\documentclass{aa}
\usepackage{psfig}
\usepackage[T1]{fontenc}
\usepackage{times}

\def\donothing{} 

\begin{document}
\thesaurus{	
20(04.01.1; 04.03.1; 08.08.1; 08.16.3; 10.07.2)
}
\title {        Photometric catalog of nearby globular clusters (I)
\thanks	{	Based on data collected at the European Southern Observatory, 
		La Silla, Chile.}
	}
\subtitle      {A large homogeneous $(V,I)$ color-magnitude diagram data-base}
\author {	A. Rosenberg \inst{1}, 
		G. Piotto \inst{2},
		I. Saviane \inst{2} \and 
		A. Aparicio    \inst{3} }
\offprints {	Alfred Rosenberg: alf@iac.es}
\institute{	
		Telescopio Nazionale Galileo, 
		vicolo dell'Osservatorio 5, I--35122 Padova, Italy
\and
		Dipartimento di Astronomia, Univ. di Padova, 
		vicolo dell'Osservatorio 5, I--35122 Padova, Italy
\and
		Instituto de Astrofisica de Canarias, 
		Via Lactea, E-38200 La Laguna, Tenerife, Spain
	}
\date{}
\titlerunning {{\it VI} photometric catalog of nearby GGC's (I)}
\authorrunning {Rosenberg A., et al.}
\maketitle

\begin{abstract}

We present the first part of the first large and homogeneous CCD
color-magnitude diagram (CMD) data base, comprising 52 nearby Galactic
globular clusters (GGC) imaged in the $V$ and $I$ bands using only two
telescopes (one for each hemisphere). The observed clusters represent
$75\%$ of the known Galactic globulars with $(m-M)_V\leq 16.15$~mag,
cover most of the globular cluster metallicity range 
($-2.2 \leq {\rm [Fe/H]}\leq -0.4$), and span Galactocentric distances
from $\sim1.2$ to $\sim18.5$ kpc.

In this paper, the CMDs for the 39 GGCs observed in the southern
hemisphere are presented. The remaining 13 northern hemisphere
clusters of the catalog are presented in a companion paper. For
four clusters (NGC~4833, NGC~5986, NGC~6543, and NGC~6638) we present
for the first time a CMD from CCD data. The typical CMD span from the
$22^{\rm nd}$ $V$ magnitude to the tip of the red giant branch. Based on a
large number of standard stars, the absolute photometric calibration
is reliable to the $\sim0.02$~mag level in both filters.

This catalog, because of its homogeneity, is expected to represent a
useful data base for the measurement of the main absolute and relative
parameters characterizing the CMD of GGCs.

\end{abstract}

\section{Introduction} 
\label{intro}

There are two main properties which make the study of the Galactic
globular clusters (GGC) particularly interesting: 1) each cluster
(with possible rare exceptions) is made up by a single population of
stars, all born at the same time, in the same place, and out of the
same material; 2) GGC stars have the oldest measurable age in the
Universe, and therefore we believe they are the oldest fossil records
of the formation history of our Galaxy.

Among the many tools we have to investigate the properties of a
stellar population, the color-magnitude diagrams (CMD) are the most
powerful ones, as they allow to recover for each individual star
its evolutionary phase, giving precious information on the age of the
entire stellar system, its chemical content, and its distance. This
information allows us to locate the system in the space, giving a base
for the distance scale, study the formation histories of the Galaxy, 
and test our knowledge of stellar evolution models.

In particular, the study of a large sample of simple stellar systems,
as the GGCs, provides important clues to the Milky Way formation
history. Recently, many studies on the relative ages of the GGCs have
been presented with results at least controversial: while some authors
find a notable age spread ($\sim5$ Gyrs) among the clusters,
others find that the bulk of GGCs is coeval. This controversy is
surely mainly due to the heterogeneity of the data used in each study,
where the combination of photographic and/or CCD data from the early
epochs of solid state detectors has been frequently used. For this
reason, a survey of both southern and northern GGCs has been started
two years ago by means of 1-m class telescopes, i.e. the 91cm European
Southern Observatory (ESO) / Dutch telescope and the 1m Isaac Newton
Group (ING) / Jacobus Kapteyn telescope (JKT). We were able to collect
the data for 52 of the 69 known GGCs with $(m-M)_V\leq16.15$.
Thirty-nine have been observed with the Dutch telescope (data that are
presented in this paper, hereafter Paper I), and the remaining ones
with the JKT (the corresponding CMDs will be presented in a companion
paper, Rosenberg et al. \cite{rosenberg00}, hereafter Paper II).

As a first exploitation of this new data base, we have conducted a GGC
relative age investigation based on the best 34 CMDs of our catalog
(Rosenberg et al. \cite{rosenberg99}, hereafter Paper III), showing
that most of the GGCs have the same age. We have also used our data
base to obtain a photometric metallicity ranking scale (Saviane et
al. \cite{saviane00}, hereafter Paper IV), based on the red giant
branch (RGB) morphology. We measured a complete set of metallicity
indices, based on the morphology and position of the RGB. Using a
grid of selected RGB fiducial points, we defined a function in the
$(V-I)_0$, $M_{\rm I}$, [Fe/H] space which is able to reproduce the
whole set of GGC RGBs in terms of a single parameter (the
metallicity). The use of this function will improve the current
determinations of metallicity and distances within the Local Group.

There are many other parameters that can be measured from a
homogeneous, well calibrated CMD data base: the horizontal branch (HB)
level, homogeneous reddening and distances, etc. We are presently
working on these problems. However, we believe it is now the time to
present to the community this data base to give to anyone interested
the opportunity to take advantage of it.

In the next section, we will describe the observations collected at
the ESO/Dutch telescope during two runs in 1997. The data reduction
and calibration is presented in Sect.~\ref{dat}, while in
Sect.~\ref{comparison} a cross check of the calibration between the two
runs is given. In order to facilitate the reader's work, we have
included the main parameters characterizing our clusters in
Sect.~\ref{parameter}. Finally, the observed fields for each
cluster, and the obtained CMDs are presented and briefly discussed in
Sect.~\ref{cmds}.

\begin{figure}
\psfig{figure=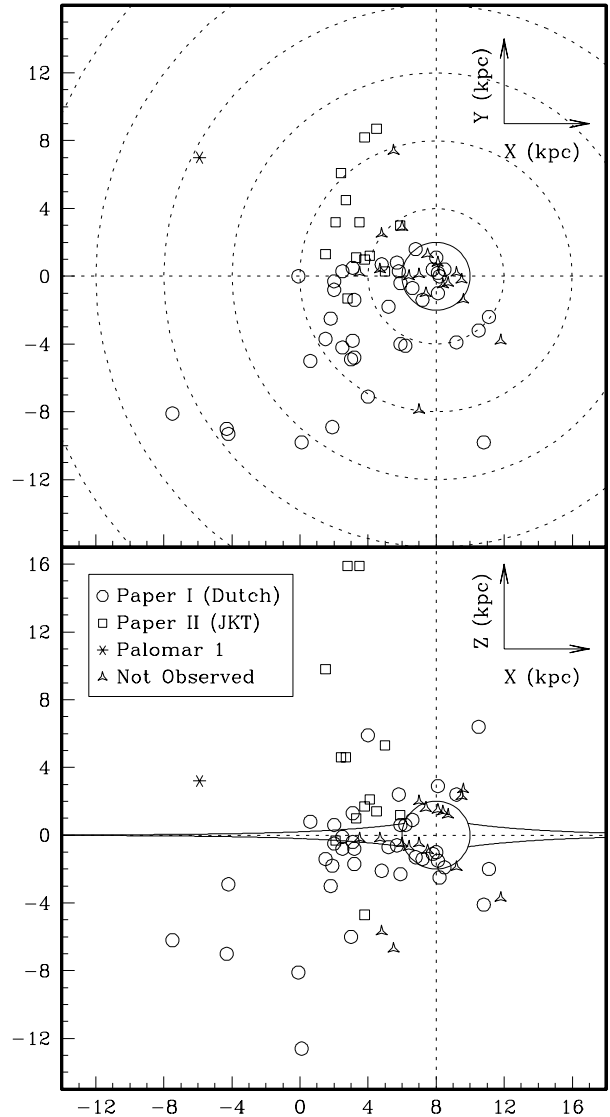,width=8cm}
\label{dist}
\caption{ Heliocentric distribution of all GGCs with
$(m-M)_V\leq16.15$~mag. In the {\it upper panel}, the GGCs projection
over the Galactic plane is presented. The {\it open circles} represent
the clusters studied in the present paper (Paper I); the {\it open
squares}, the GGCs of Paper II; and {\it the asterisk}, the GGC Pal~1
(Rosenberg et al. \cite{rosenberg98}). The clusters which are not
included in our catalog are marked by {\it open triangles}. In the
{\it lower panel}, the XZ projection is shown. The Milky Way is
schematically represented.}
\end{figure}

\begin{table}
\caption[]{Observed clusters at the DUTCH in 1997}
\label{list}
\begin{tabular}{rcccccc}
\noalign{\smallskip}
\hline
\noalign{\smallskip}
ID & Cluster & Other & Obs. & Obs.    & Seeing & Long.\\
   & (NGC)   & Name  & date & fields  & $V/I(\arcsec$) & Exp.(s)\\
\noalign{\smallskip}
\hline 
\noalign{\smallskip}
 1 & 104  & 47 Tuc 		& 23/Dec & 2 &  1.4/1.3 & 1800 \\
 2 & 288  & -       	 	& 24/Dec & 3 &  1.4/1.4 & 1800 \\
 3 & 362  & -       	 	& 26/Dec & 3 &  1.6/1.5 & 1200 \\
 4 & 1261 & -        		& 24/Dec & 3 &  1.3/1.3 & 1800 \\
\hline
 5 & 1851 & -       	 	& 23/Dec & 2 &  1.3/1.2 & 1800 \\
 6 & 1904 & M 79   	 	& 24/Dec & 3 &  1.3/1.2 & 1800 \\
 7 & 2298 & -        		& 23/Dec & 2 &  1.3/1.2 & 1800 \\
 8 & 2808 & -       	 	& 11/Apr & 2 &  1.3/1.2 & 1500 \\
\hline
 8 & 2808 & -       	 	& 26/Dec & 2 &  1.5/1.4 & 1500 \\
 9 & -    & E3     	 	& 23/Dec & 2 &  1.5/1.4 & 1800 \\
10 & 3201 & -       	 	& 12/Apr & 2 &  1.5/1.4 & 900 \\
10 & 3201 & -       	 	& 24/Dec & 3 &  1.3/1.2 & 900 \\
\hline
11 & 4372 & -       	 	& 13/Apr & 2 &  1.3/1.2 & 1500 \\
12 & 4590 & M 68         	& 14/Apr & 2 &  1.2/1.2 & 1500 \\
13 & 4833 & -            	& 15/Apr & 2 &  1.3/1.2 & 1500 \\
14 & 5139 & $\omega$ Cen	& 11/Apr & 2 &  1.2/1.2 & 900 \\
\hline
15 & 5897 & -			& 12/Apr & 1 &  1.4/1.4 & 1500 \\
16 & 5927 & -			& 13/Apr & 1 &  1.3/1.2 & 1500 \\
17 & 5986 & -			& 14/Apr & 1 &  1.3/1.3 & 1500 \\
18 & 6093 & M 80	 	& 12/Apr & 2 &  1.3/1.2 & 1500 \\
\hline
19 & 6101 & -			& 15/Apr & 1 &  1.8/1.7 & 1500 \\
20 & 6121 & M 4 	 	& 13/Apr & 2 &  1.3/1.2 & 900 \\
21 & 6171 & M 107	 	& 14/Apr & 2 &  1.4/1.3 & 1500 \\
22 & 6266 & M 62	 	& 14/Apr & 1 &  1.7/1.6 & 1500 \\
\hline
23 & 6304 & -			& 15/Apr & 1 &  1.5/1.3 & 1500 \\
24 & 6352 & -			& 11/Apr & 1 &  1.4/1.3 & 1500 \\
25 & 6362 & -			& 12/Apr & 1 &  1.4/1.3 & 1200 \\
26 & 6397 & -			& 13/Apr & 2 &  1.3/1.2 & 900 \\
\hline
27 & 6496 & -			& 14/Apr & 1 &  1.4 1.2 & 1200 \\
28 & 6541 & -			& 11/Apr & 2 &  1.3/1.2 & 1200 \\
29 & 6544 & -			& 15/Apr & 1 &  1.4/1.4 & 1500 \\
30 & 6624 & -			& 12/Apr & 1 &  1.3/1.2 & 1500 \\
\hline
31 & 6626 & M 28	 	& 13/Apr & 1 &  1.2/1.1 & 1500 \\
32 & 6637 & M 69	 	& 14/Apr & 1 &  1.2/1.1 & 1200 \\
33 & 6638 & -			& 13/Apr & 1 &  1.2/1.2 & 900 \\
34 & 6656 & M 22	 	& 15/Apr & 2 &  1.2/1.2 & 1500 \\
\hline
35 & 6681 & M 70 	 	& 11/Apr & 1 &  1.3/1.2 & 1500 \\
36 & 6717 & Pal 9	 	& 12/Apr & 1 &  1.3/1.2 & 1500 \\
37 & 6723 & -			& 13/Apr & 1 &  1.2/1.1 & 1200 \\
38 & 6752 & -			& 14/Apr & 1 &  1.3/1.2 & 1200 \\
\hline
39 & 6809 & M 55 	 	& 15/Apr & 1 &  1.3/1.1 & 900 \\
\noalign{\smallskip}
\hline
\end{tabular}
\end{table}

\section{Observations} 
\label{obs}

The data were collected during two runs in 1997: the first in April
($\rm 11^{th}-15^{th}$) and the second in December ($\rm 23^{rd},
24^{th}$ and
$\rm 26^{th}$). All nights of the first run and the first two of the
second run were photometric and had a stable seeing.

Observations were done with the ESO 91cm DUTCH telescope, at La Silla
(Chile). The same same CCD$\#33$ was used in both runs, a thinned CCD
with $512\times512$ pixels, each projecting $0.\arcsec442$ on the sky, with
a total field of view of $3.77\times3.77 (\arcmin)^2$, and the same
set of $V$ Johnson and $i$ Gunn filters.

Two short ($10-45$s), one medium ($90-120$s) and one long
($600-1800$s) exposures were taken in each band (depending on the
cluster distance modulus) for one to three fields (in order to ensure
a statistically significant sample of stars) for each of the proposed
objects. Also a large number of Landolt (\cite{landolt92}) standard
stars were
measured during each night.

In Table~\ref{list} the 39 observed GGCs are presented. Column 1 gives
an identification number adopted in this paper; cluster NGC numbers
and alternative names are given in columns 2 and 3. The observing
dates are in column 4, the number of covered fields in column 5, the
mean seeing for each filter in column 6, and the integration time for
the long exposures in column 7. In Fig.~\ref{dist} we show the
heliocentric distribution of the clusters of our entire catalog.

\section{Data reduction and calibration}
\label{dat}

The images were corrected for a constant bias, dark current, and for
spatial sensitivity variations using the respective master flats,
computed as the median of all available sky flats of the specific
run. Afterwards, photometry was performed using the
DAOPHOT/ALLSTAR/ALLFRAME software, made available to us by Dr. Stetson
(see Stetson \cite{stetson87}, \cite{stetson94}). A preliminary
photometry was carried out in order to construct a short list of stars
for each single frame. This list was used to accurately match the
different frames. With the correct coordinate transformations among the
frames, we obtained a single image, median of all the frames,
regardless of the filter. In this way we could eliminate all the
cosmic rays and obtain the highest signal/noise image for star
finding. We ran the DAOPHOT/FIND routine on the median image and
performed PSF fitting photometry in order to obtain the deepest list
of stellar objects free from spurious detections. Finally, this list
was given as input to ALLFRAME, for the simultaneous profile fitting
photometry of all the individual frames. We constructed the model PSF
for each image using typically from 60 to 120 stars.

The absolute calibration of the observations to the V-Johnson and
I-Cousins systems is based on a set of standard stars from the catalog
of Landolt (\cite{landolt92}). Specifically, the observed standard
stars were in the fields: PG0231, SA95 (41, 43, 96, 97, 98, 100, 101,
102, 112, 115), SA98 (556, 557, 563, 580, 581, L1, 614, 618, 626, 627,
634, 642), RUBIN 149, RUBIN 152, PG0918, PG0942, PG1047, PG1323,
PG1525, PG1530, PG1633, and Mark A. At least 3 exposures were taken
for each standard field, with a total of $\sim100$ standard star
measurements per night and per filter.

The reduction and aperture photometry of standard star fields were
performed in the same way as for the cluster images. The aperture
magnitudes were corrected for atmospheric extinction, assuming
$A_V=0.14$ and $A_I=0.08$ as extinction coefficients for the $V$ and
$i$ filters, respectively.

\begin{figure}
\centerline{\psfig{figure=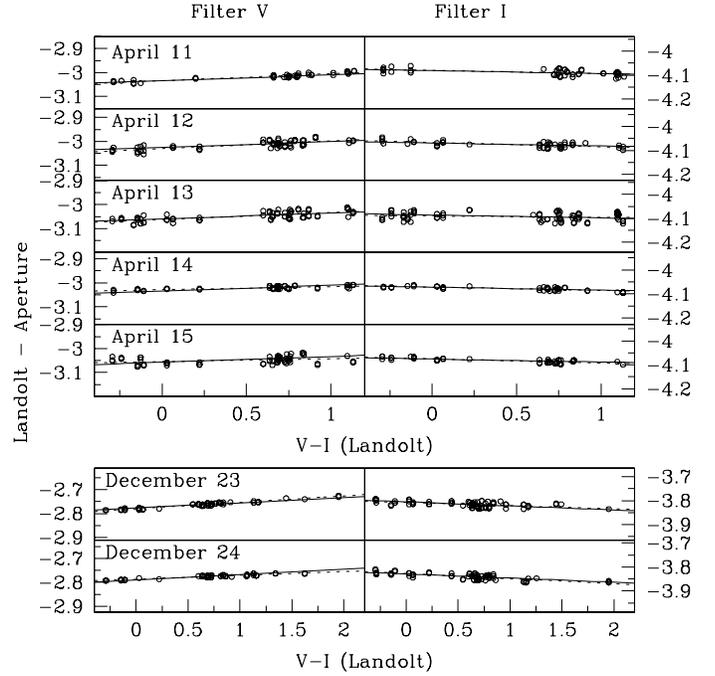,width=8.8cm}} 
\caption[]{Calibration equation for each observing night. The {\it
upper panels} refer to the run of April, while the {\it lower panels}
refer to the December run. In all cases, the $V$ filter curves are on
the {\it left side}, while the $I$ filter curves on the right.}
\label{calib} 
\end{figure} 

\begin{table}
\begin{center}
\caption[]{Calibration parameters for each observing night.}
\label{calib_par}
\begin{tabular}{ccccc}
& Filter V & & & \\
\noalign{\smallskip}
\hline
\noalign{\smallskip}
Date & $a_{\rm m}$ & error & Cons. & error\\
\noalign{\smallskip}
\hline 
\noalign{\smallskip}
11/Apr & +0.024 & $\pm0.001$ & -3.034 & $\pm0.002$ \\
12/Apr & +0.024 & $\pm0.002$ & -3.025 & $\pm0.004$ \\
13/Apr & +0.024 & $\pm0.002$ & -3.059 & $\pm0.003$ \\
14/Apr & +0.024 & $\pm0.001$ & -3.034 & $\pm0.002$ \\
15/Apr & +0.024 & $\pm0.003$ & -3.057 & $\pm0.004$ \\
23/Dec & +0.022 & $\pm0.001$ & -2.777 & $\pm0.001$ \\
24/Dec & +0.022 & $\pm0.001$ & -2.790 & $\pm0.001$ \\
\noalign{\smallskip}
\hline
\noalign{\smallskip}
& Filter I & & & \\
\noalign{\smallskip}
\hline
\noalign{\smallskip}
Date & $a_{\rm m}$ & error & Cons. & error\\
\noalign{\smallskip}
\hline 
\noalign{\smallskip}
11/Apr & -0.012 & $\pm0.003$ & -4.081 & $\pm0.004$ \\
12/Apr & -0.012 & $\pm0.002$ & -4.069 & $\pm0.003$ \\
13/Apr & -0.012 & $\pm0.003$ & -4.086 & $\pm0.004$ \\
14/Apr & -0.012 & $\pm0.001$ & -4.072 & $\pm0.002$ \\
15/Apr & -0.012 & $\pm0.001$ & -4.076 & $\pm0.004$ \\
23/Dec & -0.017 & $\pm0.002$ & -3.805 & $\pm0.002$ \\
24/Dec & -0.017 & $\pm0.002$ & -3.829 & $\pm0.002$ \\
\noalign{\smallskip}
\hline
\end{tabular}
\end{center}
\end{table}

As shown in Fig.~\ref{calib}, a straight line well reproduces the
calibration equations. As the seeing and the overall observing
conditions were stable during the run, the slopes of the calibration
equations for each observing run and for each filter have been
computed using the data from all the nights. As it can be seen in
Table~\ref{calib_par}, the standard deviations of the calibration
constants for each run and filter is $0.015$mag, corroborating our
assumption that all nights were photometric, and that we can assume a
constant slope for each filter and run.

Standard stars for which previous problems were reported (PG 1047C,
RU149A, RU149G, PG1323A; see Johnson \& Bolte \cite{johnsonbolte98})
were excluded, as well as saturated stars, those close to a cosmic ray,
etc... After this cleaning, the mean slope was computed, and finally
the different night constants were found using this slope to fit the
individual data, night by night. The adopted values are presented in
Table~\ref{calib_par}. The typical errors ($rms$) are also given.

The calibration curves are shown in Fig.~\ref{calib} for both runs. In
this figure, the {\it dotted line} represents the best fitting
equation, while the {\it continuous line} is obtained by best fitting
the data imposing the adopted mean slope. The two lines are almost
overlapping. The mean number of standard star measures used for
computing the curve per night and filter is $\sim75$. Notice the wide
color coverage for the standard stars.

The last step on the calibration is the aperture correction. As no
available bright and isolated stars exist on the cluster images, we
used DAOPHOT to subtract from the image the stars in the neighborhood
of the brightest ones, in order to compute the difference between the
aperture and the PSF-fitting magnitudes. In view of the stable seeing
conditions, we used the same aperture for calculating the aperture
photometry of the standard and cluster stars.

\section{Photometric homogeneity of the two runs}
\label{comparison}

\begin{figure} 
\centerline{\psfig{figure=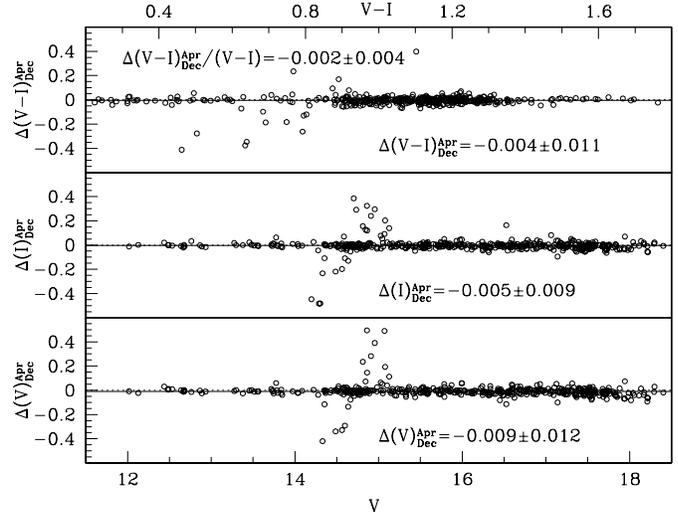,width=8.8cm}} 
\caption[]{Comparison of the magnitudes and colors of 456 stars in
common between the April and December runs, for the GGC
NGC~3201. Stars with photometric internal errors smaller than 0.02 mag
have been selected. The mean differences are given in each panel. {\it
Note:} a few NGC~3201 RR~Lyrae stars can be identified in the interval
$14.2 < V < 15.2$ (lower two panels), and between $0.4 < (V-I) <
0.9$.}
\label{compare}
\end{figure}

In order to check the photometric homogeneity of the data and of the
calibration to the standard photometric system, one cluster (NGC~3201)
was observed in both runs. Having one common field, it is possible to
analyze the individual star photometry, and test if any additional
zero point difference and/or color term exist. The latter check is
crucial when measures of the relative position of CMD features are
going to be done. The comparison between the two runs is presented in
Fig.~\ref{compare}, where 456 common stars with internal photometric
errors (as given by ALLFRAME) smaller than 0.02 mag are
used. Fig.~\ref{compare} shows that there are no systematic differences
between the two runs.

The slope of the straight lines best fitting all the points in both
the (V,$\Delta V^{\rm apr}_{\rm dec}$) plane and the (V,$\Delta I^{\rm
apr}_{\rm dec}$) plane is $\leq 0.001\pm0.002$, and $\leq
0.002\pm0.003$ in the (V-I,$\Delta (V-I)^{\rm apr}_{\rm dec}$)
plane. The zero point differences are always $\le 0.01$ mag. This
ensures the homogeneity of our database, particularly for relative
measurements within the CMDs.

\section{Parameters for the GGC sample}
\label{parameter}

\begin{table*}
\label{param01}
\caption[]{Identifications, positional data and metallicity estimates
for the observed clusters.}
\begin{center}
\begin{tabular}{rlccccccccccc}
\noalign{\smallskip}
\hline
\noalign{\smallskip}
\multicolumn{1}{c}{ID} & 
\multicolumn{1}{c}{Cluster} & 
\multicolumn{1}{c}{$RA^{ \mathrm{a}}$} & 
\multicolumn{1}{c}{$DEC^{ \mathrm{b}}$} &
\multicolumn{1}{c}{${\it l}^{ \mathrm{c}}$} & 
\multicolumn{1}{c}{${\it b}^{ \mathrm{d}}$} &
\multicolumn{1}{c}{$R_{\odot}^{ \mathrm{e}}$} &
\multicolumn{1}{c}{$R_{\rm GC}^{ \mathrm{f}}$} &
\multicolumn{1}{c}{$X^{ \mathrm{g}}$} &
\multicolumn{1}{c}{$Y^{ \mathrm{h}}$} &
\multicolumn{1}{c}{$Z^{ \mathrm{i}}$} & 
\multicolumn{2}{c}{[Fe/H]} \\
\multicolumn{1}{c}{} & 
\multicolumn{1}{c}{} & 
\multicolumn{1}{c}{($^{h \; m \; s}$)} & 
\multicolumn{1}{c}{($^{\rm o} \;$ \arcmin $\;$ \arcsec)} &
\multicolumn{1}{c}{$(^{\rm o})$} & 
\multicolumn{1}{c}{$(^{\rm o})$} &
\multicolumn{1}{c}{(kpc)} &
\multicolumn{1}{c}{(kpc)} &
\multicolumn{1}{c}{(kpc)} &
\multicolumn{1}{c}{(kpc)} &
\multicolumn{1}{c}{(kpc)} &
\multicolumn{1}{c}{$ZW84^{ \mathrm{j}}$} &
\multicolumn{1}{c}{$CG97^{ \mathrm{k}}$} \\

\noalign{\smallskip}
\hline 
\noalign{\smallskip}

 1 & NGC~104  & 00 24 05.2 & -72 04 51 & 305.90 & -44.89 & 4.3 &  7.3 &
+1.8 &  -2.5 & -3.0 & -0.71 & -0.78\\
 2 & NGC~288  & 00 52 47.5 & -26 35 24 & 152.28 & -89.38 & 8.1 & 11.4 &
-0.1 &  +0.0 & -8.1 & -1.40 & -1.14\\
 3 & NGC~362  & 01 03 14.3 & -70 50 54 & 301.53 & -46.25 & 8.3 &  9.2 &
+3.0 &  -4.9 & -6.0 & -1.33 & -1.09\\
 4 & NGC~1261 & 03 12 15.3 & -55 13 01 & 270.54 & -52.13 &16.0 & 17.9 &
+0.1 &  -9.8 &-12.6 & -1.32 & -1.08\\
\hline										  
 5 & NGC~1851 & 05 14 06.3 & -40 02 50 & 244.51 & -35.04 &12.2 & 16.8 &
-4.3 &  -9.0 & -7.0 & -1.23 & -1.03\\
 6 & NGC~1904 & 05 24 10.6 & -24 31 27 & 227.23 & -29.35 &12.6 & 18.5 &
-7.5 &  -8.1 & -6.2 & -1.67 & -1.37\\
 7 & NGC~2298 & 06 48 59.2 & -36 00 19 & 245.63 & -16.01 &10.6 & 15.6 &
-4.2 &  -9.3 & -2.9 & -1.91 & -1.71\\
 8 & NGC~2808 & 09 12 02.6 & -64 51 47 & 282.19 & -11.25 & 9.3 & 10.9 &
+1.9 &  -8.9 & -1.8 & -1.36 & -1.11\\
\hline										  
 9 & E3       & 09 20 59.3 & -77 16 57 & 292.27 & -19.02 & 4.2 &  7.6 &
+1.5 &  -3.7 & -1.4 &     - & -    \\
10 & NGC~3201 & 10 17 36.8 & -46 24 40 & 277.23 & +08.64 & 5.1 &  8.9 &
+0.6 &  -5.0 & +0.8 & -1.53 & -1.24\\
11 & NGC~4372 & 12 25 45.4 & -72 39 33 & 300.99 & -09.88 & 4.9 &  6.9 &
+2.5 &  -4.2 & -0.8 & -2.03 & -1.88\\
12 & NGC~4590 & 12 39 28.0 & -26 44 34 & 299.63 & +36.05 &10.1 & 10.0 &
+4.0 &  -7.1 & +5.9 & -2.11 & -2.00\\
\hline										  
13 & NGC~4833 & 12 59 35.0 & -70 52 29 & 303.61 & -08.01 & 5.9 &  6.9 &
+3.2 &  -4.8 & -0.8 & -1.92 & -1.71\\
14 & NGC~5139 & 13 26 45.9 & -47 28 37 & 309.10 & +14.97 & 5.1 &  6.3 &
+3.1 &  -3.8 & +1.3 &\phantom{-}-$1.62^{1}$& -\\
15 & NGC~5897 & 15 17 24.5 & -21 00 37 & 342.95 & +30.29 &12.7 &  7.6
&+10.5 &  -3.2 & +6.4 & -1.93 & -1.73\\
16 & NGC~5927 & 15 28 00.5 & -50 40 22 & 326.60 & +04.86 & 7.4 &  4.5 &
+6.2 &  -4.1 & +0.6 & -0.33 & -0.64\\
\hline										  
17 & NGC~5986 & 15 46 03.5 & -37 47 10 & 337.02 & +13.27 &10.3 &  4.7 &
+9.2 &  -3.9 & +2.4 & -1.65 & -1.35\\
18 & NGC~6093 & 16 17 02.5 & -22 58 30 & 352.67 & +19.46 & 8.7 &  3.1 &
+8.1 &  -1.0 & +2.9 & -1.75 & -1.47\\
19 & NGC~6121 & 16 23 35.5 & -26 31 31 & 350.97 & +15.97 & 2.2 &  6.0 &
+2.0 &  -0.3 & +0.6 & -1.27 & -1.05\\
20 & NGC~6101 & 16 25 48.6 & -72 12 06 & 317.75 & -15.82 &15.1 & 11.0
&+10.8 &  -9.8 & -4.1 & -1.95 & -1.76\\
\hline										  
21 & NGC~6171 & 16 32 31.9 & -13 03 13 & 003.37 & +23.01 & 6.3 &  3.3 &
+5.8 &  +0.3 & +2.4 & -1.09 & -0.95\\
22 & NGC~6266 & 17 01 12.6 & -30 06 44 & 353.58 & +07.32 & 6.7 &  1.8 &
+6.6 &  -0.7 & +0.9 & -1.23 & -1.02\\
23 & NGC~6304 & 17 14 32.5 & -29 27 44 & 355.83 & +05.38 & 6.0 &  2.2 &
+5.9 &  -0.4 & +0.6 & -0.38 & -0.66\\
24 & NGC~6352 & 17 25 29.2 & -48 25 22 & 341.42 & -07.17 & 5.6 &  3.3 &
+5.2 &  -1.8 & -0.7 & -0.50 & -0.70\\
\hline										  
25 & NGC~6362 & 17 31 54.8 & -67 02 53 & 325.55 & -17.57 & 7.5 &  5.1 &
+5.9 &  -4.0 & -2.3 & -1.18 & -0.99\\
26 & NGC~6397 & 17 40 41.3 & -53 40 25 & 338.17 & -11.96 & 2.2 &  6.0 &
+2.0 &  -0.8 & -0.5 & -1.94 & -1.76\\
27 & NGC~6496 & 17 59 02.0 & -44 15 54 & 348.02 & -10.01 &11.6 &  4.4
&+11.1 &  -2.4 & -2.0 & -0.50 & -0.70\\
28 & NGC~6544 & 18 07 20.6 & -24 59 51 & 005.84 & -02.20 & 2.5 &  5.5 &
+2.5 &  +0.3 & -0.1 & -1.48 & -1.20\\
\hline										  
29 & NGC~6541 & 18 08 02.2 & -43 42 20 & 349.29 & -11.18 & 7.4 &  2.1 &
+7.2 &  -1.4 & -1.4 & -1.79 & -1.53\\
30 & NGC~6624 & 18 23 40.5 & -30 21 40 & 002.79 & -07.91 & 7.9 &  1.2 &
+7.8 &  +0.4 & -1.1 & -0.50 & -0.70\\
31 & NGC~6626 & 18 24 32.9 & -24 52 12 & 007.80 & -05.58 & 5.7 &  2.5 &
+5.7 &  +0.8 & -0.6 & -1.23 & -1.03\\
32 & NGC~6638 & 18 30 56.2 & -25 29 47 & 007.90 & -07.15 & 8.2 &  1.5 &
+8.0 &  +1.1 & -1.0 & -1.00 & -0.90\\
\hline										  
33 & NGC~6637 & 18 31 23.2 & -32 20 53 & 001.72 & -10.27 & 8.2 &  1.5 &
+8.1 &  +0.2 & -1.5 & -0.72 & -0.78\\
34 & NGC~6656 & 18 36 24.2 & -23 54 12 & 009.89 & -07.55 & 3.2 &  5.0 &
+3.1 &  +0.5 & -0.4 &\phantom{-}-$1.64^{1}$& -\\
35 & NGC~6681 & 18 43 12.7 & -32 17 31 & 002.85 & -12.51 & 8.7 &  2.0 &
+8.5 &  +0.4 & -1.9 & -1.64 & -1.35\\
36 & NGC~6717 & 18 55 06.2 & -22 42 03 & 012.88 & -10.90 & 7.1 &  2.4 &
+6.8 &  +1.6 & -1.3 & -1.33 & -1.09\\
\hline										  
37 & NGC~6723 & 18 59 33.2 & -36 37 54 & 000.07 & -17.30 & 8.6 &  2.6 &
+8.2 &  +0.0 & -2.5 & -1.12 & -0.96\\
38 & NGC~6752 & 19 10 51.8 & -59 58 55 & 336.50 & -25.63 & 3.9 &  5.3 &
+3.2 &  -1.4 & -1.7 & -1.54 & -1.24\\
39 & NGC~6809 & 19 39 59.4 & -30 57 44 & 008.80 & -23.27 & 5.3 &  3.9 &
+4.8 &  +0.7 & -2.1 & -1.80 & -1.54\\
\noalign{\smallskip}
\hline
\end{tabular}\\
In the following cases, the [Fe/H] values were taken directly from
($^1$) ZW84.
\end{center}

\begin{tabular}{lrl}
$^{\mathrm{a}}$ Right Ascension (2000)&
Sun-Centered coordinates: & 
$^{\mathrm{g}}$ X: Toward the Galactic Center\\
$^{\mathrm{b}}$ Declination (2000) &
 &
$^{\mathrm{h}}$ Y: in direction of Galactic rotation \\
$^{\mathrm{c}}$ Galactic Longitude &
 &
$^{\mathrm{i}}$ Z: Towards North Galactic Plane \\
$^{\mathrm{d}}$ Galactic Latitude &
 & 
 \\
$^{\mathrm{e}}$ Heliocentric Distance &
[Fe/H] (From Rutledge et al. 1997):&
$^{\mathrm{j}}$ in the ZW84 scale\\
$^{\mathrm{f}}$ Galactocentric Distance &
 & 
$^{\mathrm{k}}$ in the CG97 scale\\
\end{tabular}

\end{table*}

\begin{table*}
\begin{center}
\caption[]{Photometric Parameters.}
\label{param2}
\begin{tabular}{rlclcrrccccrrc}
\noalign{\smallskip}
\hline
\noalign{\smallskip}

\multicolumn{1}{r}{ID} & 
\multicolumn{1}{c}{Cluster} & 
\multicolumn{1}{c}{${\rm E(B-V)^{ \mathrm{a}}}$} & 
\multicolumn{1}{c}{$V_{\rm HB} ^{ \mathrm{b}}$} & 
\multicolumn{1}{c}{${\rm (m-M)_V^{ \mathrm{c}}}$} & 
\multicolumn{1}{r}{${\rm V_t}^{ \mathrm{d}}$} & 
\multicolumn{1}{r}{${\rm Mv_t}^{ \mathrm{e}}$} &
\multicolumn{1}{c}{\rm $U-B^{ \mathrm{f}}$} & 
\multicolumn{1}{c}{\rm $B-V^{ \mathrm{f}}$} & 
\multicolumn{1}{c}{\rm $V-R^{ \mathrm{f}}$} & 
\multicolumn{1}{c}{\rm $V-I^{ \mathrm{f}}$} & 
\multicolumn{1}{r}{$S_{\rm RR} ^{ \mathrm{g}}$} & 
\multicolumn{1}{r}{$HBR^{ \mathrm{h}}$} \\

\noalign{\smallskip}
\hline 
\noalign{\smallskip}
 1 & NGC~104  & 0.05 & 14.05* & 13.32 &  3.95 & -9.37 & 0.37 & 0.88 &
0.53 & 1.14 &  0.4 & -0.99\\
 2 & NGC~288  & 0.03 & 15.40* & 14.64 &  8.09 & -6.55 & 0.08 & 0.65 &
0.45 & 0.94 &  4.8 &  0.98\\
 3 & NGC~362  & 0.05 & 15.43  & 14.75 &  6.40 & -8.35 & 0.16 & 0.77 &
0.49 & 1.01 &  5.9 & -0.87\\
 4 & NGC~1261 & 0.01 & 16.68* & 16.05 &  8.29 & -7.76 & 0.13 & 0.72 &
0.45 & 0.93 & 14.9 & -0.71\\
\hline			     								       
 5 & NGC~1851 & 0.02 & 16.18* & 15.49 &  7.14 & -8.35 & 0.17 & 0.76 &
0.49 & 1.01 & 10.1 & -0.36\\
 6 & NGC~1904 & 0.01 & 16.15* & 15.53 &  7.73 & -7.80 & 0.06 & 0.65 &
0.44 & 0.91 &  5.3 &  0.89\\
 7 & NGC~2298 & 0.13 & 16.11  & 15.54 &  9.29 & -6.25 & 0.17 & 0.75 &
0.54 & 1.11 &  9.5 &  0.93\\
 8 & NGC~2808 & 0.23 & 16.30* & 15.55 &  6.20 & -9.35 & 0.28 & 0.92 &
0.57 & 1.18 &  0.4 & -0.49\\
\hline			     								       
 9 & E 3      & 0.30 & 14.80  & 14.07 & 11.35 & -2.72 & -    & -    &
-    & -    &  0.0 &     -\\
10 & NGC~3201 & 0.21 & 14.75* & 14.17 &  6.75 & -7.42 & 0.38 & 0.96 &
0.62 & 1.23 & 91.3 &  0.08\\
11 & NGC~4372 & 0.42 & 15.30  & 14.76 &  7.24 & -7.52 & 0.31 & 1.10 &
0.72 & 1.50 &  0.0 &  1.00\\
12 & NGC~4590 & 0.04 & 15.75* & 15.14 &  7.84 & -7.30 & 0.04 & 0.63 &
0.46 & 0.94 & 49.3 &  0.17\\
\hline			     								       
13 & NGC~4833 & 0.33 & 15.45  & 14.87 &  6.91 & -7.96 & 0.29 & 0.93 &
0.63 & 1.33 & 11.8 &  0.93\\
14 & NGC~5139 & 0.12 & 14.53  & 13.92 &  3.68 &-10.24 & 0.19 & 0.78 &
0.51 & 1.05 & 12.2 &   -  \\
15 & NGC~5897 & 0.08 & 16.35  & 15.77 &  8.53 & -7.24 & 0.08 & 0.74 &
0.50 & 1.04 &  8.9 &  0.86\\
16 & NGC~5927 & 0.47 & 16.60  & 15.81 &  8.01 & -7.80 & 0.85 & 1.31 &
0.79 & 1.63 &  0.0 & -1.00\\
\hline			     								       
17 & NGC~5986 & 0.27 & 16.50  & 15.90 &  7.52 & -8.38 & 0.30 & 0.90 &
0.58 & 1.22 &  4.4 &  0.97\\
18 & NGC~6093 & 0.18 & 16.25* & 15.25 &  7.33 & -7.92 & 0.21 & 0.84 &
0.56 & 1.11 &  4.1 &  0.93\\
19 & NGC~6121 & 0.36 & 13.36* & 12.78 &  5.63 & -7.15 & 0.43 & 1.03 &
0.69 & 1.42 & 70.4 & -0.06\\
20 & NGC~6101 & 0.04 & 16.60  & 16.02 &  9.16 & -6.86 & 0.06 & 0.68 &
0.50 &    - & 19.8 &  0.84\\
\hline			     								       
21 & NGC~6171 & 0.33 & 15.65* & 15.01 &  7.93 & -7.08 & 0.69 & 1.10 &
0.72 & 1.45 & 32.5 & -0.73\\
22 & NGC~6266 & 0.47 & 16.25  & 15.59 &  6.45 & -9.14 & 0.52 & 1.19 &
0.74 & 1.58 & 19.1 &  0.32\\
23 & NGC~6304 & 0.52 & 16.25  & 15.49 &  8.22 & -7.27 & 0.82 & 1.31 &
0.77 & 1.70 &  0.0 & -1.00\\
24 & NGC~6352 & 0.21 & 15.25* & 14.39 &  7.96 & -6.43 & 0.64 & 1.06 &
0.66 & 1.50 &  0.0 & -1.00\\
\hline			     								       
25 & NGC~6362 & 0.09 & 15.35* & 14.65 &  7.73 & -6.92 & 0.29 & 0.85 &
0.56 & 1.14 & 56.4 & -0.58\\
26 & NGC~6397 & 0.18 & 12.95* & 12.31 &  5.73 & -6.58 & 0.12 & 0.73 &
0.49 & 1.03 &  0.0 &  0.98\\
27 & NGC~6496 & 0.13 & 16.47  & 15.72 &  8.54 & -7.18 & 0.45 & 0.98 &   
- &    - &  0.0 & -1.00\\
28 & NGC~6544 & 0.74 & 14.90  & 14.28 &  7.77 & -6.51 & 0.73 & 1.46 &
0.98 & 1.92 &   -  &  1.00\\
\hline			     								       
29 & NGC~6541 & 0.12 & 15.30  & 14.72 &  6.30 & -8.42 & 0.13 & 0.76 &
0.49 & 1.01 &  0.0 &  1.00\\
30 & NGC~6624 & 0.27 & 16.11  & 15.32 &  7.87 & -7.45 & 0.60 & 1.11 &
0.67 & 1.42 &  1.0 & -1.00\\
31 & NGC~6626 & 0.41 & 15.70  & 15.07 &  6.79 & -8.28 & 0.46 & 1.08 &
0.69 & 1.41 &  6.4 &  0.90\\
32 & NGC~6638 & 0.40 & 16.50  & 15.80 &  9.02 & -6.78 & 0.56 & 1.15 &
0.72 & 1.50 & 38.9 & -0.30\\
\hline			     								       
33 & NGC~6637 & 0.17 & 15.85  & 15.11 &  7.64 & -7.47 & 0.48 & 1.01 &
0.62 & 1.28 &  0.0 & -1.00\\
34 & NGC~6656 & 0.34 & 14.25* & 13.55 &  5.10 & -8.45 & 0.28 & 0.98 &
0.68 & 1.42 &  7.5 &  0.91\\
35 & NGC~6681 & 0.07 & 15.70* & 14.93 &  7.87 & -7.06 & 0.12 & 0.72 &
0.47 & 0.99 &  3.0 &  0.96\\
36 & NGC~6717 & 0.21 & 15.56  & 14.90 &  9.28 & -5.62 & 0.35 & 1.00 &
0.65 & 1.37 &  5.6 &  0.98\\
\hline			     								       
37 & NGC~6723 & 0.05 & 15.45* & 14.82 &  7.01 & -7.81 & 0.21 & 0.75 &
0.50 & 1.05 & 21.8 & -0.08\\
38 & NGC~6752 & 0.04 & 13.80* & 13.08 &  5.40 & -7.68 & 0.07 & 0.66 &
0.43 & 0.93 &  0.0 &  1.00\\
39 & NGC~6809 & 0.07 & 14.45* & 13.82 &  6.32 & -7.50 & 0.11 & 0.72 &
0.48 & 1.00 & 10.0 &  0.87\\
\noalign{\smallskip}
\hline
\end{tabular}
The HB levels (column 4) with an asterisk have been measured directly on
our CMDs.
\end{center}

\begin{tabular}{llll}
$^{\mathrm{a}}$ Foreground reddening &
$^{\mathrm{e}}$ Absolute visual magnitude \\
$^{\mathrm{b}}$ HB Level &
$^{\mathrm{f}}$ Integrated color indices \\
$^{\mathrm{c}}$ Apparent visual distance modulus &
$^{\mathrm{g}}$ Specific frequency of RR Lyrae variables \\
$^{\mathrm{d}}$ Integrated $V$ mag. of clusters &
$^{\mathrm{h}}$ HB ratio: $HBR=(B-R)/(B+V+R)$ \\
\end{tabular}
\end{table*}

\begin{table*}
\begin{center}
\caption[]{Kinematics, and Structural Parameters}
\label{param3}
\begin{tabular}{rlrrccccccc}
\noalign{\smallskip}
\hline
\noalign{\smallskip}
ID & 
Cluster & 
$V_{\rm r}^{ \mathrm{a}}$ & 
$V_{\rm LSR}^{ \mathrm{b}}$ & 
$c^{ \mathrm{c}}$ & 
$r_{\rm c}^{ \mathrm{d}}$ & 
$r_{\rm h}^{ \mathrm{e}}$ & 
$lg(t_{\rm c})^{ \mathrm{f}}$ & 
$lg(t_{\rm h})^{ \mathrm{g}}$ & 
$\mu_{\rm V}^{ \mathrm{h}}$ & 
$\rho_{\rm 0}^{ \mathrm{i}}$ \\
\noalign{\smallskip}
\hline 
\noalign{\smallskip}
01 & NGC~104  &   $-18.7\pm0.2$ &  $-28.0$ &  2.04 & 0.37 & 2.79 & 7.99
&  9.24 &  14.43 & 4.87 \\
02 & NGC~288  &   $-46.6\pm0.4$ &  $-53.9$ &  0.96 & 1.42 & 2.22 & 9.09
&  8.99 &  19.95 & 1.84 \\
03 & NGC~362  &  $+223.5\pm0.5$ & $+213.3$ &  1.94c& 0.17 & 0.81 & 7.79
&  8.43 &  14.88 & 4.74 \\
04 & NGC~1261 &   $+68.2\pm4.6$ &  $+53.4$ &  1.27 & 0.39 & 0.75 & 8.79
&  8.81 &  17.65 & 2.97 \\
\hline		       
05 & NGC~1851 &  $+320.9\pm1.0$ & $+302.1$ &  2.24 & 0.08 & 0.52 & 7.41
&  8.50 &  14.15 & 5.17 \\
06 & NGC~1904 &  $+207.5\pm0.5$ & $+188.3$ &  1.72 & 0.16 & 0.80 & 7.87
&  8.66 &  16.23 & 4.01 \\
07 & NGC~2298 &  $+148.9\pm1.2$ & $+129.8$ &  1.28 & 0.34 & 0.78 & 8.02
&  8.36 &  18.79 & 2.90 \\
08 & NGC~2808 &  $ +93.6\pm2.4$ &  $+80.1$ &  1.77 & 0.26 & 0.76 & 8.35
&  8.77 &  15.17 & 4.62 \\
\hline		       
09 & E3       &     -           &    -     &  -    & 0.75 & 1.87 & 2.06
&  -    &  23.10 & 1.12 \\
10 & NGC~3201 &  $+494.0\pm0.2$ & $+481.9$ &  1.31 & 1.45 & 2.68 & 8.82
&  8.79 &  18.77 & 2.69 \\
11 & NGC~4372 &  $ +72.3\pm1.3$ &  $+63.8$ &  1.30 & 1.75 & 3.90 & 8.88
&  9.23 &  20.51 & 2.19 \\
12 & NGC~4590 &  $ -95.2\pm0.4$ &  $-97.1$ &  1.64 & 0.69 & 1.55 & 8.60
&  8.90 &  18.67 & 2.53 \\
\hline		       
13 & NGC~4833 &  $+200.2\pm1.2$ & $+192.7$ &  1.25 & 1.00 & 2.41 & 8.79
&  8.77 &  18.45 & 3.07 \\
14 & NGC~5139 &  $+232.3\pm0.5$ & $+229.4$ &  1.24 & 2.58 & 4.18 & 9.76
&  9.72 &  16.77 & 3.13 \\
15 & NGC~5897 &  $+101.5\pm1.0$ & $+110.0$ &  0.79 & 1.96 & 2.11 & 9.78
&  9.31 &  20.49 & 1.38 \\
16 & NGC~5927 &  $-115.7\pm3.1$ & $-114.5$ &  1.60 & 0.42 & 1.15 & 8.53
&  8.71 &  17.45 & 3.90 \\
\hline		       
17 & NGC~5986 &  $ +88.9\pm3.7$ &  $+94.3$ &  1.22 & 0.63 & 1.05 & 8.97
&  8.78 &  17.56 & 3.31 \\
18 & NGC~6093 &  $  +9.3\pm3.1$ &  $+19.7$ &  1.95 & 0.15 & 0.65 & 7.60
&  8.32 &  15.19 & 4.82 \\
19 & NGC~6121 &  $ +70.2\pm0.3$ &  $+79.8$ &  1.59 & 0.83 & 3.65 & 7.57
&  8.64 &  17.88 & 3.83 \\
20 & NGC~6101 &  $+361.4\pm1.7$ & $+357.2$ &  0.80 & 1.15 & 1.71 & 9.44
&  9.22 &  20.12 & 1.63 \\
\hline		       
21 & NGC~6171 &  $ -33.8\pm0.3$ &  $-20.6$ &  1.51 & 0.54 & 2.70 & 8.10
&  8.75 &  18.84 & 3.14 \\
22 & NGC~6266 &  $ -65.8\pm2.5$ &  $-56.3$ &  1.70c& 0.18 & 1.23 & 7.54
&  8.55 &  15.35 & 5.15 \\
23 & NGC~6304 &  $-107.3\pm3.6$ &  $-97.4$ &  1.80 & 0.21 & 1.41 & 7.45
&  8.56 &  17.34 & 4.40 \\
24 & NGC~6352 &  $-120.9\pm3.0$ & $-116.7$ &  1.10 & 0.83 & 2.00 & 8.64
&  8.71 &  18.42 & 3.05 \\
\hline		       
25 & NGC~6362 &  $ -13.1\pm0.6$ &  $-15.1$ &  1.10 & 1.32 & 2.18 & 9.09
&  8.83 &  19.19 & 2.27 \\
26 & NGC~6397 &  $ +18.9\pm0.1$ &  $+21.4$ &  2.50c& 0.05 & 2.33 & 4.93
&  8.35 &  15.65 & 5.69 \\
27 & NGC~6496 &  $-112.7\pm5.7$ & $-107.0$ &  0.70 & 1.05 & 1.87 & 8.46
&  8.46 &  20.10 & 1.94 \\
28 & NGC~6544 &  $ -27.3\pm3.9$ &  $-15.7$ &  1.63c& 0.05 & 1.77 & 5.23
&  7.82 &  17.13 & 5.78 \\
\hline		       
29 & NGC~6541 &  $-156.2\pm2.7$ & $-150.3$ &  2.00c& 0.30 & 1.19 & 8.04
&  8.58 &  15.58 & 4.36 \\
30 & NGC~6624 &  $ +53.9\pm0.6$ &  $+63.9$ &  2.50c& 0.06 & 0.82 & 6.71
&  8.50 &  15.42 & 5.24 \\
31 & NGC~6626 &  $ +17.0\pm1.0$ &  $+28.5$ &  1.67 & 0.24 & 1.56 & 7.73
&  8.78 &  16.08 & 4.73 \\
32 & NGC~6638 &  $ +18.1\pm3.9$ &  $+29.4$ &  1.40 & 0.26 & 0.66 & 8.00
&  8.02 &  17.27 & 4.06 \\
\hline		       
33 & NGC~6637 &  $ +39.9\pm2.8$ &  $+49.3$ &  1.39 & 0.34 & 0.83 & 8.40
&  8.69 &  16.83 & 3.83 \\
34 & NGC~6656 &  $-148.9\pm0.4$ & $-137.2$ &  1.31 & 1.42 & 3.26 & 8.62
&  8.86 &  17.32 & 3.65 \\
35 & NGC~6681 &  $+218.6\pm1.2$ & $+227.9$ &  2.50c& 0.03 & 0.93 & 5.82
&  8.40 &  15.28 & 5.42 \\
36 & NGC~6717 &  $ +22.8\pm3.4$ &  $+34.6$ &  2.07c& 0.08 & 0.68 & 6.61
&  8.14 &  16.48 & 4.68 \\
\hline		       
37 & NGC~6723 &  $ -94.5\pm3.6$ &  $-86.7$ &  1.05 & 0.94 & 1.61 & 9.02
&  8.94 &  17.92 & 2.82 \\
38 & NGC~6752 &  $ -24.5\pm1.9$ &  $-24.4$ &  2.50c& 0.17 & 2.34 & 6.95
&  8.65 &  15.20 & 4.92 \\
39 & NGC~6809 &  $+174.8\pm0.4$ & $+183.4$ &  0.76 & 2.83 & 2.89 & 9.40
&  8.89 &  19.13 & 2.15 \\
\noalign{\smallskip}
\hline
\end{tabular}
\smallskip

\begin{tabular}{lll}

$^{\mathrm{a}}$Heliocentric radial velocity &
$^{\mathrm{d}}$The core radii &
$^{\mathrm{g}}$Log. of core relaxation time at $r_{\rm h}$ \\
$^{\mathrm{b}}$Radial velocity relative to the(LSR) &
$^{\mathrm{e}}$The core median radii &
$^{\mathrm{h}}$Central surface brightness \\
$^{\mathrm{c}}$Concentration parameter $[c=log(r_{\rm t}/r_{\rm c})]$ &
$^{\mathrm{f}}$Log. of relaxation time in years &
$^{\mathrm{i}}$Log. of central luminosity density \\
\end{tabular}
\end{center}
\end{table*}

In order to facilitate the readers work, we present in Tables
3\donothing{}, \ref{param2} and \ref{param3} the basic parameters
available for our GGCs sample\footnote {Unless otherwise stated, the
data presented in these tables are taken from the McMaster catalog
described by Harris (\cite{harris96}).}.

In Table~3\donothing{} we give the coordinates, the position, and the
metallicity of the clusters: right ascension and declination (epoch
J2000, columns 3 and 4); Galactic longitude and latitude (columns 5
and 6); Heliocentric (column 7) and Galactocentric (column 8)
distances (assuming $R_{\sun}$=8.0 kpc); spatial components (X,Y,Z)
(columns 9, 10 and 11) in the Sun-centered coordinate system (X
pointing toward the Galactic center, Y in direction of Galactic
rotation, Z toward North Galactic Pole) and, finally, the metallicity
given in Rutledge et al. (\cite{rutledge97}), on both the Zinn \& West
(\cite{zinnwest84}) and Carretta \& Gratton (\cite{carretagratton97})
scales.

In Table~\ref{param2}, the photometric parameters are given. Column 3
lists the foreground reddening; column 4, the $V$ magnitude level of
the horizontal branch; column 5, the apparent visual distance modulus;
integrated $V$ magnitudes of the clusters are given in column 6;
column 7 gives the absolute visual magnitude. Columns 8 to 11 give the
integrated color indices (uncorrected for reddening). Column 12 gives
the specific frequency of RR Lyrae variables, while column 13 list the
horizontal-branch morphological parameter (Lee \cite{lee90}).

In Table~\ref{param3}, we present the kinematical and structural
parameters for the observed clusters. Column 3 gives the heliocentric
radial velocity (km/s) with the observational (internal) uncertainty;
column 4, the radial velocity relative to the local standard of rest;
column 5, the concentration parameter ($c = \log (r_{\rm t}/r_{\rm
c})$); a 'c' denotes a core-collapsed cluster; columns 6 and 7, the
core and the half mass radii in arcmin; column 8, the logarithm of the
core relaxation time, in years; and column 9 the logarithm of the
relaxation time at the half mass radius. Column 10, the central
surface brightness in $V$; and column 11, the logarithm of central
luminosity density (Solar luminosities per cubic parsec).

\section{The Color-Magnitude Diagrams}
\label{cmds}

In this section the $V$  vs. $(V-I)$ CMDs for the 39 GGCs and the
covered fields are presented.

The same color and magnitude scale has been used in plotting the CMDs,
so that differential measures can be done directly using the plots.
Two dot sizes have been used, with the bigger ones corresponding to
the better measured stars, normally selected on the basis of error
($\leq0.1$) and sharpness parameter (Stetson 1987). In some
exceptional cases, a selection on radius is also done in order to make
evident the cluster stars over the field stars, or to show
differential reddening effects. The smaller size dots show all the
measured stars with errors (as calculated by DAOPHOT) smaller than
0.15 mag.

The images of the fields are oriented with the North at the top and
East on the left side. As explained in Sect.~\ref{obs}, each field
covers $3.77\times3.77(\arcmin)^2$, and the overlaps between fields of
the same object are about $20-25\%$ of the area. For some clusters,
only short exposures were obtained for the central fields.

In the next subsections, we present the single CMDs and clusters, and
give some references to the best existing CMDs. This is by no means a
complete bibliographical catalog: a large number of CMDs are available
in the literature for many of the clusters of this survey, but we will
concentrate just on the best CCD photometric works. The tables with
the position and photometry of the measured stars will be available
via a web interface at IAC and Padova in the near future.

\begin{figure*}
\begin{tabular}{c@{}c}
\raisebox{-4cm}{
\psfig{figure=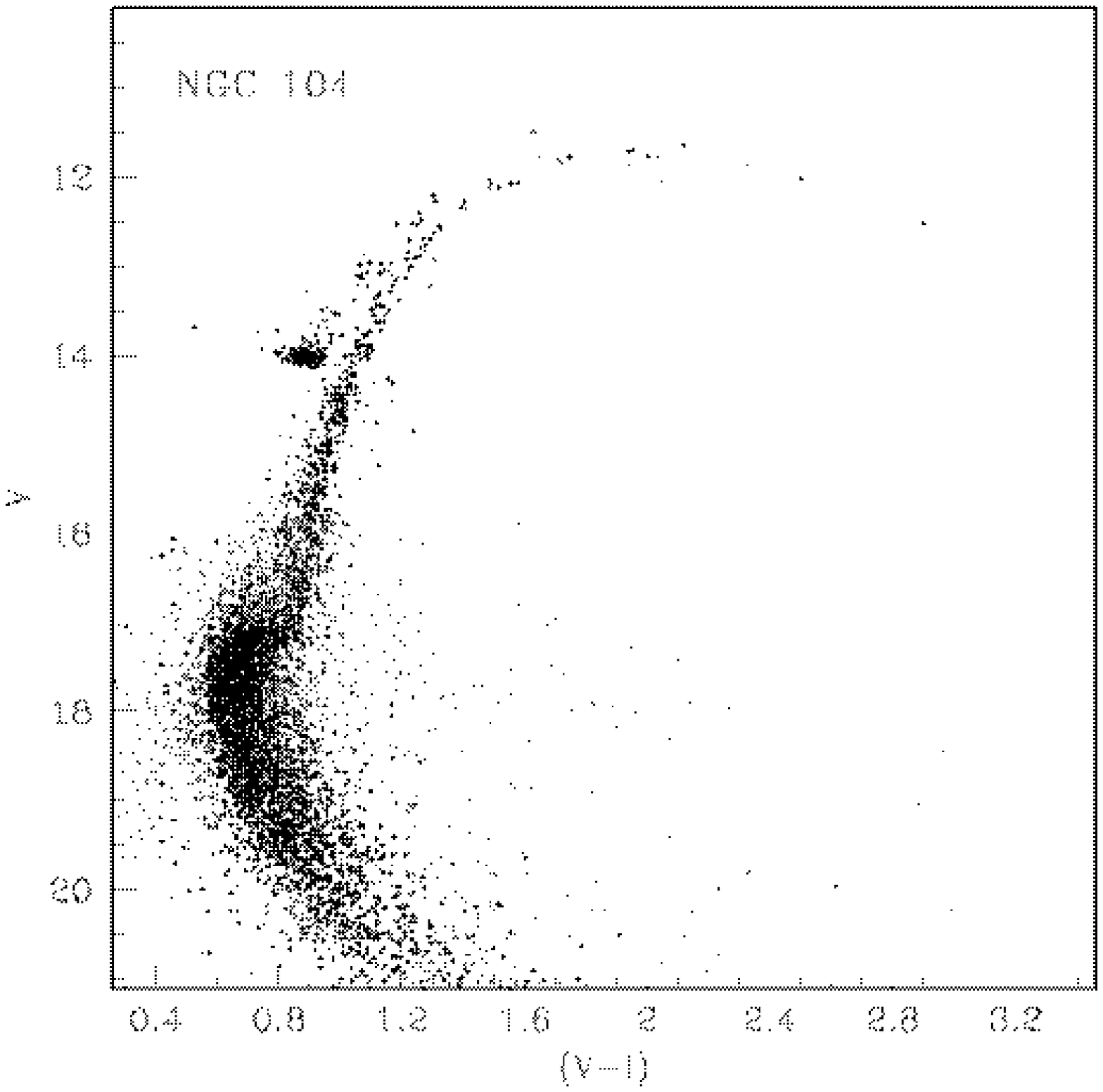,width=10.9cm}
} &
\begin{minipage}[t]{8.8cm}
\begin{tabular}{c@{}c}
\hspace{-2cm}
\fbox{\psfig{figure=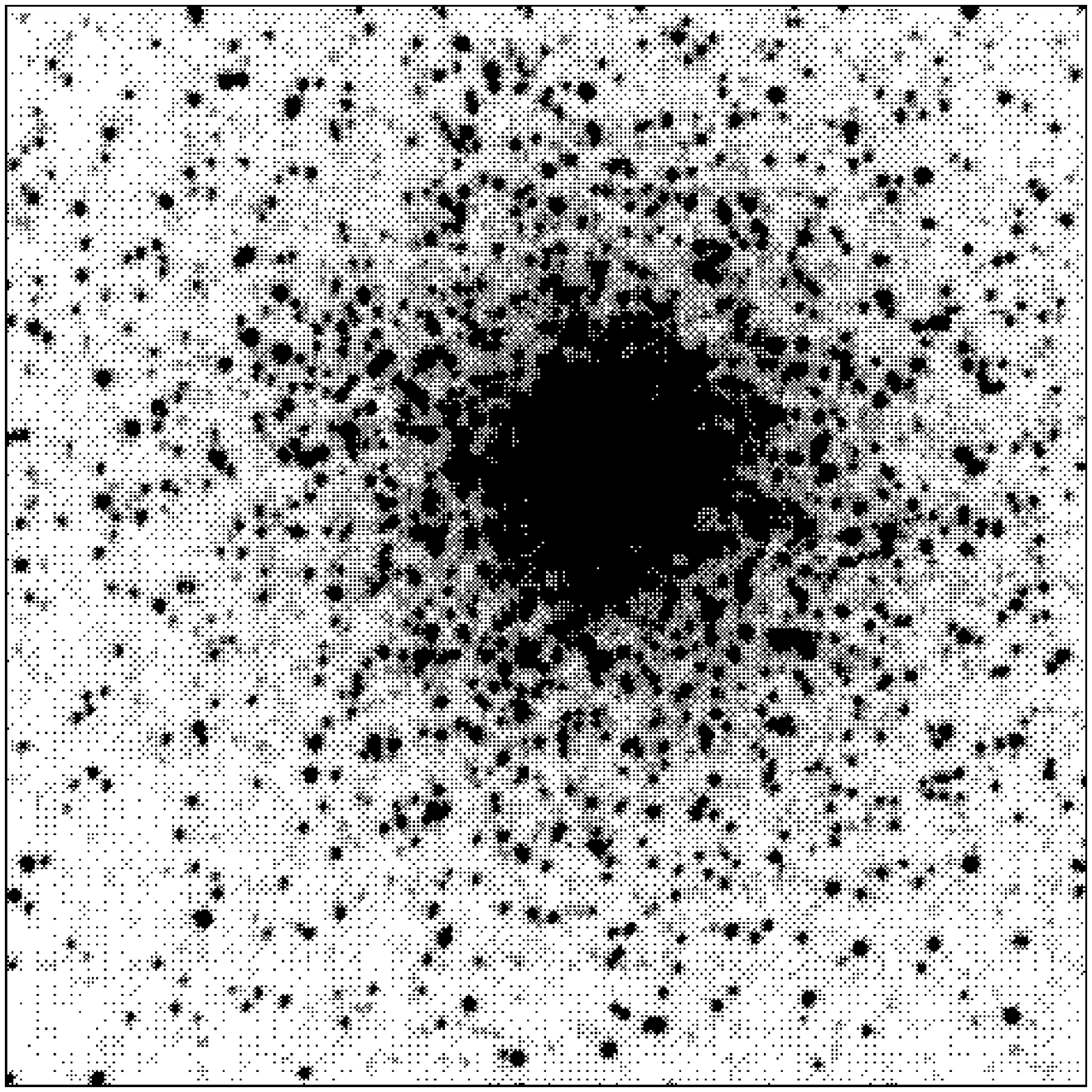,width=4cm}} &
\fbox{\psfig{figure=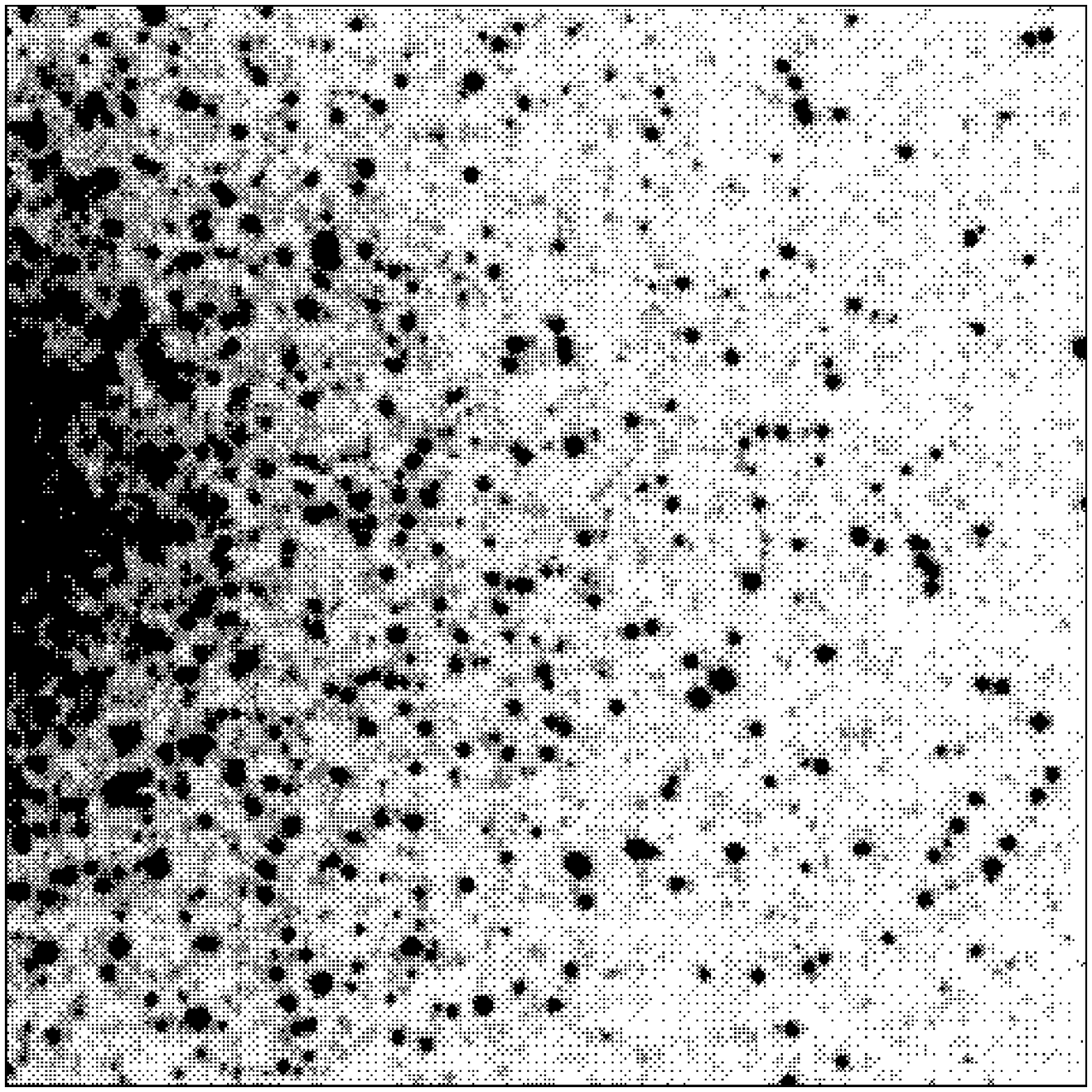,width=4cm}} 
\end{tabular}
\end{minipage}
\end{tabular}
\caption[]{CMD and covered fields for NGC~104 (47 Tucanae)}
\label{ngc104}
\end{figure*}

\paragraph {\bf NGC 104 (47 Tucanae).}(Fig.~\ref{ngc104})

The cluster 47 Tucanae is (after $\omega$ Centauri) the second
brightest globular cluster in the sky, and consequently a lot of work
has been done on this object. 47 Tucanae has been often indicated as
the prototype of the metal-rich GGCs, characterized by a well
populated red HB (RHB) clump and an extended RGB that also in our CMD
spans $\sim$2~mag in color from the RHB to the reddest stars at the
tip.

A classical CMD of 47~Tucanae is that presented by Hesser et
al. (\cite{hesser87}) where a composite CMD was obtained from the
superposition of $B$ and $V$ CCD photometry for the main sequence (MS)
and photographic data for the evolved part of the diagram. The same
year, Alcaino \& Liller (\cite{alcainoliller87a}) published a BVI CCD
photometry. One year later, Armandroff (\cite{armandroff88}) presented
the RGB $V$ and $I$ bands photometry for this cluster (together with
other five). In 1994, Sarajedini \& Norris (\cite{sarajedininorris94})
presented a study of the RGB and HB stars in the $B$ and $I$ bands.
Sosin et al. (\cite{sosin97a}) and Rich et al. (\cite{rich97}) have
published a $B$,$V$ photometry based on HST data.

A recent work in the $V$ and $I$ bands has been presented by Kaluzny
et al. (\cite{kaluzny98}), who focussed their study on the variable
stars. They do not find any RR-Lyrae, but many other variables (mostly
located in the BSS region), identified as binary stars. As already
stated by these authors, a small difference is found between their and
our photometry. Indeed, their magnitudes coincide with ours at
$\sim 12.5$~mag in both bands, but there is a small deviation from
linearity of $\sim -0.015$ magnitudes per magnitude (with the Kaluzny
et al. stars brighter than ours), in both bands (computed from 90
common stars with small photometric errors) within a magnitude range
of $\sim 3$ mag. We are confident that our calibration, within the
quoted errors, is correct, as further confirmed by the comparison
with other authors for other objects, as discussed below. Although
small, these differences could be important in relative measures, if
they appear randomly in different CMDs. For example, in this case, the
$\Delta V_{\rm TO}^{\rm HB}$ parameter is $\sim0.05$ mag smaller in
Kaluzny et al. (\cite{kaluzny98}) CMD than in our one, implying, for
the 47 Tuc metallicity, an age difference of $\sim0.8$ \rm Gyrs. We
want to stress the importance of a homogeneous database for a
reliable measurement of differential parameters on the CMDs.

\begin{figure*}
\begin{tabular}{c@{}c}
\raisebox{-6cm}{
\psfig{figure=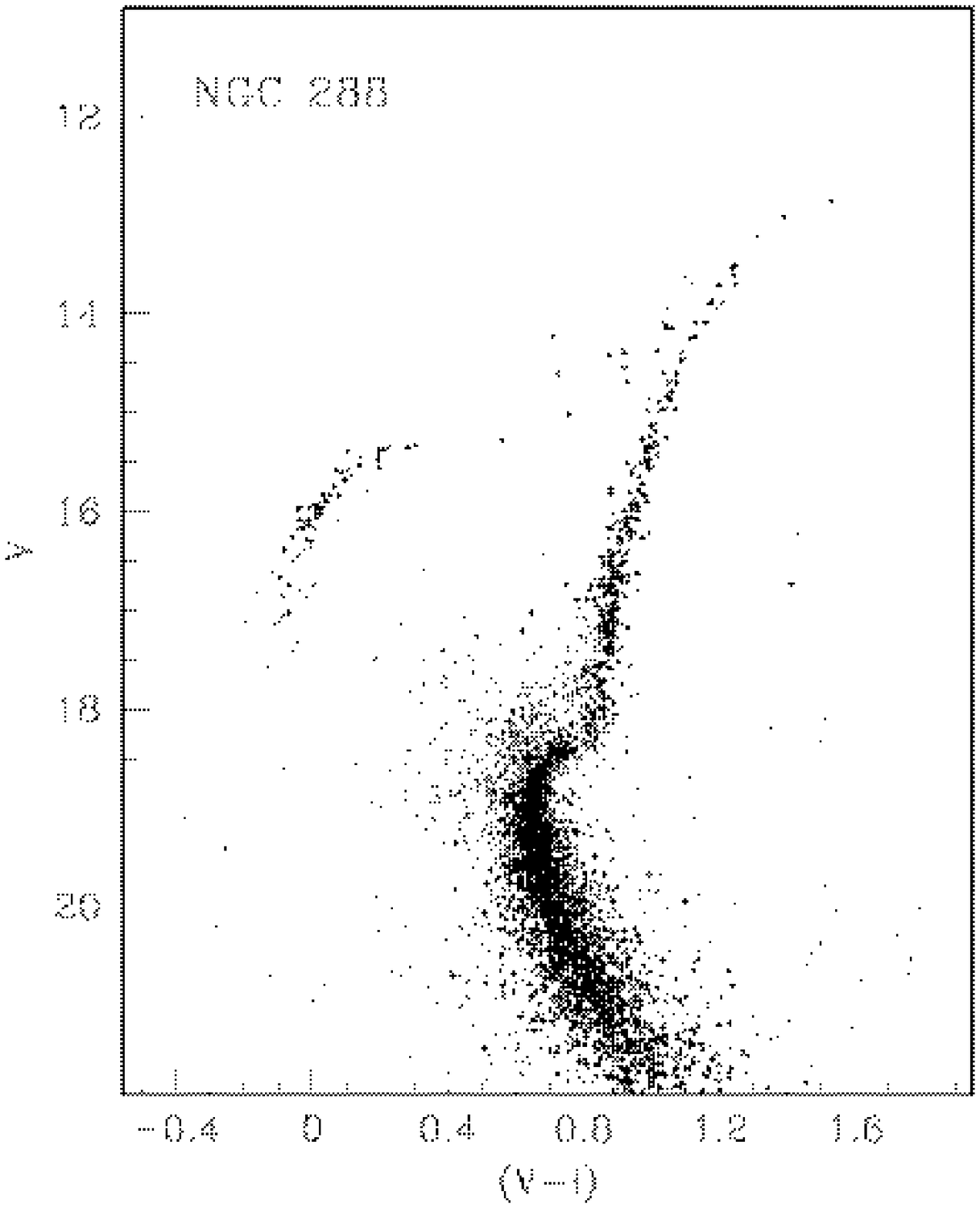,width=8.8cm}
} &
\begin{minipage}[t]{8.8cm}
\begin{tabular}{c@{}c}
\fbox{\psfig{figure=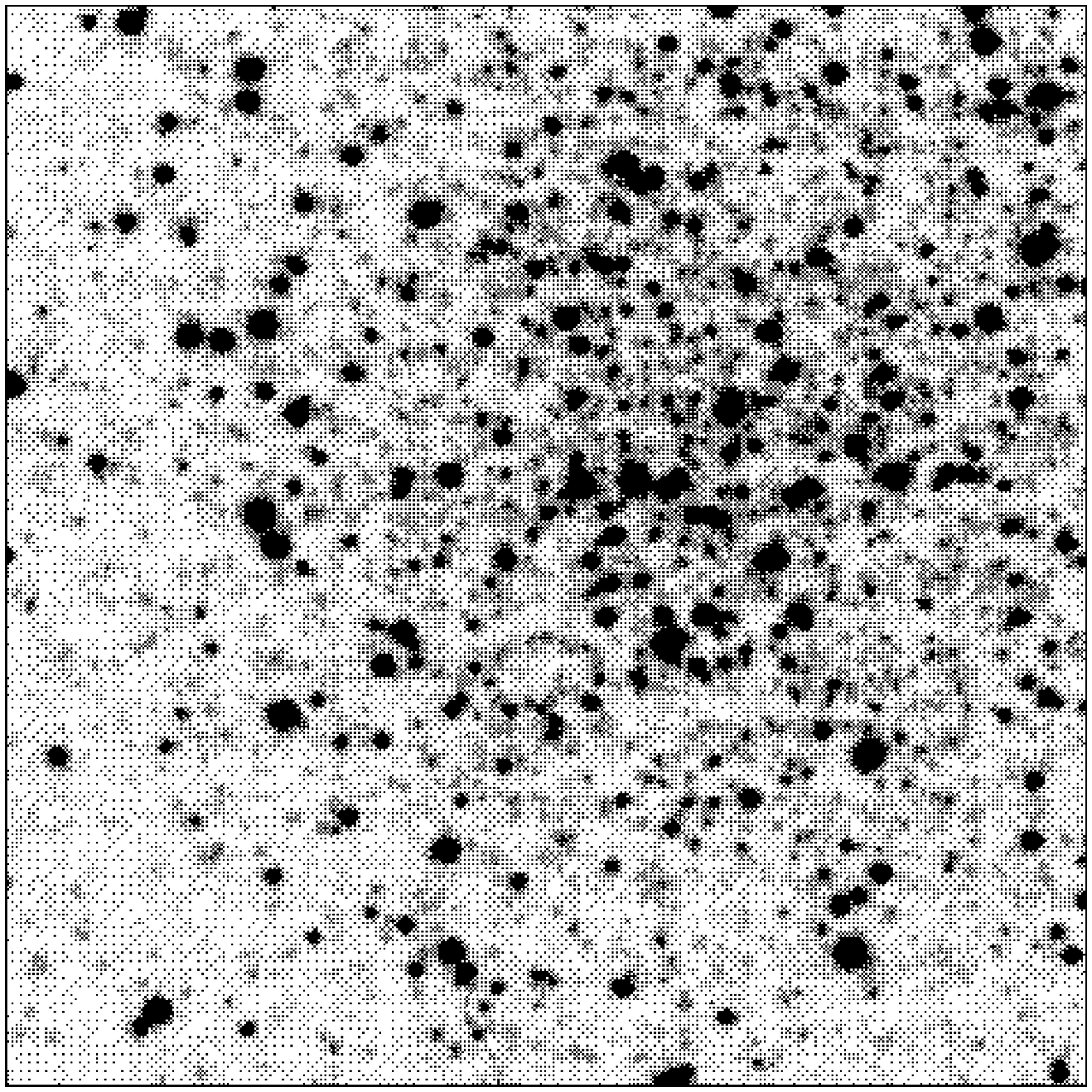,width=4cm}} &
\fbox{\psfig{figure=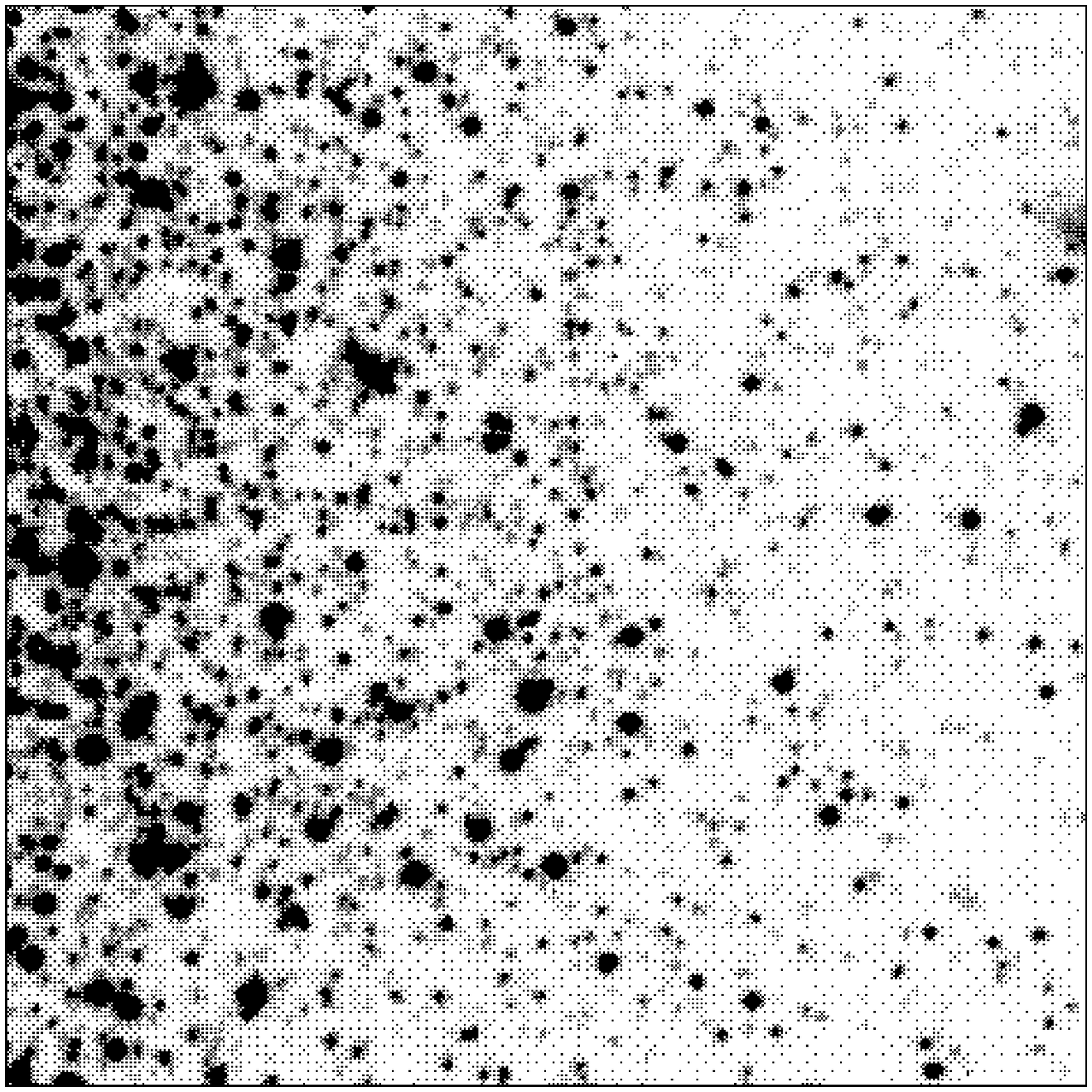,width=4cm}} \\
\fbox{\psfig{figure=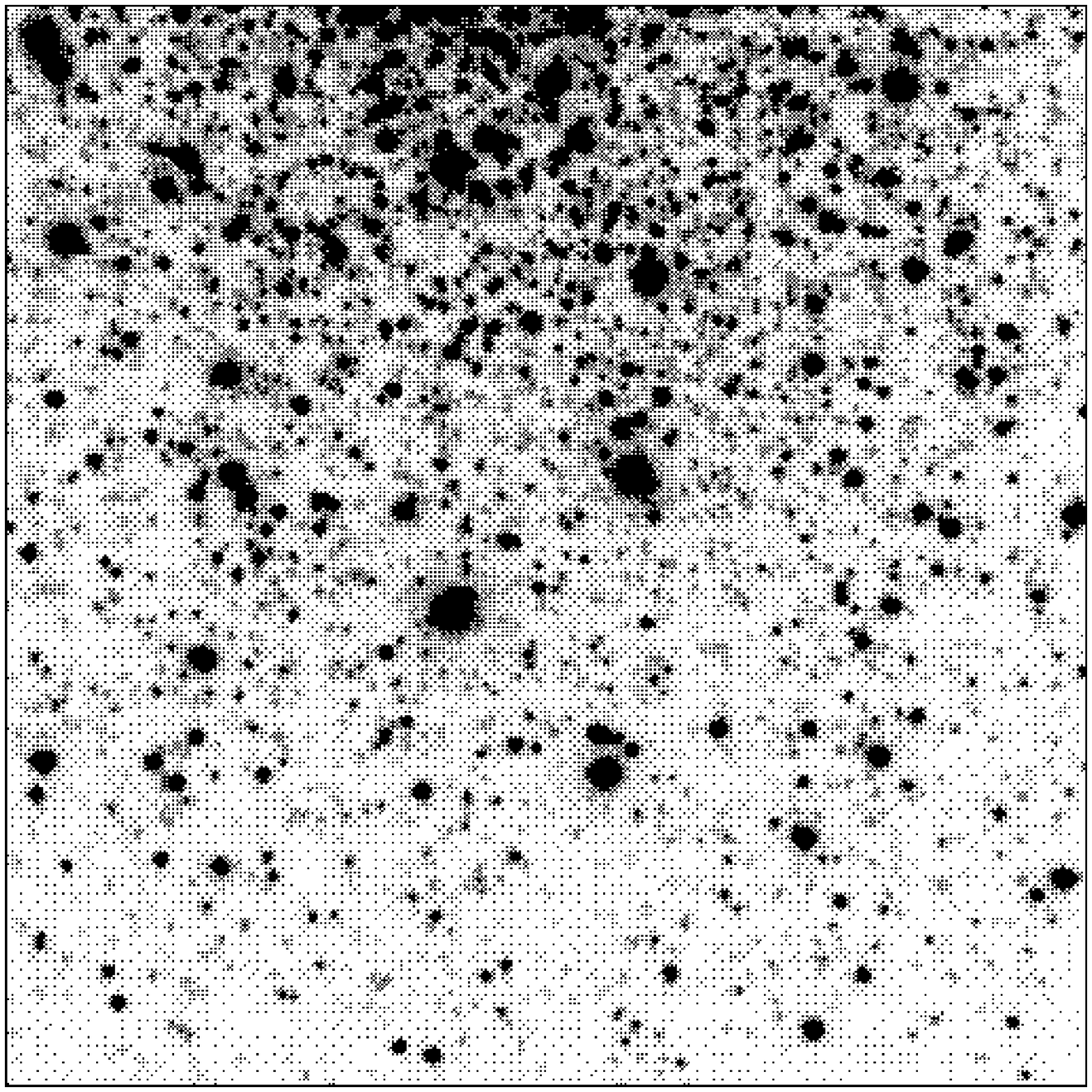,width=4cm}}
\end{tabular}
\end{minipage}
\end{tabular}
\caption[]{CMD and covered fields for NGC~288}
\label{ngc288}
\end{figure*}

\begin{figure*}
\begin{tabular}{c@{}c}
\raisebox{-6cm}{
\psfig{figure=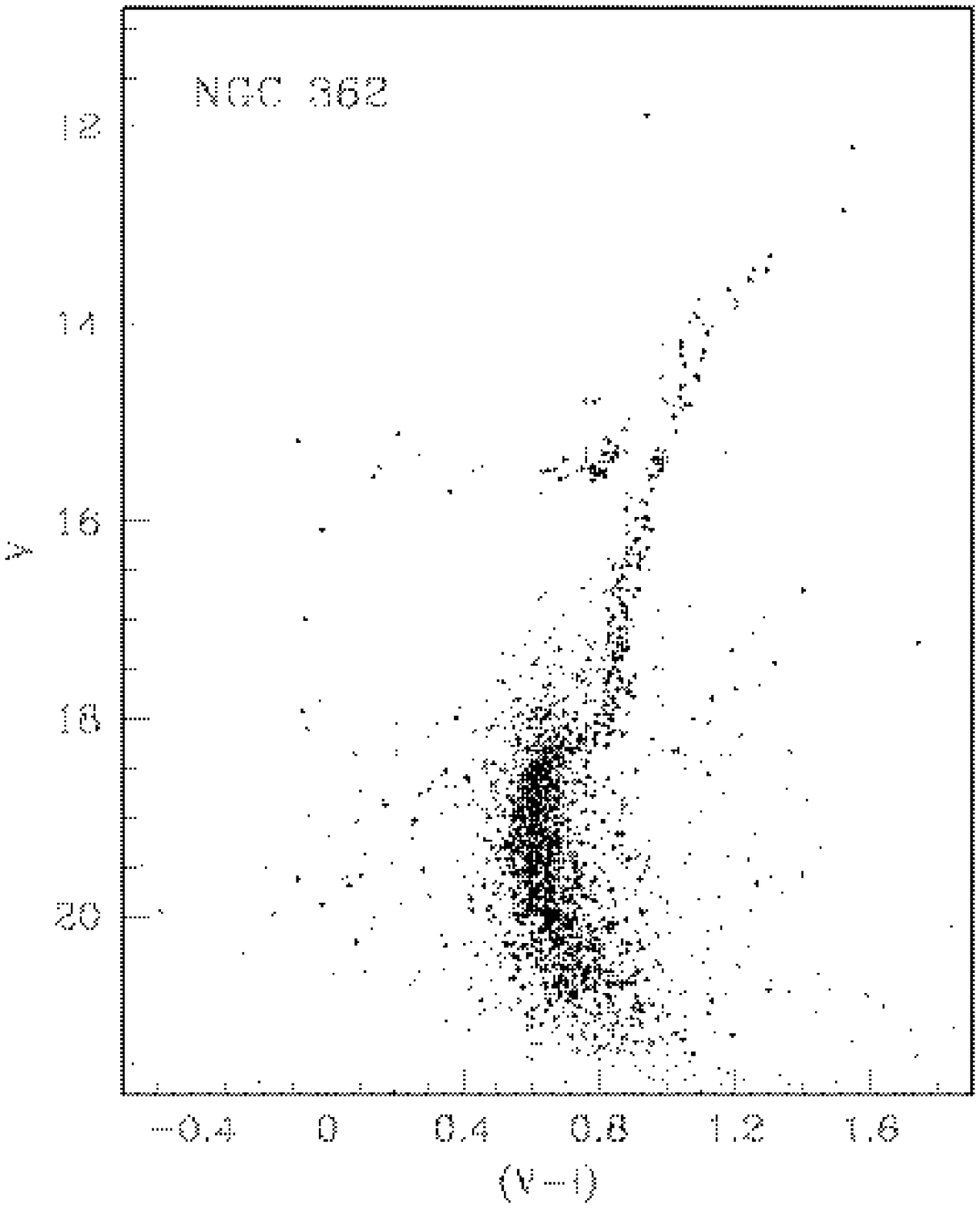,width=8.8cm}
} &
\begin{minipage}[t]{8.8cm}
\begin{tabular}{c@{}c}
\fbox{\psfig{figure=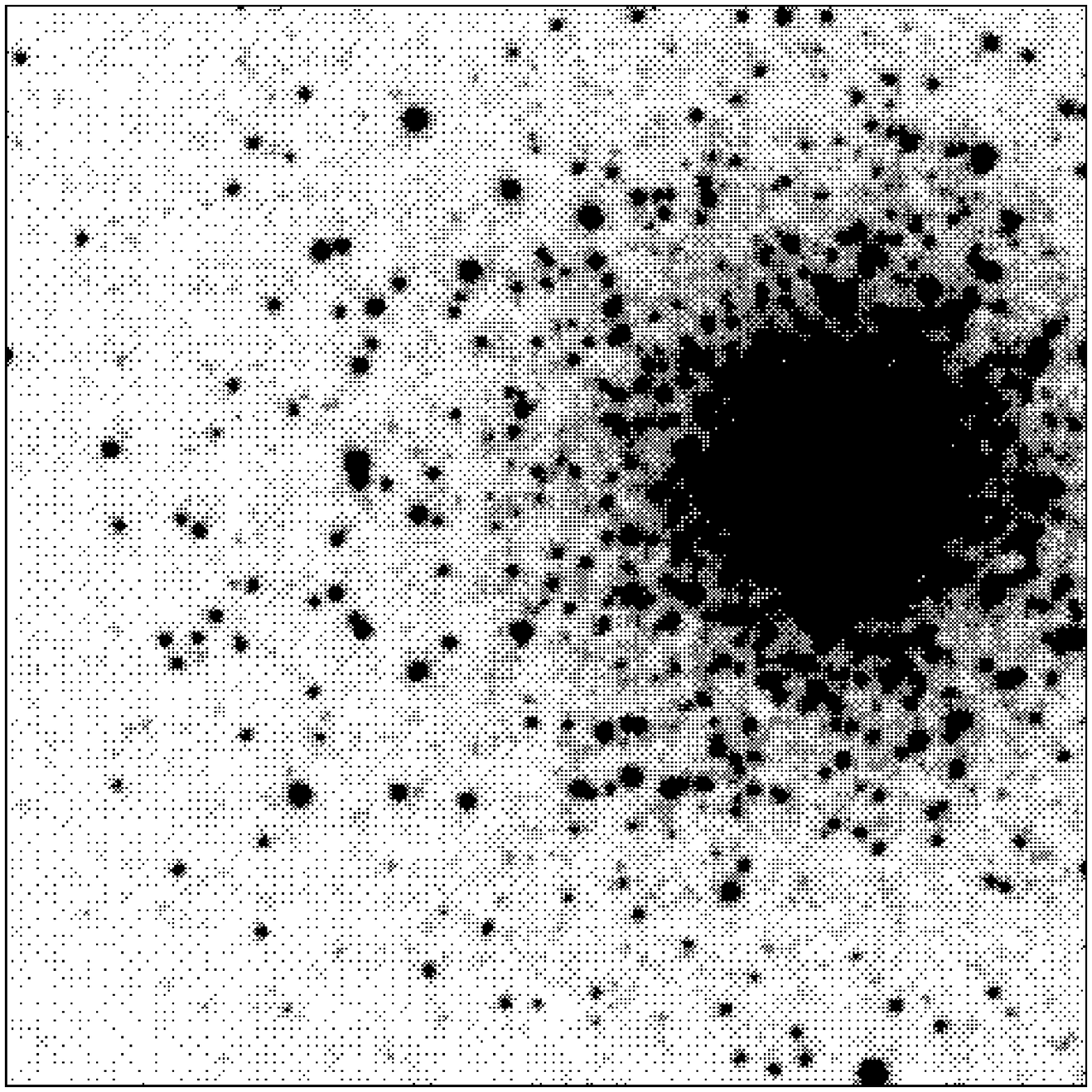,width=4cm}} &
\fbox{\psfig{figure=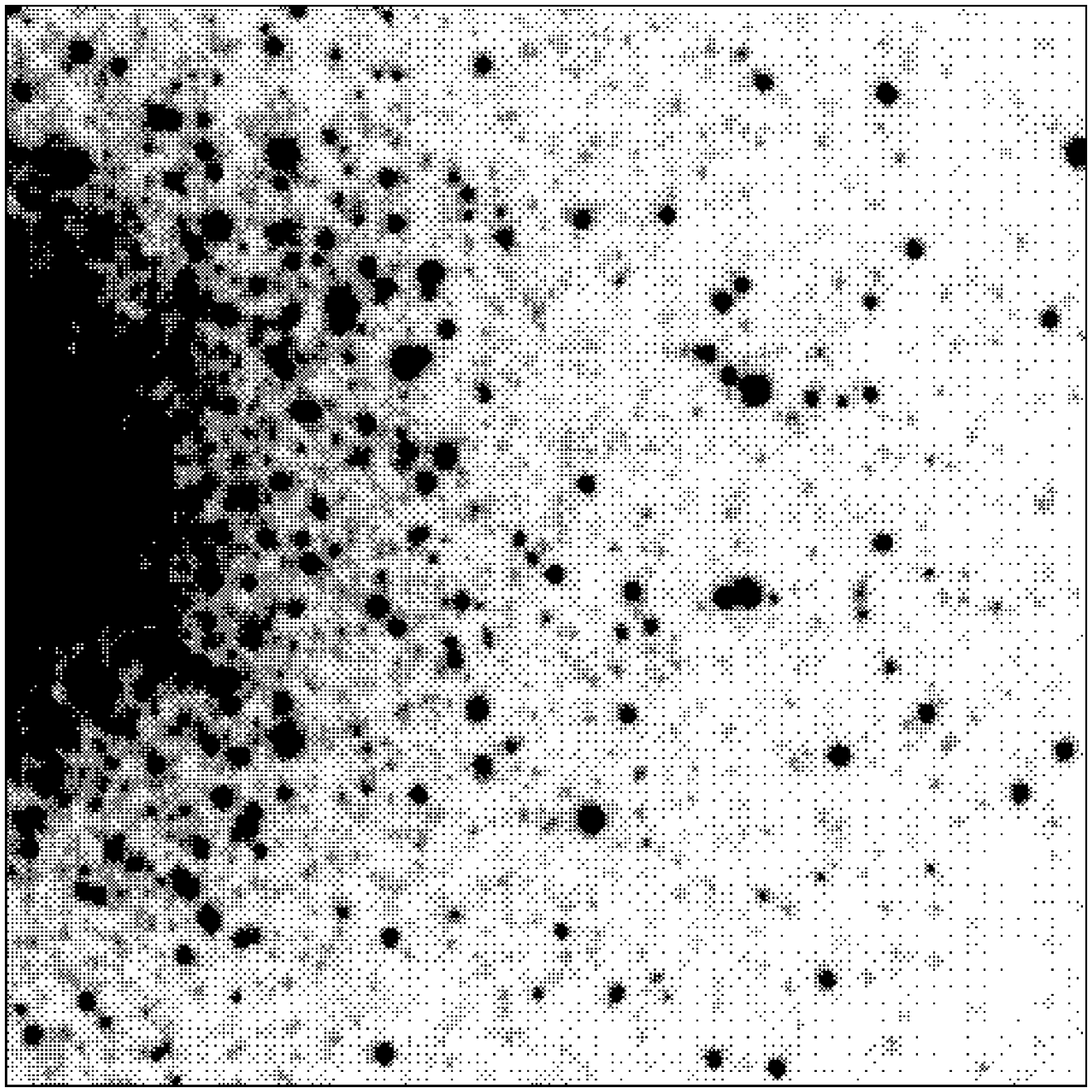,width=4cm}} \\
\fbox{\psfig{figure=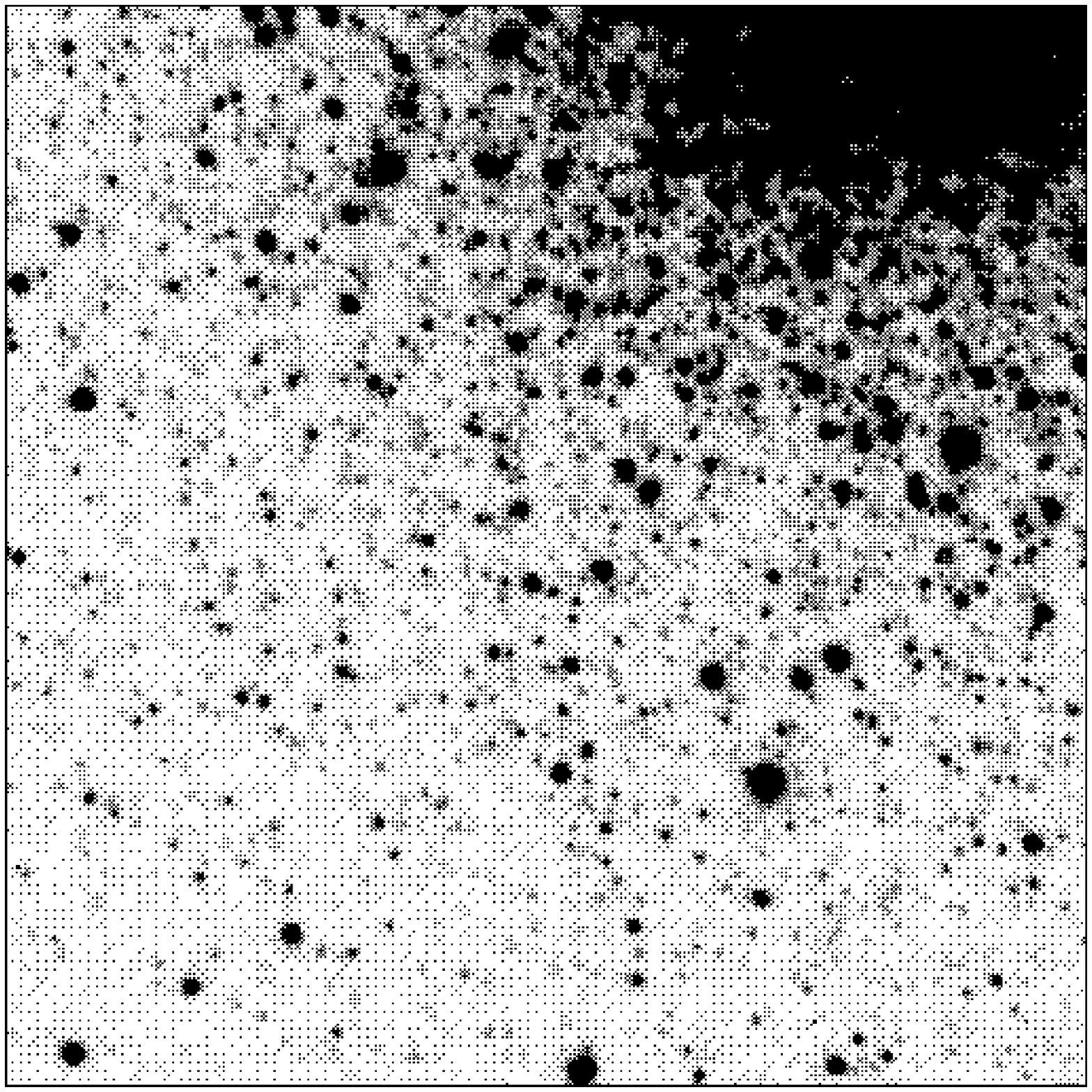,width=4cm}}
\end{tabular}
\end{minipage}
\end{tabular}
\caption[]{CMD and covered fields for NGC~362}
\label{ngc362}
\end{figure*}

\paragraph {\bf NGC~288 and NGC~362.} 
(Figs.~\ref{ngc288} and \ref{ngc362})

The diagram of NGC~288 is well defined and presents an extended blue
horizontal branch (EBHB) which extends from the blue side of the
RR-Lyrae region, to just above the TO. Conversely,
NGC~362 has a populated RHB with just a few blue HB stars.

These two clusters define one of the most studied second parameter
couple: despite their similar metallicities, their HB morphologies are
different. Much work have been done on both clusters in order to try
to understand the origin of such differences: Bolte (\cite{bolte89})
and Sarajedini \& Demarque (\cite{sarajedinidemarque90}) in the $B$
and $V$ bands, and Green \& Norris (\cite{greennorris90}) in the $B$
and $R$ bands, based on homogeneous CCD photometry, obtain an age
difference of $\sim3$ \rm Gyrs, NGC~288 being older than NGC~362. A
similar conclusion is obtained in our study (Paper III), where NGC~362
is found $\sim20\%$ younger than NGC~288. It has also been proposed
(e.g. Green \& Norris \cite{greennorris90}) that these age differences
might be responsible of the HB differences between the two clusters.
On the other side, Buonanno et al. (\cite{buonanno98}) and Salaris \&
Weiss (\cite{salarisweiss98}) do not find significant age
differences. Another $B$,$V$ photometry of NGC~362 based on HST data
is in Sosin et al. (\cite{sosin97a}).

It might be worth to remark here that, as it will be discussed in
Paper II, there are clusters with different HB morphologies, though
with the same metallicities and ages (within errors). This means that
the analysis of a single couple of GGCs can not be considered
conclusive for understanding the second parameter problem, while a
large scale study (as that feasible with this catalog) can be of more
help.

\begin{figure*}
\begin{tabular}{c@{}c}
\raisebox{-6cm}{
\psfig{figure=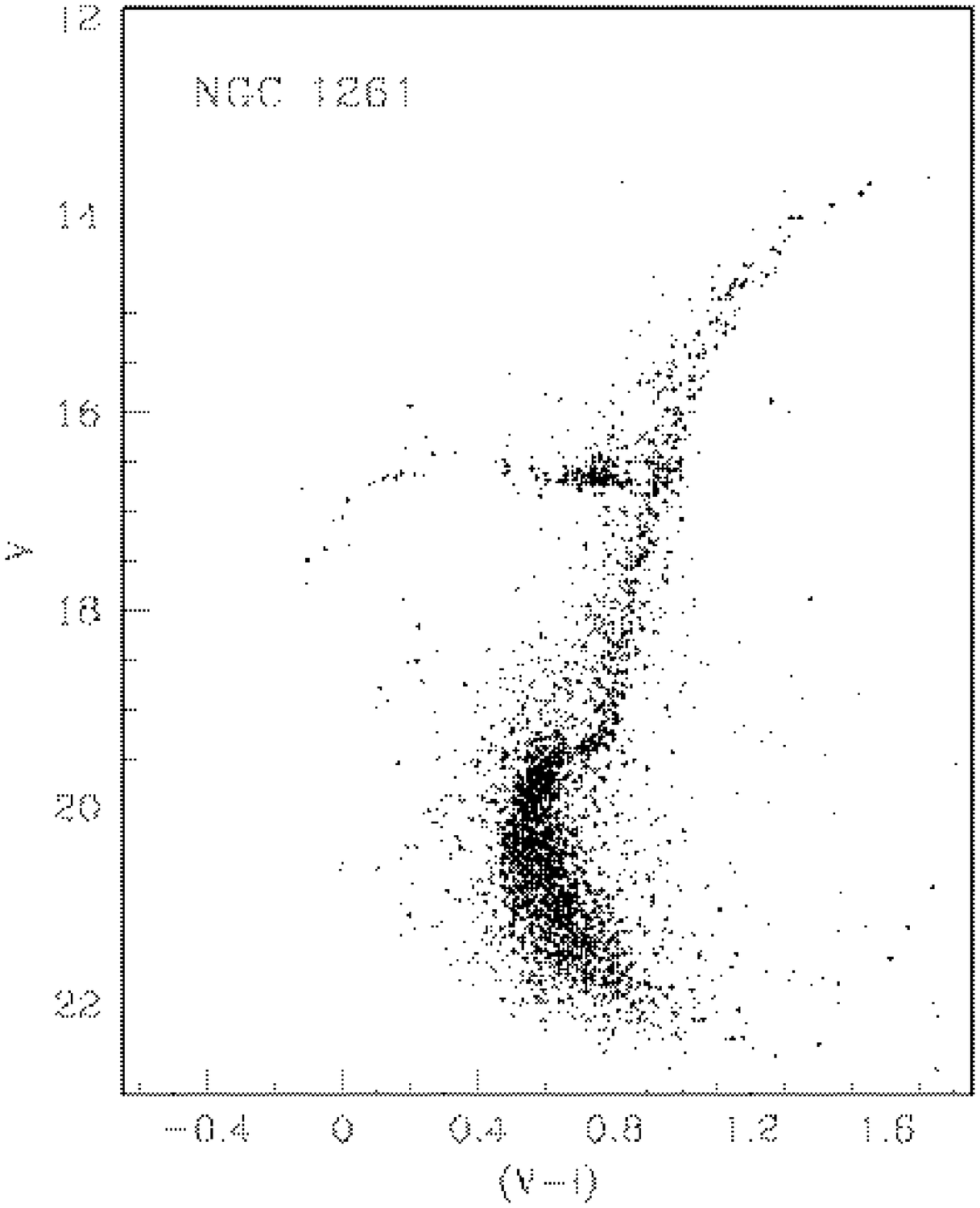,width=8.8cm}
} &
\begin{minipage}[t]{8.8cm}
\begin{tabular}{c@{}c}
\fbox{\psfig{figure=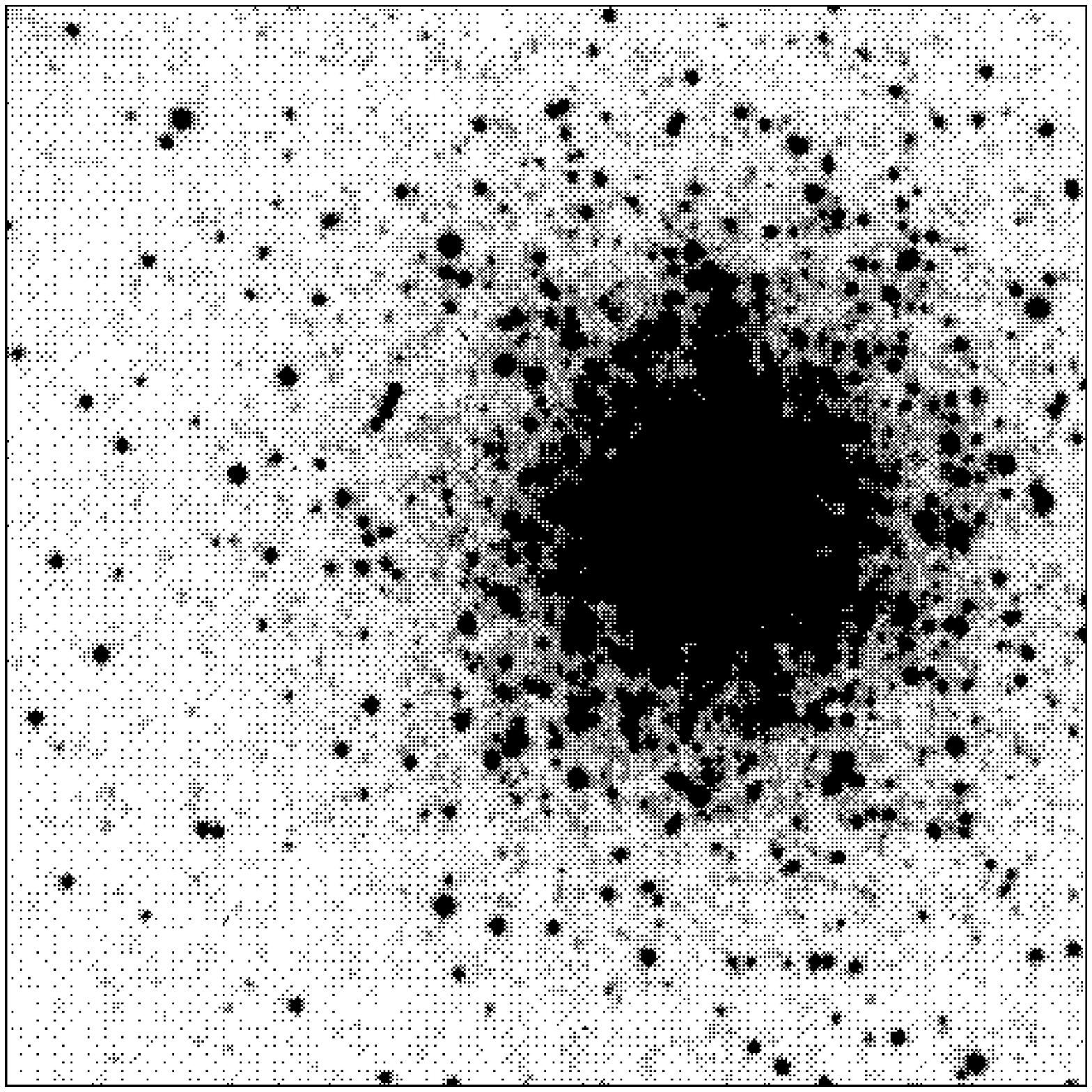,width=4cm}} &
\fbox{\psfig{figure=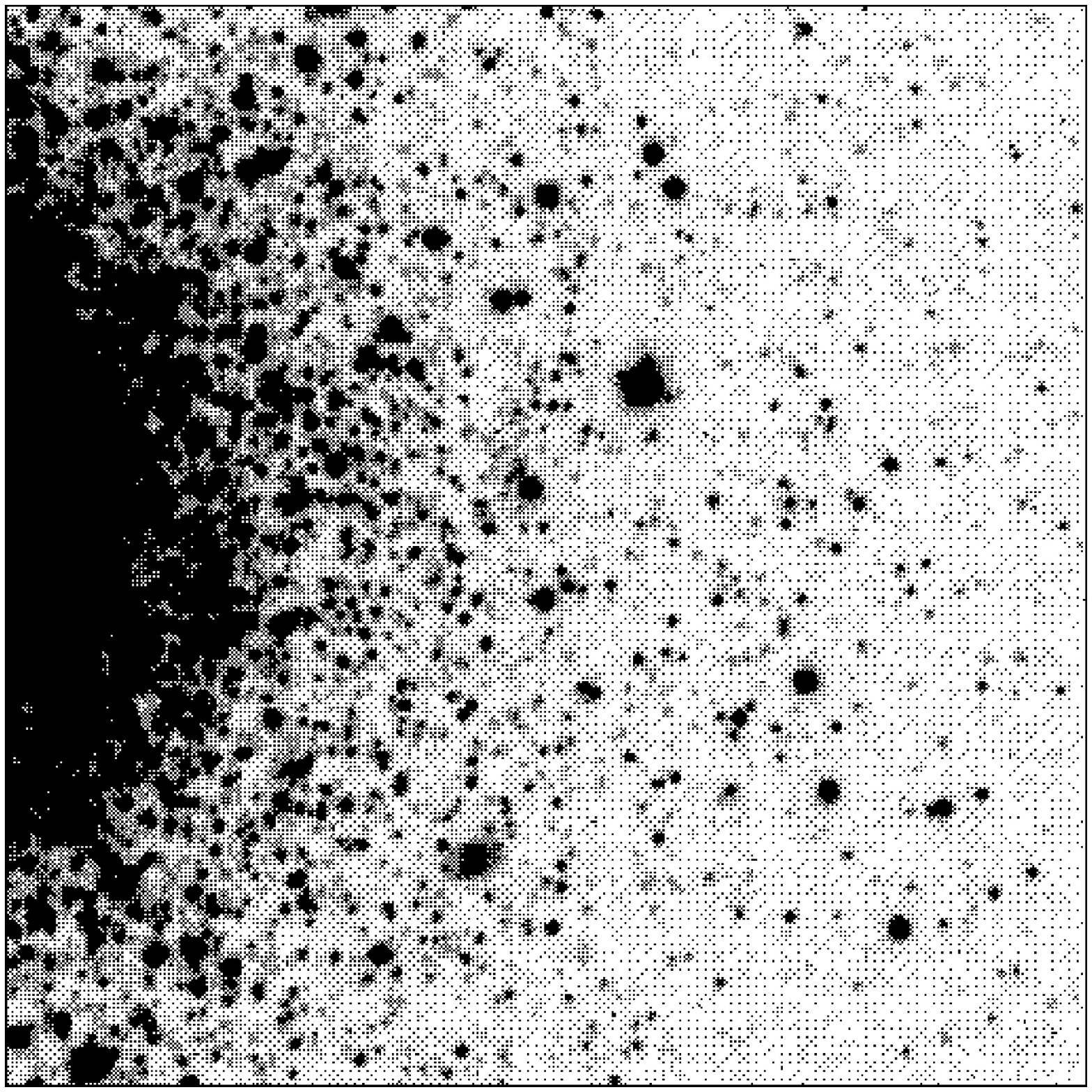,width=4cm}} \\
\fbox{\psfig{figure=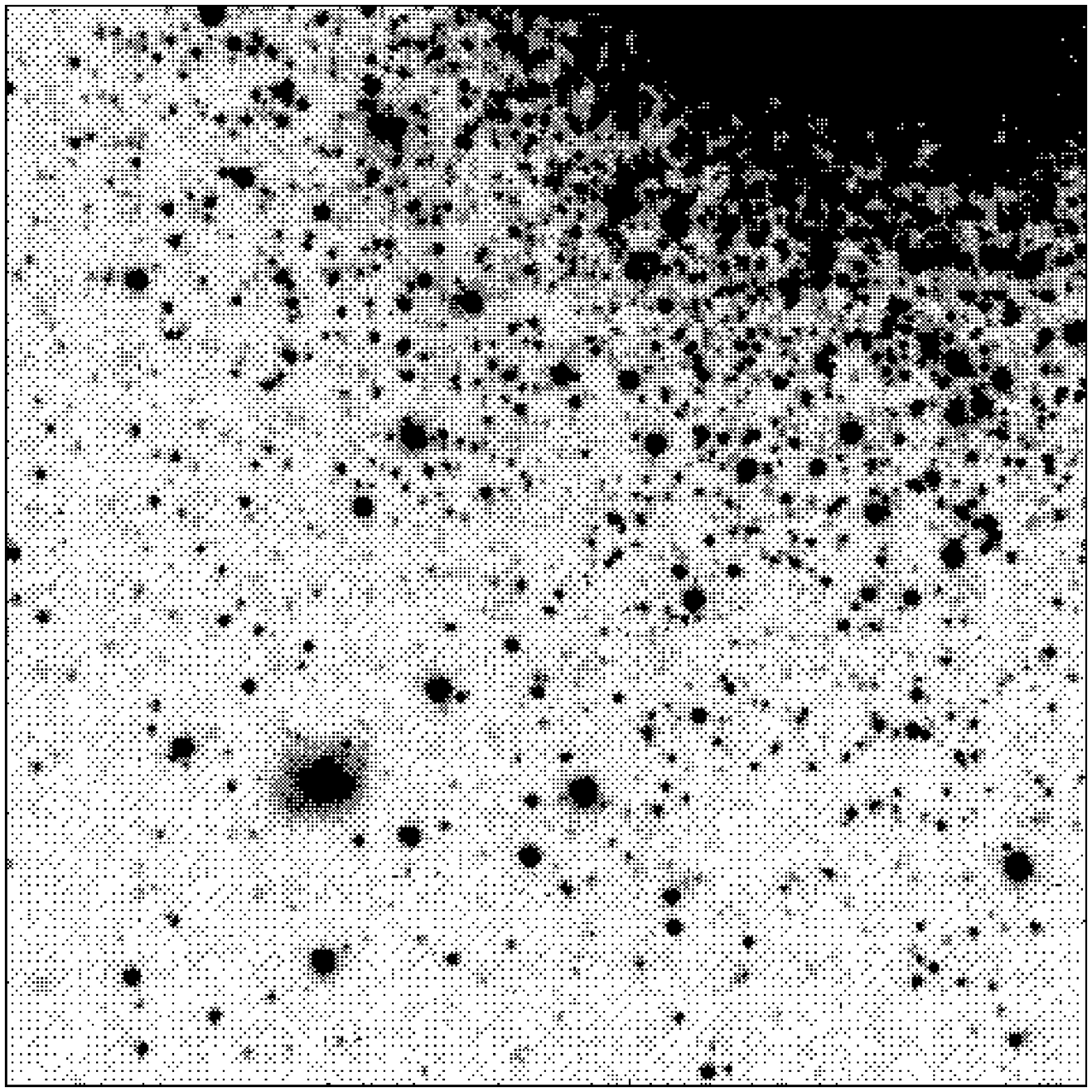,width=4cm}} 
\end{tabular}
\end{minipage}
\end{tabular}
\caption[]{CMD and covered fields for NGC~1261}
\label{ngc1261}
\end{figure*}

\paragraph {\bf NGC 1261.} (Fig.~\ref{ngc1261})

This cluster is the object with the largest distance in our southern
hemisphere sample. It is located at $\sim 16$~kpc from the Sun.

Three major CCD CMDs have been published for NGC~1261: Bolte \&
Marleau (\cite{boltemarleau89}) in $B,V$, Alcaino et
al. (\cite{alcaino92}) in $B,V,R,I$, and Ferraro et
al. (\cite{ferraro93b}) in the $B$ and $V$ bands.

The CMD is characterized by an HB which is similar to the HB of
NGC~1851. From here on, clusters with an HB well populated both on the
red and blue side of the RR-Lyrae gap will be named bimodal HB
clusters, though a more objective classification would require taking
into account the color distribution of stars along the HB including
the RR Lyrae (Catelan et al. \cite{catelan98}). NGC 1261 has a
metallicity very close to that of the previous couple; Chaboyer et
al. (\cite{chaboyer96}), Richer et al. (\cite{richer96}) and Rosenberg
et al. (Paper III) find that it is younger (similar in age to NGC~362)
than the bulk of GGCs. A blue straggler (BS) is clearly visible in
Fig.~\ref{ngc1261}.

\begin{figure*}
\begin{tabular}{c@{}c}
\raisebox{-6cm}{
\psfig{figure=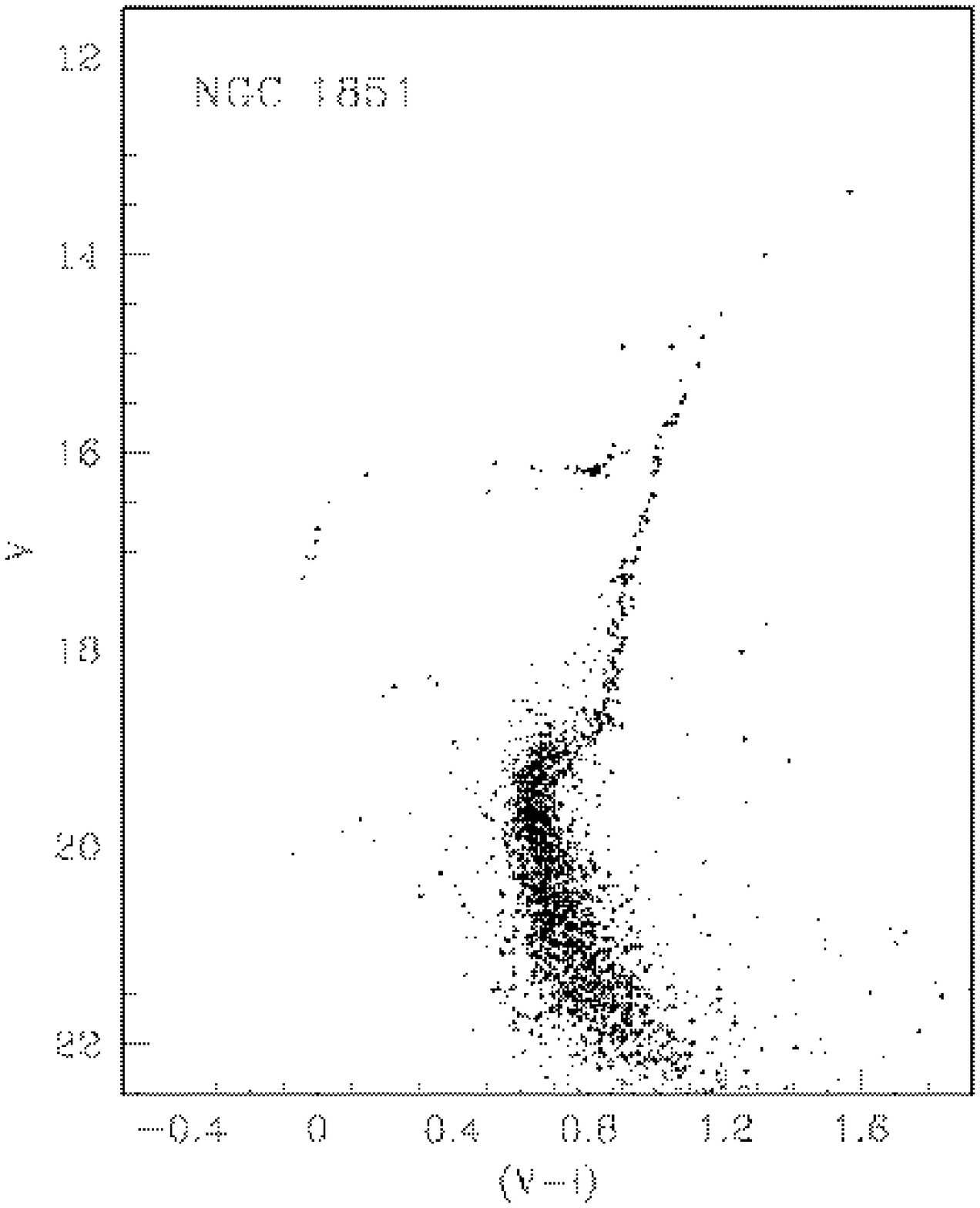,width=8.8cm}
} &
\begin{minipage}[t]{8.8cm}
\begin{tabular}{c@{}c}
\fbox{\psfig{figure=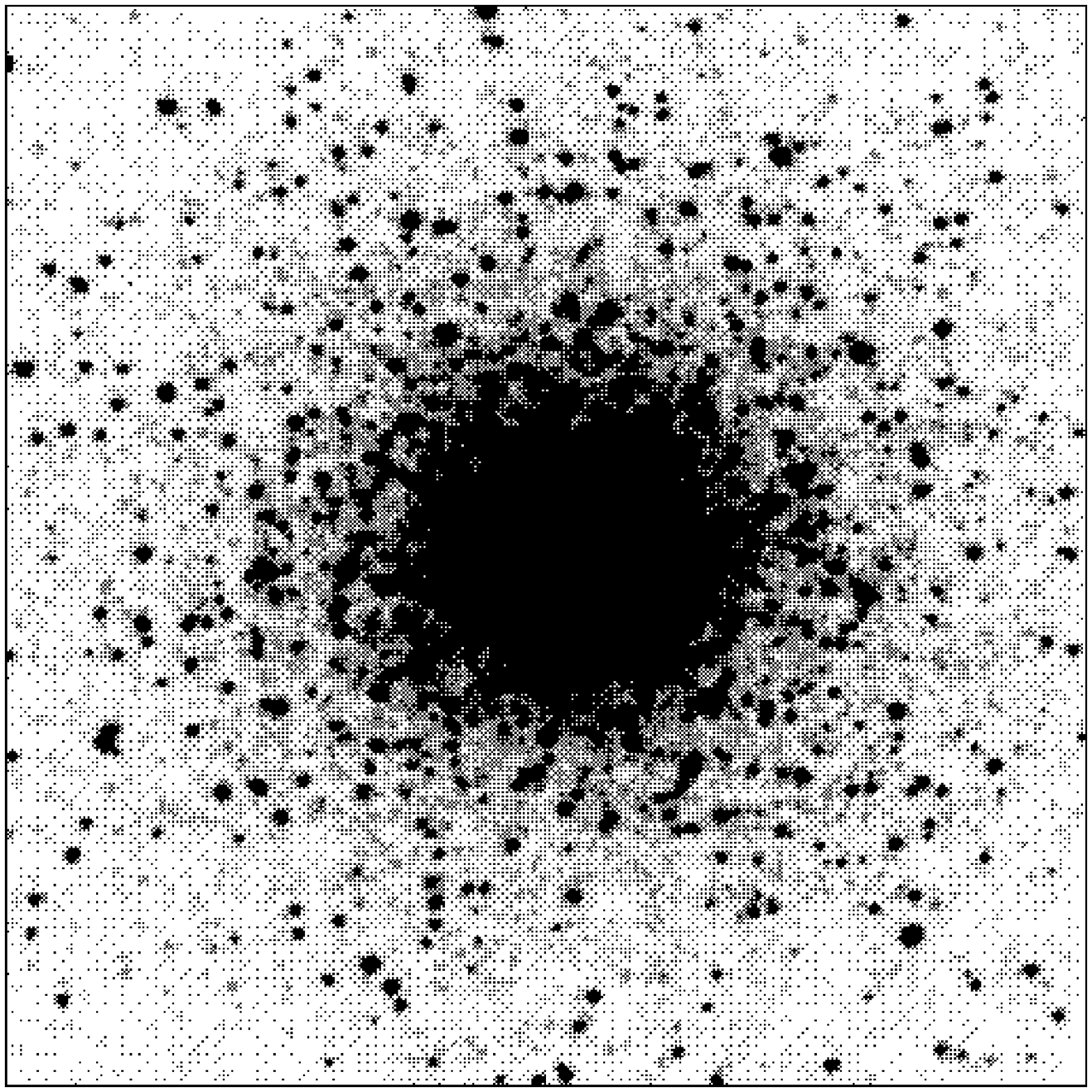,width=4cm}} &
\fbox{\psfig{figure=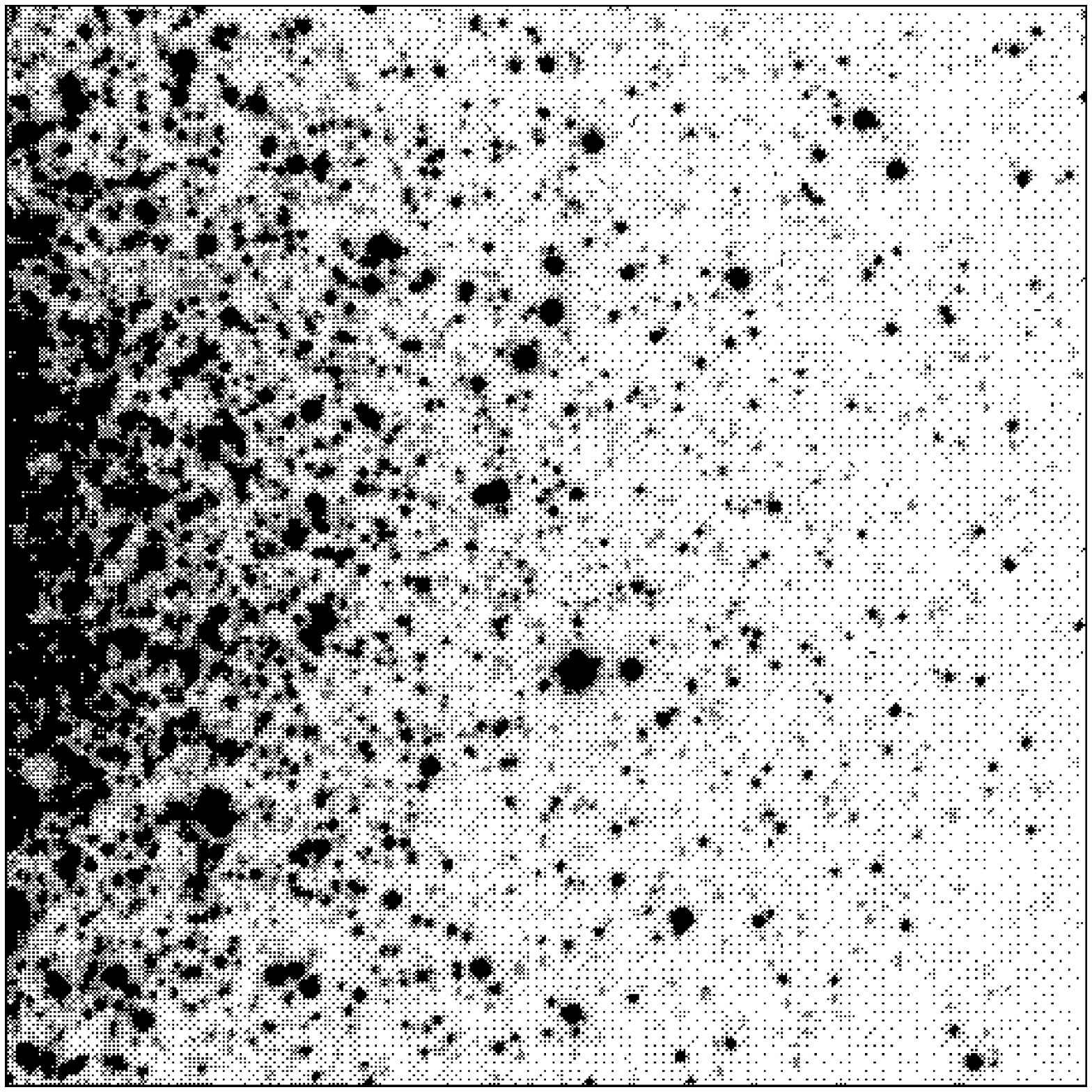,width=4cm}}
\end{tabular}
\end{minipage}
\end{tabular}
\caption[]{CMD and covered fields for NGC~1851}
\label{ngc1851}
\end{figure*}

\paragraph {\bf NGC 1851.} (Fig.~\ref{ngc1851})

This cluster has a bimodal HB, with very well defined RHB and blue HB
(BHB). Also in this case, a BS sequence is visible in
Fig.~\ref{ngc1851}. It is curious that, again, a bimodal cluster
results to be younger than the GGCs bulk. From the 34 clusters in the
present catalog, only 4 result to be surely younger, i.e. the already
described NGC~362 and NGC~1261, this cluster, and NGC~2808: three of
them have a bimodal HB (cf. Rosenberg et al. \cite{rosenberg99} for a
detailed discussion). There exist other two recent $(V,I)$ CCD
photometries of NGC~1851 by Walker (\cite{walker98}) and Saviane et
al. (\cite{saviane98}). The three photometries are all in agreement
within the errors, confirming our calibration to the standard
system. A CMD of NGC~1851 in the $B$,$V$ bands from HST is in Sosin et
al. (\cite{sosin97a}). Older CCD photometries are found in Alcaino et
al (\cite{alcaino90a}) ($B,V,I$ bands) and Walker (\cite{walker92})
($B,V$ bands)

\begin{figure*}
\begin{tabular}{c@{}c}
\raisebox{-6cm}{
\psfig{figure=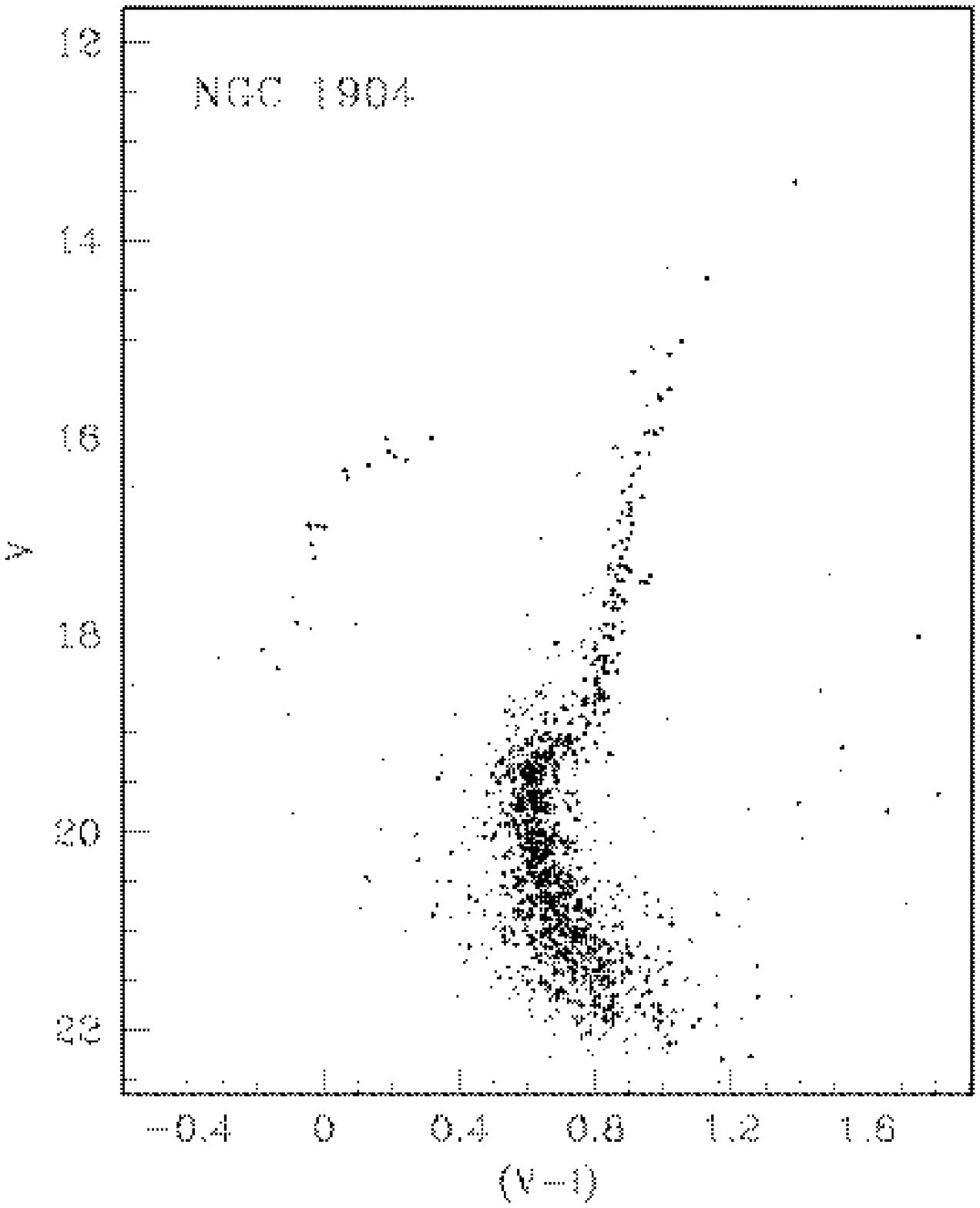,width=8.8cm}
} &
\begin{minipage}[t]{8.8cm}
\begin{tabular}{c@{}c}
\fbox{\psfig{figure=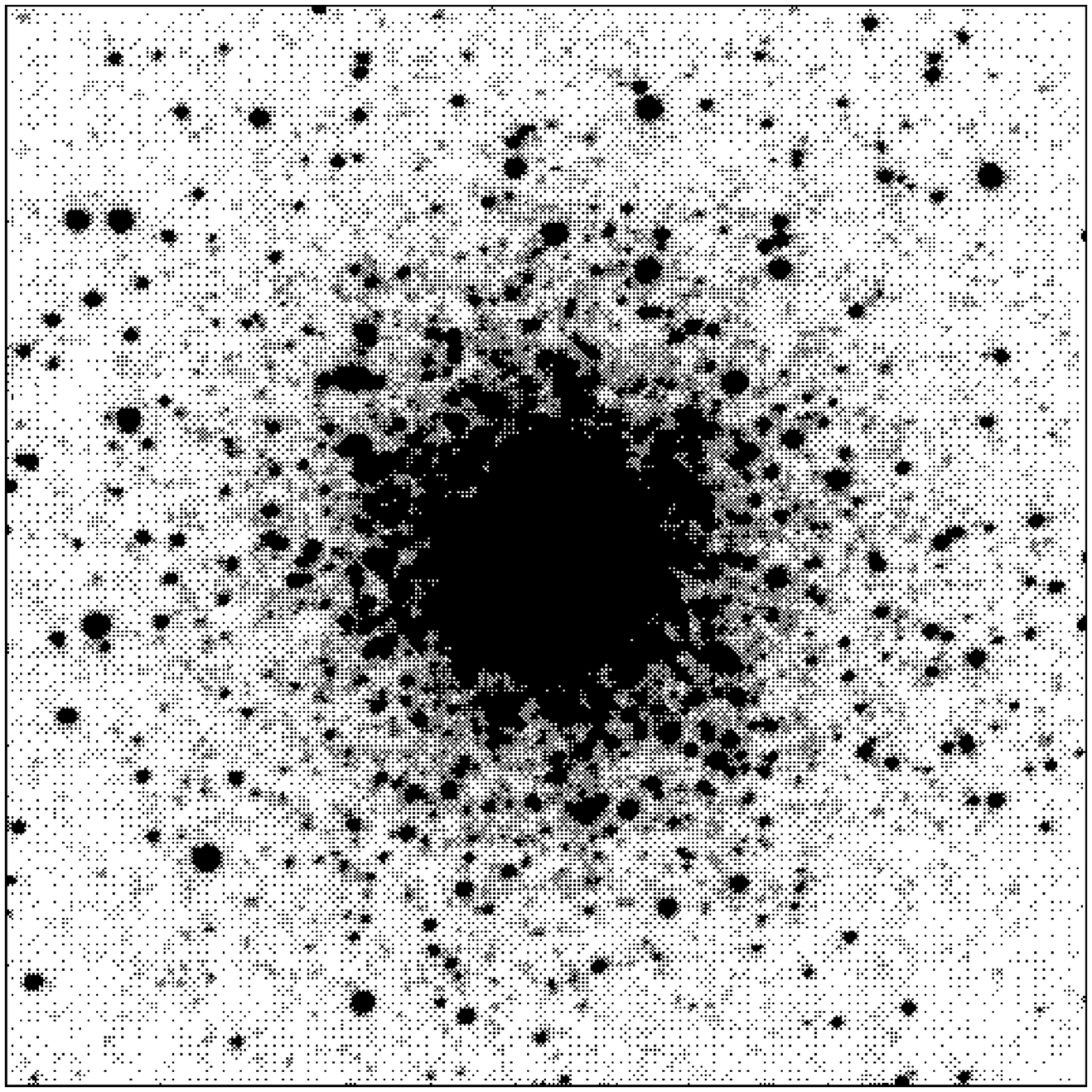,width=4cm}} &
\fbox{\psfig{figure=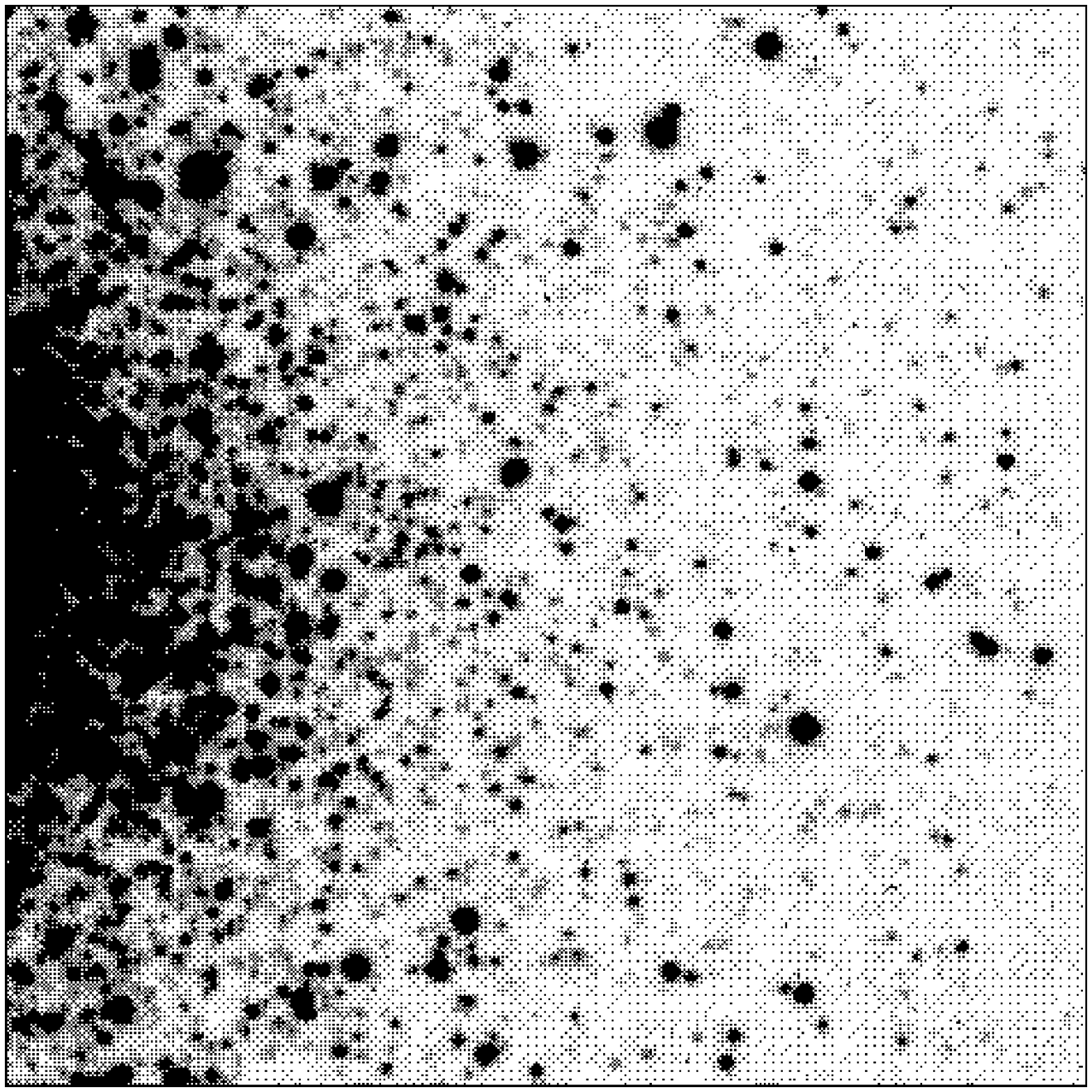,width=4cm}} \\ 
&
\fbox{\psfig{figure=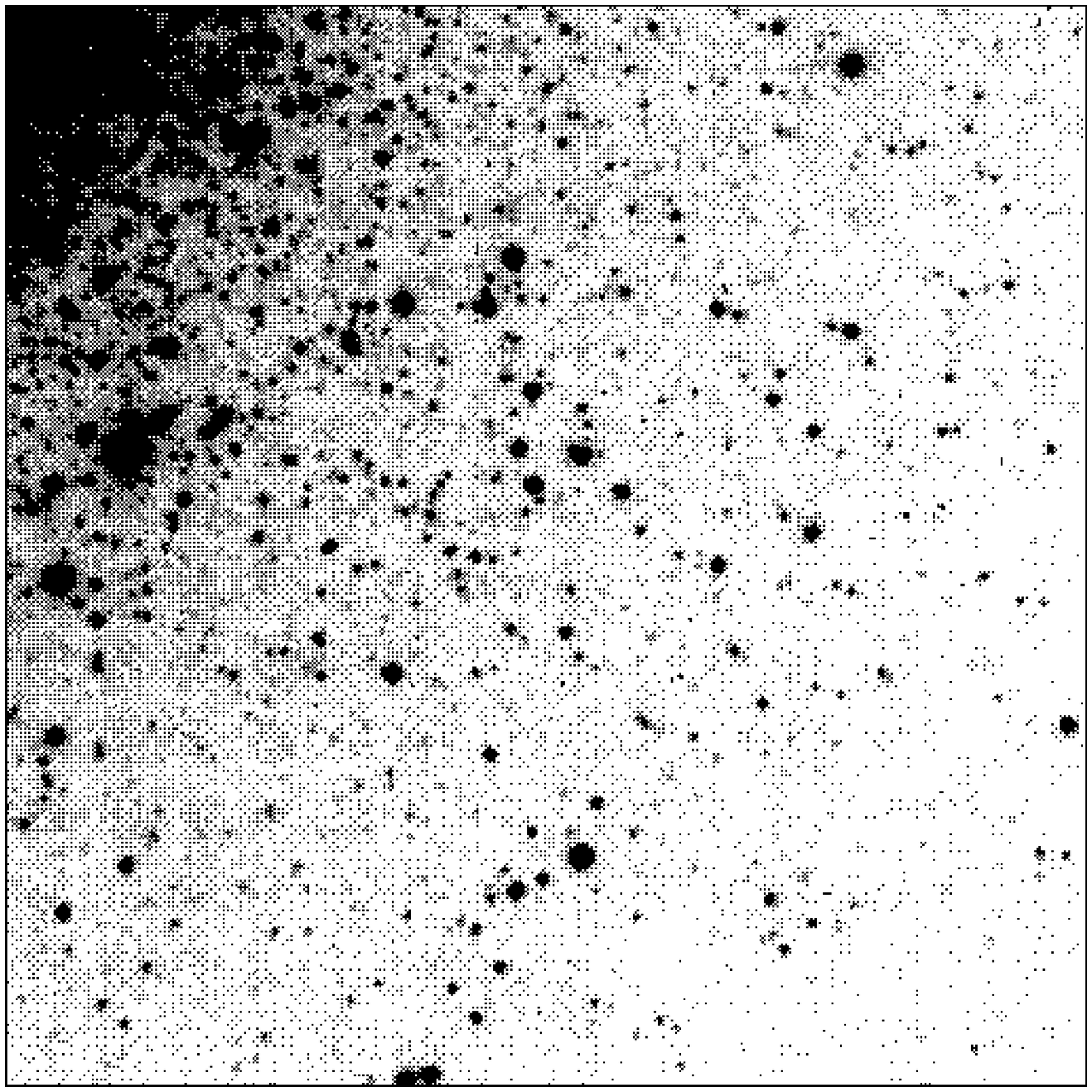,width=4cm}}
\end{tabular}
\end{minipage}
\end{tabular}
\caption[]{CMD and covered fields for NGC~1904 (M~79)}
\label{ngc1904}
\end{figure*}

\paragraph {\bf NGC 1904 (M~79).} (Fig.~\ref{ngc1904})

M~79 is the farthest cluster ($R_{\rm GC}=18.5$ kpc) from the Galactic
center in our sample. The main feature in the CMD of
Fig.~\ref{ngc1904} is the EBHB. Previous CMDs from CCD photometry are
in Heasley et al. (\cite{heasley83}) ($U,B,V$ bands), Gratton \&
Ortolani (\cite{grattonortolani86}) ($B,V$ bands), Ferraro et
al. (\cite{ferraro93a}) ($B,V$ bands), Alcaino et
al. (\cite{alcaino94}) ($B,V,R,I$ bands), and Kravtsov et
al. (\cite{kravstov97}) ($U,B,V$ bands),
and the $B$,$V$ photometry from HST in Sosin et al. (\cite{sosin97a}). 

\begin{figure*}
\begin{tabular}{c@{}c}
\raisebox{-6cm}{
\psfig{figure=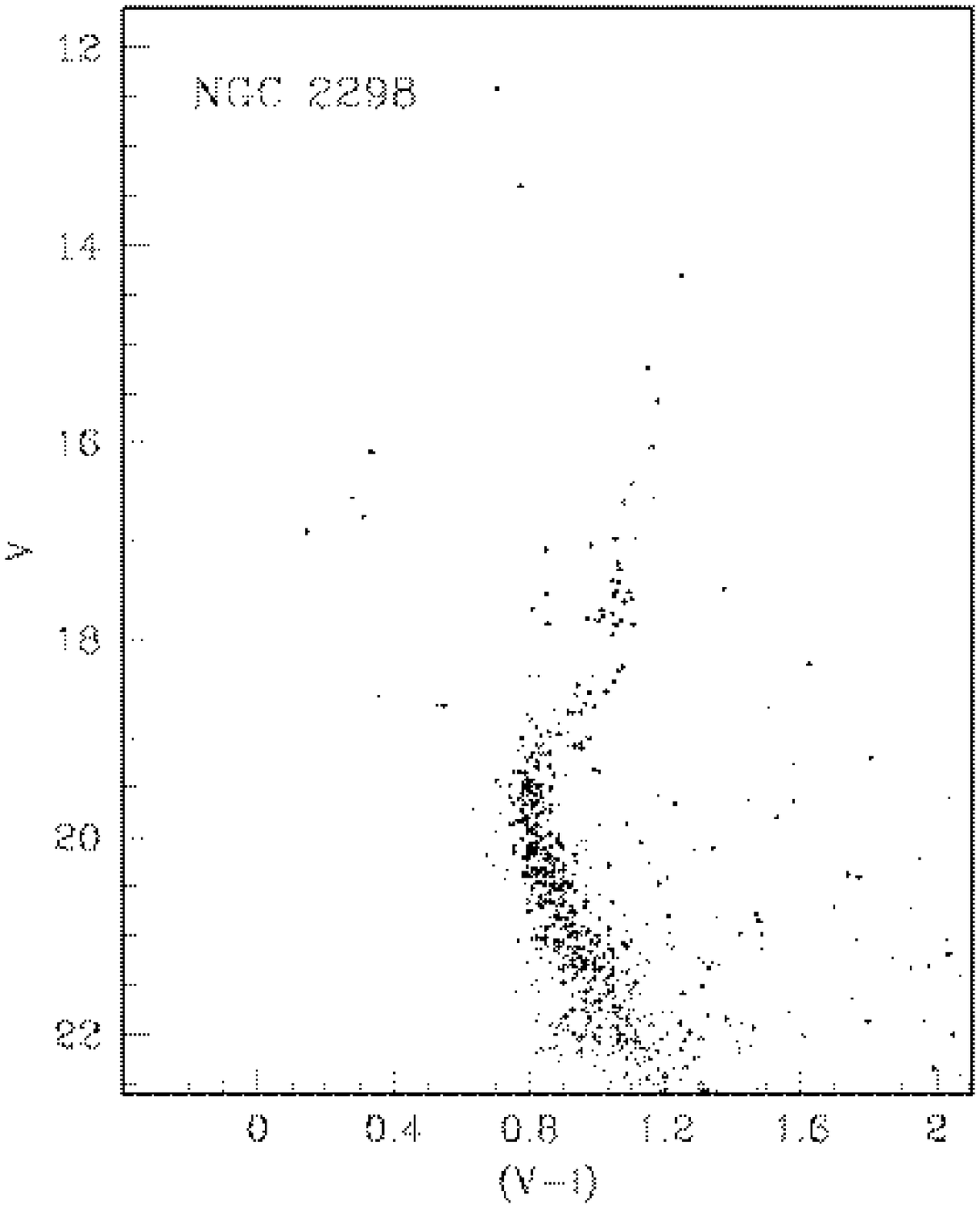,width=8.8cm}
} &
\begin{minipage}[t]{8.8cm}
\begin{tabular}{c@{}c}
\fbox{\psfig{figure=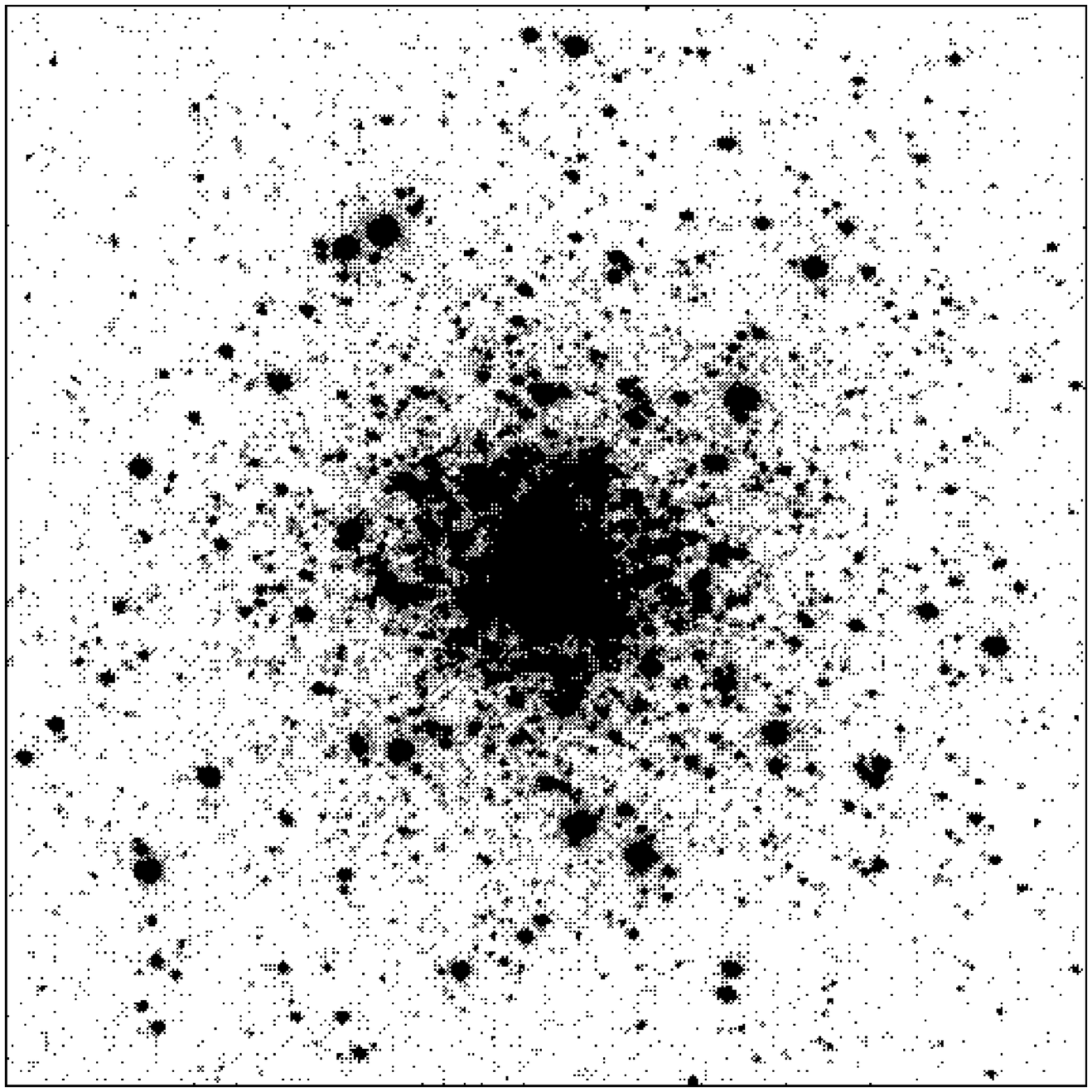,width=4cm}} &
\fbox{\psfig{figure=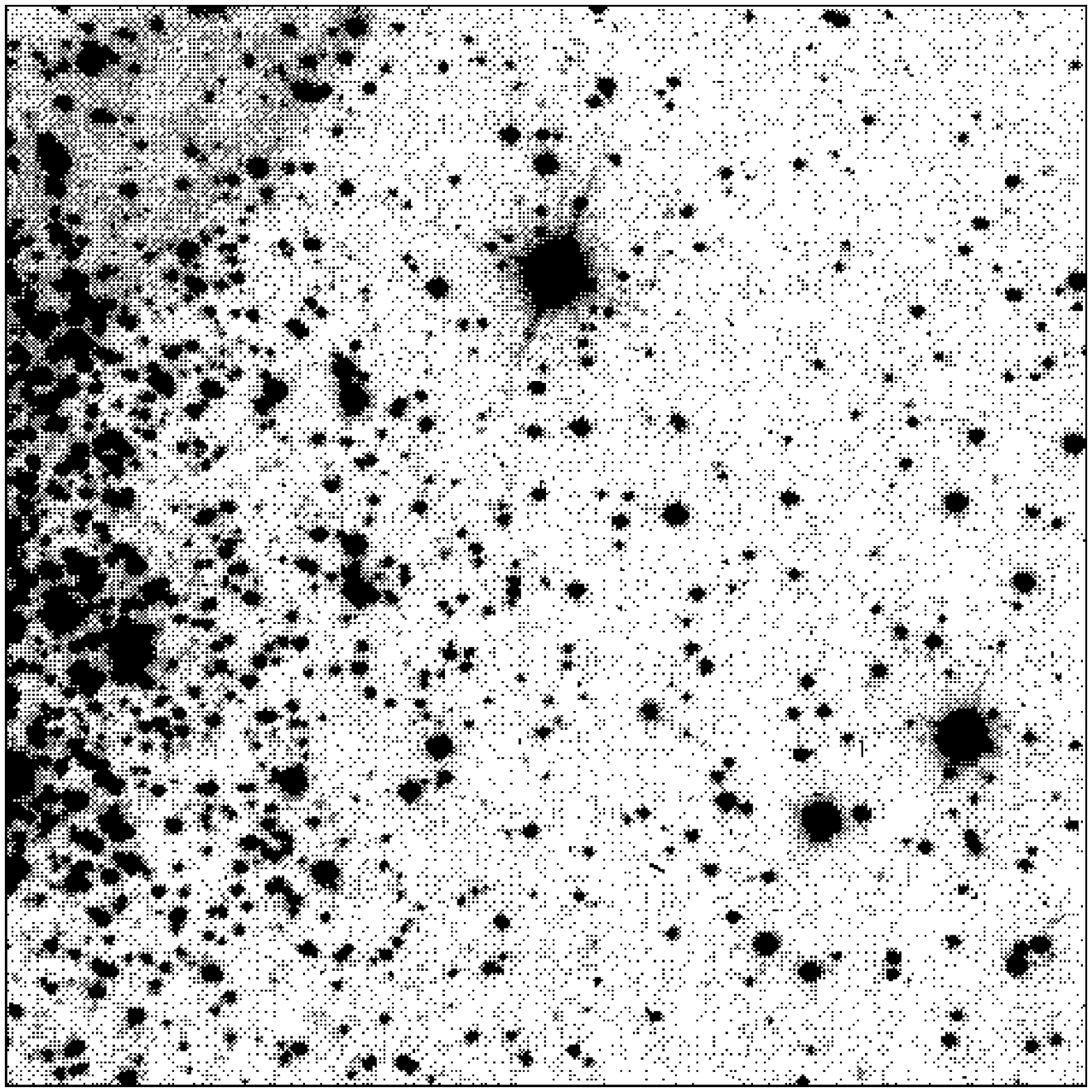,width=4cm}} 
\end{tabular}
\end{minipage}
\end{tabular}
\caption[]{CMD and covered fields for NGC~2298}
\label{ngc2298}

\vspace {-1.5cm}

\begin{tabular}{c@{}c}
\raisebox{-6.5cm}{
\psfig{figure=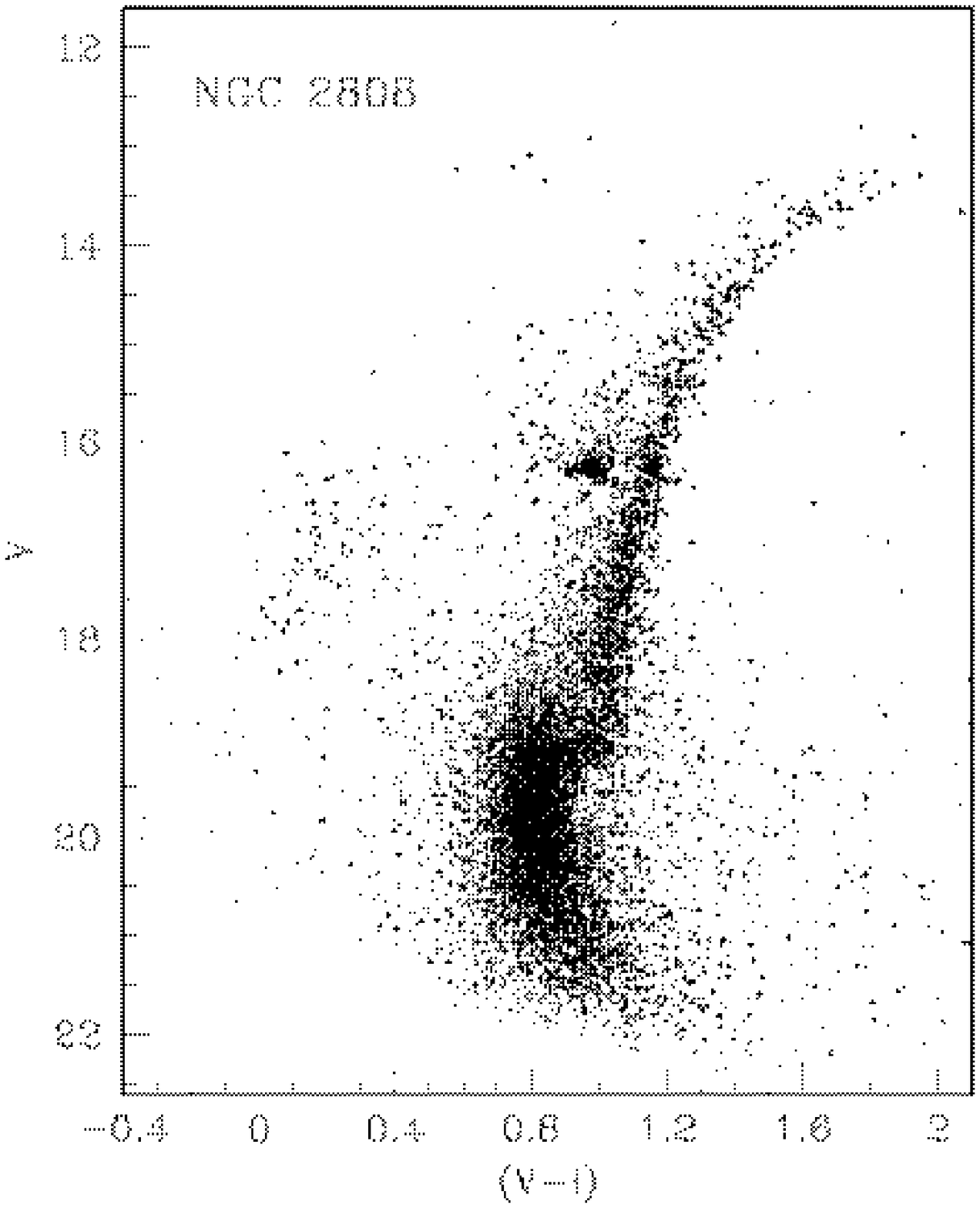,width=8.8cm}
} &
\begin{minipage}[t]{8.8cm}
\begin{tabular}{c@{}c}
&
\fbox{\psfig{figure=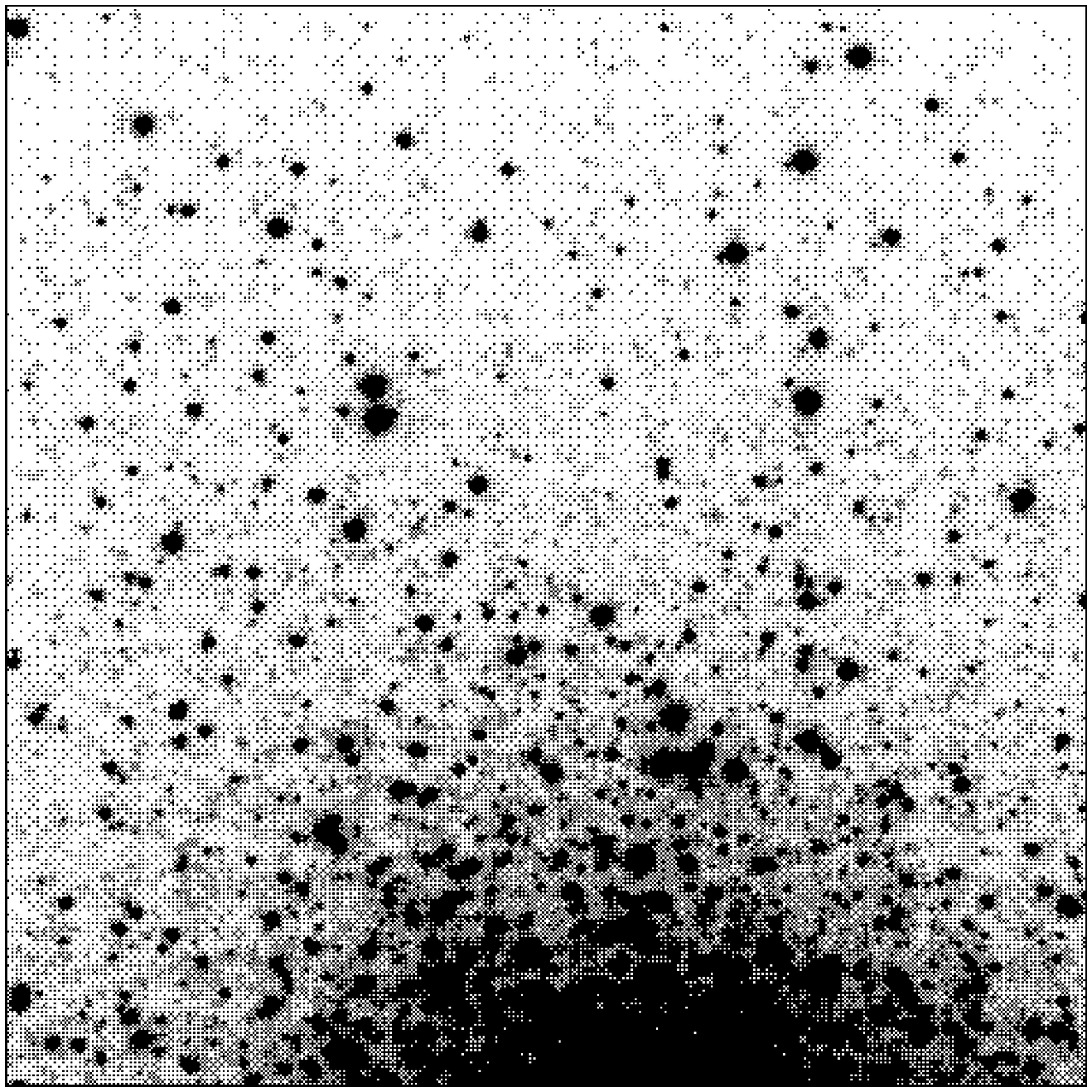,width=4cm}} \\
\fbox{\psfig{figure=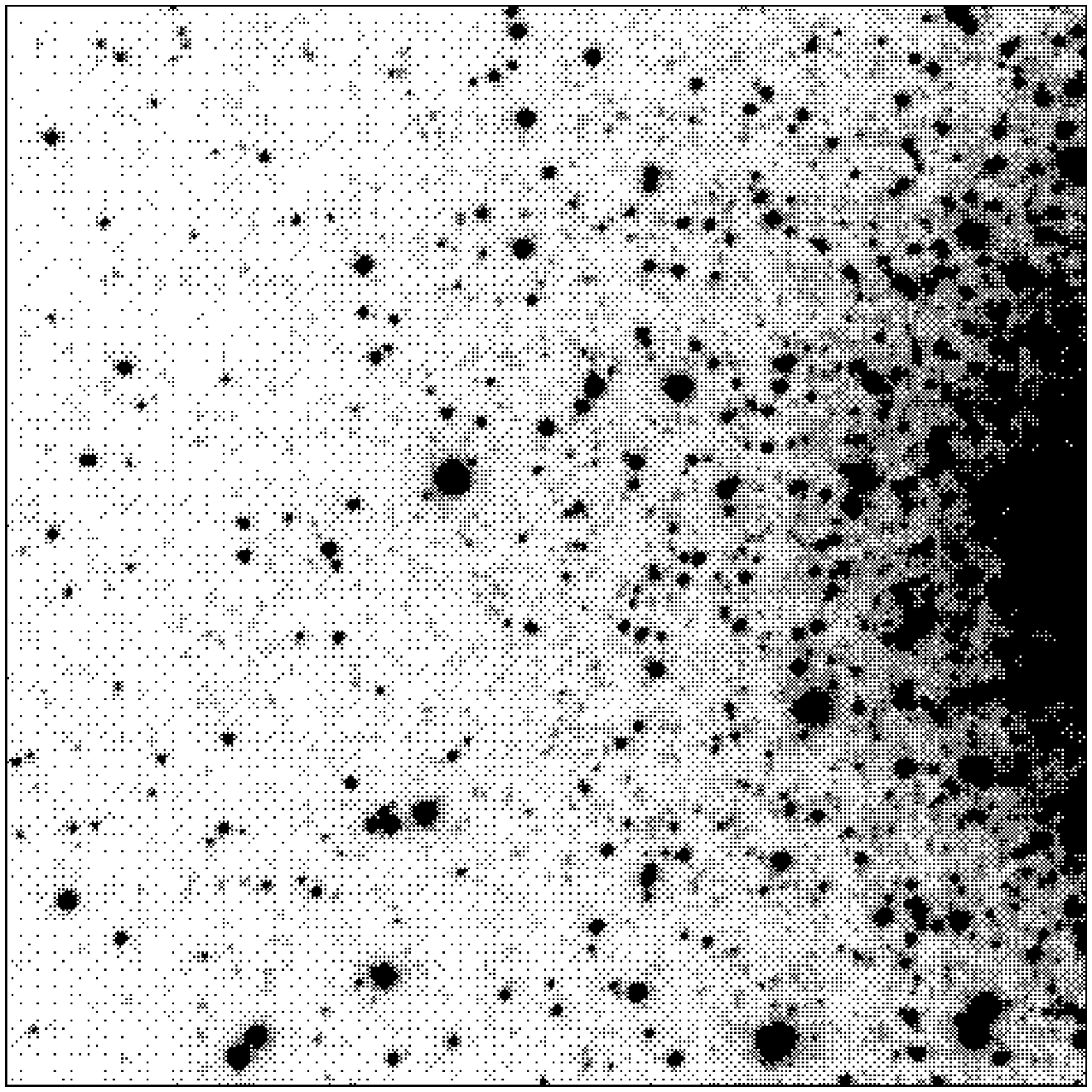,width=4cm}} &
\fbox{\psfig{figure=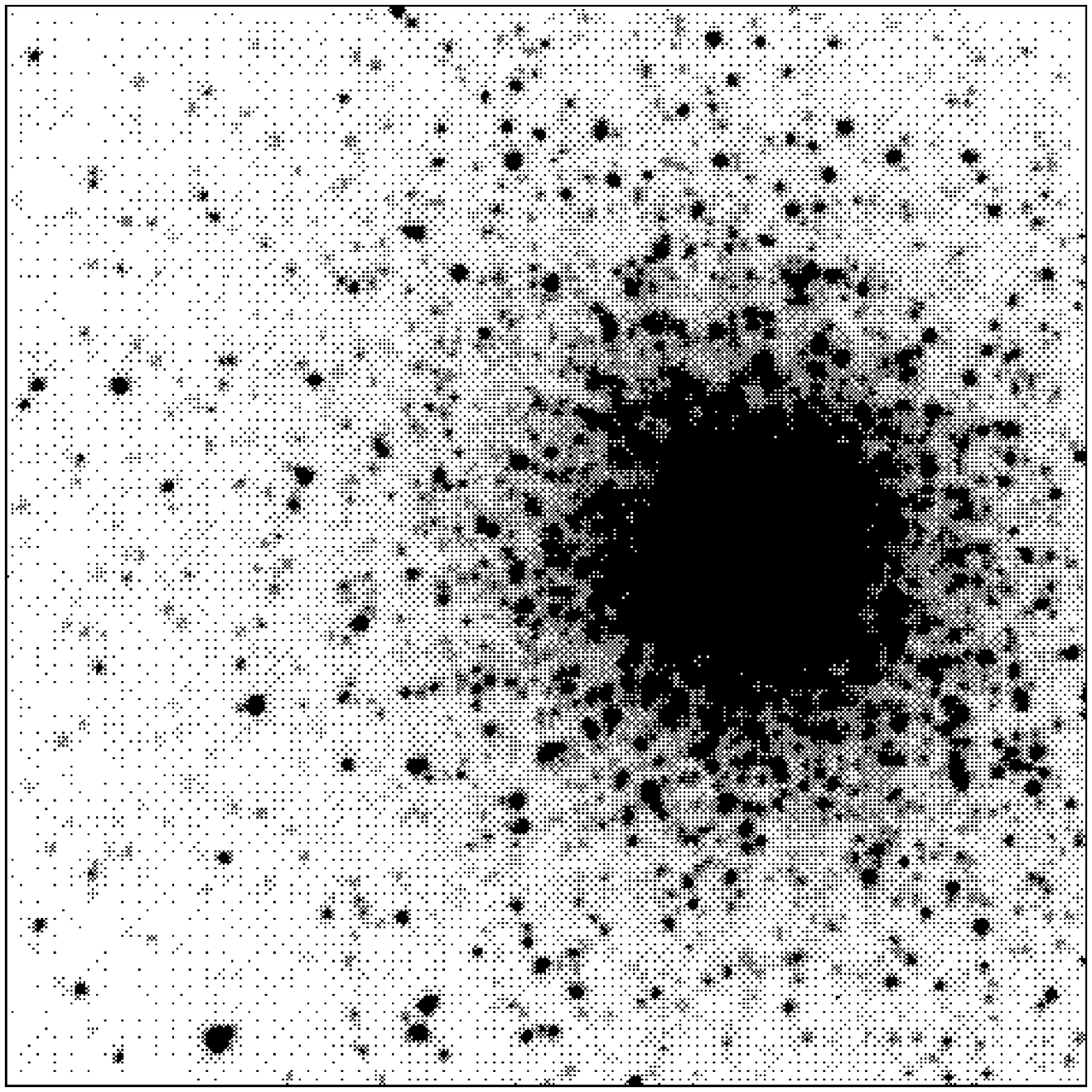,width=4cm}} \\
&
\fbox{\psfig{figure=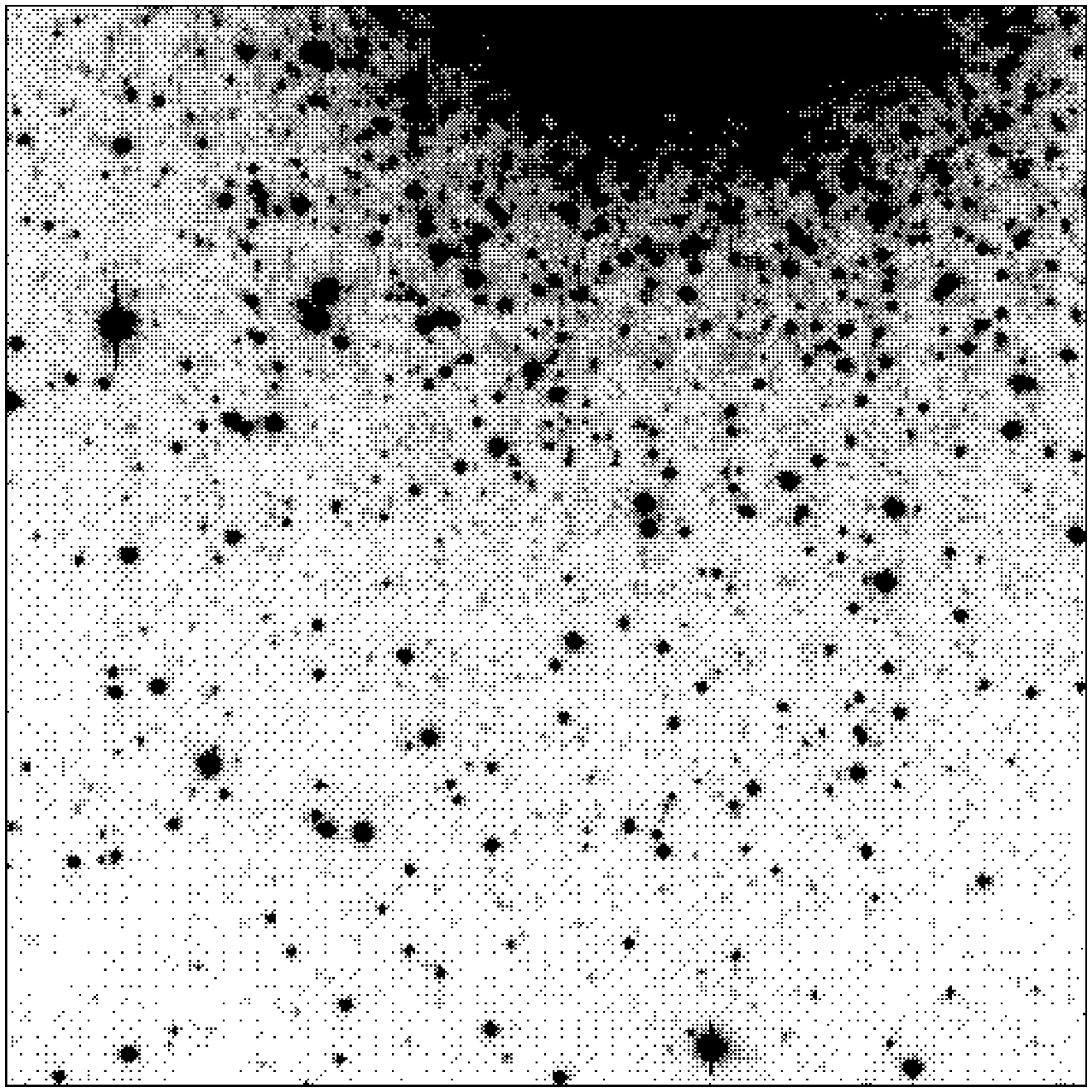,width=4cm}}
\end{tabular}
\end{minipage}
\end{tabular}
\caption[]{CMD and covered fields for NGC~2808}
\label{ngc2808}
\end{figure*}

\paragraph {\bf NGC 2298.} (Fig.~\ref{ngc2298})

This cluster is poorly sampled, particularly for the bright part of
the diagram (due to problems with a short exposure). Only four BHB
stars are present in the HB region. Recent photometric works on this
object are in Gratton \& Ortolani (\cite{grattonortolani86}) ($B,V$
bands), Alcaino \& Liller (\cite{alcainoliller86a}) ($B,V,R,I$ bands),
Janes \& Heasley (\cite{janesheasley88}) ($U,B,V$ bands), and Alcaino
et al. (\cite{alcaino90b}) ($B,V,R,I$ bands).

\paragraph {\bf NGC 2808.} (Fig.~\ref{ngc2808})

This cluster has  some differential reddening (Walker 1999), as it
can be inferred also from the broadening of the sequences in the CMD
of Fig.~\ref{ngc2808}, and a moderate field contamination. The most
interesting features of the CMD are the bimodal HB and the EBHB tail
with other two gaps, as extensively discussed in Sosin et al.
(\cite{sosin97b}). As previously discussed, NGC~2808 is another bimodal
HB cluster at intermediate metallicity with a younger age (Rosenberg
et al. \cite{rosenberg99}). Apart from the already quoted $B$, and
$V$ band photometry from HST data by Sosin et al. (\cite{sosin97b}),
there are many other CCD photometries: 
Gratton \& Ortolani (\cite{grattonortolani86}) ($B,V$ bands), 
Buonanno et al. (\cite{buonanno89}) ($B,V$ bands),
Ferraro et al. (\cite{ferraro90}) ($B,V$ bands), 
Alcaino et al. (\cite{alcaino90c}) ($B,V,R,I$ bands),
Byun \& Lee (\cite{byunlee93}),
Ferraro et al. (\cite{ferraro97}) ($V,I$ bands), and
more recently Walker (\cite{walker99}) ($B,V$ bands).

\begin{figure*}
\begin{tabular}{c@{}c}
\raisebox{-6cm}{
\psfig{figure=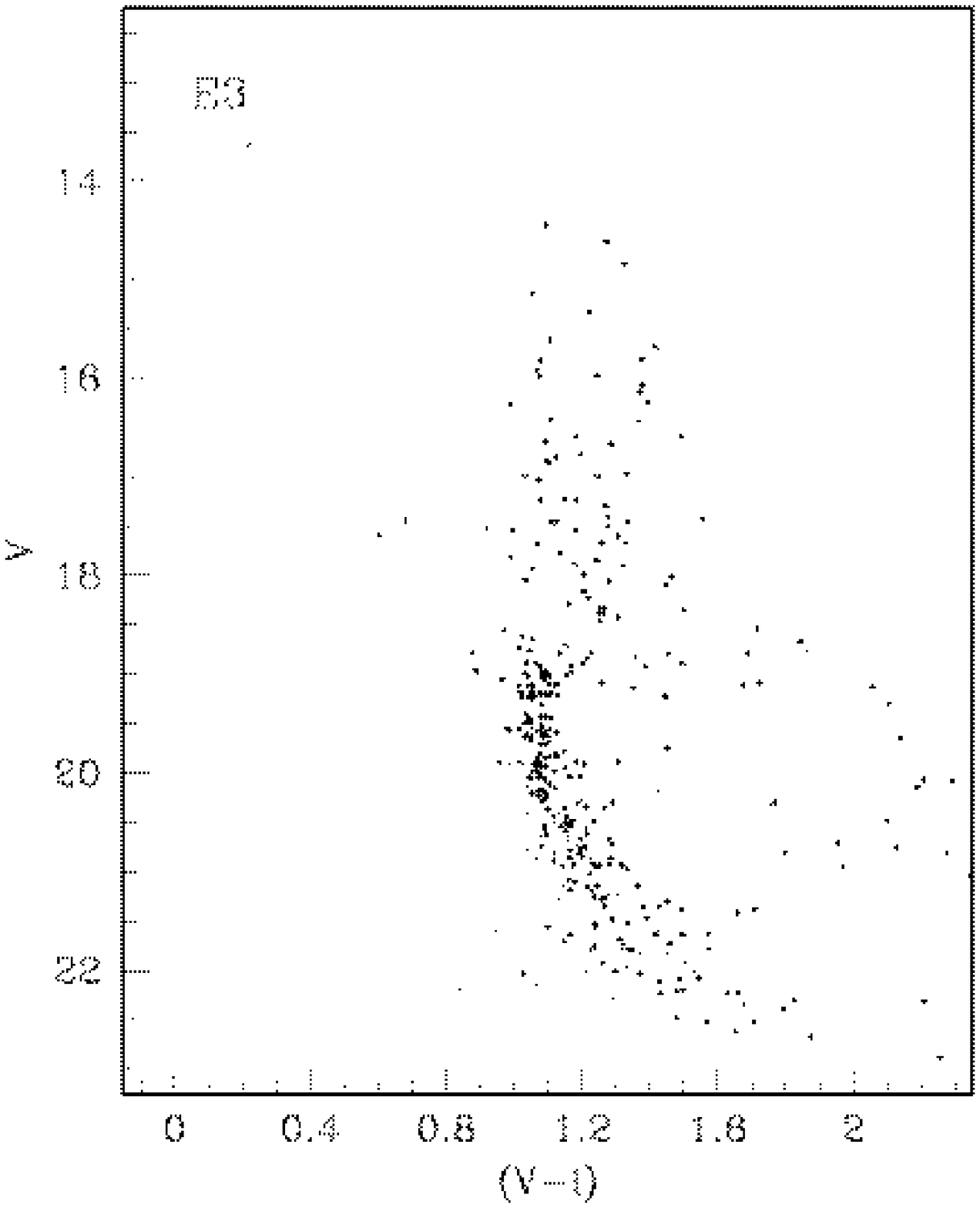,width=8.8cm}
} &
\begin{minipage}[t]{8.8cm}
\begin{tabular}{c@{}c}
\fbox{\psfig{figure=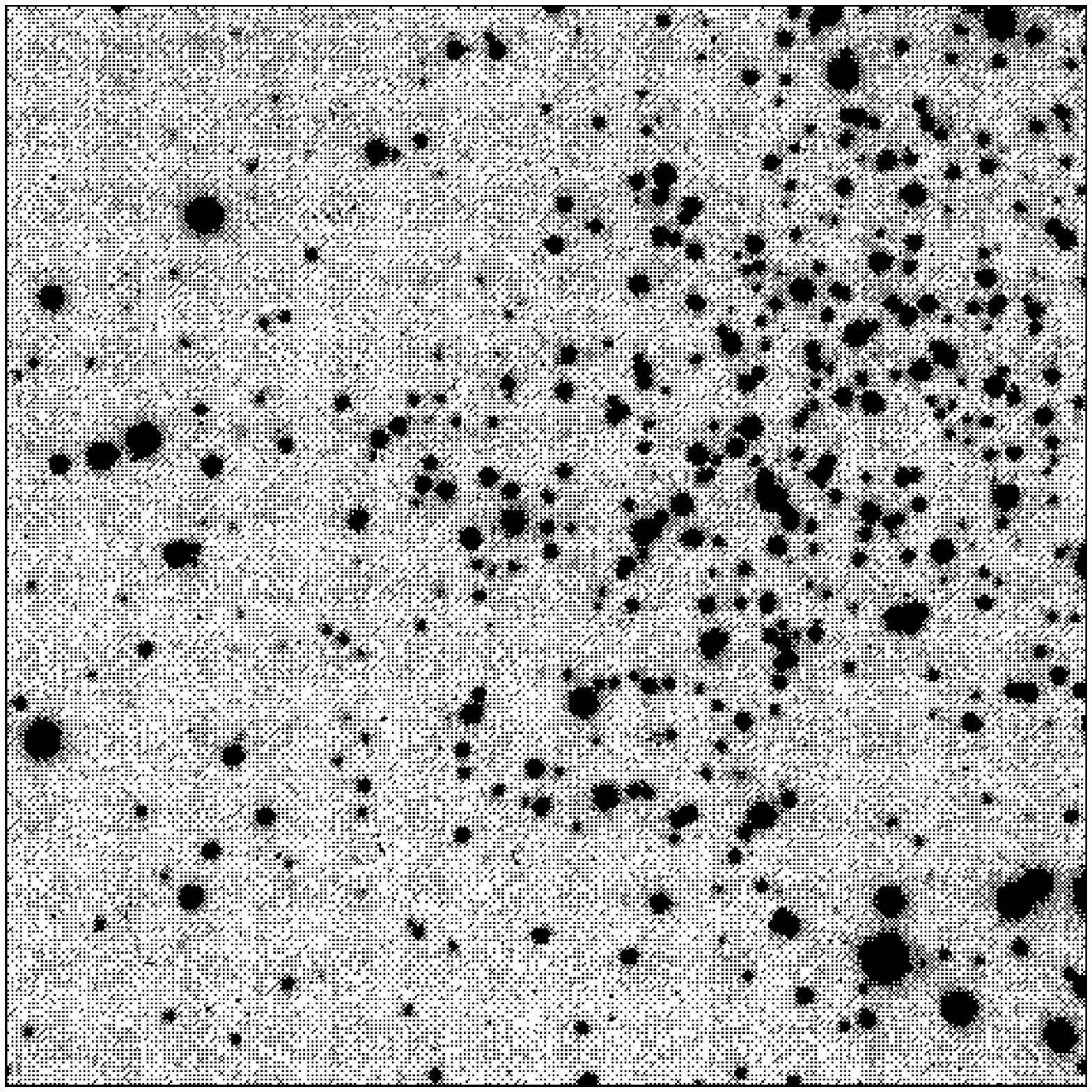,width=4cm}} &
\fbox{\psfig{figure=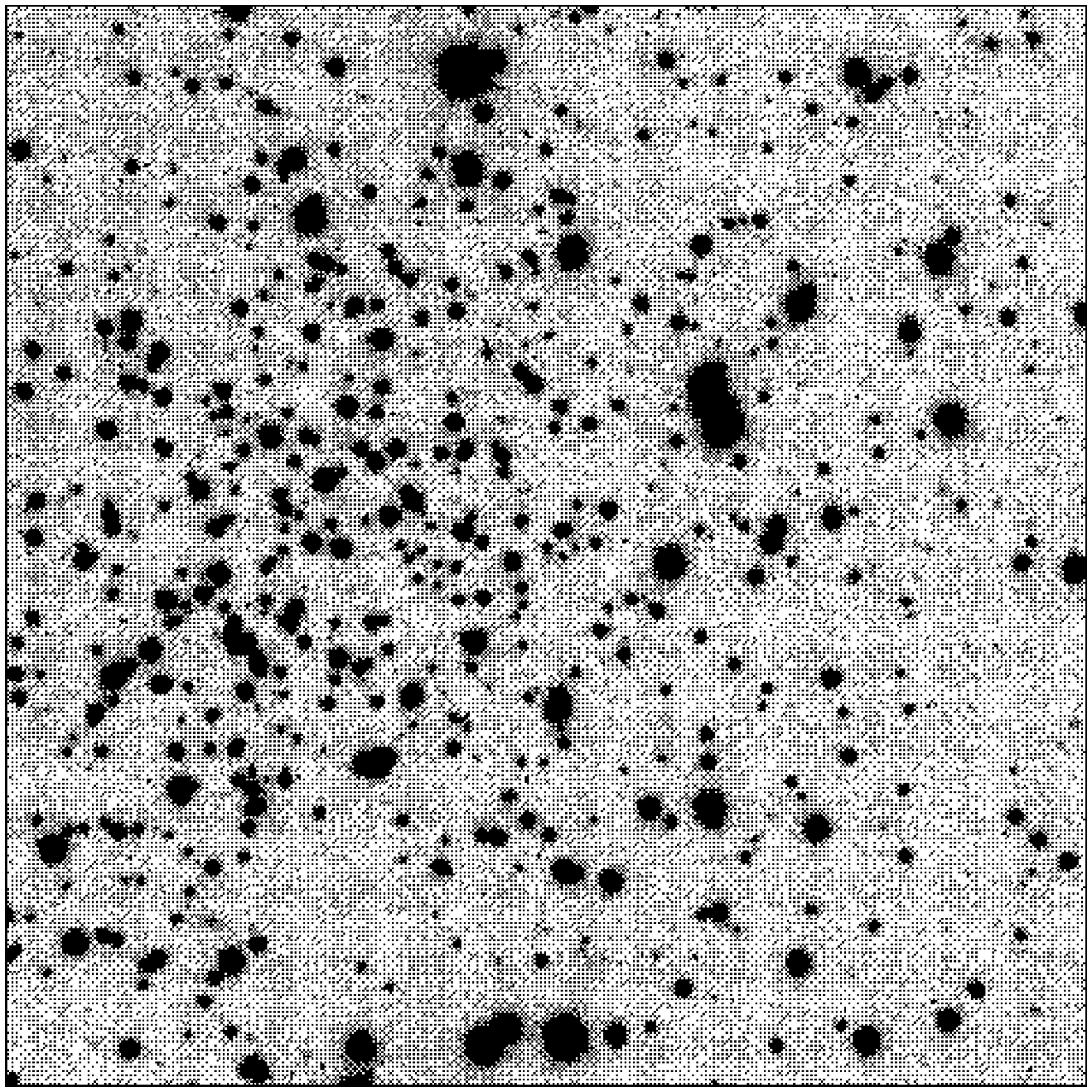,width=4cm}} 
\end{tabular}
\end{minipage}
\end{tabular}
\caption[]{CMD and covered fields for E3}
\label{e3}
\end{figure*}

\paragraph {\bf E3.} (Fig.~\ref{e3})

This cluster is one of the less populated clusters in our Galaxy,
resembling some Palomar-like globular as Pal~1 (Rosenberg et
al. \cite{rosenberg98}). As in Pal~1, there are no HB stars in the CMD, and
the entire population of observed stars is smaller than 1000 objects. E3 is
suspected to have a metallicity close to that of Pal~1.  From the $\delta
(V-I)_{\rm @2.5}$ (Paper III) measured on Fig.~\ref{e3}, E3 is coeval with
the other GGCs of similar metallicity, though the result is necessarily
very uncertain, due to the high contamination and the small number of RGB
stars. E3 is the cluster with the better defined MS binary sequence
(Veronesi et al. \cite{veronesi96}), which can be also seen in
Fig.~\ref{e3}. Previous CCD CMDs are in McClure et al. (\cite{mcclure85})
($B,V$ bands), Gratton \& Ortolani (\cite{grattonortolani87}) ($B,V$
bands), and Veronesi et al. (\cite{veronesi96}) ($B,V,R,I$ bands).

\begin{figure*}
\begin{tabular}{c@{}c}
\raisebox{-6cm}{
\psfig{figure=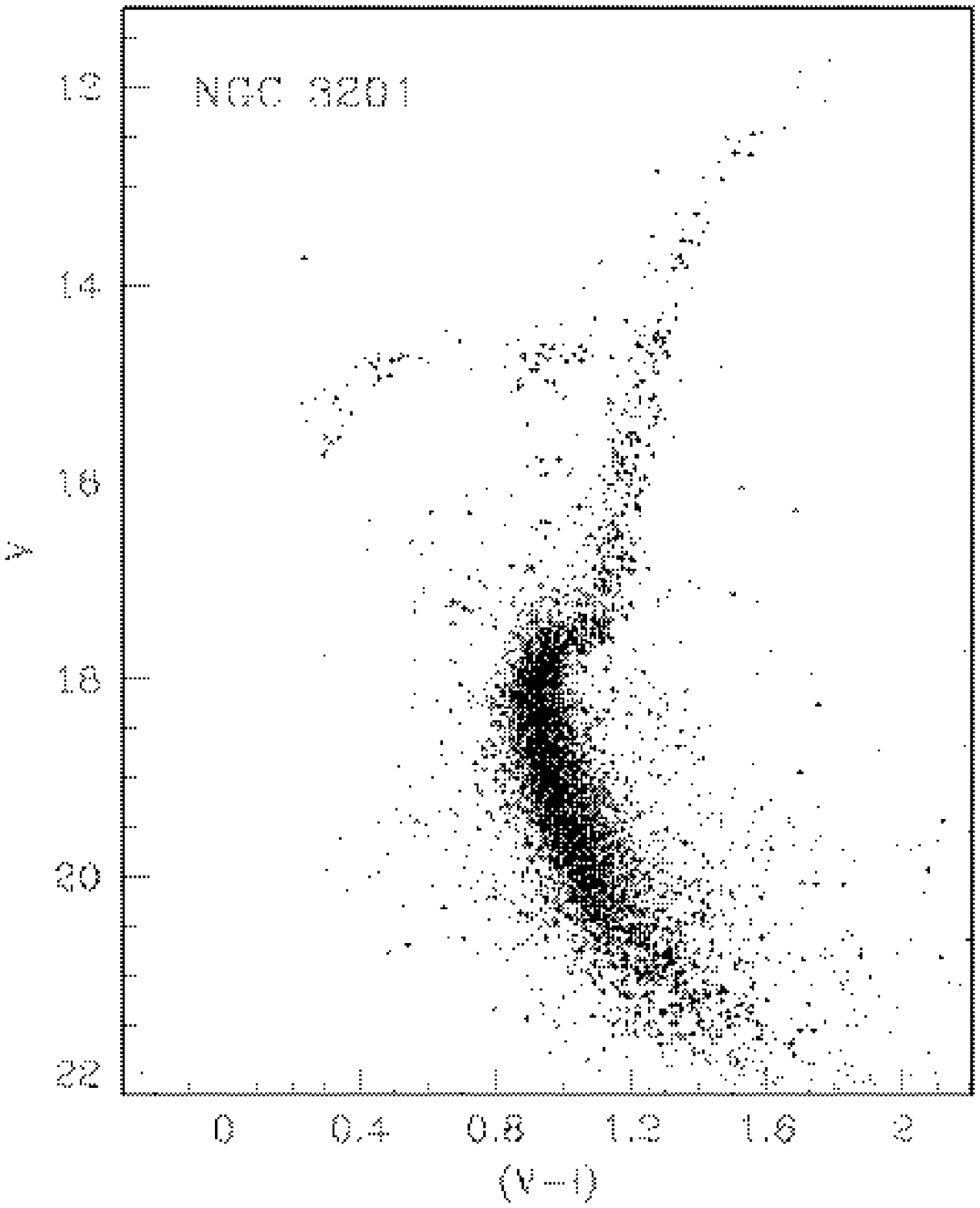,width=8.8cm}
} &
\begin{minipage}[t]{8.8cm}
\begin{tabular}{c@{}c}
\fbox{\psfig{figure=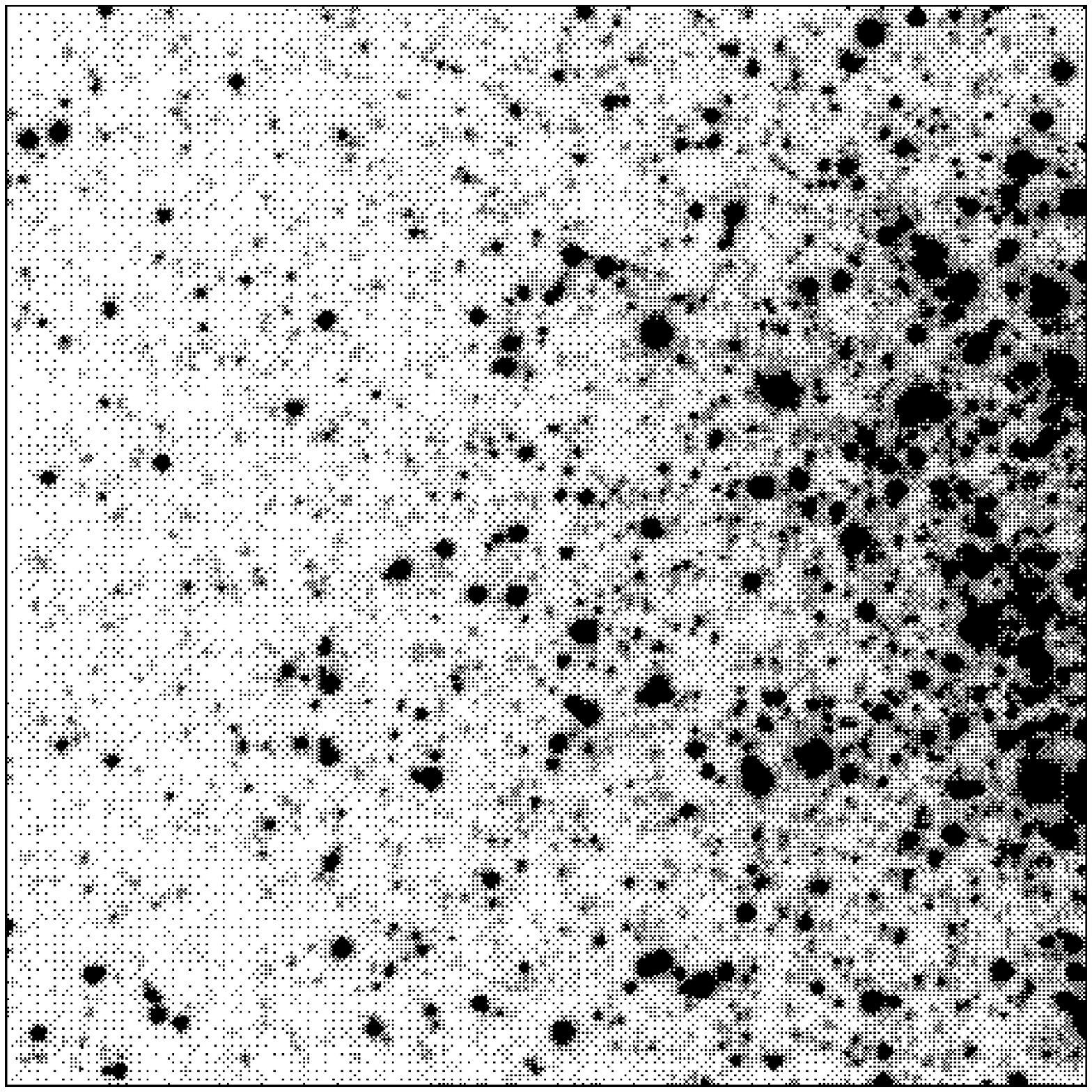,width=4cm}} &
\fbox{\psfig{figure=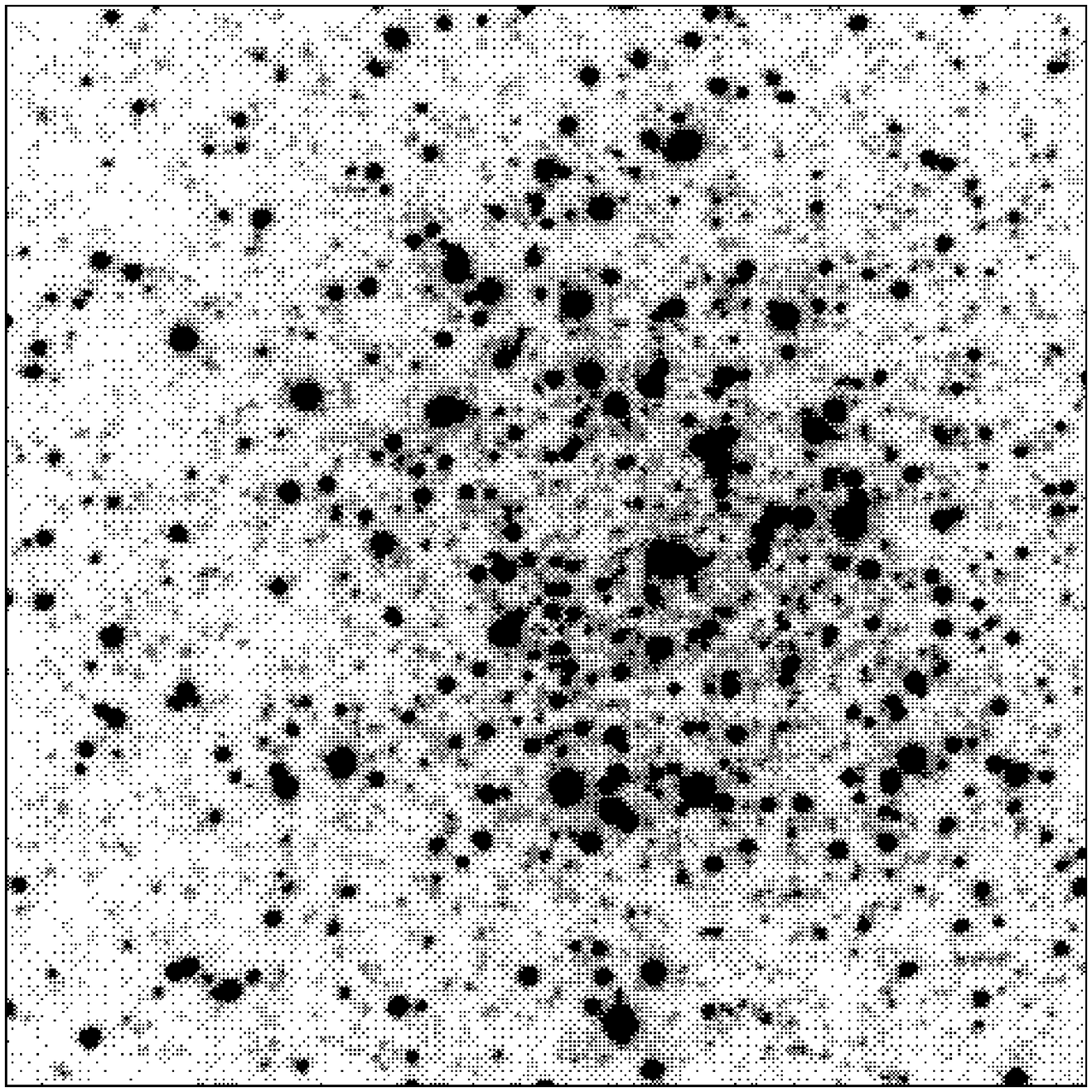,width=4cm}} \\
&
\fbox{\psfig{figure=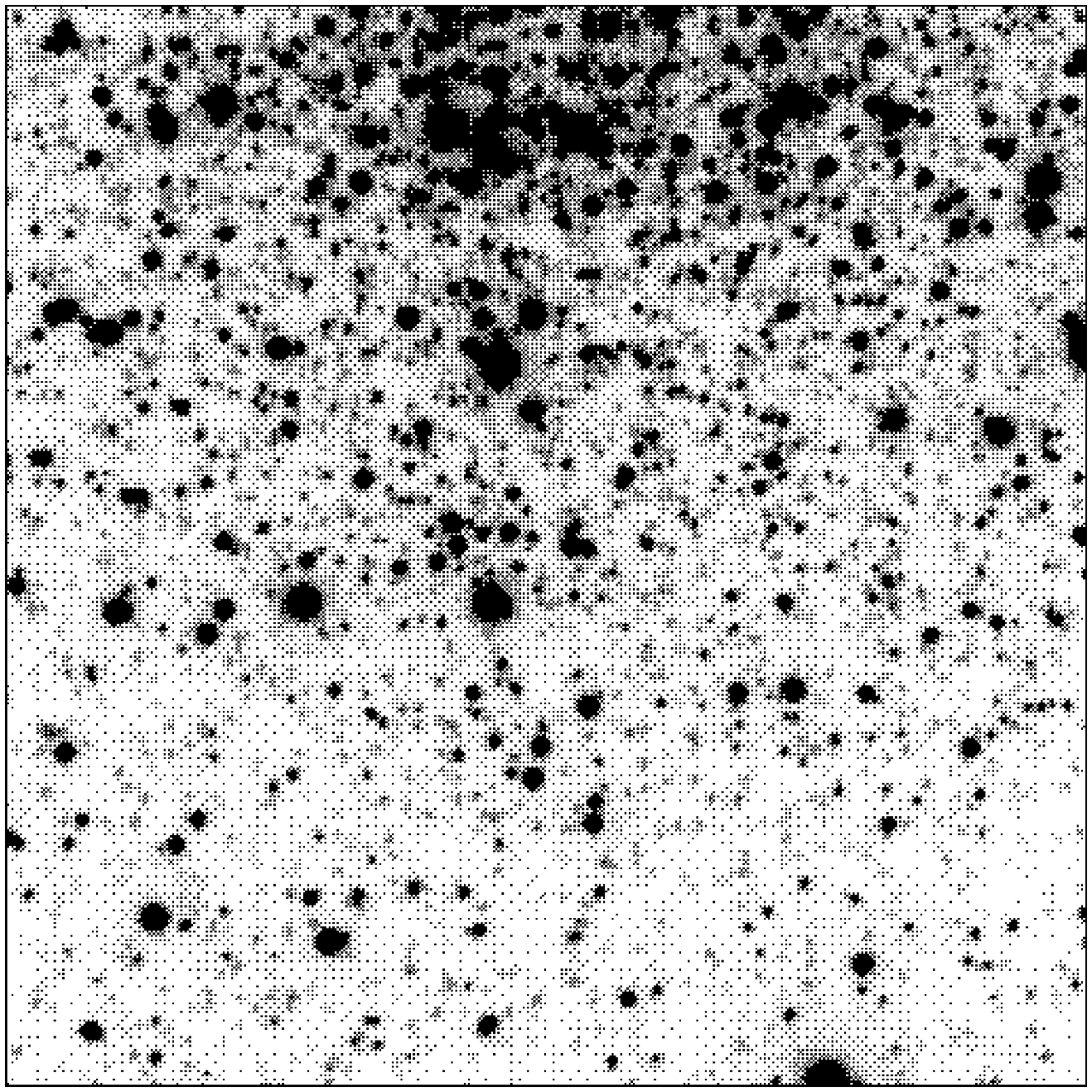,width=4cm}}
\end{tabular}
\end{minipage}
\end{tabular}
\caption[]{CMD and covered fields for NGC~3201}
\label{ngc3201}
\end{figure*}

\paragraph {\bf NGC 3201.}  (Fig.~\ref{ngc3201})

The two lateral fields presented in Fig.~\ref{ngc3201} were observed
in both runs, in order to test the homogeneity of the data and
instrumentation (see Sect.~\ref{comparison}). The HB of NGC~3201 has
a bimodal appearance, though it is not younger than the bulk of GGCs
of the same metallicity group, at variance with the previously
discussed cases. It has a small differential reddening. A blue
straggler (BS) sequence is visible in Fig.~\ref{ngc3201}. Previous
CCD studies of this cluster include Penny (\cite{penny84}) ($B,V,I$
bands), Alcaino et al. (\cite{alcaino89}) ($B,V,R,I$ bands), Brewer et
al. (\cite{brewer93}) ($U,B,V,I$ bands) and Covino \& Ortolani
(\cite{covinoortolani97}) ($B,V$ bands).

\begin{figure*}
\begin{tabular}{c@{}c}
\raisebox{-6cm}{
\psfig{figure=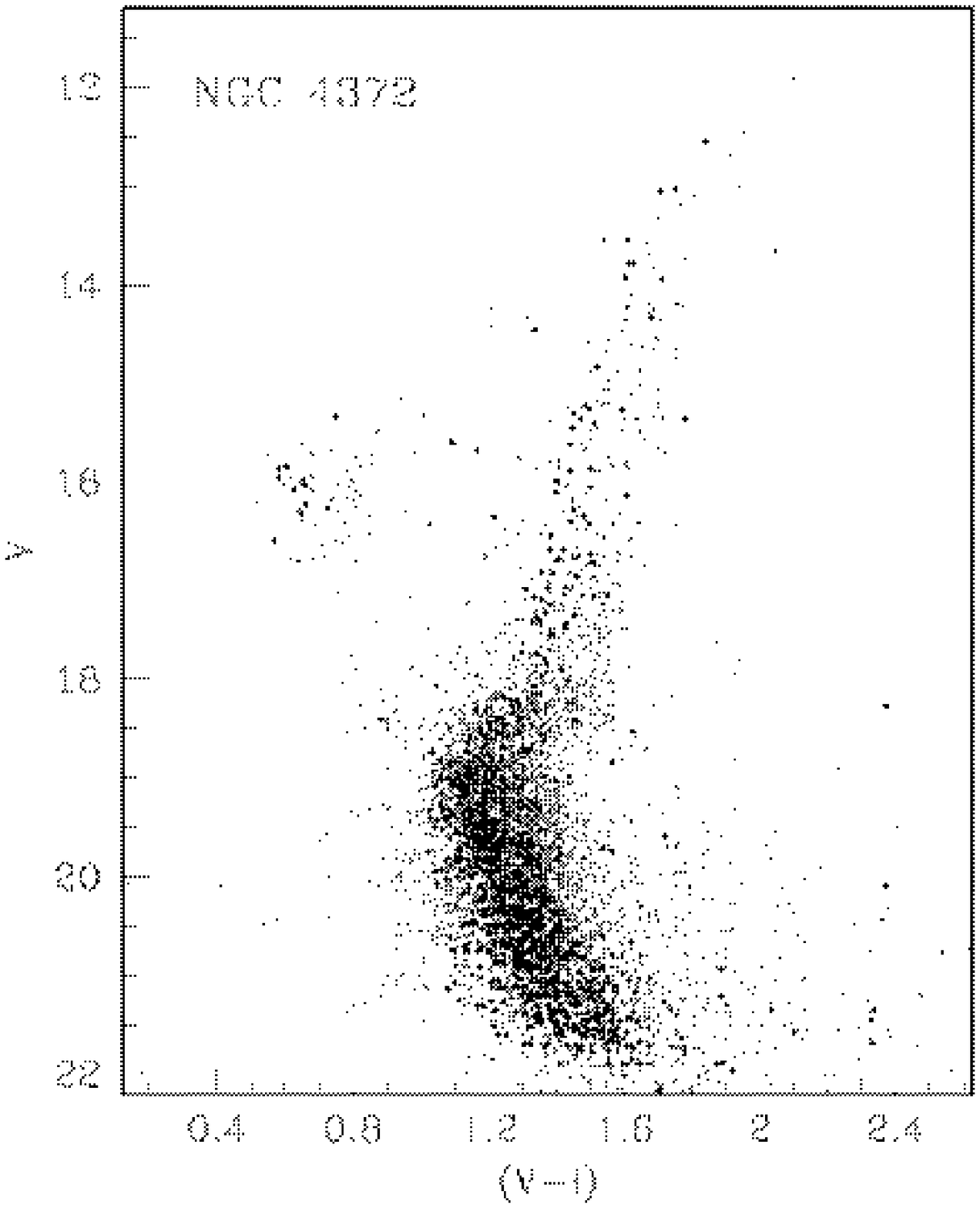,width=8.8cm}
} &
\begin{minipage}[t]{8.8cm}
\begin{tabular}{c@{}c}
&
\fbox{\psfig{figure=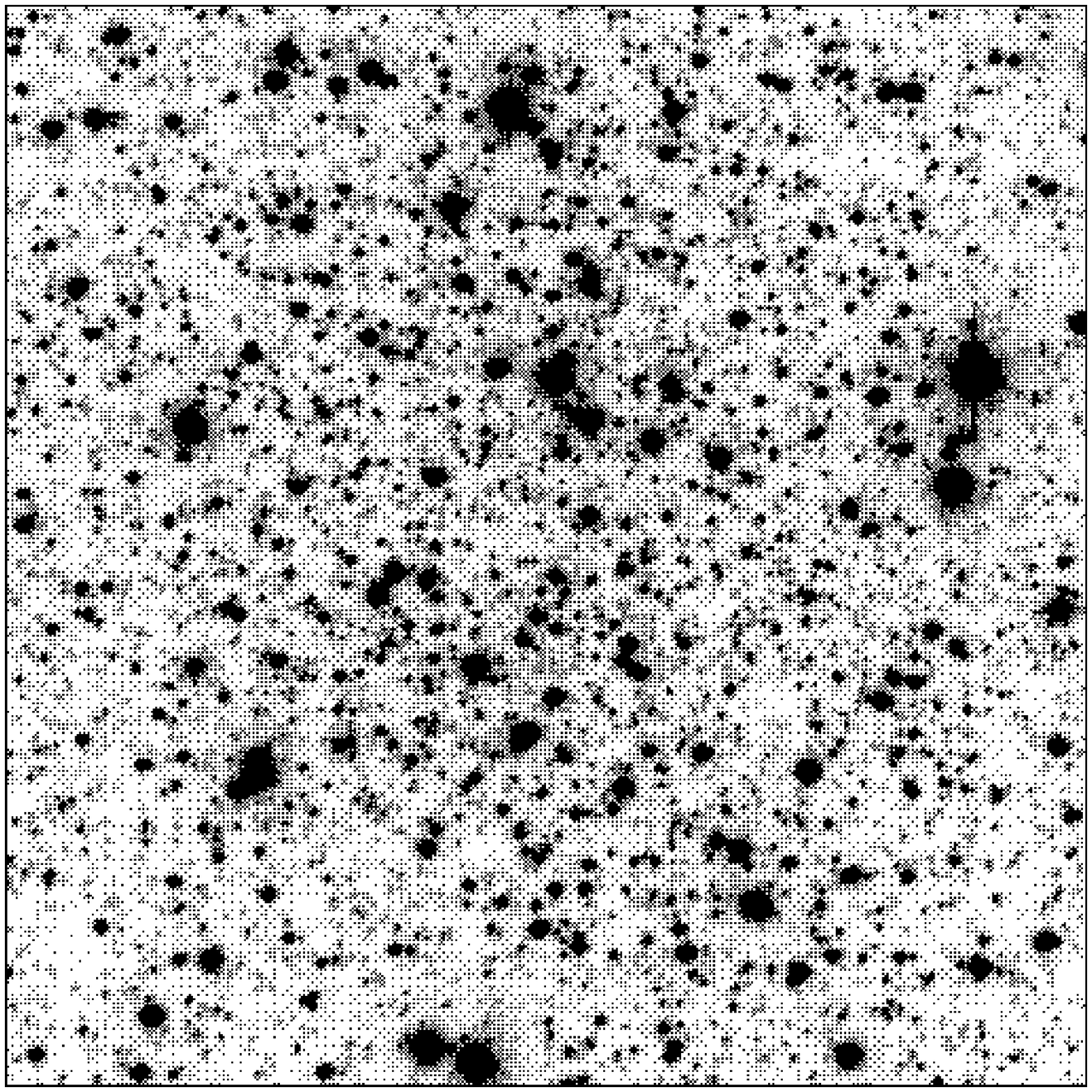,width=4cm}}
\\
\fbox{\psfig{figure=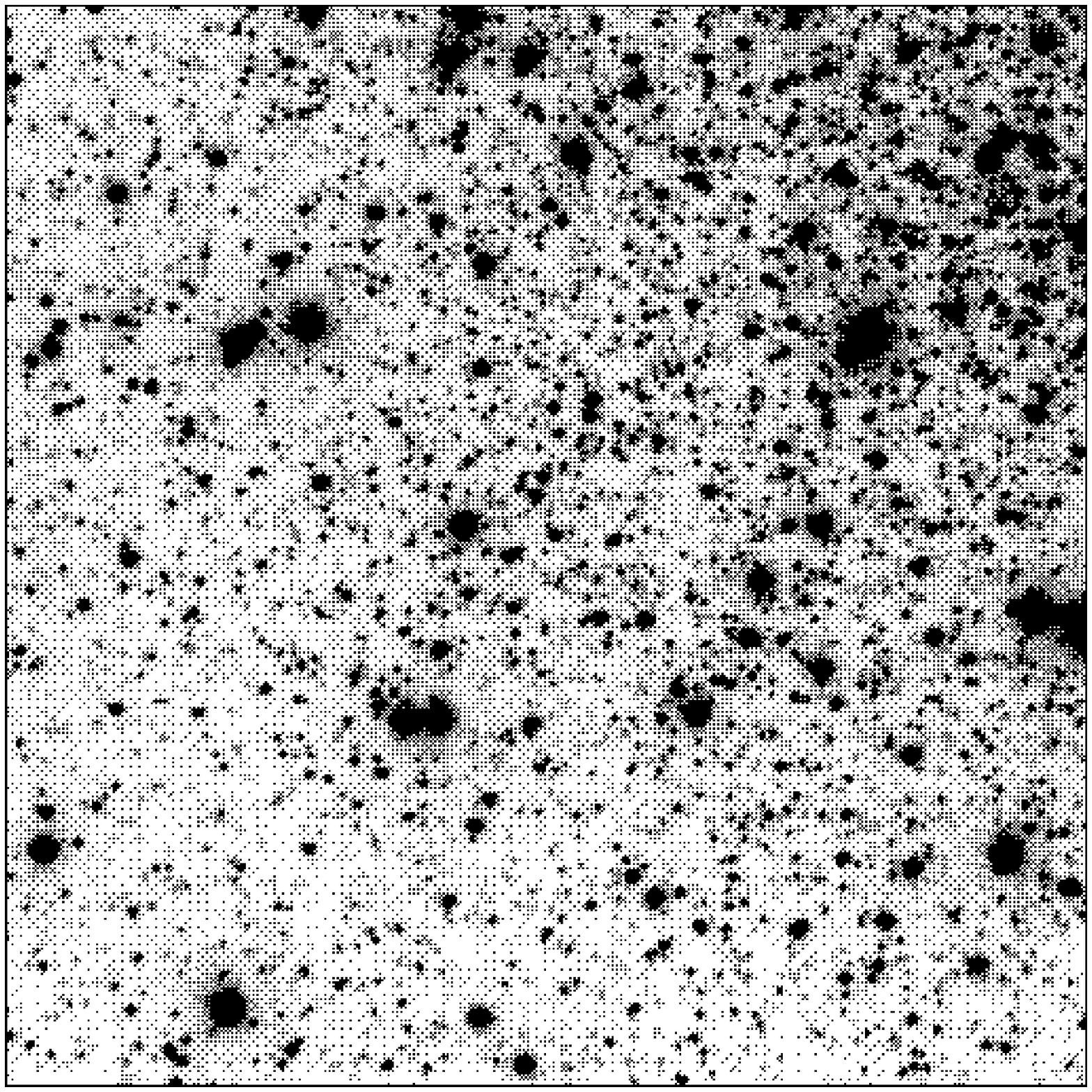,width=4cm}} 
\end{tabular}
\end{minipage}
\end{tabular}
\caption[]{CMD and covered fields for NGC~4372}
\label{ngc4372}
\end{figure*}

\paragraph {\bf NGC 4372.}  (Fig.~\ref{ngc4372})

The principal characteristic of the CMD of this cluster is the
broadening of all the sequences, consequence of the high differential
reddening, probably due to the Coal-sack Nebulae. In the CMD of
Fig.~\ref{ngc4372} the darker dots are from the stars in the lowest
reddening region (south east) of the observed fields. We have computed
the reddening field for this cluster from the shift of the CMDs
obtained in different positions, finding that it is homogeneously
distributed in space and quite easy to correct by a second order
polynomial surface. Two previous CCD photometries can be found in
Alcaino et al. (\cite{alcaino91}) ($B,V,R,I$ bands) and Brocato et
al. (\cite{brocato96}) ($B,V$ bands).

\begin{figure*}
\begin{tabular}{c@{}c}
\raisebox{-6cm}{
\psfig{figure=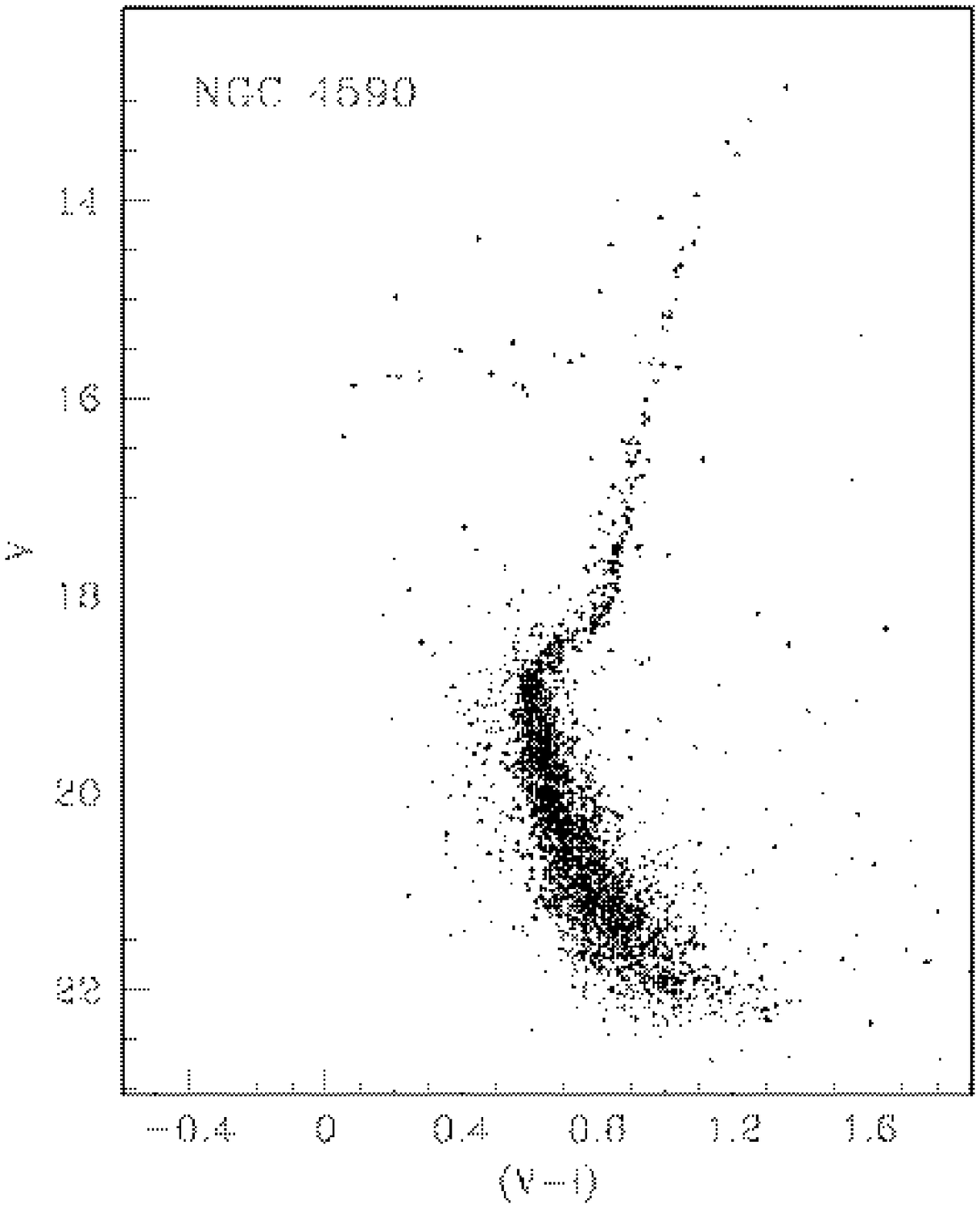,width=8.8cm}
} &
\begin{minipage}[t]{8.8cm}
\begin{tabular}{c@{}c}
\fbox{\psfig{figure=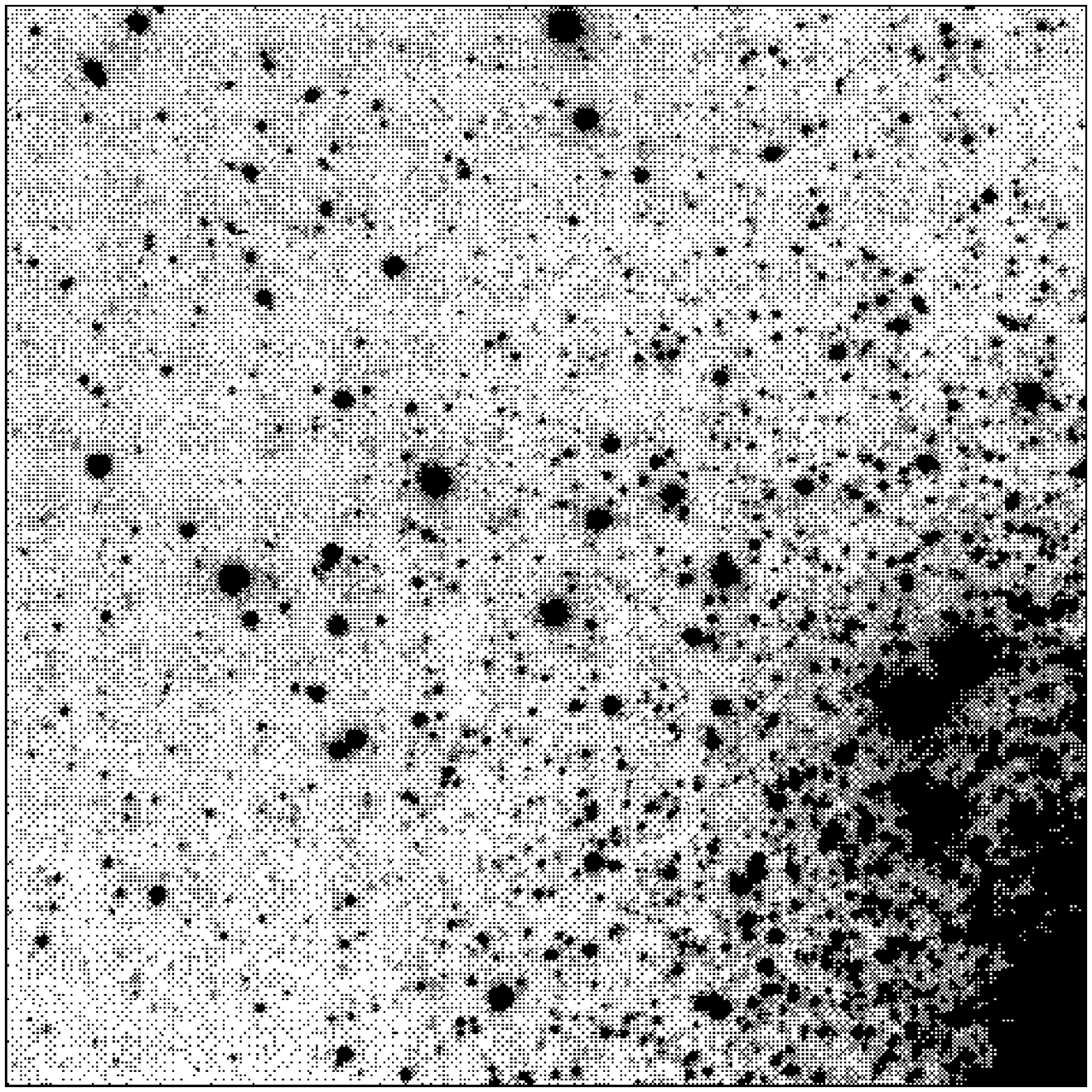,width=4cm}} &
\\
\fbox{\psfig{figure=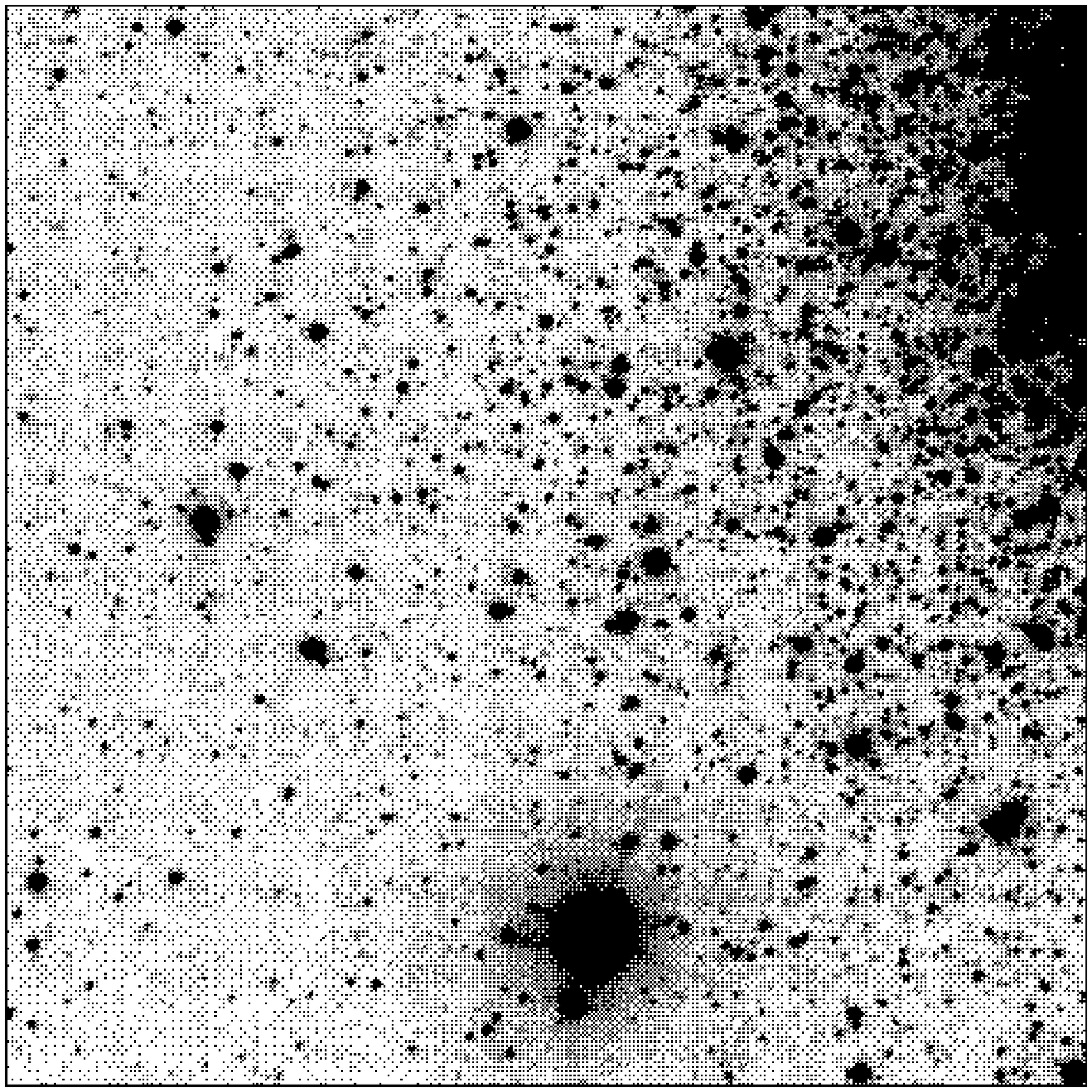,width=4cm}} 
\end{tabular}
\end{minipage}
\end{tabular}
\caption[]{CMD and covered fields for NGC~4590 (M~68)}
\label{ngc4590}
\end{figure*}

\paragraph {\bf NGC 4590 (M~68).} (Fig.~\ref{ngc4590})

This cluster is probably the lowest metallicity cluster of the present
sample. It has a well defined CMD, with an HB populated on both sides
of the instability strip, and including some RR-Lyrae stars. It has
sometimes been classified as one of the oldest GGCs (Salaris et al
\cite{salaris97}), and, in fact, we find that M68 is old, though
coeval with the rest of the metal poor clusters (Paper III). Other
CCD CMDs for this cluster are in McClure et al. (\cite{mcclure87})
($B,V$ bands), Alcaino et al. (\cite{alcaino90d}) ($B,V,R,I$ bands)
and Walker (\cite{walker94}) ($B,V,I$ bands).

\begin{figure*}
\begin{tabular}{c@{}c}
\raisebox{-6cm}{
\psfig{figure=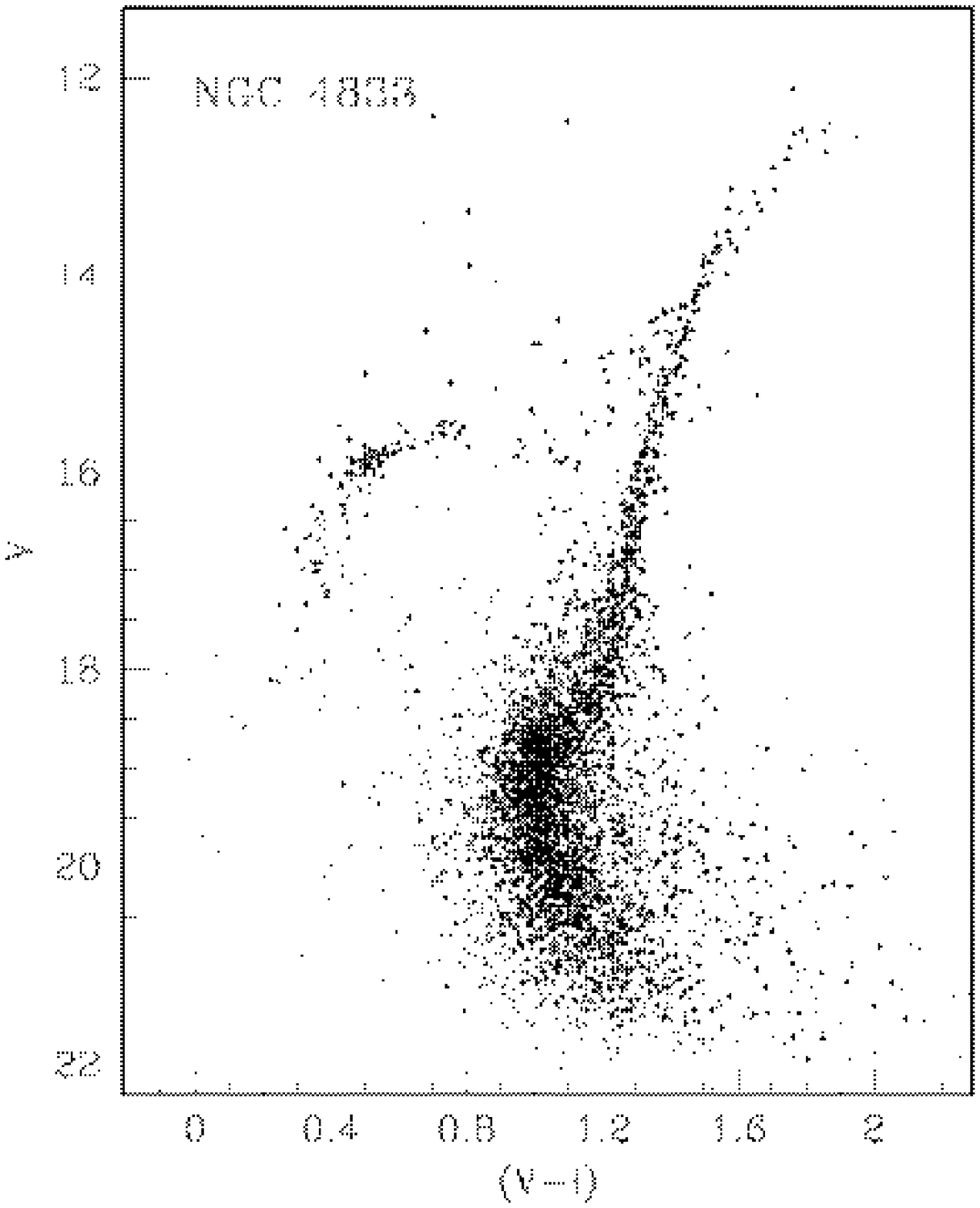,width=8.8cm}
} &
\begin{minipage}[t]{8.8cm}
\begin{tabular}{c@{}c}
\fbox{\psfig{figure=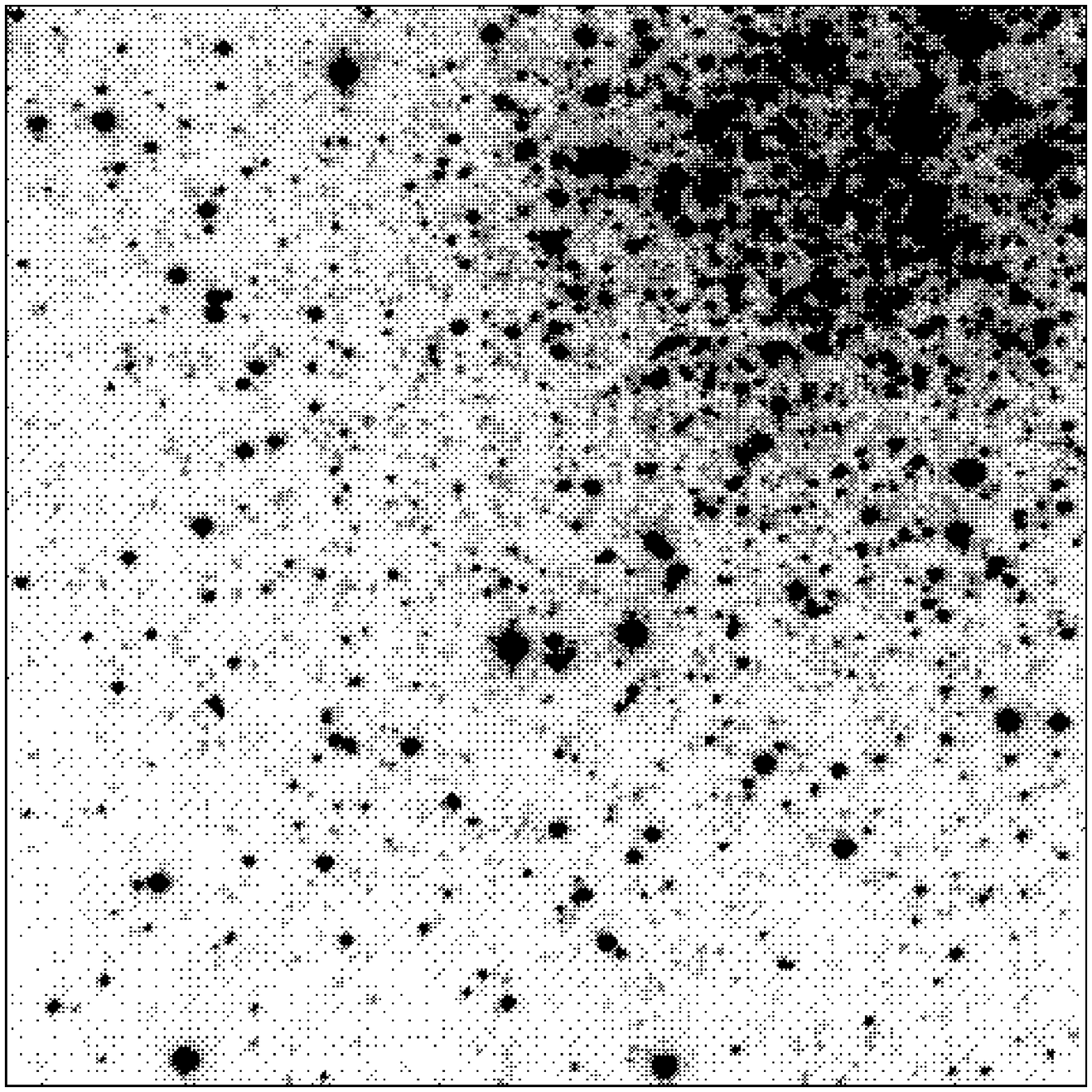,width=4cm}} &
\fbox{\psfig{figure=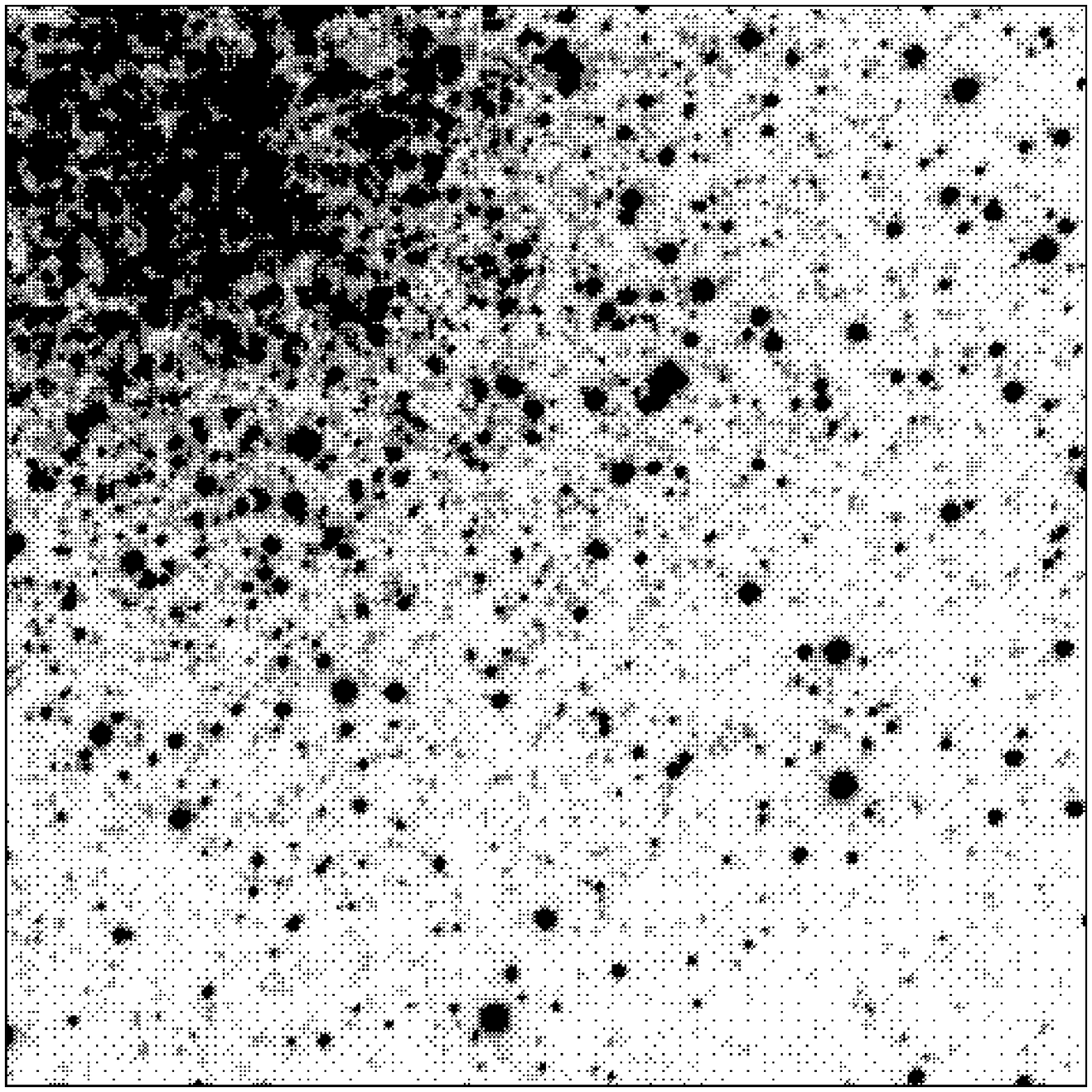,width=4cm}} 
\end{tabular}
\end{minipage}
\end{tabular}
\caption[]{CMD and covered fields for NGC~4833}
\label{ngc4833}
\end{figure*}

\paragraph {\bf NGC 4833.} (Fig.~\ref{ngc4833})

NGC~4833 is another metal-poor cluster, with an extended BHB, likely
with gaps, for which we have not found any previous CCD photometry.

\begin{figure*}
\begin{tabular}{c@{}c}
\raisebox{-6cm}{
\psfig{figure=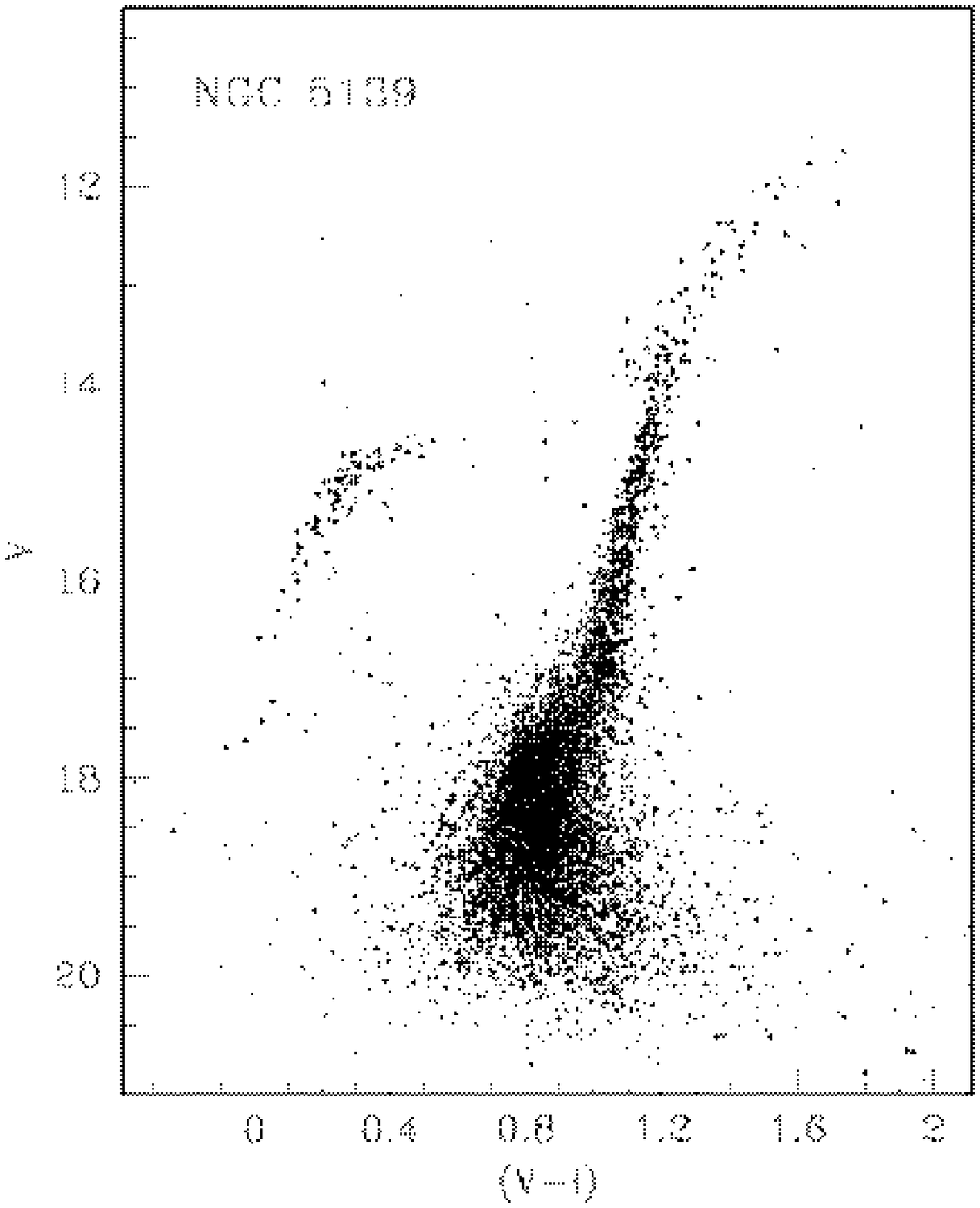,width=8.8cm}
} &
\begin{minipage}[t]{8.8cm}
\begin{tabular}{c@{}c}
\fbox{\psfig{figure=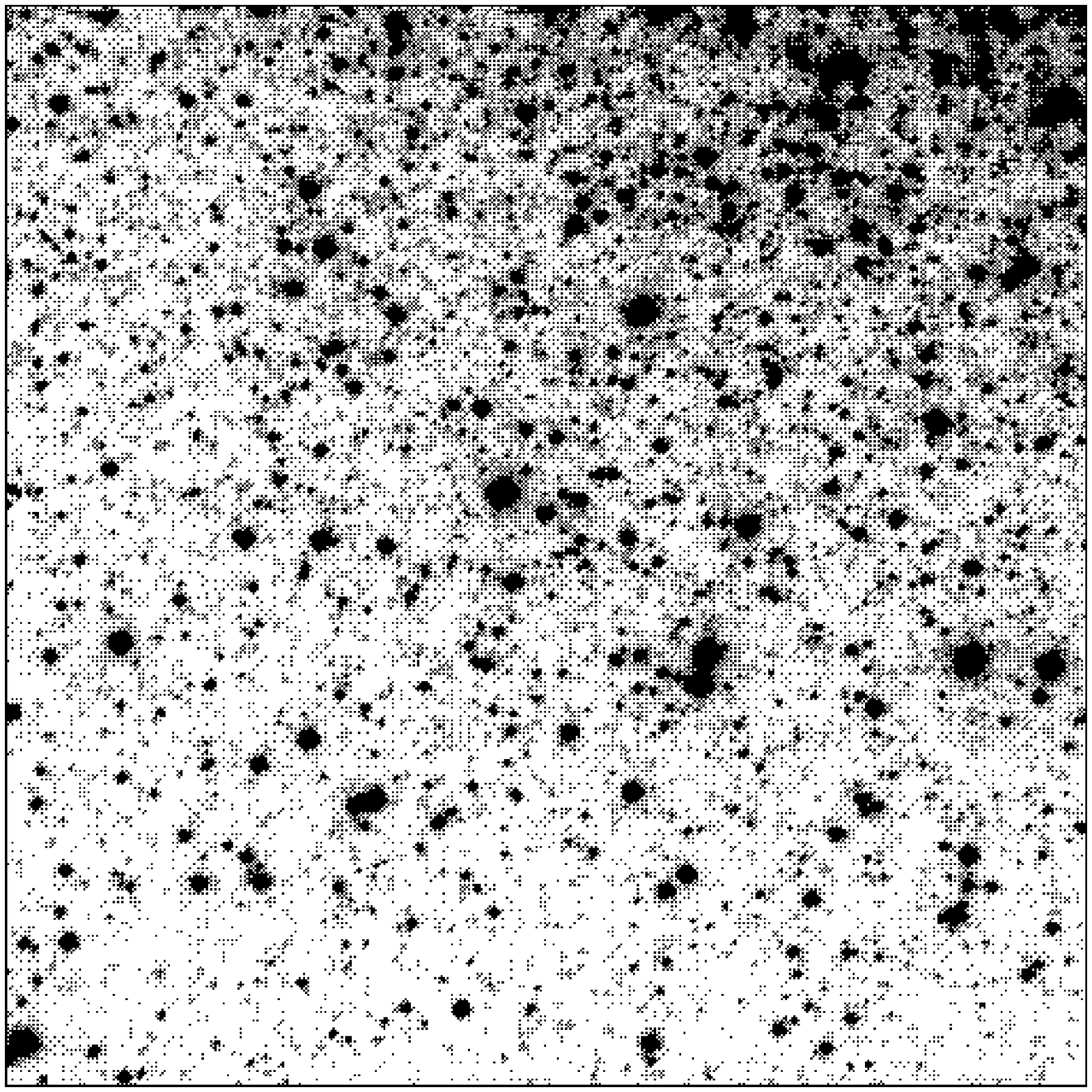,width=4cm}} &
\fbox{\psfig{figure=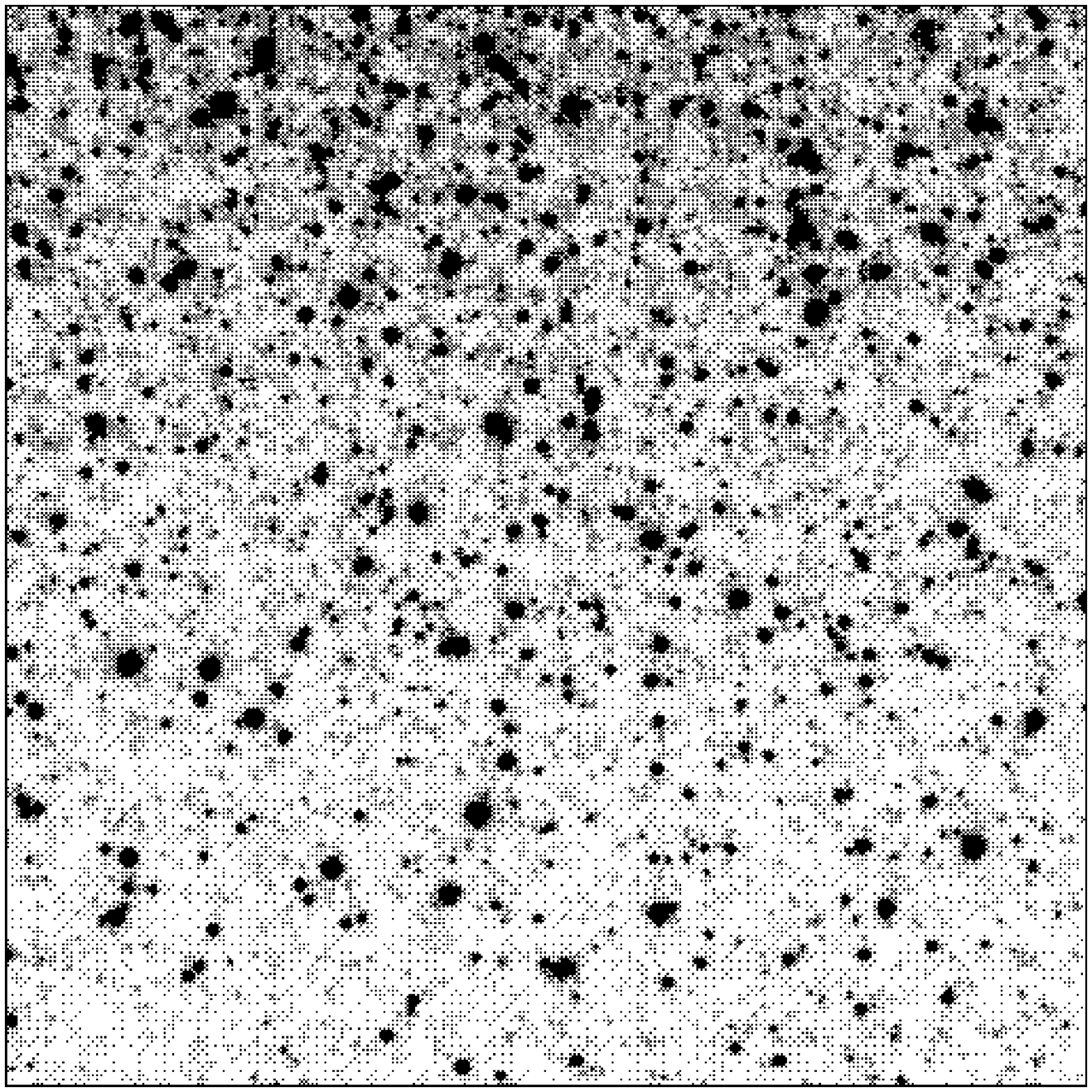,width=4cm}} 
\end{tabular}
\end{minipage}
\end{tabular}
\caption[]{CMD and covered fields for NGC~5139 ($\omega$ Centauri)}
\label{ngc5139}
\end{figure*}

\paragraph {\bf NGC 5139 ($\omega$ Centauri).} (Fig.~\ref{ngc5139}) 

NGC~5139 is the intrinsically brightest cluster in our Galaxy. Apart
from this, there are many other properties of $\omega$ Centauri which
make it a very particular object. Its stellar population shows
metallicity variations as large as $\sim1.5$~dex from star to star
(Norris et al. \cite{norris96}). Its overall properties suggest that
this clusters could have a different origin from the bulk of GGCs. It
has an extended BHB and probably numerous BSS. The broad sequences in
the CMD are mainly due to the metallicity variations though likely
there is some differential reddening in the field of $\omega$
Centauri. Due to its peculiarities, $\omega$ Centauri has been (and
is!)  extensively studied; there is a large number of photometries,
and we cannot cite all of them. The most recent and interesting CCD
CMDs are in: Alcaino \& Liller (\cite{alcainoliller87b}), who present
a multi-band ($B,V,R,I$) photometry, but poorly sampled, specially
for the evolved part of the diagram; Noble et al. (\cite{noble91})
present a deep $B,V$ diagram, where the MS is well sampled, but the RGB
is not so clear and only 3-5 stars are present in the HB; Elson et al.
(\cite{elson95}) present a HST $V,I$ photometry of the MS; Lynga
(\cite{lynga96}) presents a $BVRI$ study of the evolved part of the
diagram ($\sim 2$ mag below the HB); Kaluzny et al. (\cite{kaluzny96},
\cite{kaluzny97a}) present a $V, I$ CMD covering more than $10^5$ stars.

\begin{figure*}
\begin{tabular}{c@{}c}
\raisebox{-6cm}{
\psfig{figure=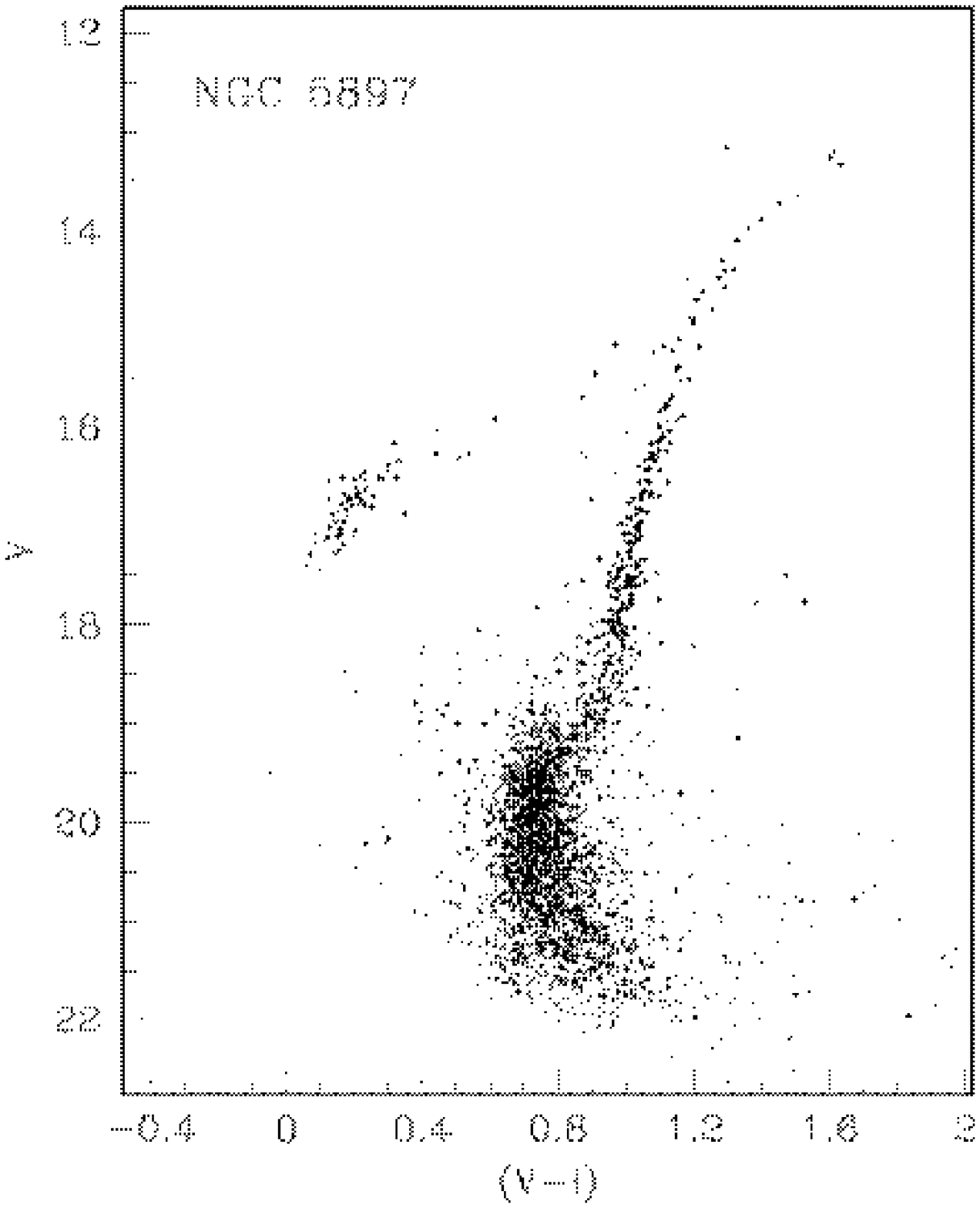,width=8.8cm}
} &
\begin{minipage}[t]{8.8cm}
\begin{tabular}{c@{}c}
\fbox{\psfig{figure=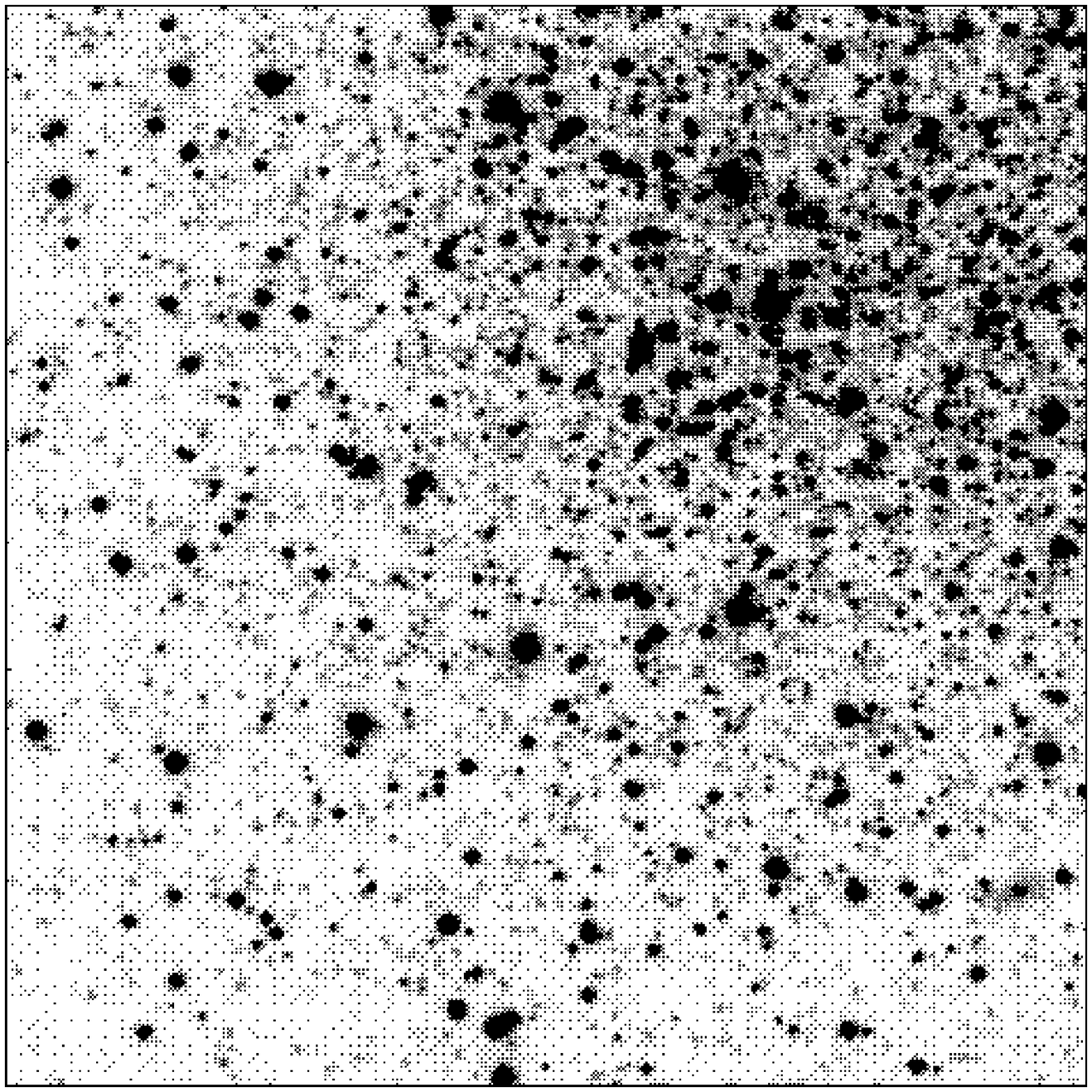,width=4cm}}
\end{tabular}
\end{minipage}
\end{tabular}
\caption[]{CMD and covered fields for NGC~5897}
\label{ngc5897}
\end{figure*}

\paragraph {\bf NGC 5897.} (Fig.~\ref{ngc5897}) 

NGC~5897 is a metal poor cluster with a blue, not extended HB, typical
for its metallicity. All the sequences of Fig.~\ref{ngc5897} are well
defined and populated, including a BS sequence. Two CCD photometric
studies exist for this cluster: Sarajedini (\cite{sarajedini92})
($B,V$ bands) and Ferraro et al. (\cite{ferraro92}) ($U,B,V,I$ bands).

\begin{figure*}
\begin{tabular}{c@{}c}
\raisebox{-6cm}{
\psfig{figure=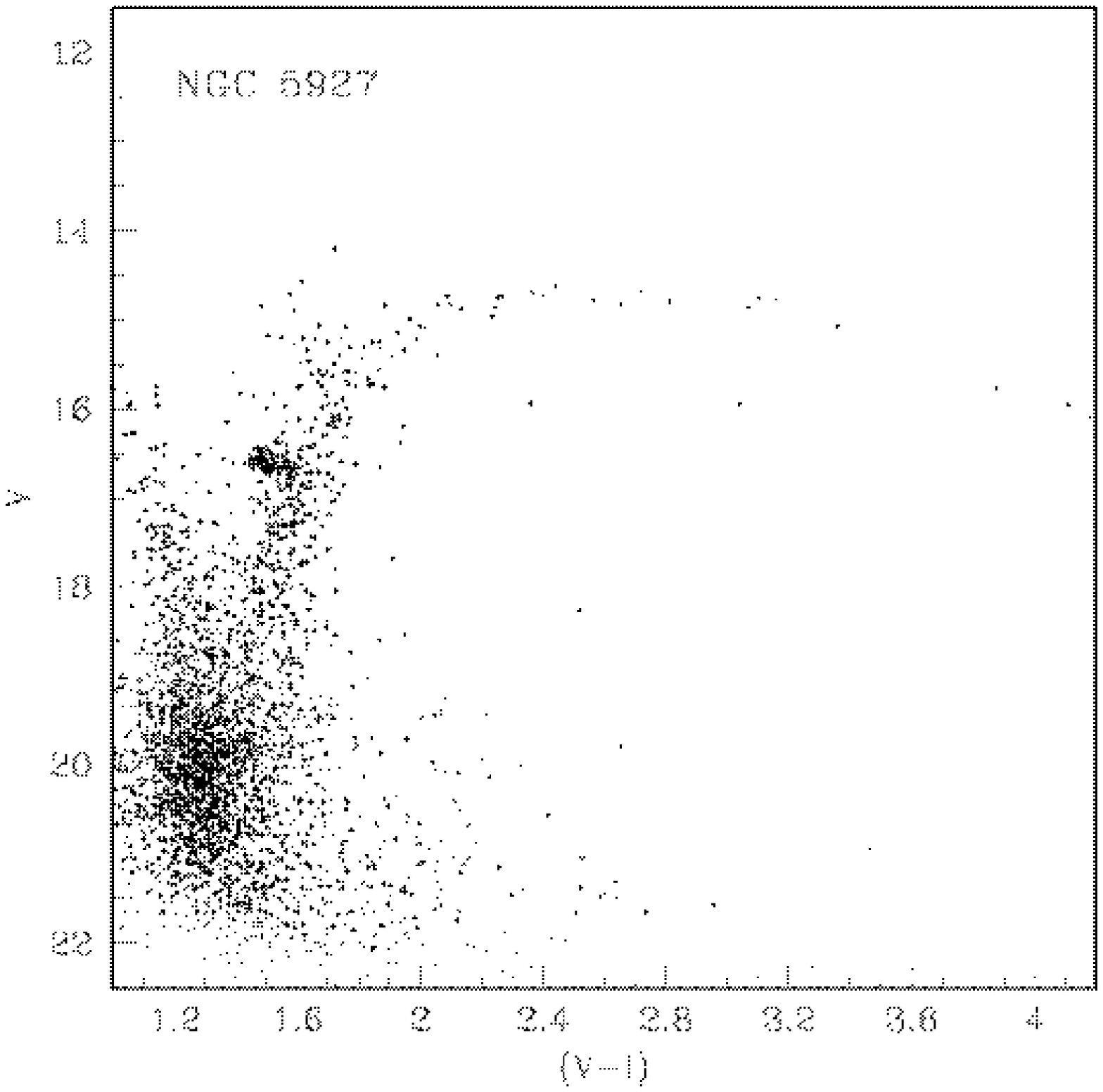,width=10.9cm}
} &
\begin{minipage}[t]{8.8cm}
\begin{tabular}{c@{}c}
\fbox{\psfig{figure=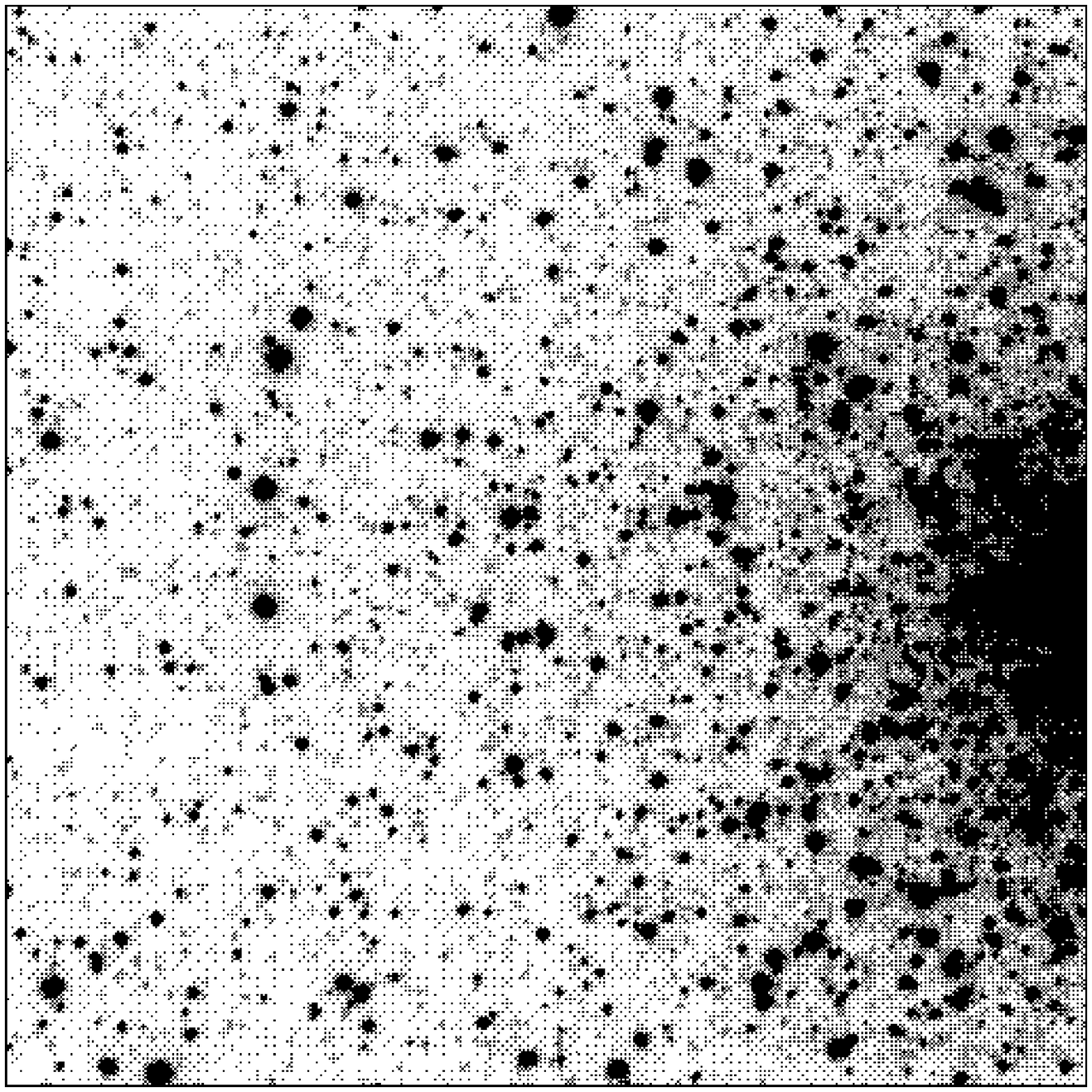,width=4cm}}
\end{tabular}
\end{minipage}
\end{tabular}
\caption[]{CMD and covered field for NGC~5927}
\label{ngc5927}
\end{figure*}

\paragraph {\bf NGC 5927.}  (Fig.~\ref{ngc5927})

NGC~5927 has the highest metallicity among the objects of our
catalog. It has, as most of the GGCs with [Fe/H]$>-0.8$, a well
populated red horizontal branch (RHB), and an extended RGB, which, in
our CMD, covers more than $\sim$2.5~mag in ($V-I$), from the RHB
(partially overlapped with the RGB) to the reddest stars of the RGB
tip. It has a high reddening, possibly differential, judging from the
broadening of the RGB, and, due to its location (projected towards the
Galactic center), the field object contamination (disk and bulge
stars) is very high. Previous CCD photometries are in 
Friel \& Geisler (\cite{frielgeisler91}) (Washington photometry), 
Sarajedini \& Norris (\cite{sarajedininorris94}) ($B,V$ bands), 
Samus et al. (\cite{samus96}) ($B,V,I$ bands), Sosin et al.
(\cite{sosin97a}), and Rich et al. (\cite{rich97}) (HST $B,V$ bands).

\begin{figure*}
\begin{tabular}{c@{}c}
\raisebox{-6cm}{
\psfig{figure=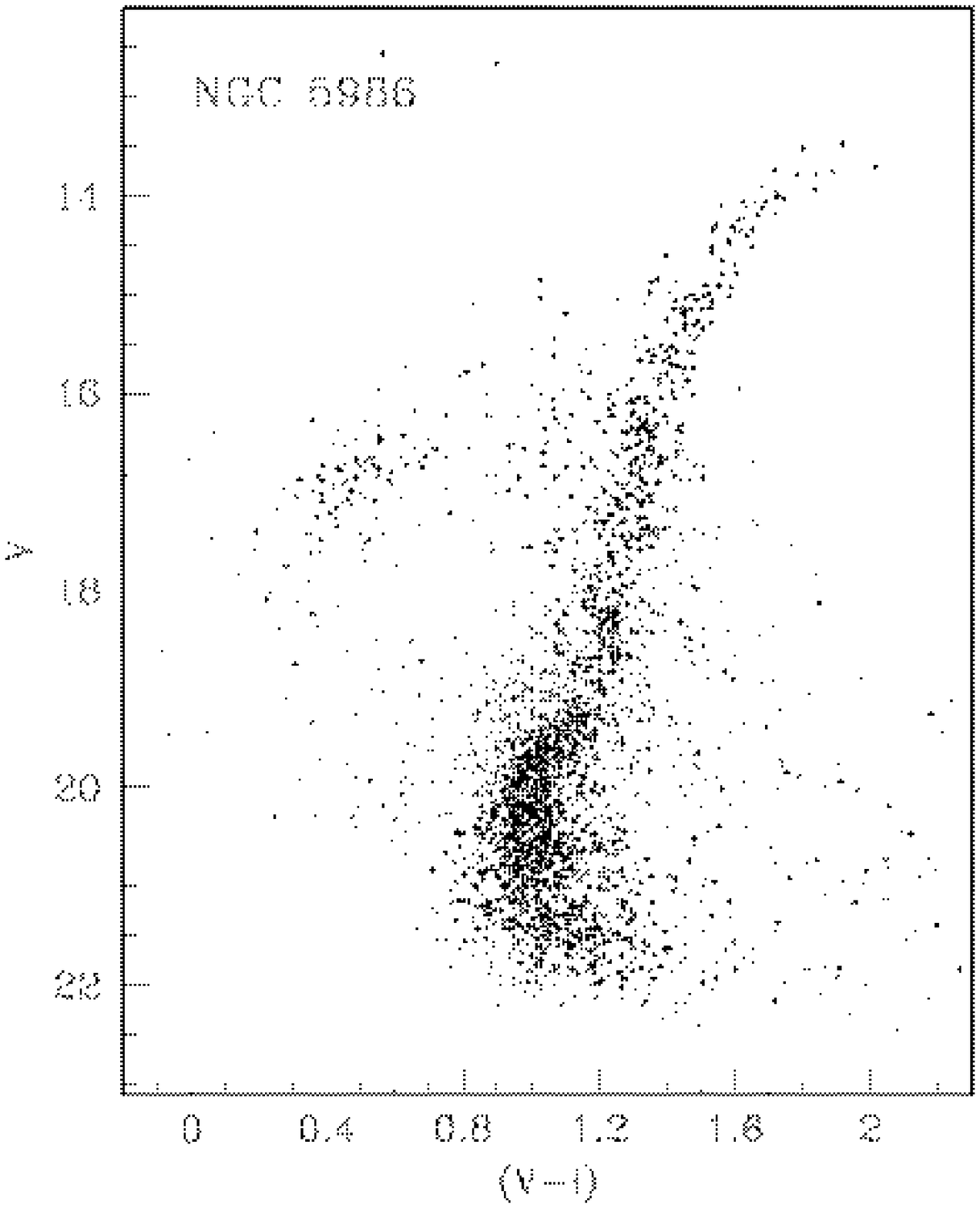,width=8.8cm}
} &
\begin{minipage}[t]{8.8cm}
\begin{tabular}{c@{}c}
\fbox{\psfig{figure=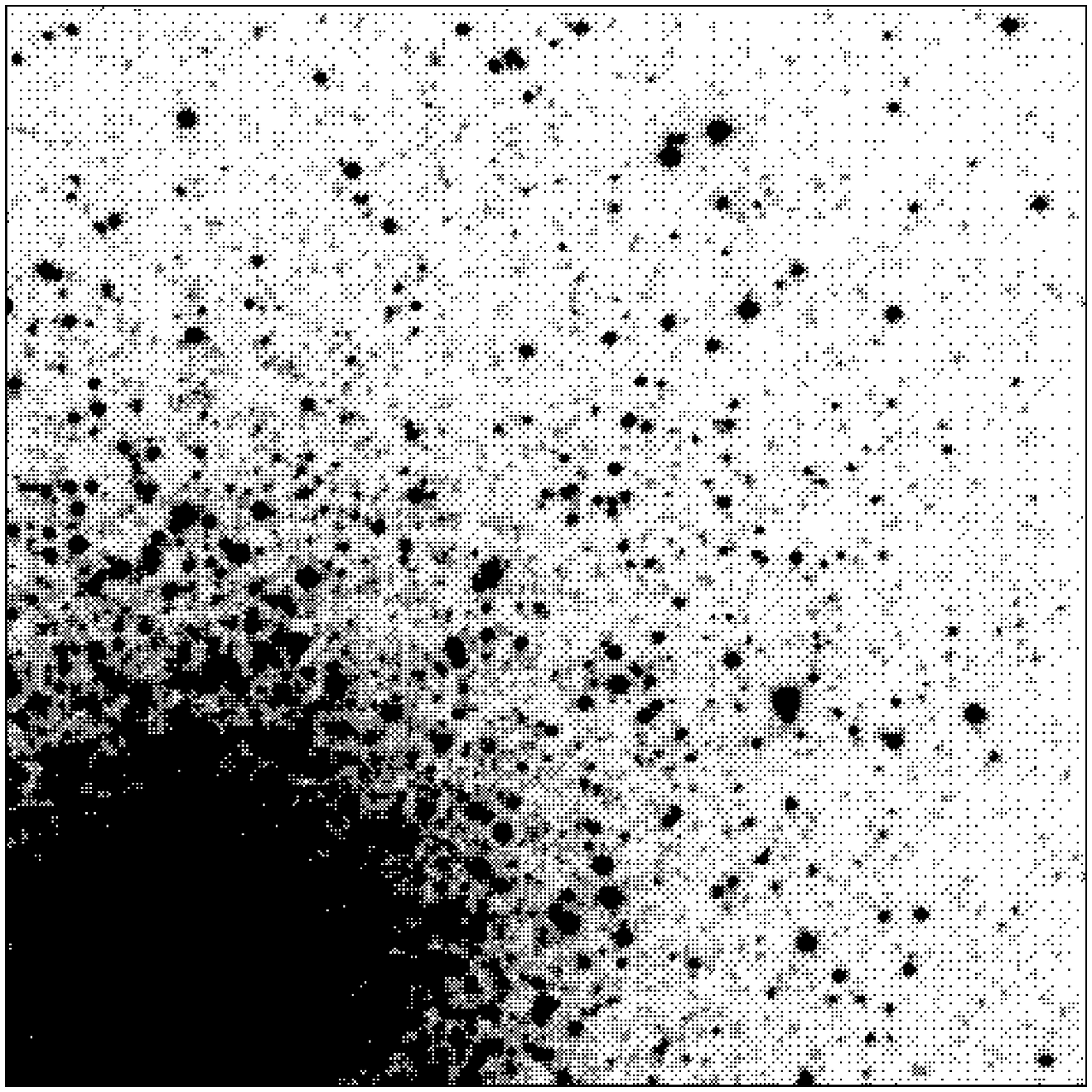,width=4cm}}
\end{tabular}
\end{minipage}
\end{tabular}
\caption[]{CMD and covered field for NGC~5986}
\label{ngc5986}
\end{figure*}

\paragraph {\bf NGC 5986.}  (Fig.~\ref{ngc5986})

To our knowledge, this is the first CCD photometry for this
cluster. NGC~5986 is an intermediate metallicity cluster, but with a
metal-poor like HB. The broadening of the CMD suggests some
differential reddening. Contamination by field stars is clearly
visible, as expected on the basis of the position within the Galaxy of
this cluster.

\begin{figure*}
\begin{tabular}{c@{}c}
\raisebox{-6cm}{
\psfig{figure=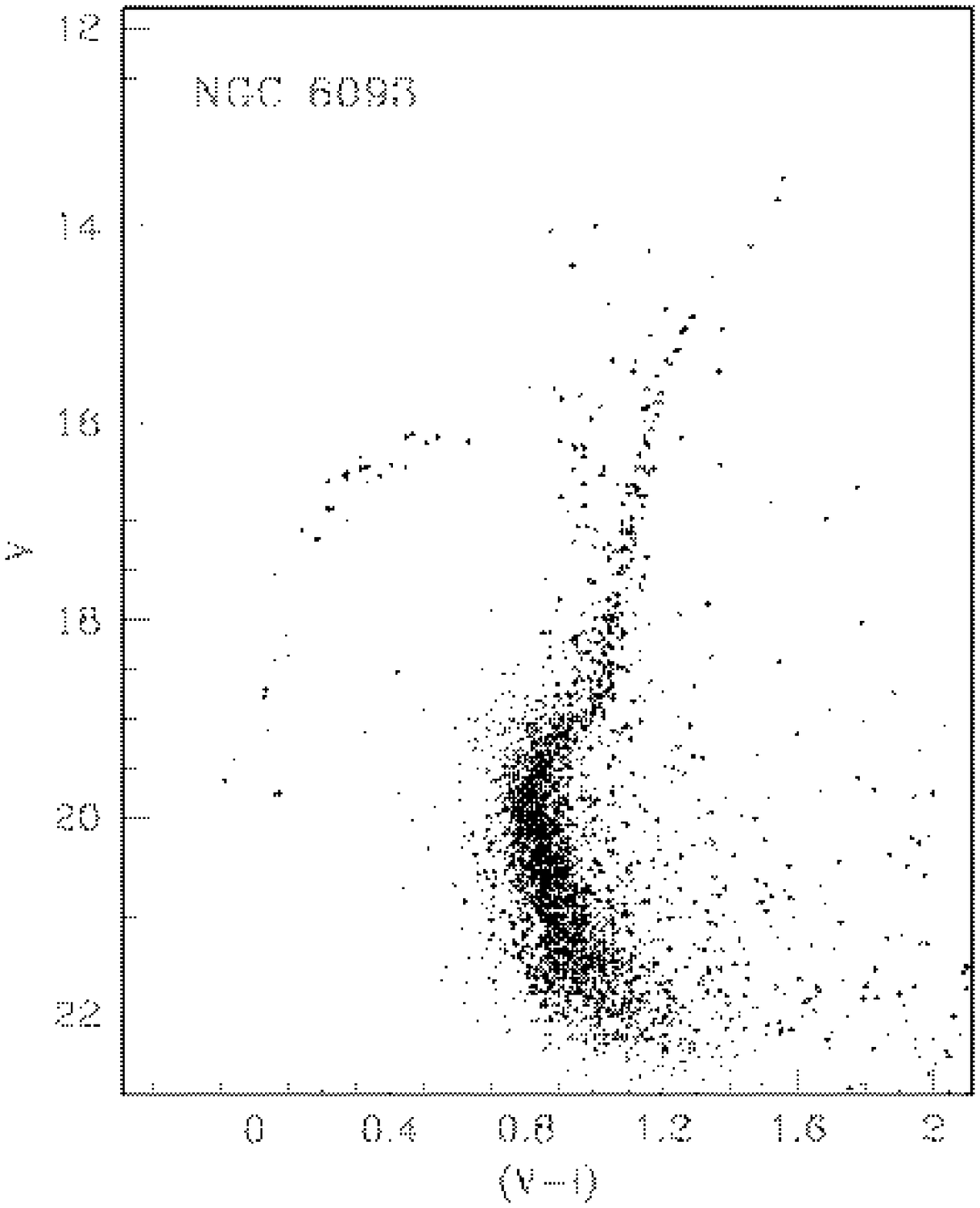,width=8.8cm}
} &
\begin{minipage}[t]{8.8cm}
\begin{tabular}{c@{}c}
\fbox{\psfig{figure=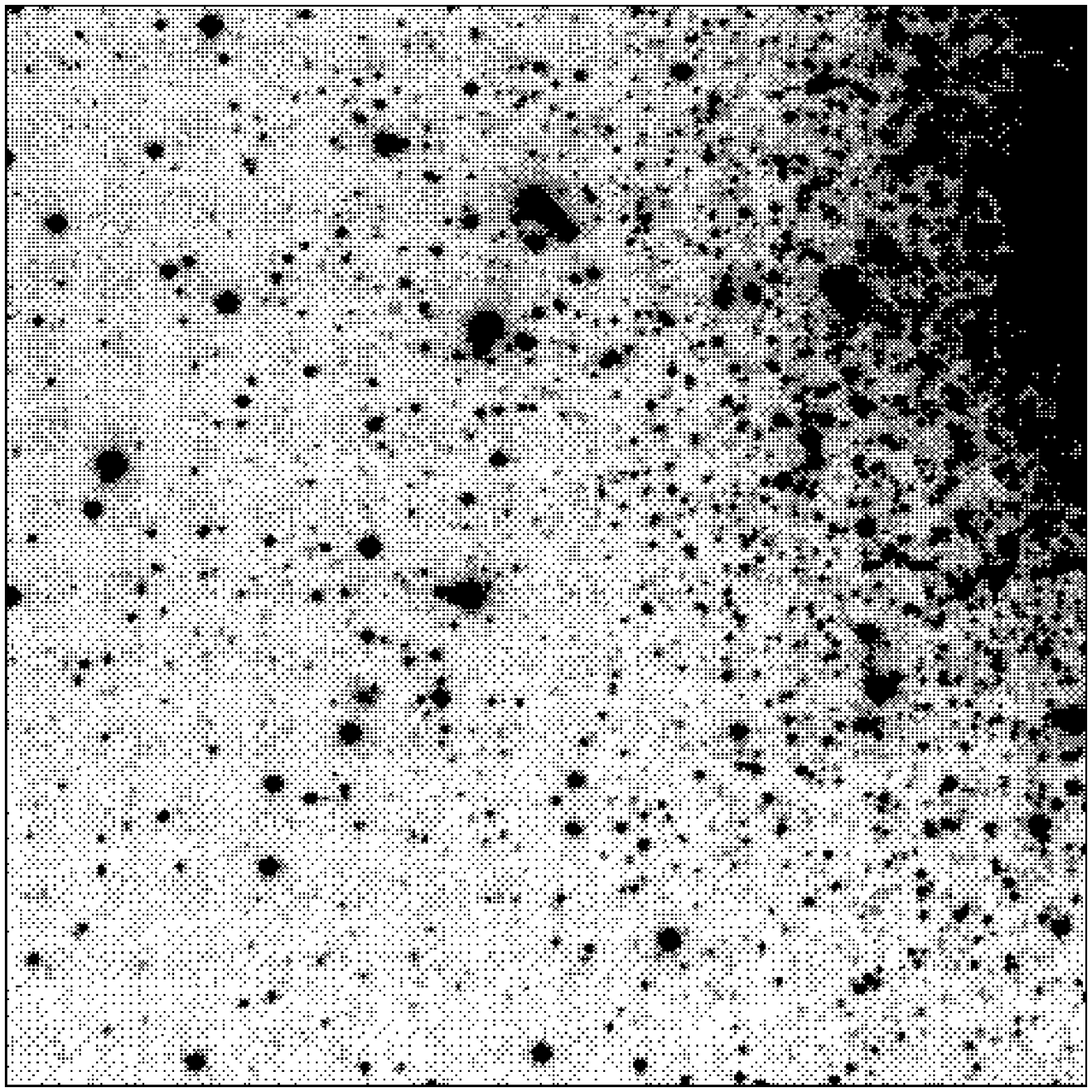,width=4cm}}\\
&
\fbox{\psfig{figure=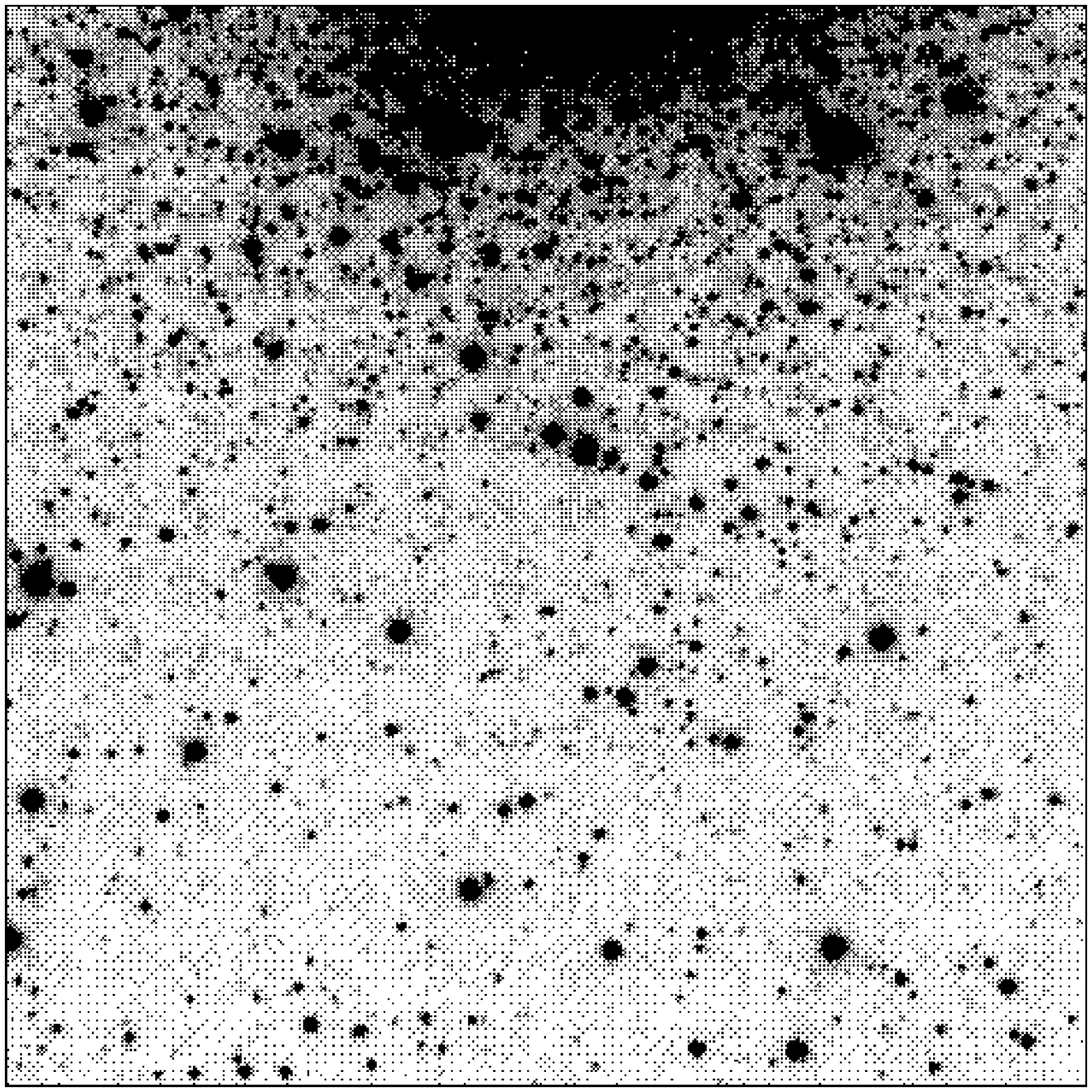,width=4cm}} 
\end{tabular}
\end{minipage}
\end{tabular}
\caption[]{CMD and covered fields for NGC~6093 (M~80)}
\label{ngc6093}
\end{figure*}

\paragraph {\bf NGC 6093 (M~80).}  (Fig.~\ref{ngc6093})

NGC~6093 is a bright and moderately metal poor cluster, and one of the
densest globular clusters in the Galaxy. It has an EBHB, which extends
well below the TO as clearly visible also in the CMD of
Fig.~\ref{ngc6093}, with gaps (Ferraro et al. 1998). Three recent CCD
photometries that cover the entire object, with CMD from the brightest
stars to above the TO exist for this cluster: Brocato et
al. (\cite{brocato98}) ($B,V$ bands) and Ferraro et al.
(\cite{ferraro98}) (HST $U,V$, and far-UV (F160BW) bands). A
ground-based multicolor $U,B,V,I$ CCD CMD has been published also by
Alcaino et al. (\cite{alcaino98}).

\begin{figure*}
\begin{tabular}{c@{}c}
\raisebox{-6cm}{
\psfig{figure=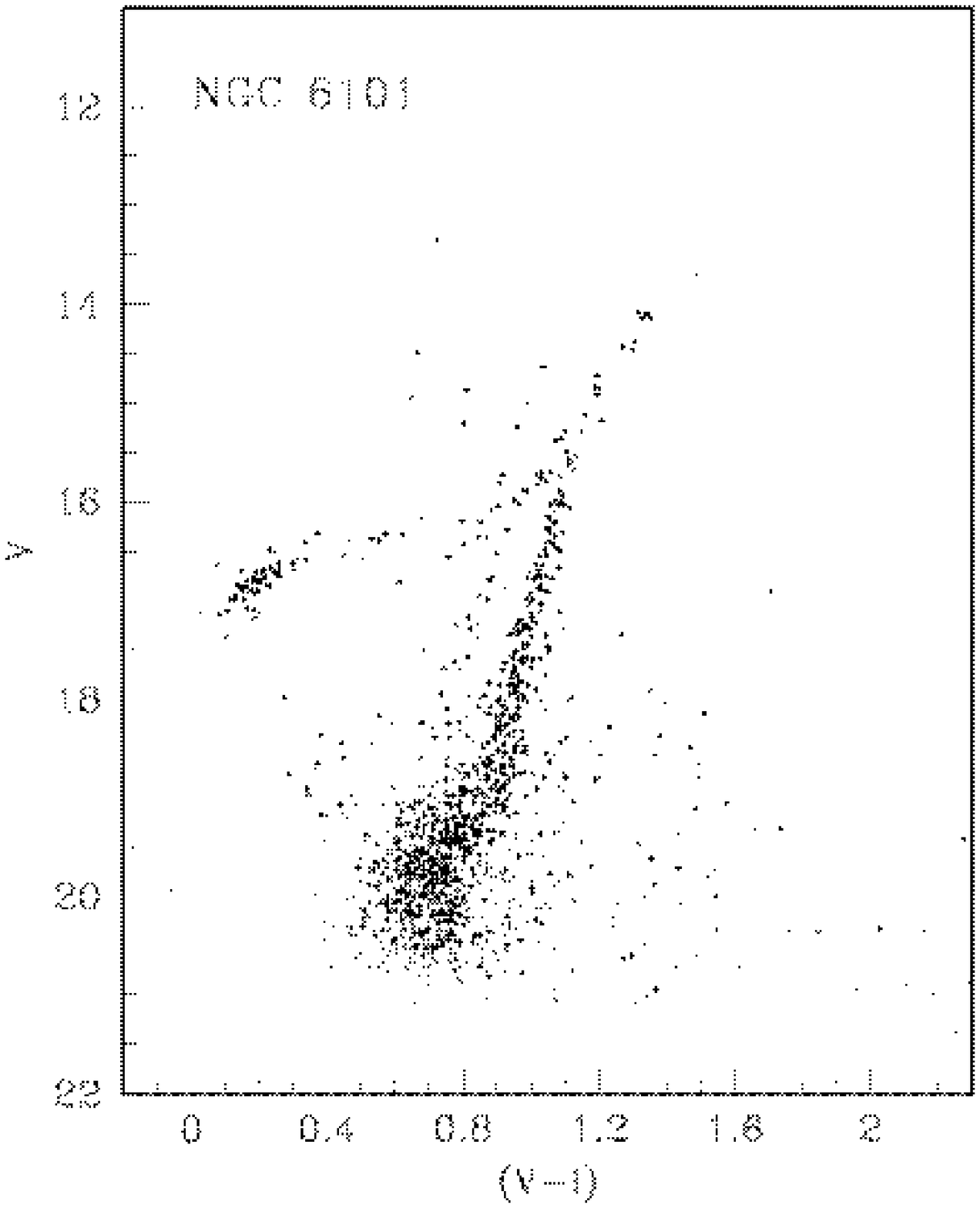,width=8.8cm}
} &
\begin{minipage}[t]{8.8cm}
\begin{tabular}{c@{}c}
\fbox{\psfig{figure=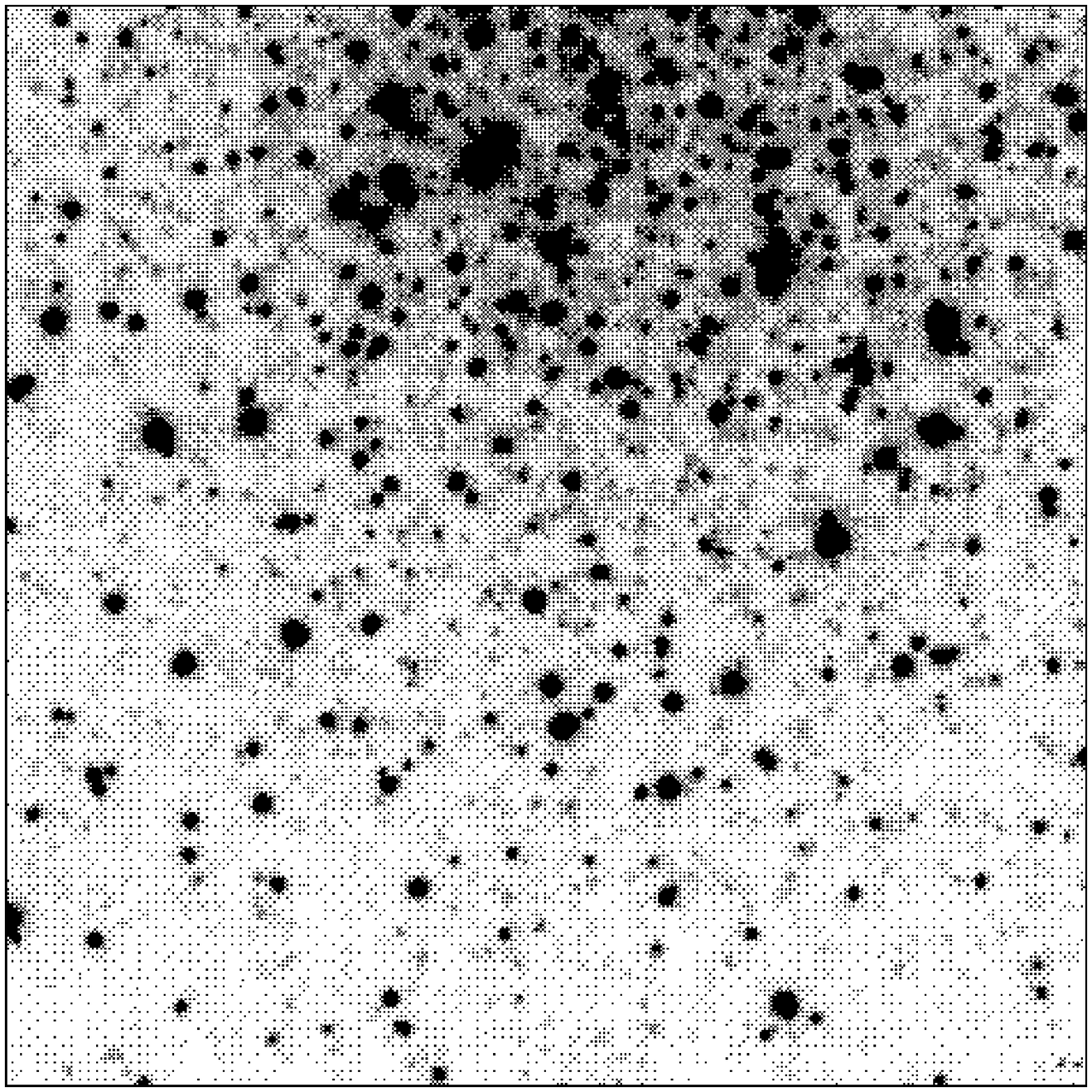,width=4cm}}
\end{tabular}
\end{minipage}
\end{tabular}
\caption[]{CMD and covered field for NGC~6101}
\label{ngc6101}
\end{figure*}

\paragraph {\bf NGC 6101.}  (Fig.~\ref{ngc6101})

NGC~6101 was observed under not very good seeing conditions, and this
is the reason for the brighter limiting magnitude. Its CMD has the
morphology expected for a metal-poor cluster: the HB is predominantly
blue, and the giant branch is steep. In Fig.~\ref{ngc6101} we note
that, starting from the BSS sequence, there is a sequence of stars
parallel to the RGB on its blue side. In view of the position of the
cluster ($l,b$)=(318,$-16$) these can unlikely be bulge stars; it is
possible that on the same line of sight there is an open cluster,
though the slope of the two RGBs are quite similar, implying an
unlikely similar metallicity. A larger field coverage of NGC~6101 is
desirable. 
The only previous CCD photometry that exists for
this cluster is the $B$ and $V$ study by Sarajedini \& Da Costa
(\cite{sarajedinidacosta91}), which shows these stars in the same CMD
location.  However, being the background-foreground stellar contamination
heavier, the sequences we discussed can hardly be seen.

\begin{figure*}
\begin{tabular}{c@{}c}
\raisebox{-6cm}{
\psfig{figure=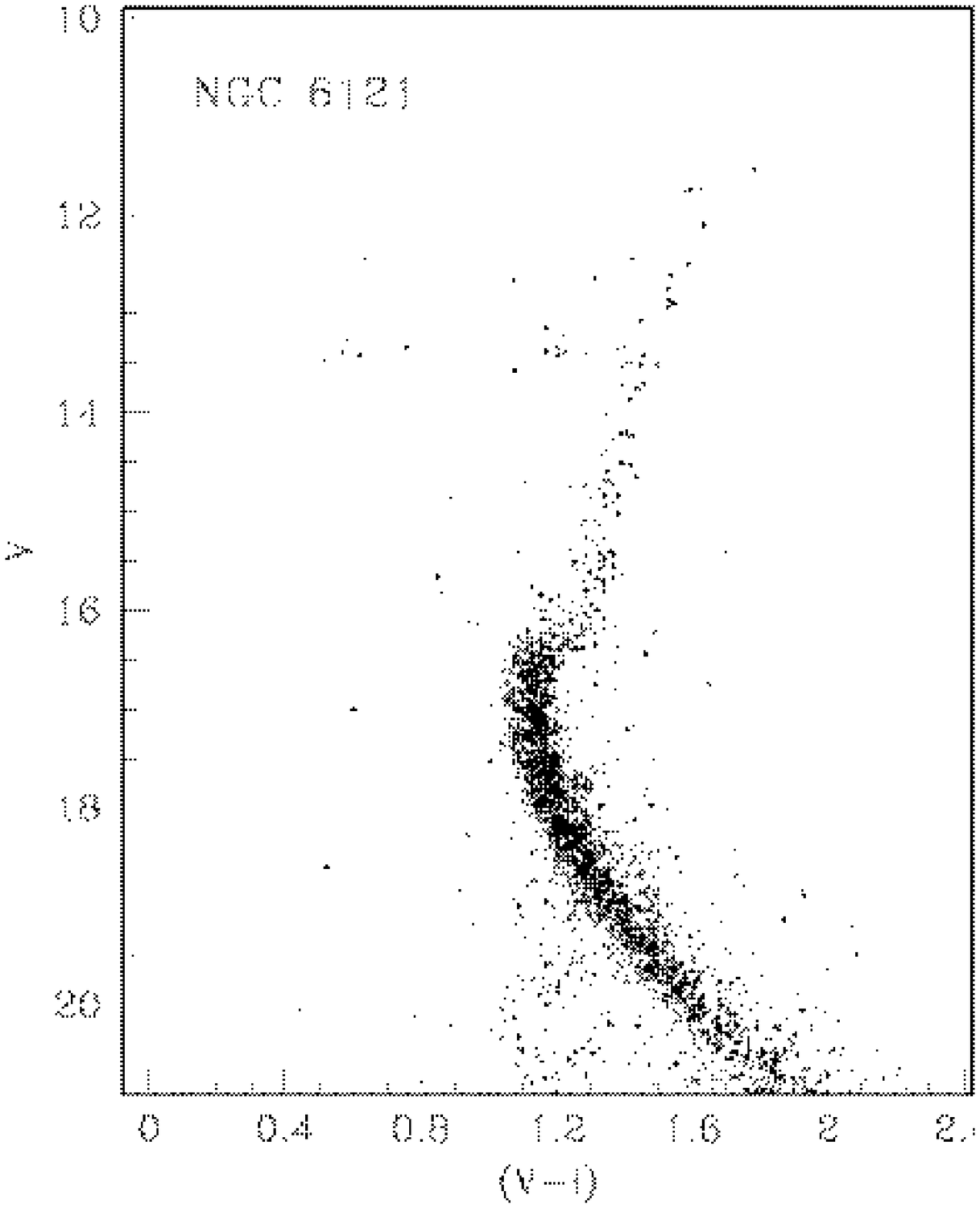,width=8.8cm}
} &
\begin{minipage}[t]{8.8cm}
\begin{tabular}{c@{}c}
\fbox{\psfig{figure=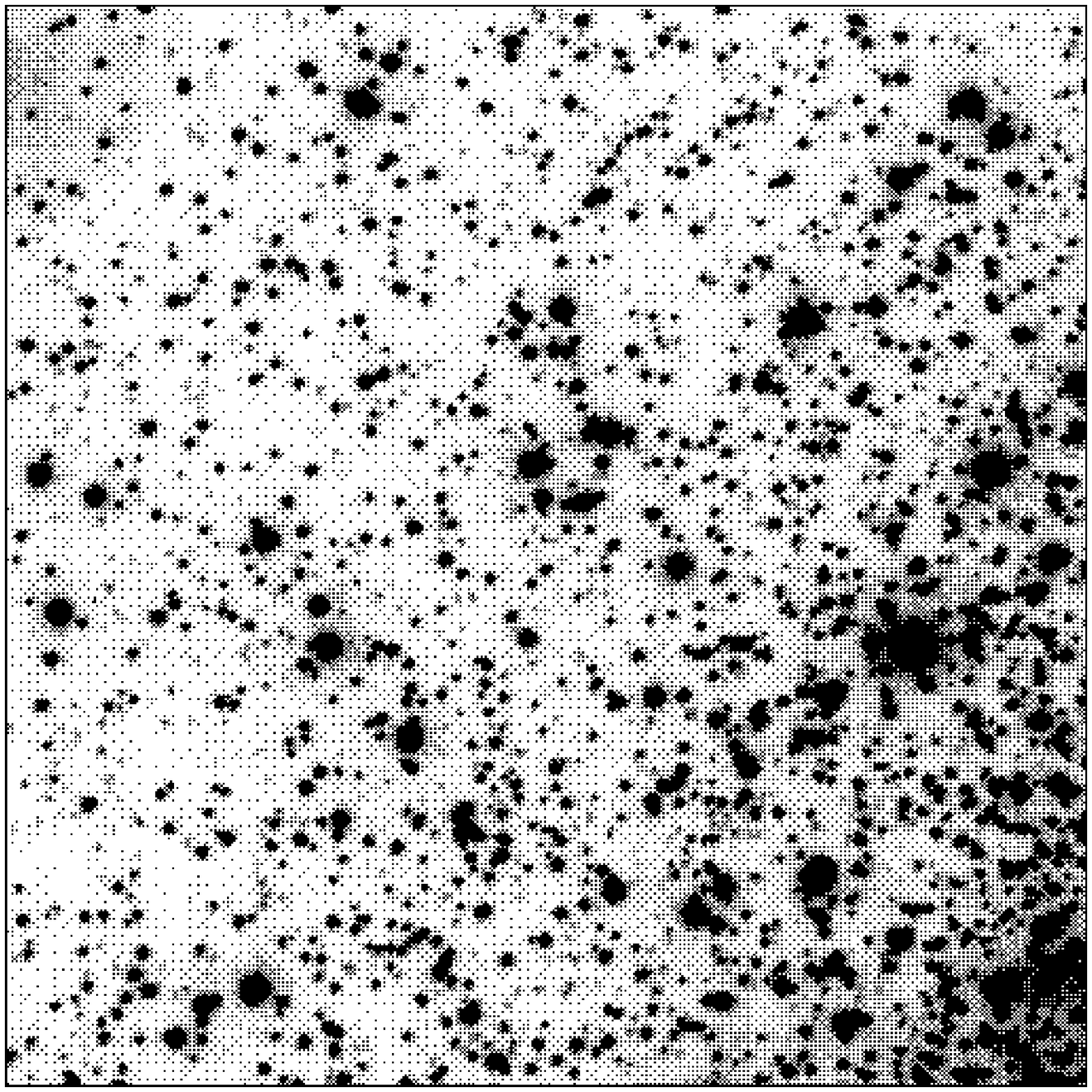,width=4cm}} \\
\fbox{\psfig{figure=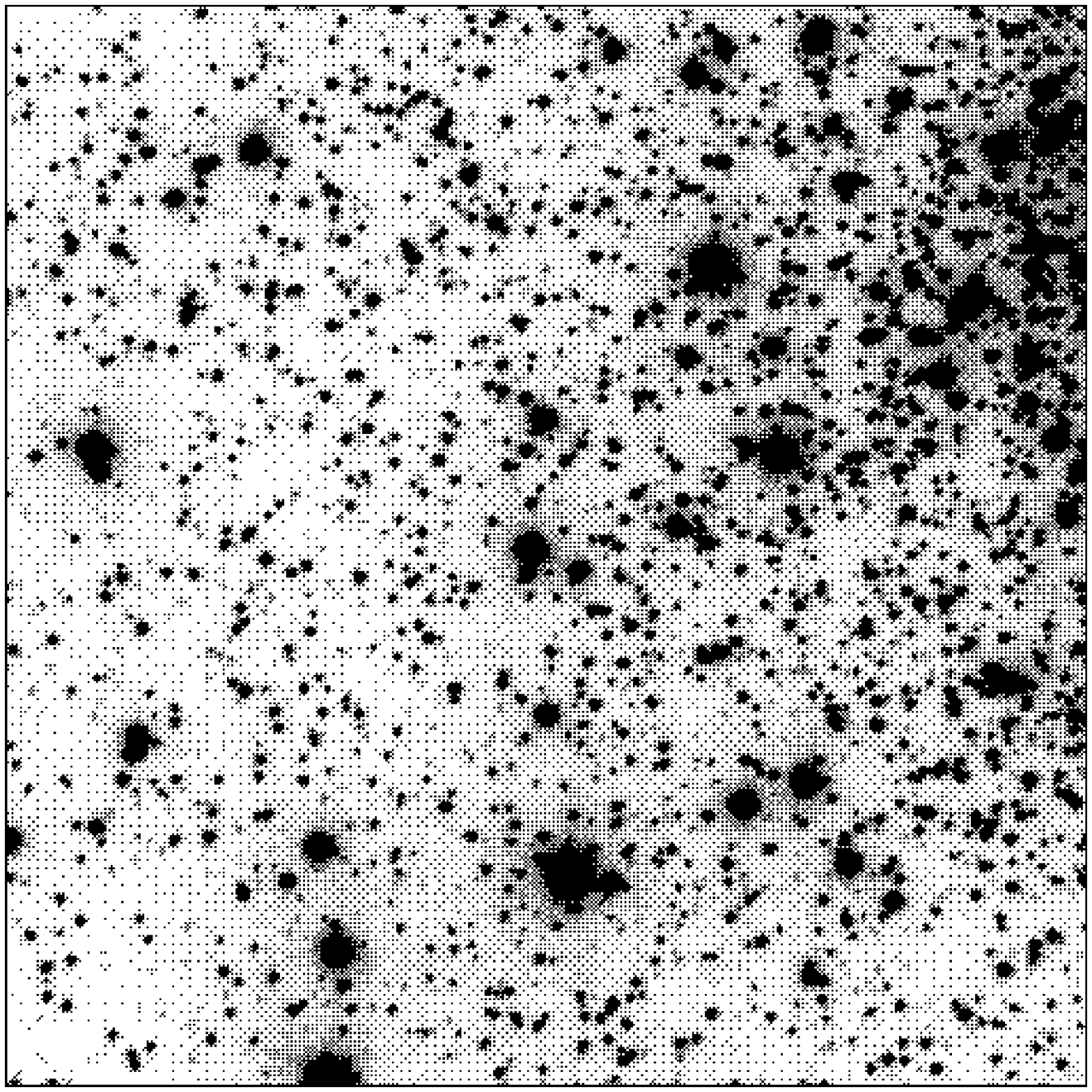,width=4cm}} 
\end{tabular}
\end{minipage}
\end{tabular}
\caption[]{CMD and covered fields for NGC~6121 (M~4)}
\label{ngc6121}
\end{figure*}

\paragraph {\bf NGC 6121 (M~4).} (Fig.~\ref{ngc6121})

This cluster is the closest GGC, located approximately at
$\sim2.2$~kpc from the Sun, though, due to the large reddening caused
by the nebulosity in Scorpio-Ophiuchus, it has an apparent visual
distance modulus larger than NGC~6397. The reddening is differential,
though (as in the case of NGC~4372) it is homogeneously distributed in
space. The mean regions of the CMD can be improved using an
appropriate second order polynomial fit to the reddening distribution,
at least on the two fields shown in Fig.~\ref{ngc6121}. The stars from
the southern field have been plotted as darker dots; they are located
on the redder (more reddened) part of the CMD. The two most recent CMD
of M4 are in Ibata et al. (\cite{ibata99}) ($V,I,U$ filters) and Pulone et
al. (\cite{pulone99}), who present (near IR) HST studies of the faint part of
the MS and of the WD sequence. Other recent CMDs from the RGB tip to
below the MSTO are in Alcaino et al. (\cite{alcaino97a}), who
presented an $UBVI$ CCD photometry, and Kanatas et
al. (\cite{kanatas95}), who obtained a composite ($B,V$) CMD from
$V\sim12$ to $V\sim25$.

\begin{figure*}
\begin{tabular}{c@{}c}
\raisebox{-6cm}{
\psfig{figure=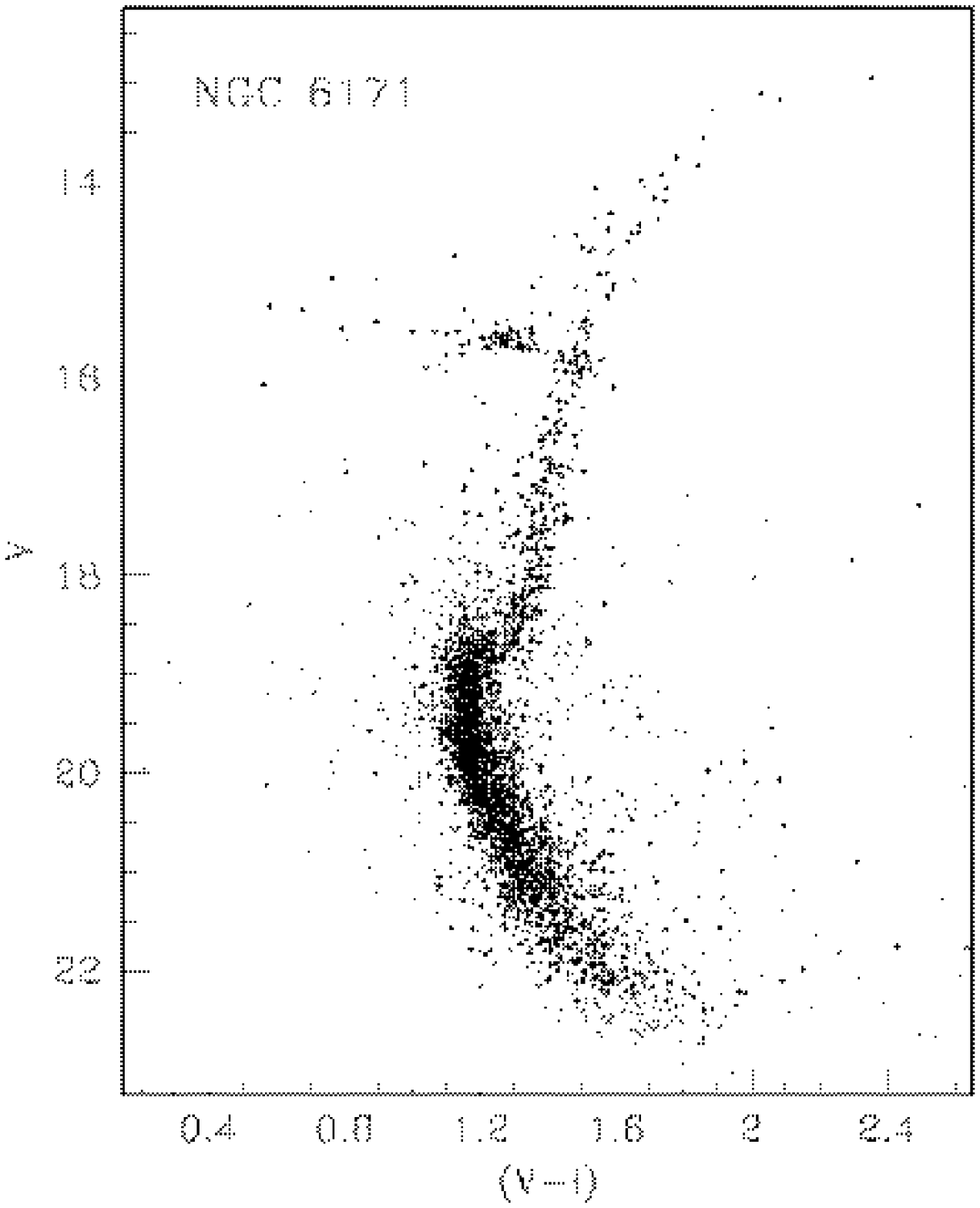,width=8.8cm}
} &
\begin{minipage}[t]{8.8cm}
\begin{tabular}{c@{}c}
\fbox{\psfig{figure=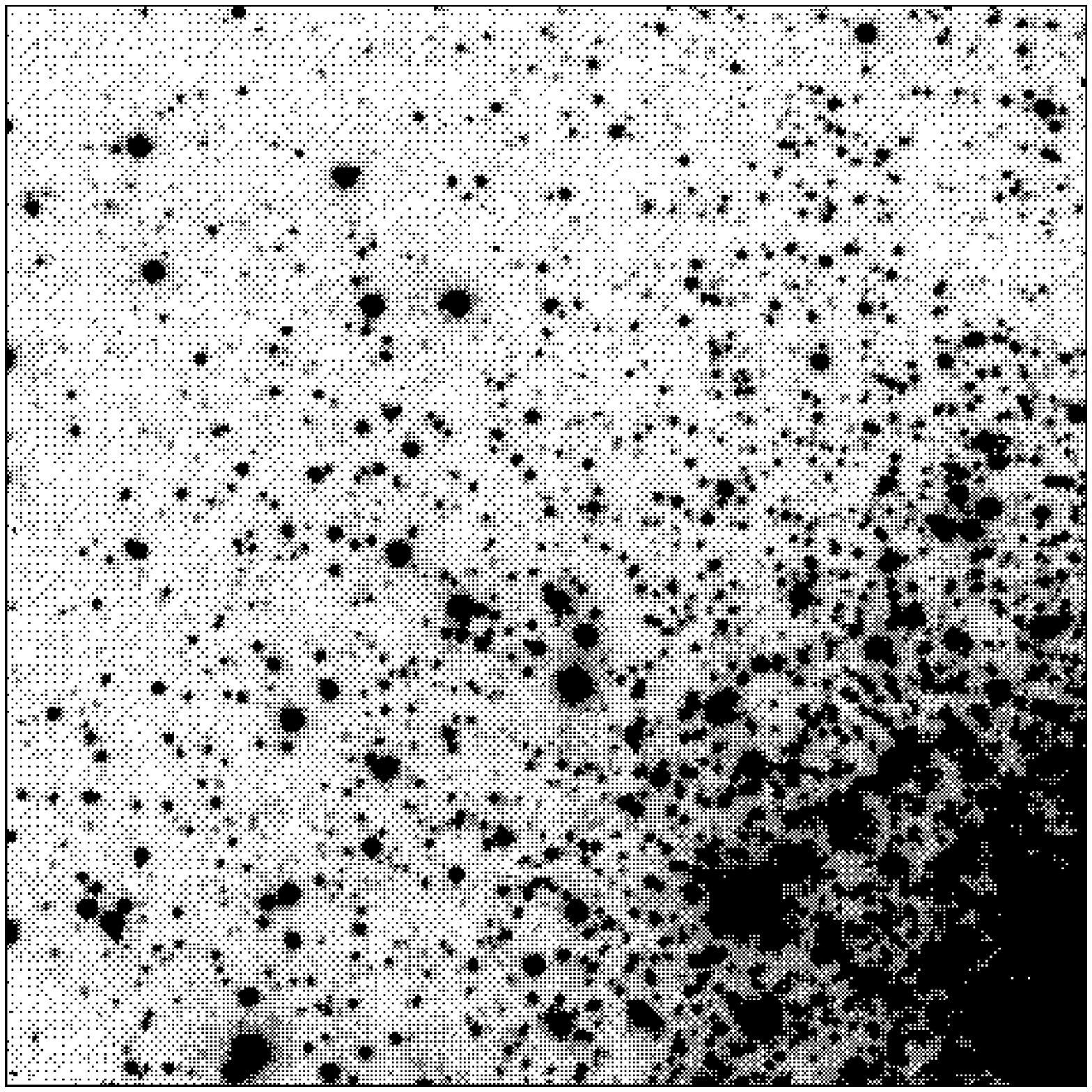,width=4cm}} &
\fbox{\psfig{figure=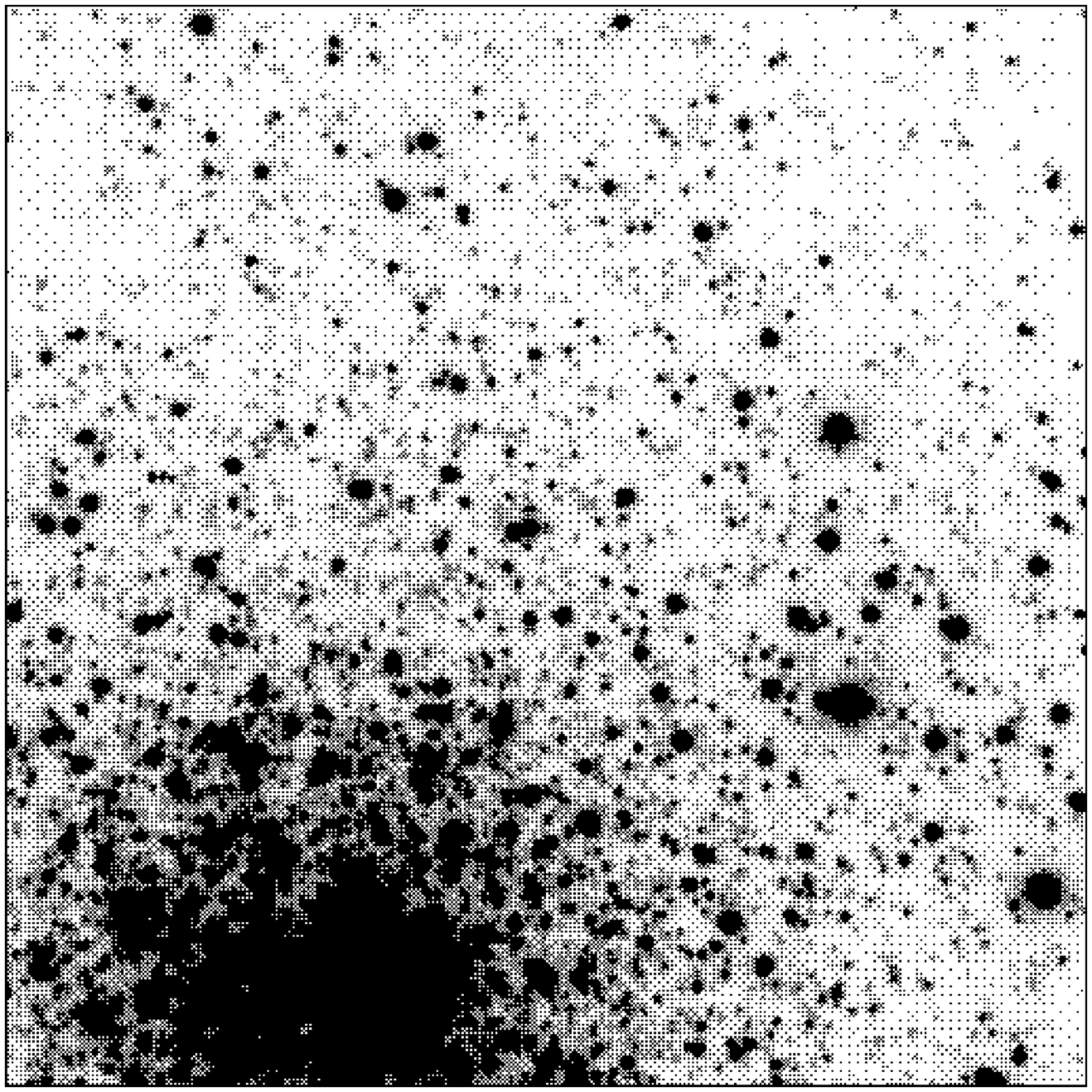,width=4cm}} 
\end{tabular}
\end{minipage}
\end{tabular}
\caption[]{CMD and covered fields for NGC~6171 (M~107)}
\label{ngc6171}
\end{figure*}

\paragraph {\bf NGC 6171 (M~107).} (Fig.~\ref{ngc6171})

Previous CCD studies of NGC~6171 are the $(J,K)$ and $(B,V)$
photometry by Ferraro et al. (\cite{ferraro95} and \cite{ferraro91},
respectively). This cluster is affected by a moderate reddening, which
could be slightly differential. It has a RHB, with a few stars bluer
than the instability strip blue edge.

\begin{figure*}
\begin{tabular}{c@{}c}
\raisebox{-6cm}{
\psfig{figure=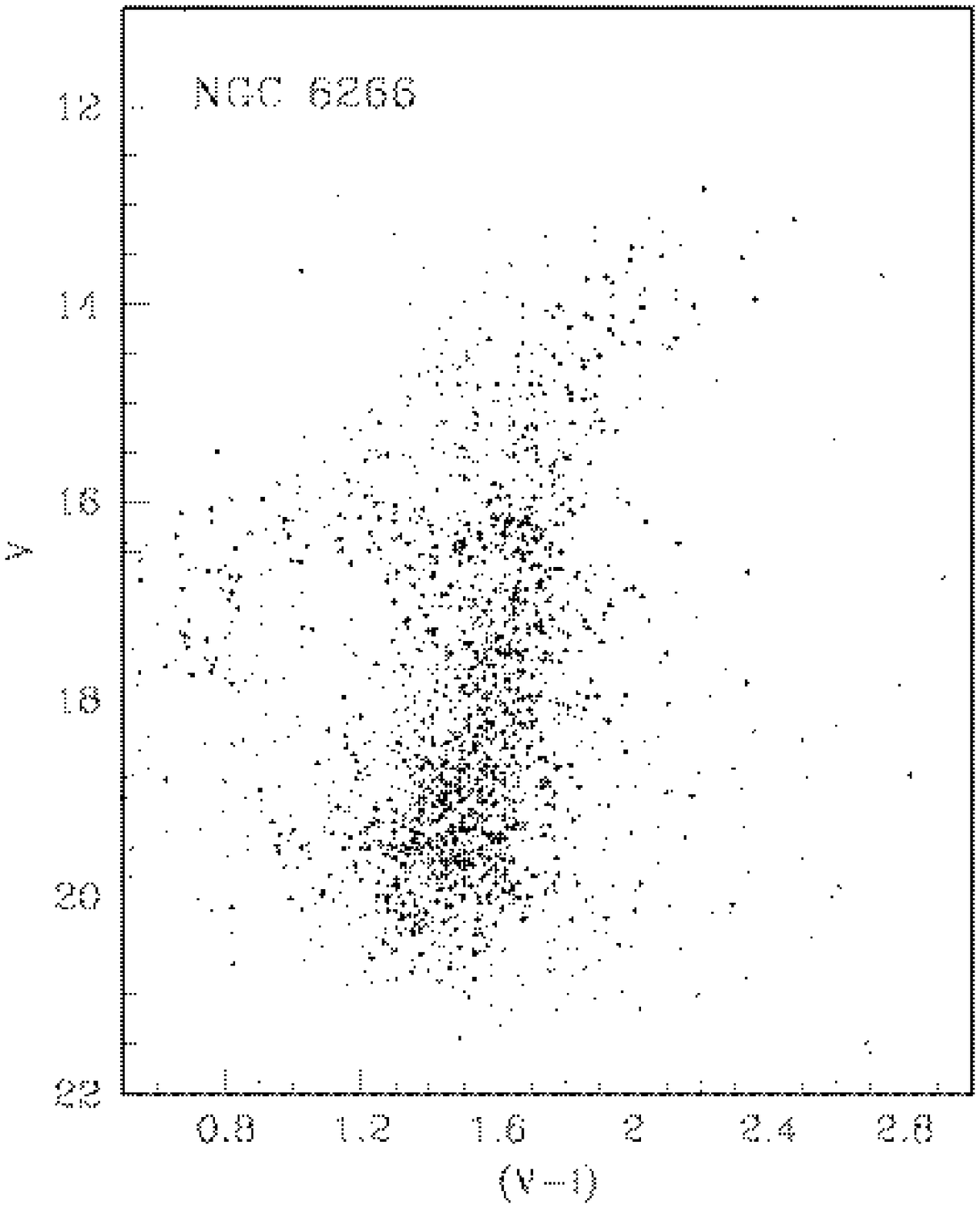,width=8.8cm}
} &
\begin{minipage}[t]{8.8cm}
\begin{tabular}{c@{}c}
\fbox{\psfig{figure=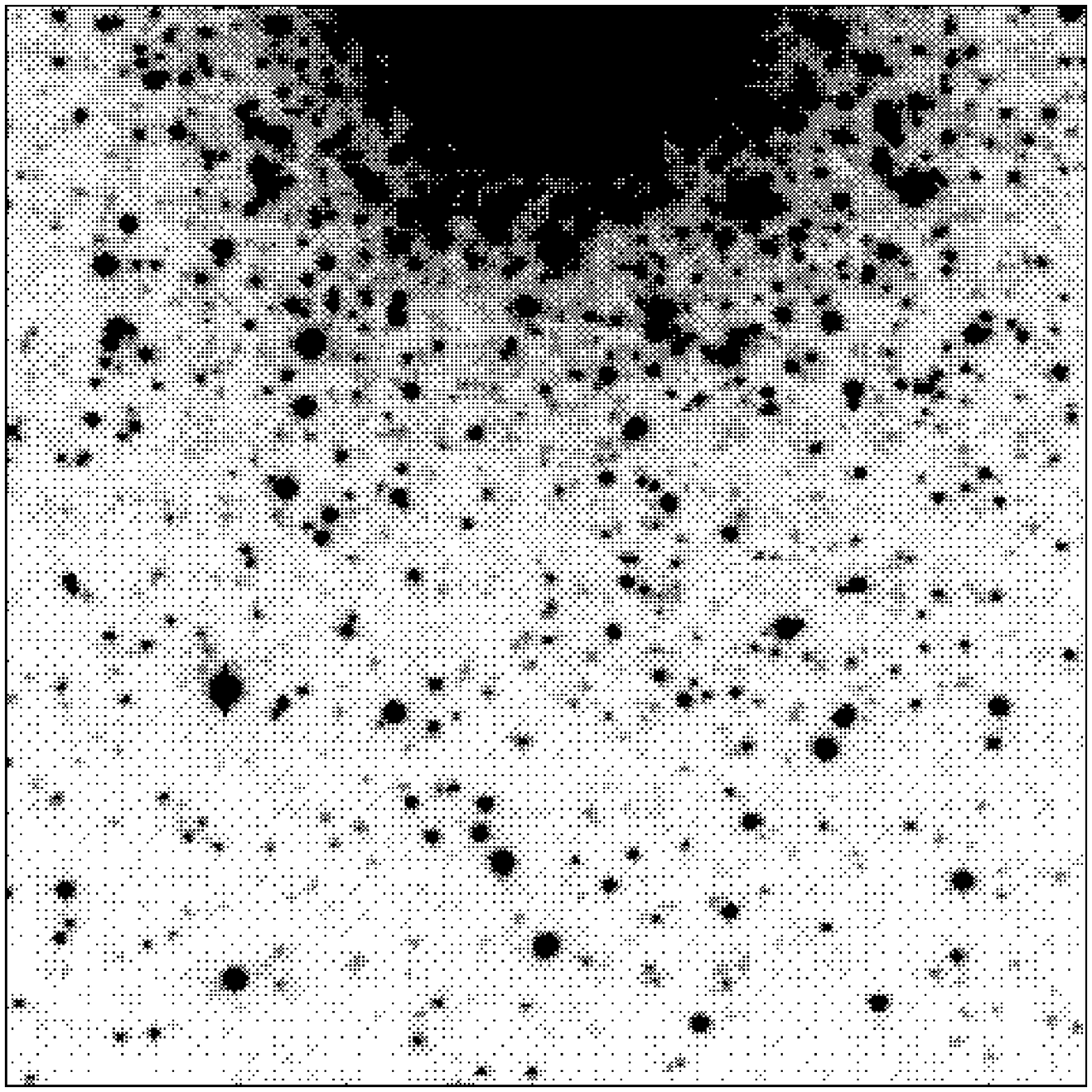,width=4cm}}
\end{tabular}
\end{minipage}
\end{tabular}
\caption[]{CMD and covered field for NGC~6266}
\label{ngc6266}
\end{figure*}

\paragraph {\bf NGC 6266 (M~62).} (Fig.~\ref{ngc6266})

This cluster is located very close to the Galactic center, and it has
a high differential reddening. It seems to have both a RHB and a BHB
resembling the HB of NGC~1851. Previous $B,V$ bands CCD works are in
Caloi et al. (\cite{caloi87}), and Brocato et
al. (\cite{brocato96}). A de-reddened CMD and RR-Lyrae stars are also
studied in Malakhova et al. (\cite{malakhova97}).

\begin{figure*}
\begin{tabular}{c@{}c}
\raisebox{-8cm}{
\psfig{figure=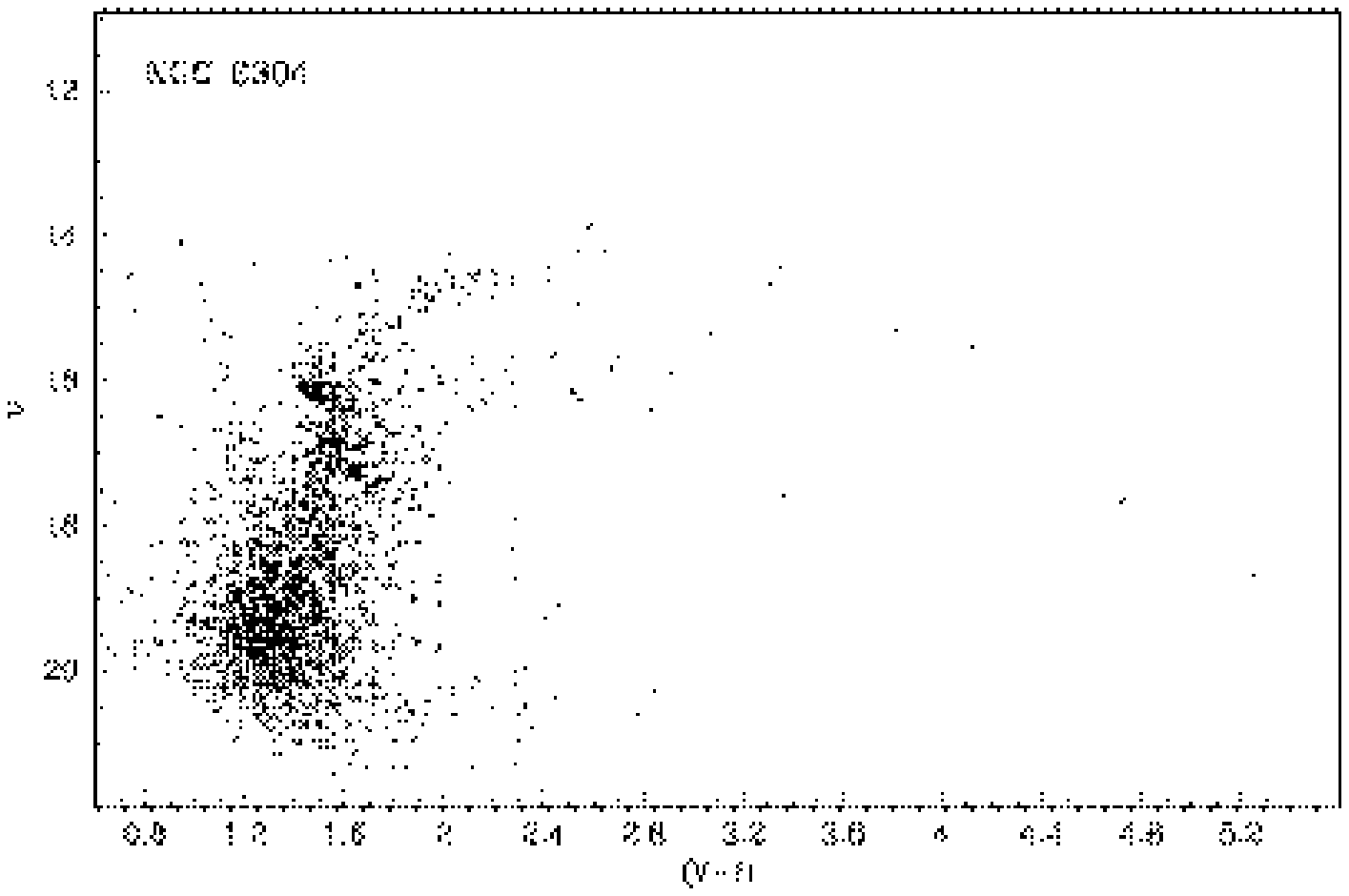,width=16.5cm}
} &
\begin{minipage}[t]{8.8cm}
\begin{tabular}{c@{}c}
\hspace{-4cm}
\fbox{\psfig{figure=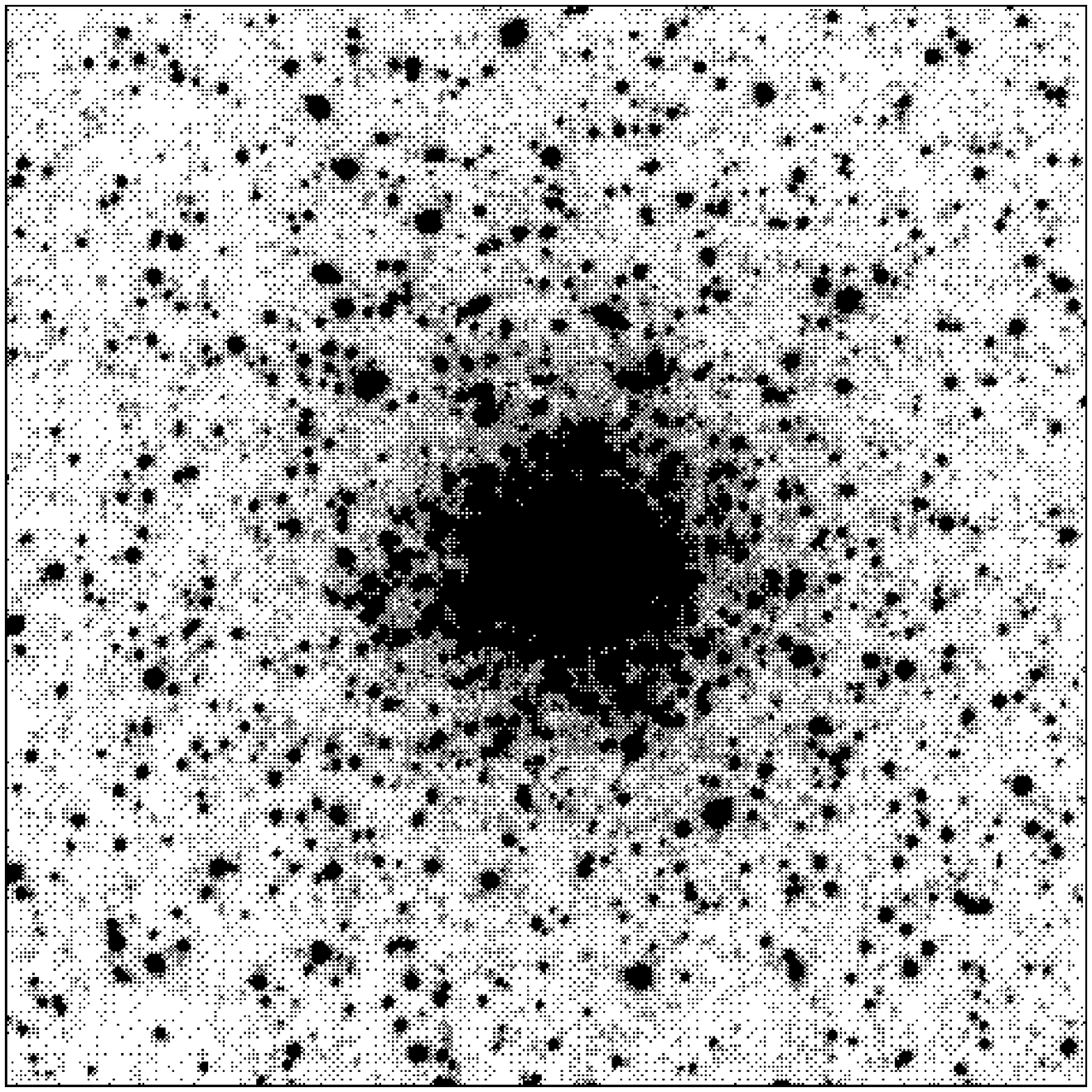,width=4cm}}
\end{tabular}
\end{minipage}
\end{tabular}
\caption[]{CMD and covered field for NGC~6304 (M~62)}
\label{ngc6304}
\end{figure*}

\paragraph {\bf NGC 6304.}  (Fig.~\ref{ngc6304})

NGC~6304 is a high metallicity cluster very close to the Galactic
center, and has one of the highest reddenings in our sample. It has
some disk and bulge star contamination. There is a second RGB fainter
and redder than the main RGB (bulge star contamination or a more
absorbing patch?), but the most noticeable feature is the extremely
long RGB. The reddest star of its RGB is located $\sim$3.7~mag redward
from the RHB! To our knowledge, this is the most extended RGB known
for a GGC. The most recent CCD CMD for this cluster comes from the $V$
and $K$ photometry by Davidge et al. (\cite{davidge92}) which covers
the hottest RGB stars and the HB.

\begin{figure*}
\begin{tabular}{c@{}c}
\raisebox{-6cm}{
\psfig{figure=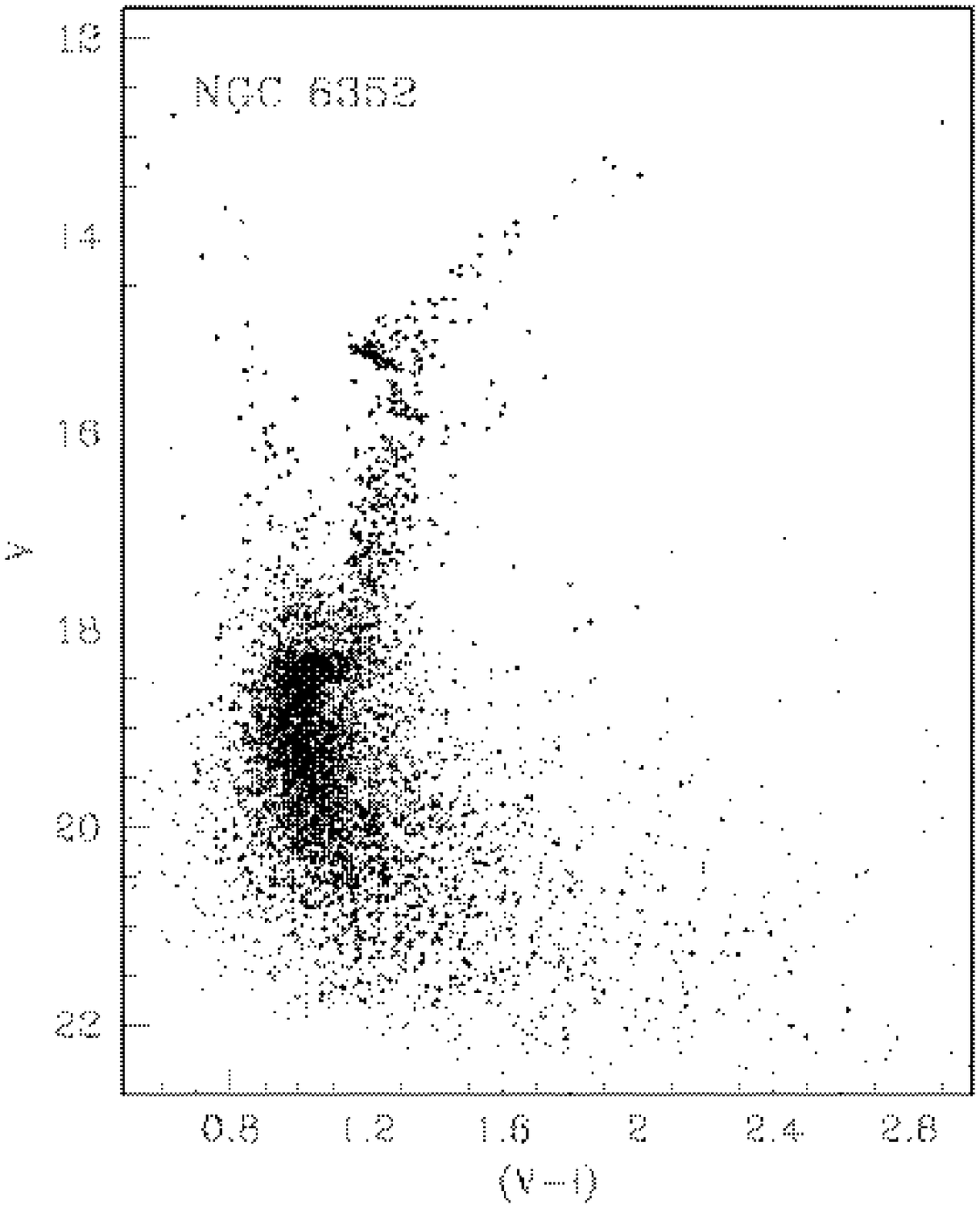,width=8.8cm}
} &
\begin{minipage}[t]{8.8cm}
\begin{tabular}{c@{}c}
\fbox{\psfig{figure=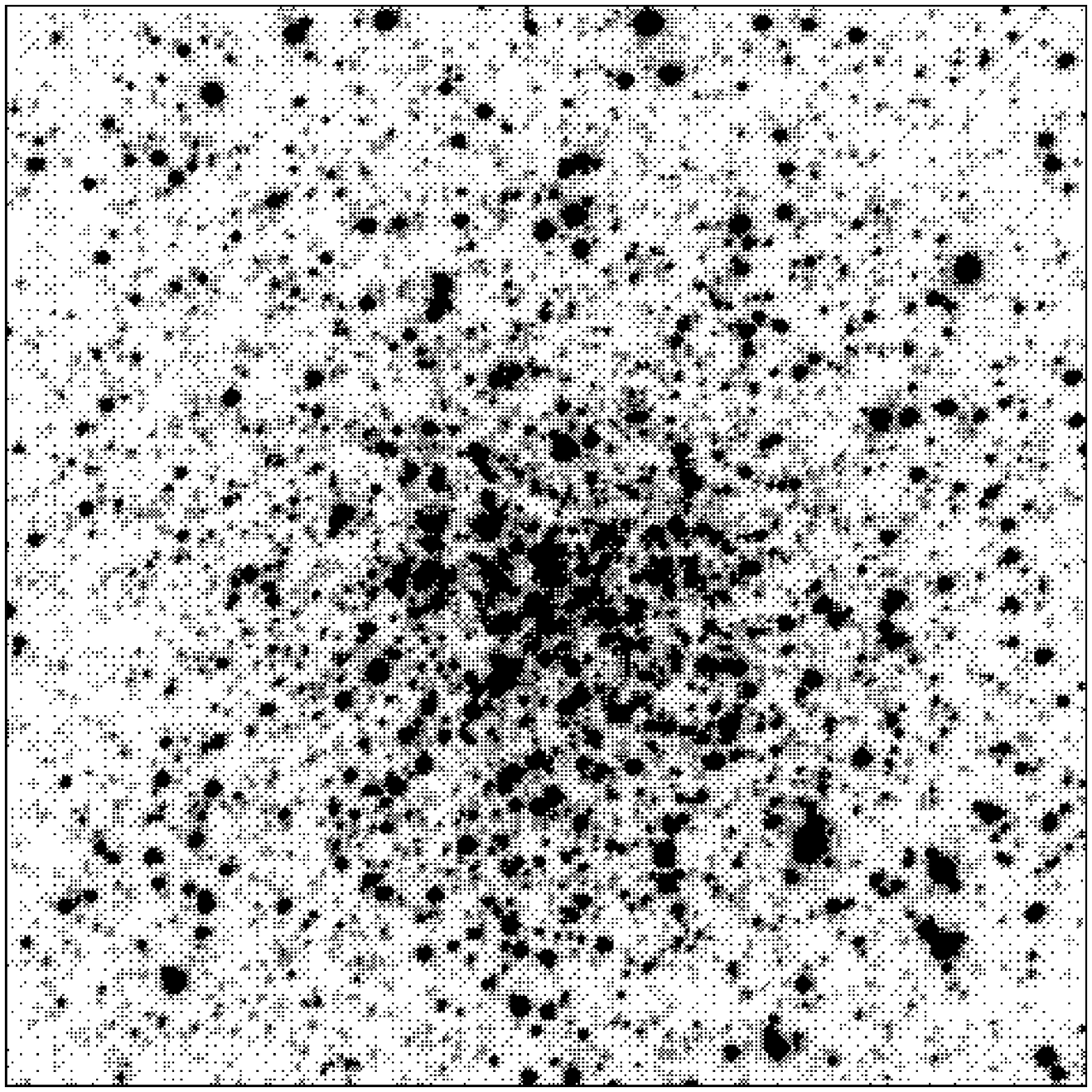,width=4cm}}
\end{tabular}
\end{minipage}
\end{tabular}
\caption[]{CMD and covered field for NGC~6352}
\label{ngc6352}
\end{figure*}

\paragraph {\bf NGC 6352.}  (Fig.~\ref{ngc6352})

NGC~6352 is another high metallicity bulge GGC, with a CMD typical of
a cluster with this metal content. The most recent CCD study on this
cluster is in Fullton et al. (\cite{fullton95}), where a $VI_{\rm c}$ CMD
from HST data combined with ground-based observations is
presented. Another study of the RGB and HB regions of this cluster is
presented by Sarajedini \& Norris (\cite{sarajedininorris94}) in the
$B,V$ bands.

\begin{figure*}
\begin{tabular}{c@{}c}
\raisebox{-6cm}{
\psfig{figure=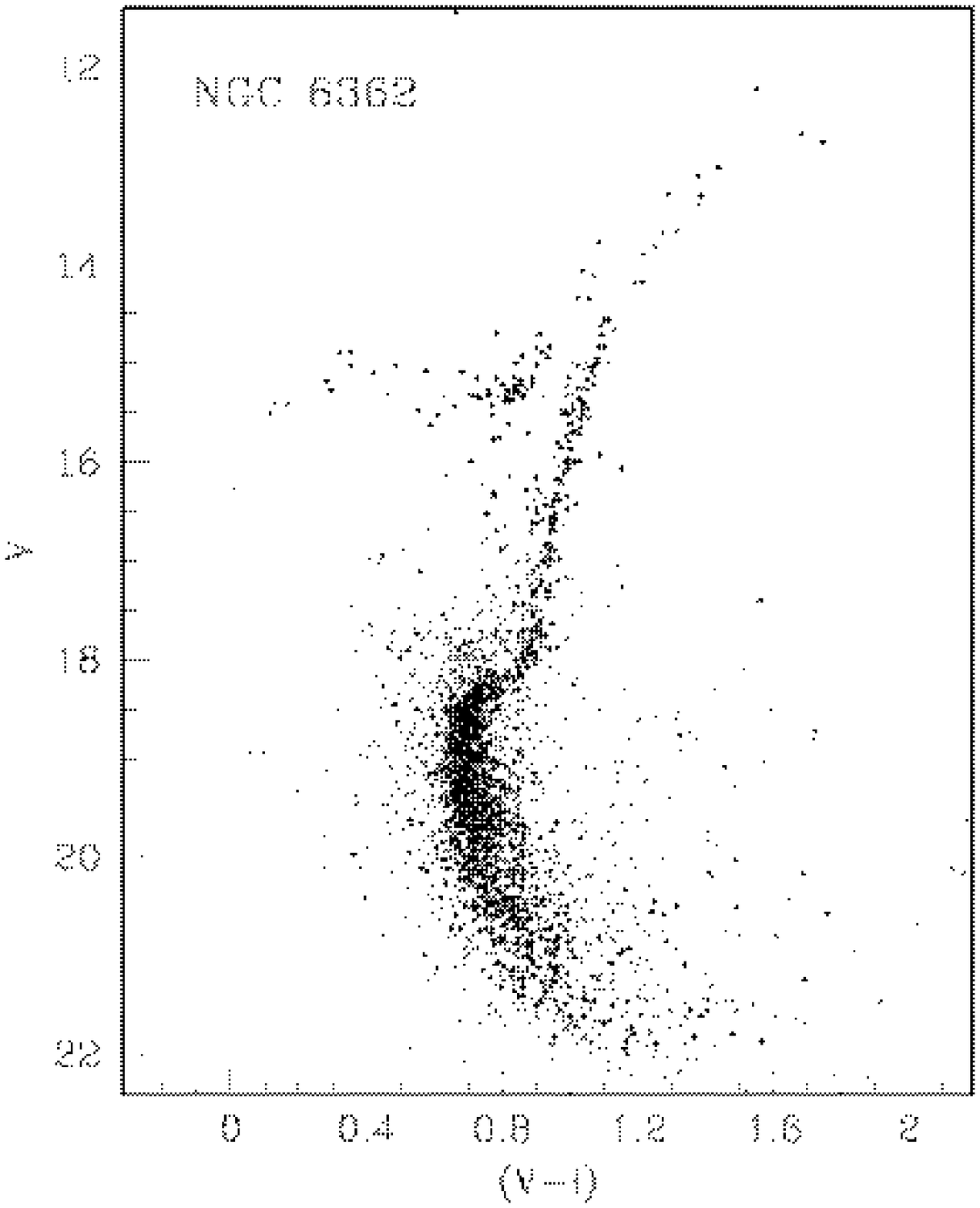,width=8.8cm}
} &
\begin{minipage}[t]{8.8cm}
\begin{tabular}{c@{}c}
\fbox{\psfig{figure=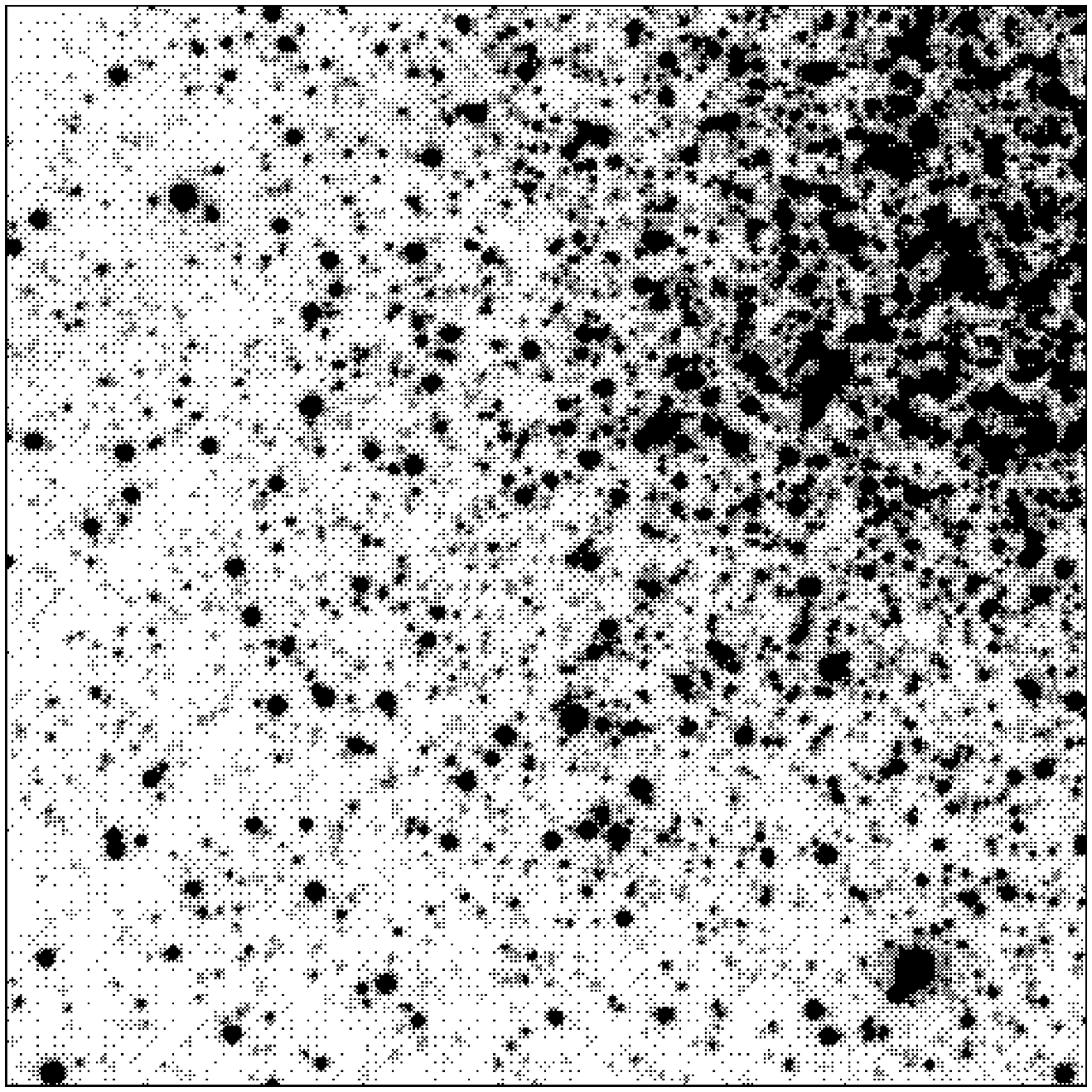,width=4cm}}
\end{tabular}
\end{minipage}
\end{tabular}
\caption[]{CMD and covered field for NGC~6362}
\label{ngc6362}
\end{figure*}

\paragraph {\bf NGC 6362.}  (Fig.~\ref{ngc6362})

NGC~6362 presents a well defined CMD with a bimodal HB. The most
recent CMD on this cluster is given by Piotto et al.
(\cite{piotto99}), who present observations of the center of the
cluster obtained with the HST/WFPC2 camera in the $B$ and $V$
bands. The only previous ground-based CCD photometry is in Alcaino \&
Liller (\cite{alcainoliller86b}). Our field has been also observed in
the same filters by Walker (priv. comm.), who made available to us his
data for a cross-check of the photometric calibration. We find that
the two photometries agree within the errors. In particular, we found
a zero point difference of 0.02~mag for the $V$ band and 0.01~mag for
the $I$ band, with a negligible -0.001 color term difference between
Walker and our data. These discrepancies are well within the
uncertainties, and allow to further confirm our calibration to the
standard (Landolt \cite{landolt92}) system.

\begin{figure*}
\begin{tabular}{c@{}c}
\raisebox{-6cm}{
\psfig{figure=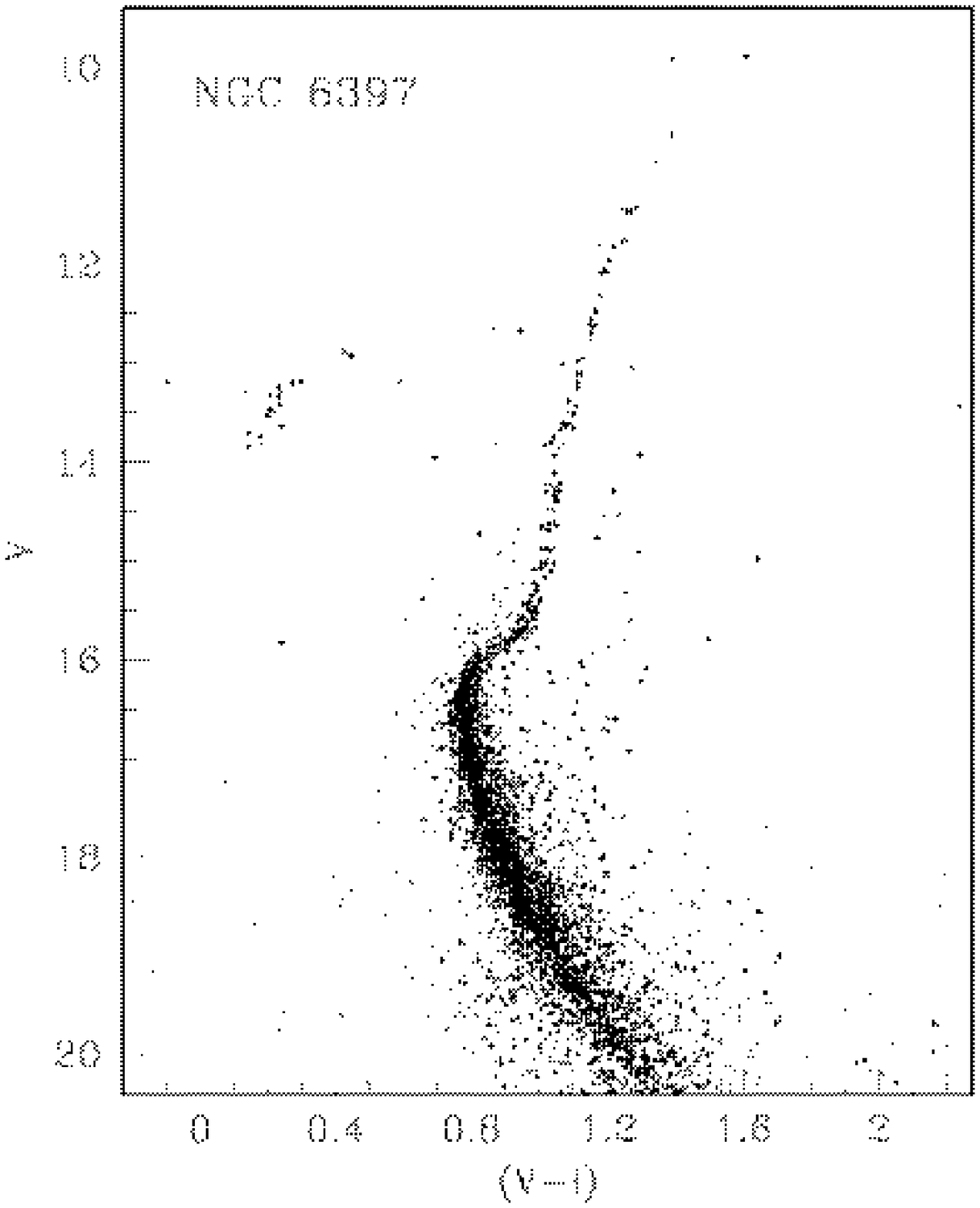,width=8.8cm}
} &
\begin{minipage}[t]{8.8cm}
\begin{tabular}{c@{}c}
\fbox{\psfig{figure=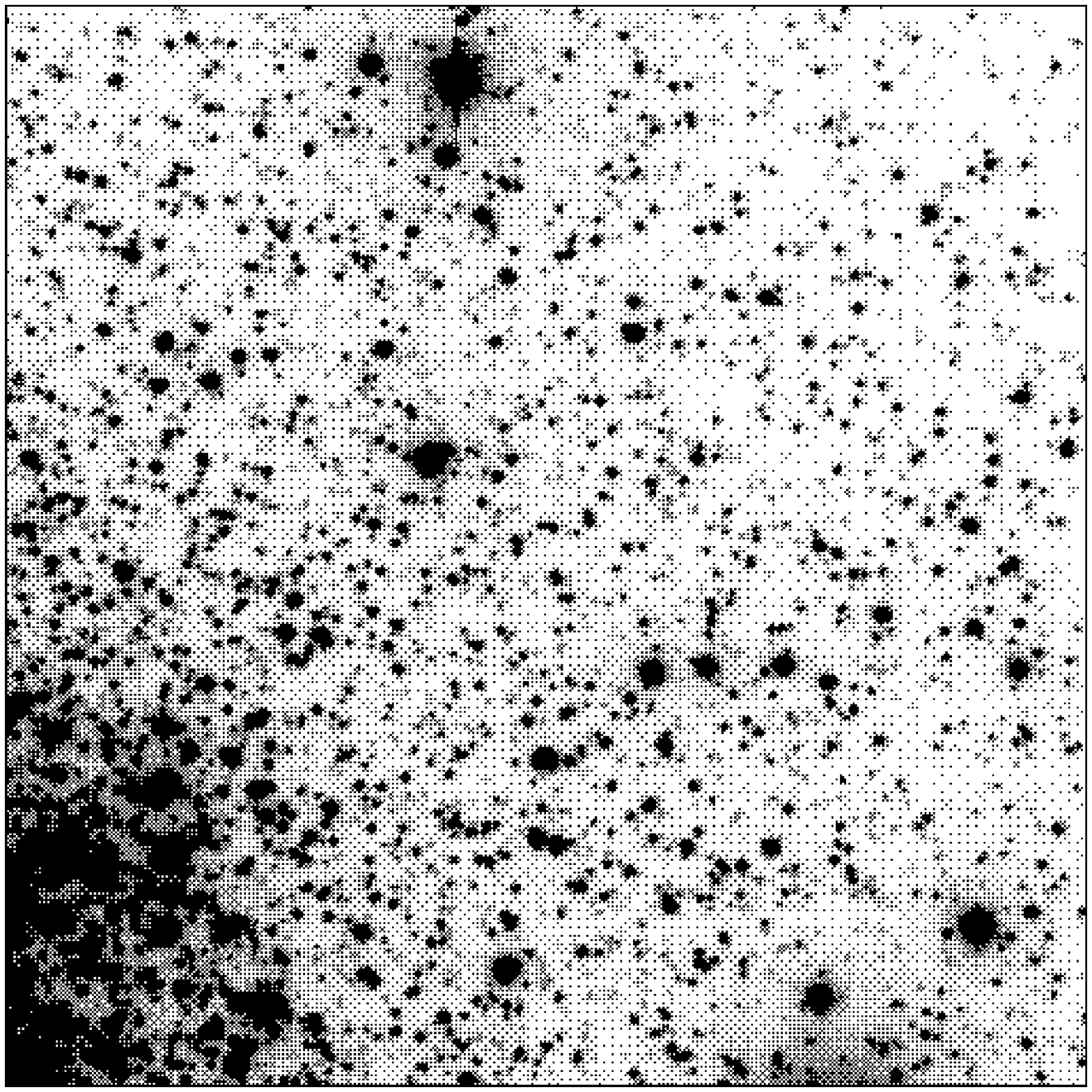,width=4cm}} \\
\fbox{\psfig{figure=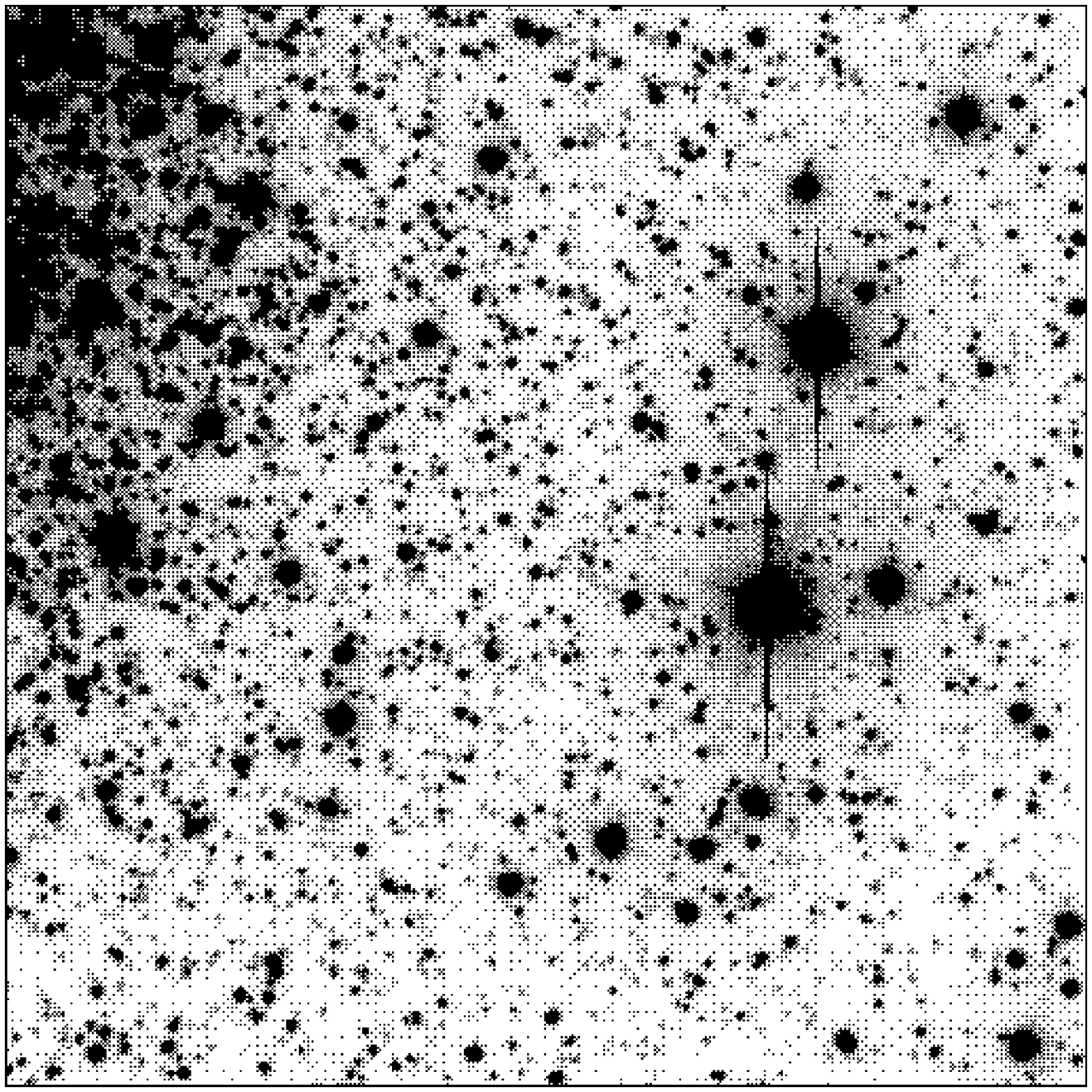,width=4cm}} 
\end{tabular}
\end{minipage}
\end{tabular}
\caption[]{CMD and covered fields for NGC~6397}
\label{ngc6397}
\end{figure*}

\paragraph {\bf NGC 6397.}  (Fig.~\ref{ngc6397})

This cluster is the GGC with the smallest apparent distance
modulus. Cool et al. (\cite{cool96}) and King et al.
(\cite{king98}) present an extremely well defined 
CMD of the main sequence of this cluster, from HST
data, from just below the TO down to $I=24.5$, which correspond to a
mass of less than $0.1 M_\odot$. Other HST studies on this cluster have
been presented by Burgarella et al. (\cite{burgarella94}), De Marchi
\& Paresce (\cite{demarchiparesce94}), Cool et al. (\cite{cool95}) and
King et al. (\cite{king95}). Many ground-based CCD data have also
been published: Auriere et al. (\cite{auriere90}), Anthony-Twarog et
al. (\cite{anthonytwarog92}) (Stromgren photometry), Lauzeral et
al. (\cite{lauzeral92}, \cite{lauzeral93}), Kaluzny (\cite{kaluzny97b})
($B,V$ bands) and Alcaino et al. (\cite{alcaino87}: $B,V$ bands;
\cite{alcaino97b}: $U,B,V,I$ bands).

\begin{figure*}
\begin{tabular}{c@{}c}
\raisebox{-6cm}{
\psfig{figure=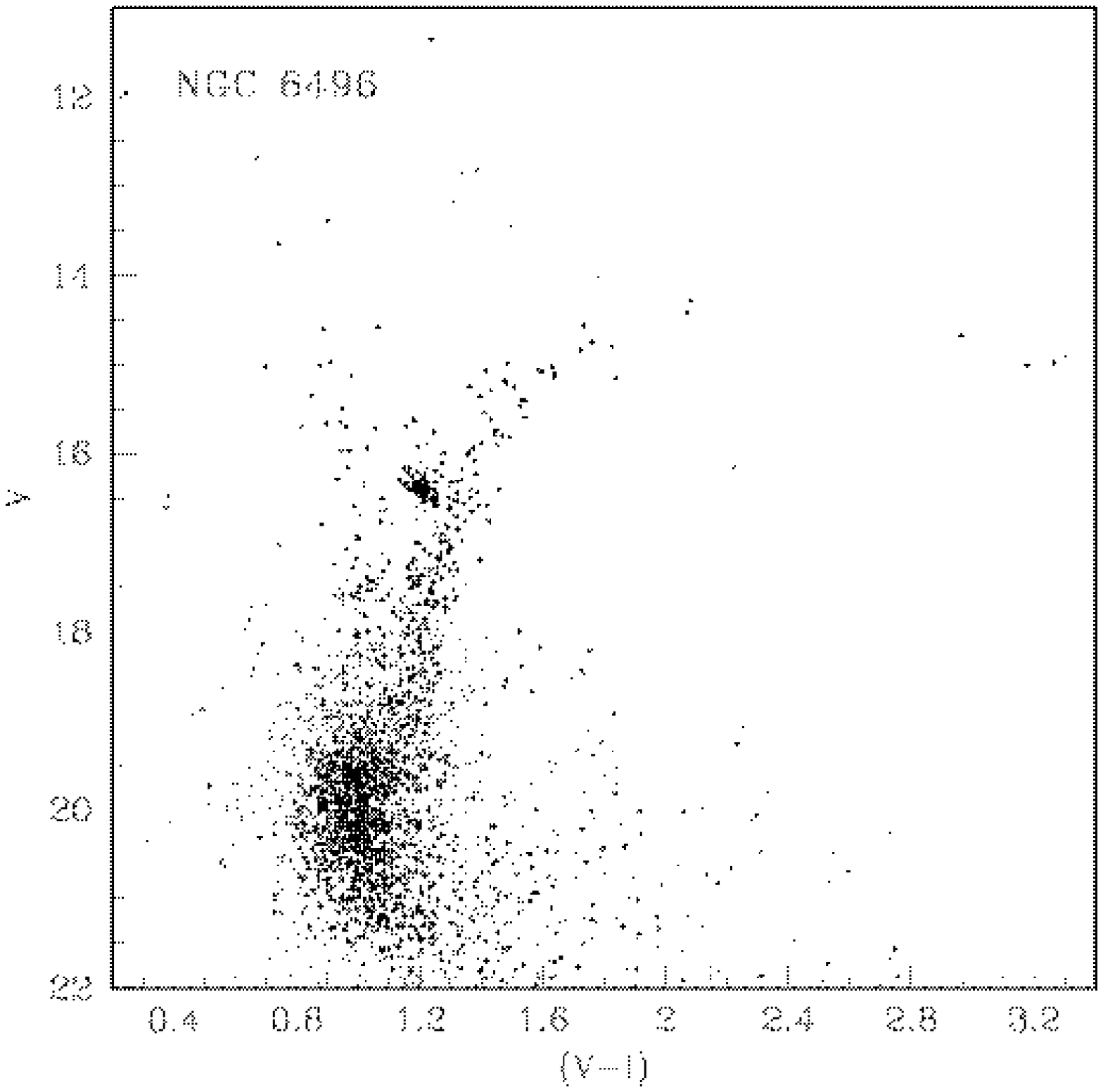,width=10.9cm}
} &
\begin{minipage}[t]{8.8cm}
\begin{tabular}{c@{}c}
\fbox{\psfig{figure=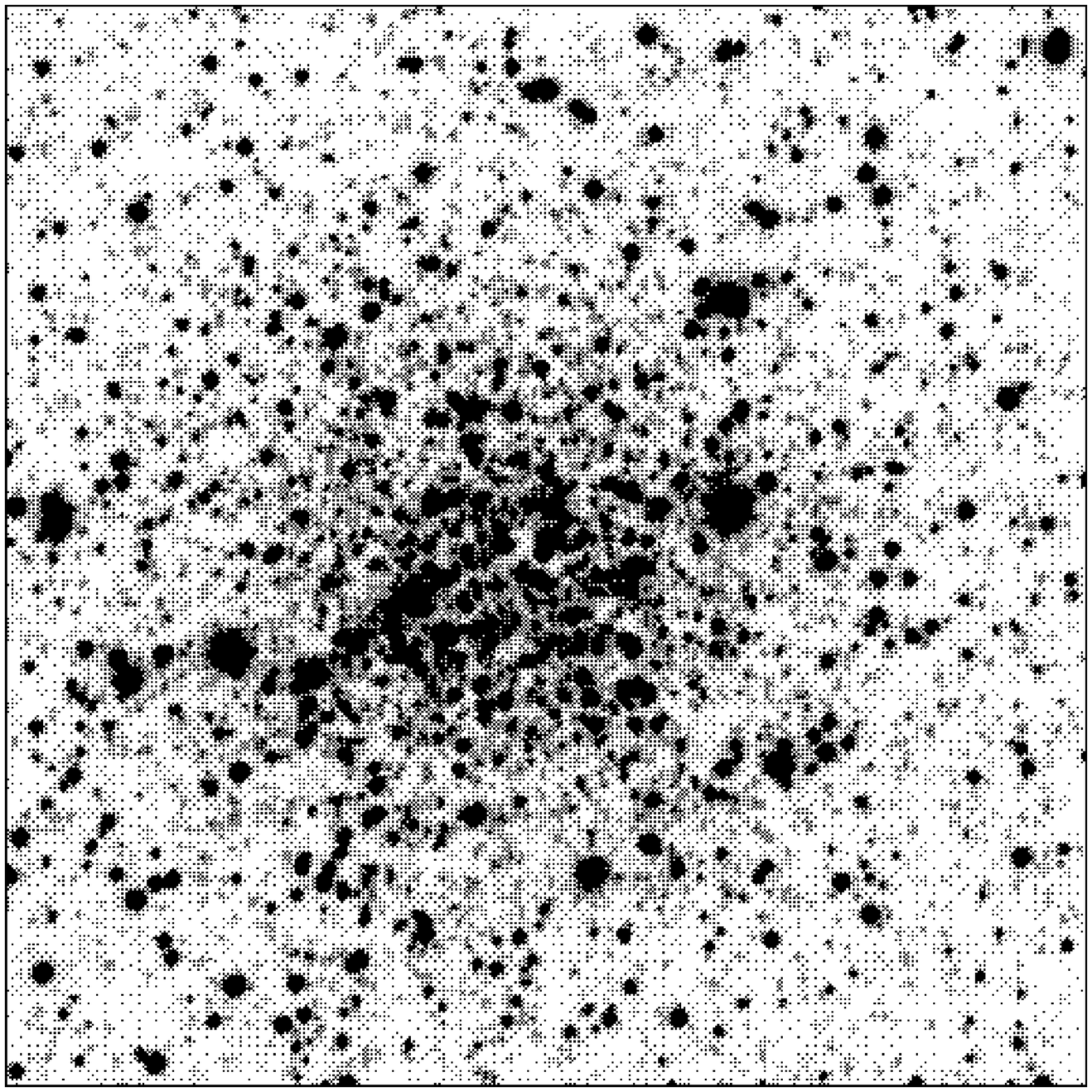,width=4cm}}
\end{tabular}
\end{minipage}
\end{tabular}
\caption[]{CMD and covered field for NGC~6496}
\label{ngc6496}
\end{figure*}

\paragraph {\bf NGC 6496.}  (Fig.~\ref{ngc6496})

NGC~6496 is another metal rich GGC which presents an extended RGB. In
this case, the reddest stars are $\sim$ 2~mag redder than the RHB. It
has also a remarkably tilted RHB, already noted by Richtler et al.
(\cite{richtler94}), who present a CCD $(B,V)$ photometry of this
cluster; Armandroff (\cite{armandroff88}) gives $(V,I)$ CCD
photometry. A tilted RHB can be noted not only in this CMD, but also
in the CMDs of most of the very metal-rich clusters of our
sample. Such a feature is usually not present in the canonical models.
The RHB is well populated, and there are two stars located on the BHB
region. This is quite unusual considering the metallicity of NGC~6496,
and it would be interesting to study the membership and to obtain a
CMD on a larger field. Another CCD photometry of this cluster is in
Friel \& Geisler (\cite{frielgeisler91}) in the Washington
system. Sarajedini \& Norris (\cite{sarajedininorris94}) present a $B$
and $V$ photometry for the RGB and HB region.

\begin{figure*}
\begin{tabular}{c@{}c}
\raisebox{-6cm}{
\psfig{figure=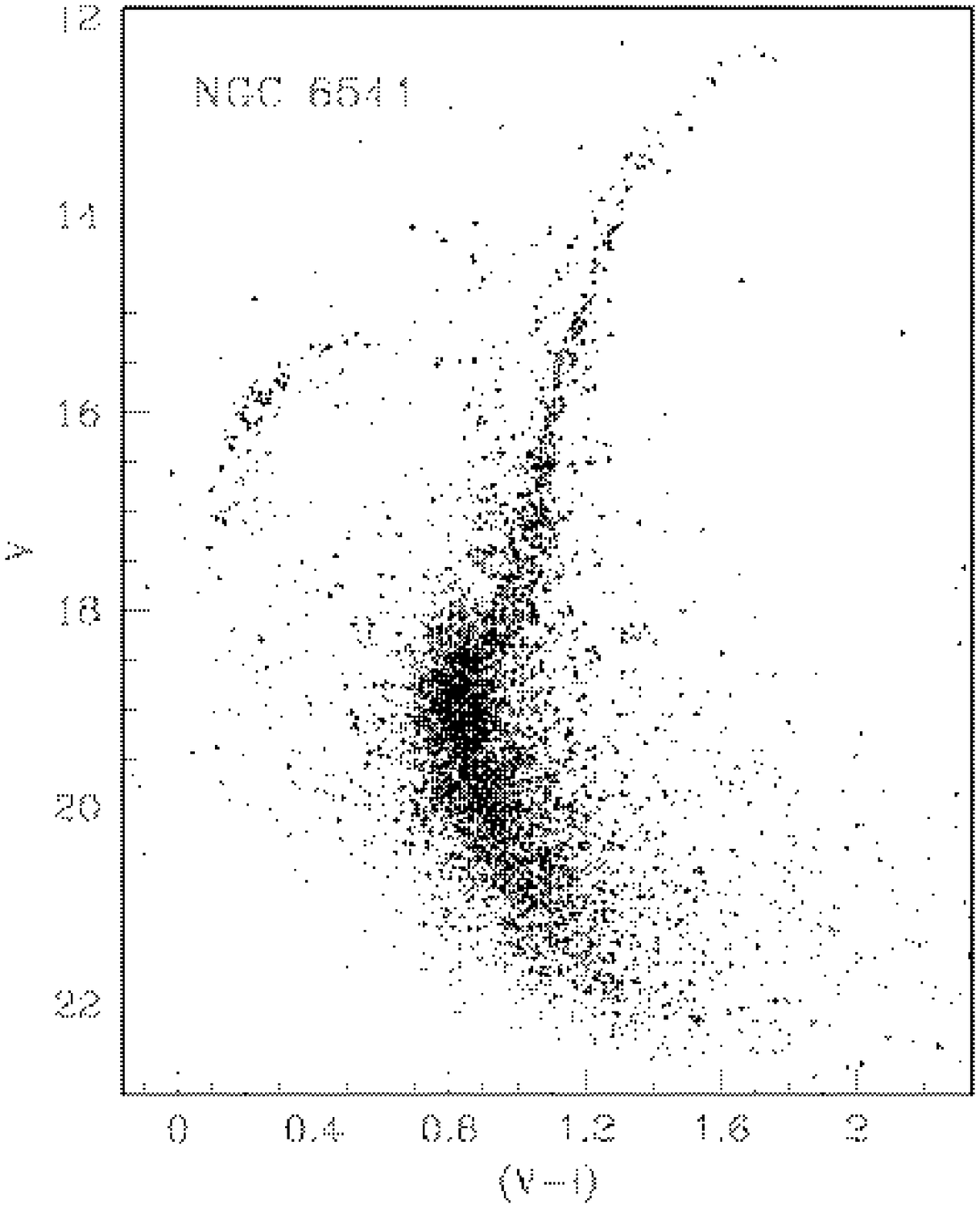,width=8.8cm}
} &
\begin{minipage}[t]{8.8cm}
\begin{tabular}{c@{}c}
\fbox{\psfig{figure=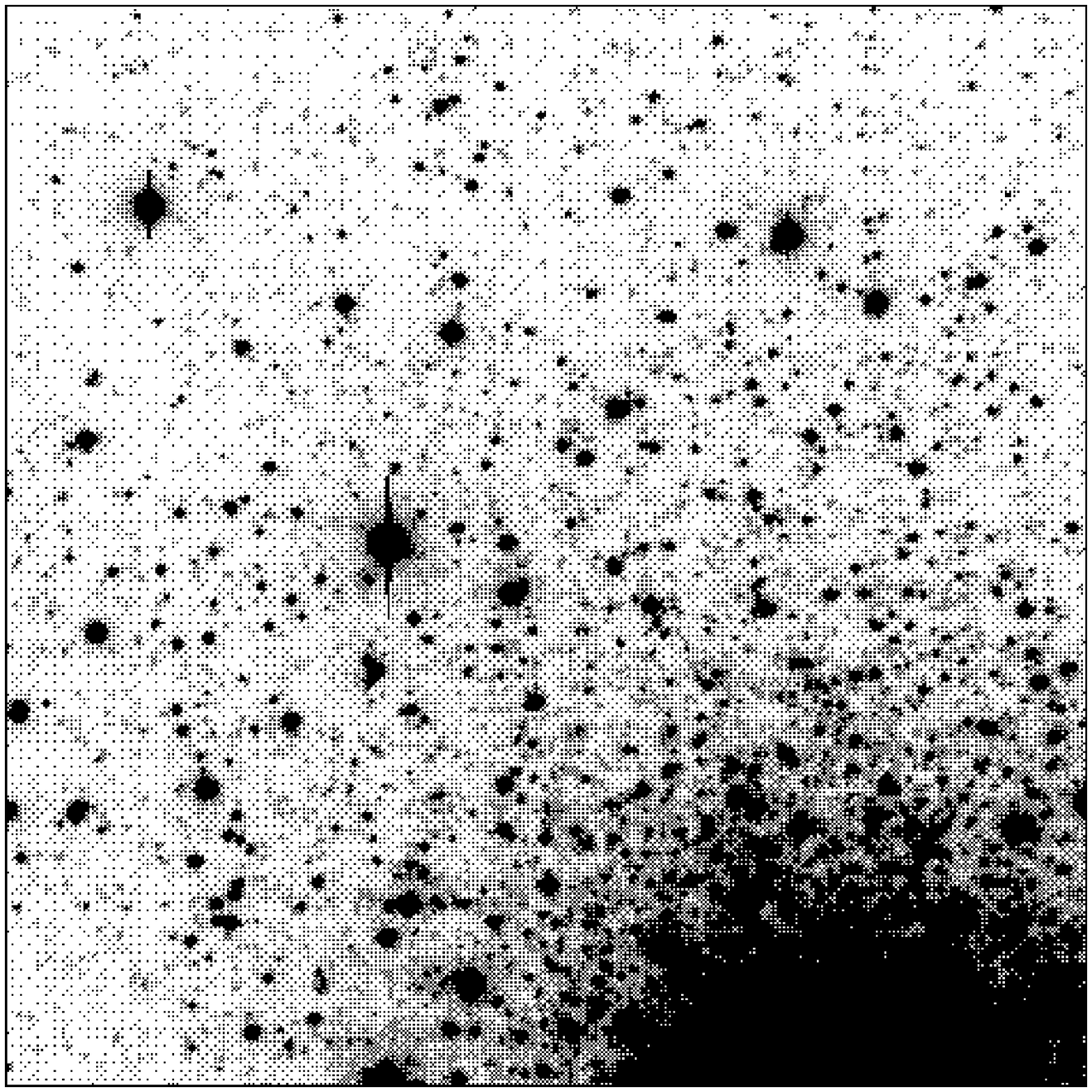,width=4cm}} &
\fbox{\psfig{figure=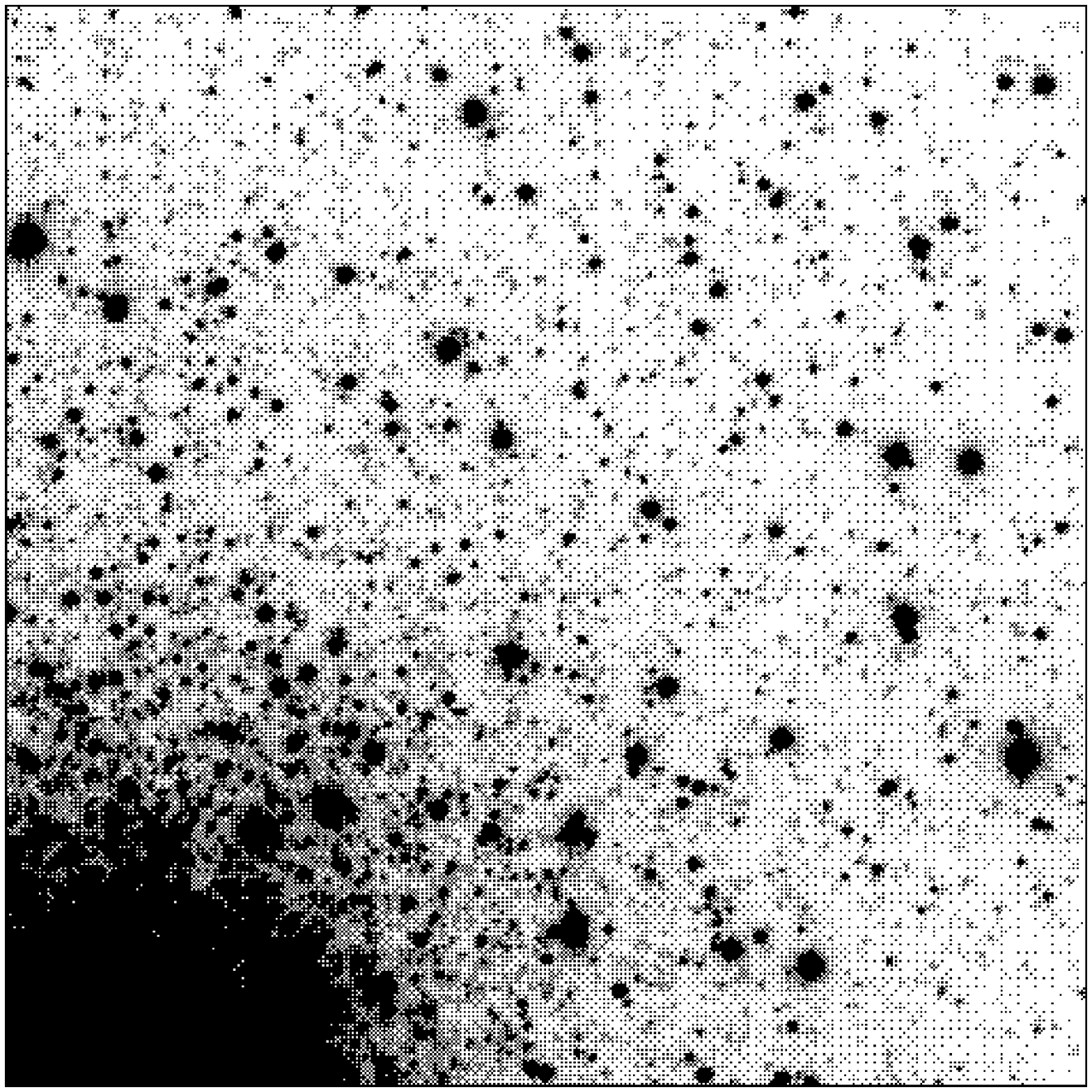,width=4cm}} 
\end{tabular}
\end{minipage}
\end{tabular}
\caption[]{CMD and covered fields for NGC~6541}
\label{ngc6541}
\end{figure*}

\paragraph {\bf NGC 6541.} (Fig.~\ref{ngc6541})

NGC 6541 is located rather close to the Galactic center, and this
explains the high field star contamination of the CMD. It has a BHB, as
expected from its metal content. The only previous CCD study of this
cluster is the multicolor photometry by Alcaino et
al. (\cite{alcaino97c}).

\begin{figure*}
\begin{tabular}{c@{}c}
\raisebox{-6cm}{
\psfig{figure=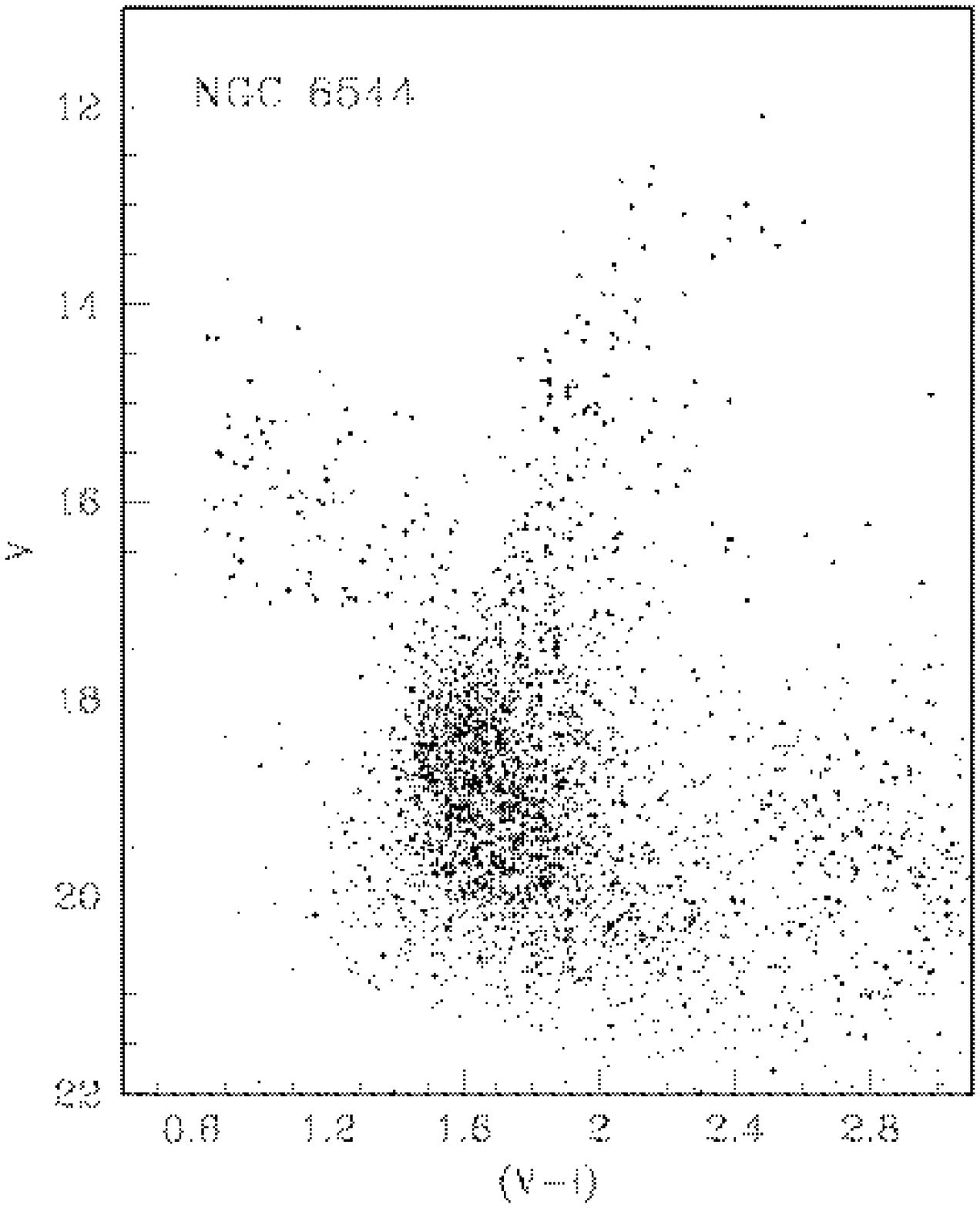,width=8.8cm}
} &
\begin{minipage}[t]{8.8cm}
\begin{tabular}{c@{}c}
\fbox{\psfig{figure=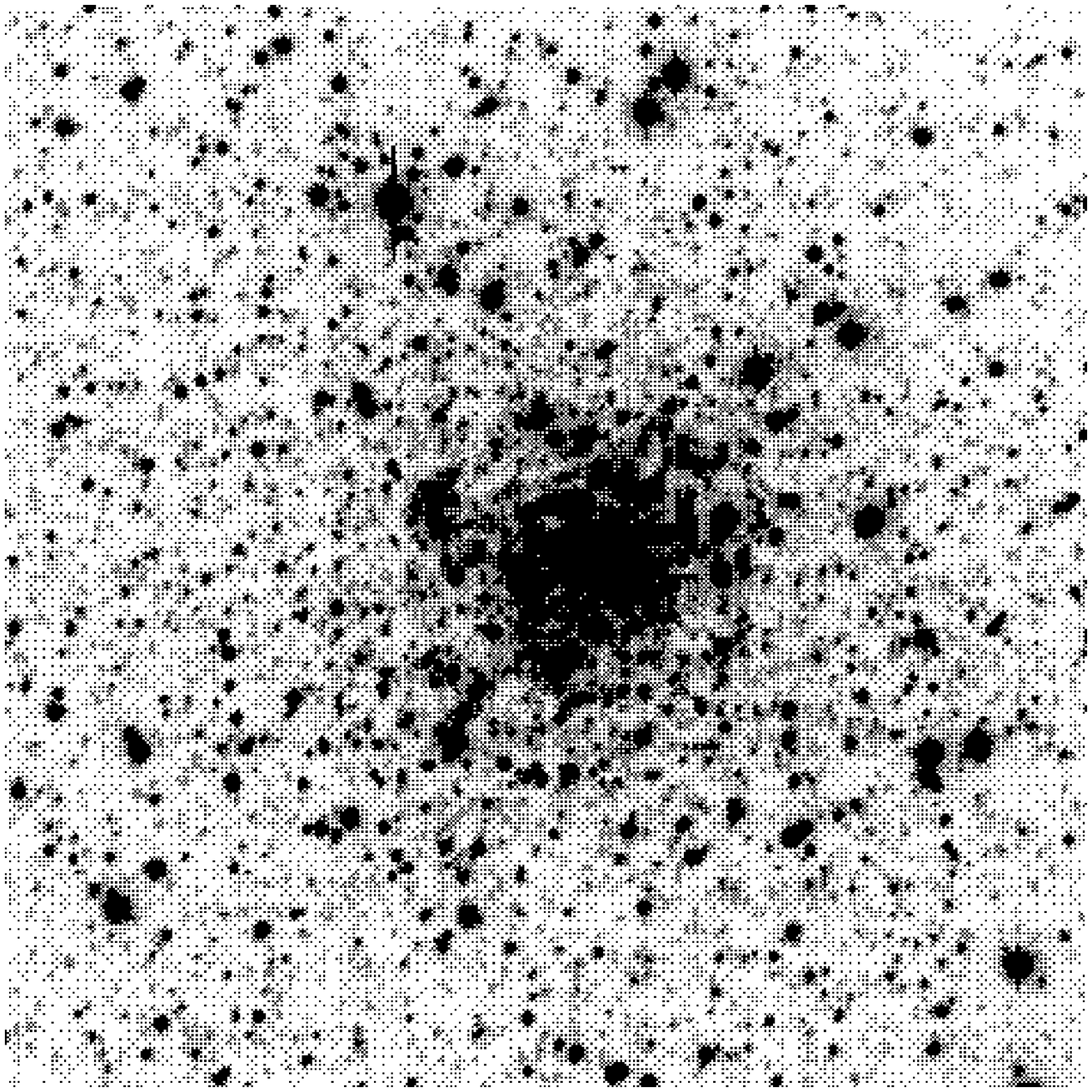,width=4cm}}
\end{tabular}
\end{minipage}
\end{tabular}
\caption[]{CMD and covered field for NGC~6544}
\label{ngc6544}
\end{figure*}

\paragraph {\bf NGC 6544.} (Fig.~\ref{ngc6544})

This is an example of a terrible ``spotty'' field with a high (the
highest in our sample) and highly differential reddening, due to the
location of NGC~6544, which is very close to the Galactic plane and
projected towards the Galactic center. Interestingly enough, despite
its intermediate metallicity, there are only BHB stars. Probably, the
use of the HST in this case is almost inevitable if we want to
estimate the age of this kind of clusters. We have not found any
previous CCD photometry of this cluster.

\begin{figure*}
\begin{tabular}{c@{}c}
\raisebox{-6cm}{
\psfig{figure=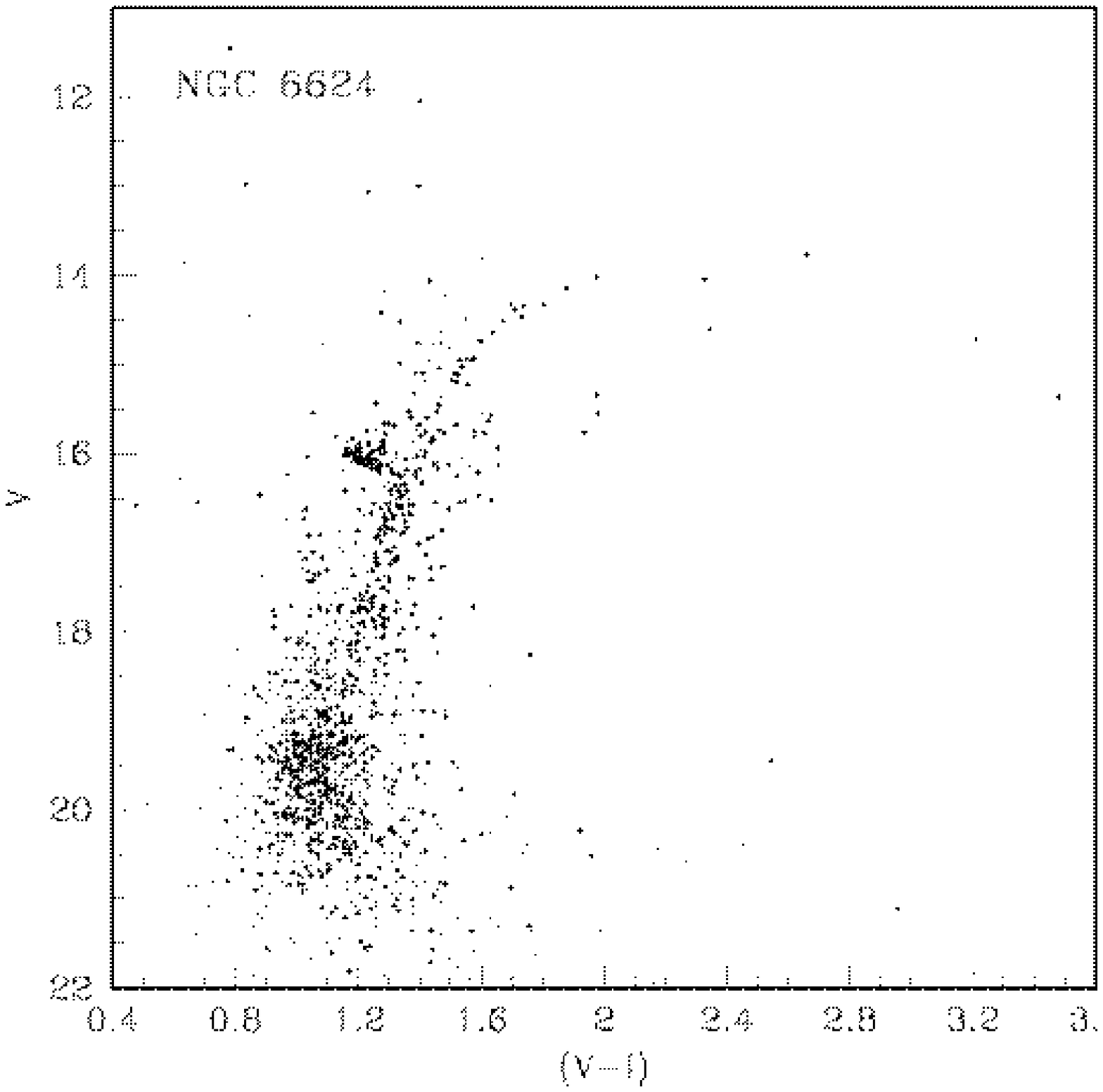,width=10.9cm}
} &
\begin{minipage}[t]{8.8cm}
\begin{tabular}{c@{}c}
\fbox{\psfig{figure=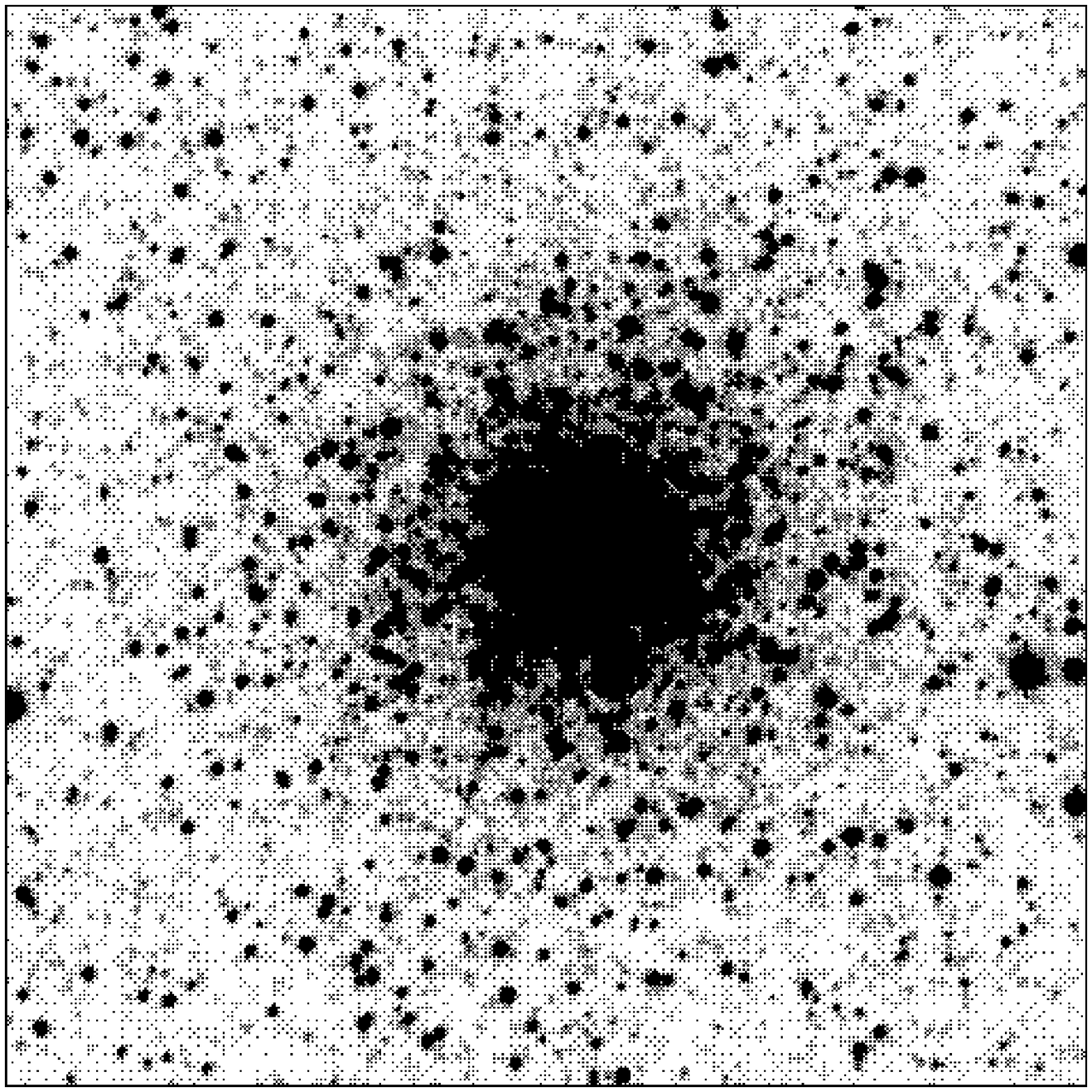,width=4cm}}
\end{tabular}
\end{minipage}
\end{tabular}
\caption[]{CMD and covered field for NGC~6624}
\label{ngc6624}
\end{figure*}

\paragraph {\bf NGC 6624.} (Fig.~\ref{ngc6624})

Another member of the metal-rich group is presented in
Fig.~\ref{ngc6624}. Despite of being the cluster closest to the Galactic
center, NGC~6624 has a moderate field star contamination, and a very
well
defined RGB and RHB. The reddest stars of the RGB are in this case
$\sim2.2$~mag redder than the RHB. 

Richtler et al. (\cite{richtler94}) present a $B$ and $V$ CCD CMD of
this cluster extending well below the TO, while Sarajedini \& Norris
(\cite{sarajedininorris94}) present a photometric study of the RGB
and HB in the same bands. A $B,V$ CMD from HST data is in Sosin \& King
(\cite{sosinking95}) and Sosin et al. (\cite{sosin97a}).

\begin{figure*}
\begin{tabular}{c@{}c}
\raisebox{-6cm}{
\psfig{figure=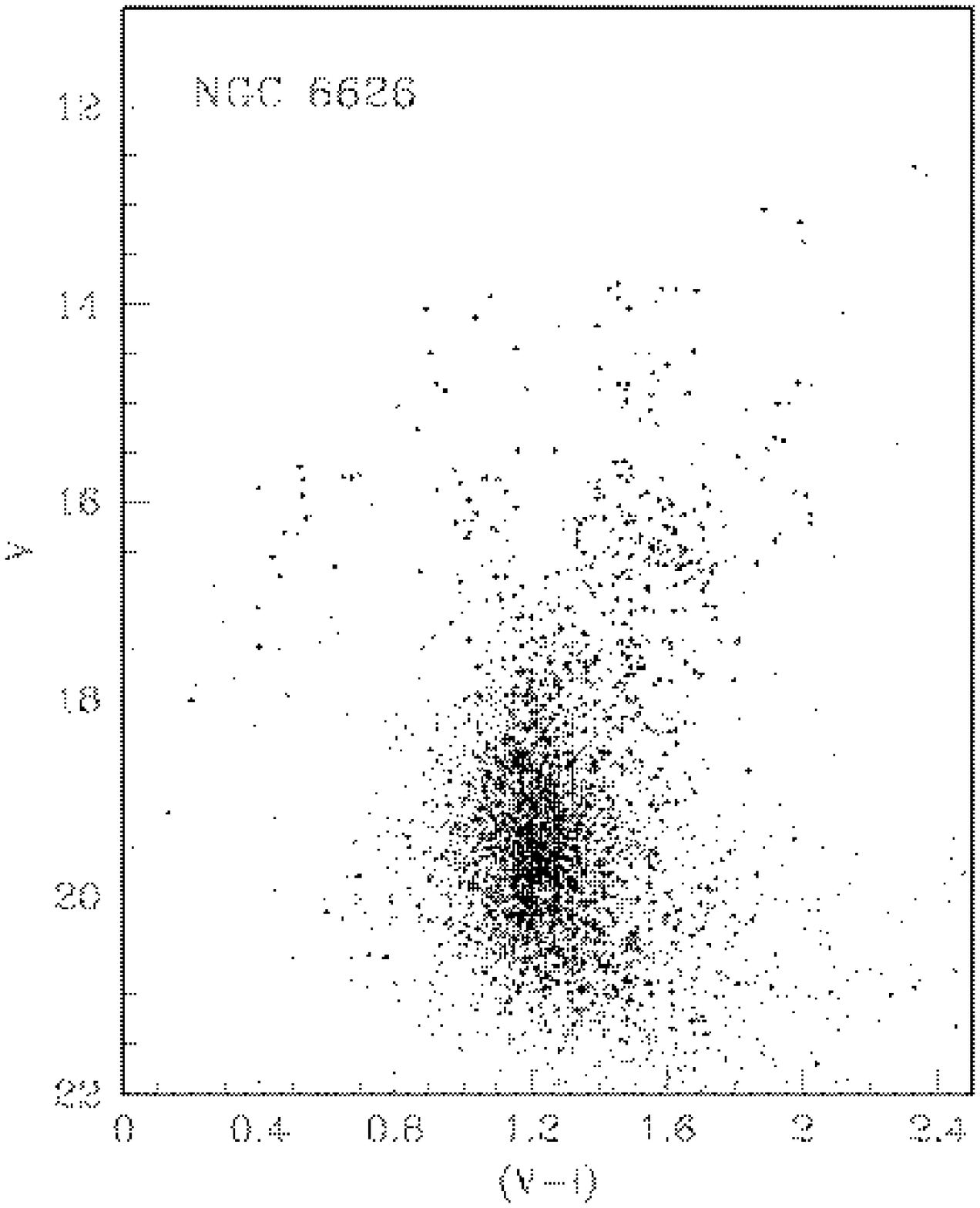,width=8.8cm}
} &
\begin{minipage}[t]{8.8cm}
\begin{tabular}{c@{}c}
\fbox{\psfig{figure=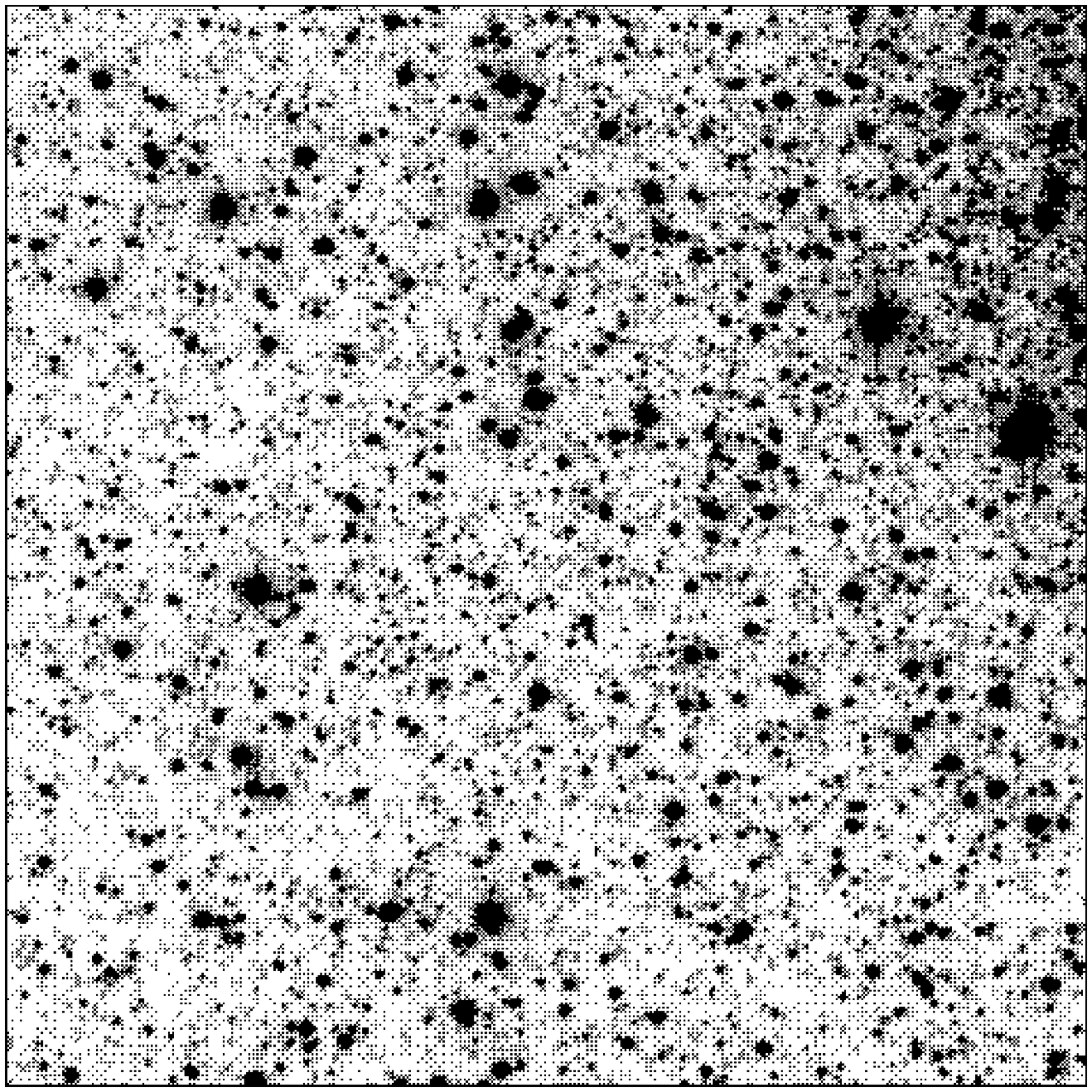,width=4cm}}
\end{tabular}
\end{minipage}
\end{tabular}
\caption[]{CMD and covered field for NGC~6626 (M~28)}
\label{ngc6626}
\end{figure*}

\paragraph {\bf NGC 6626 (M~28).}  (Fig.~\ref{ngc6626})

Again a high differential reddening is present in the field of NGC
6626, which is located close to the Galactic center. NGC 6626 seems to
have an extended BHB, and maybe a few RHB stars, though the field star
contamination makes it rather difficult to see them. Previous CCD photometry
is given by Davidge et al. (\cite{davidge96}), who present a deep
near infrared photometry.

\begin{figure*}
\begin{tabular}{c@{}c}
\raisebox{-6cm}{
\psfig{figure=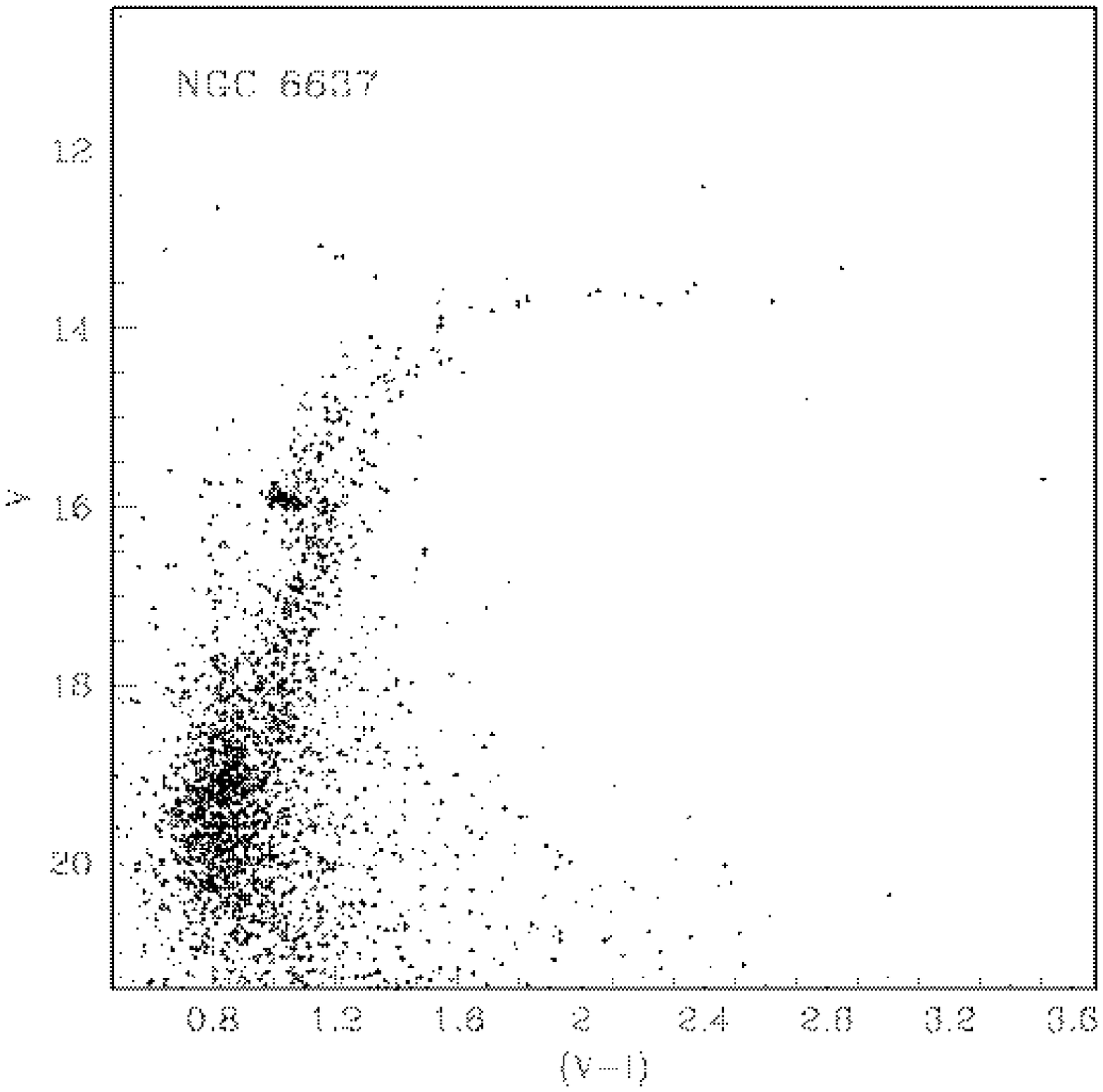,width=10.9cm}
} &
\begin{minipage}[t]{8.8cm}
\begin{tabular}{c@{}c}
\fbox{\psfig{figure=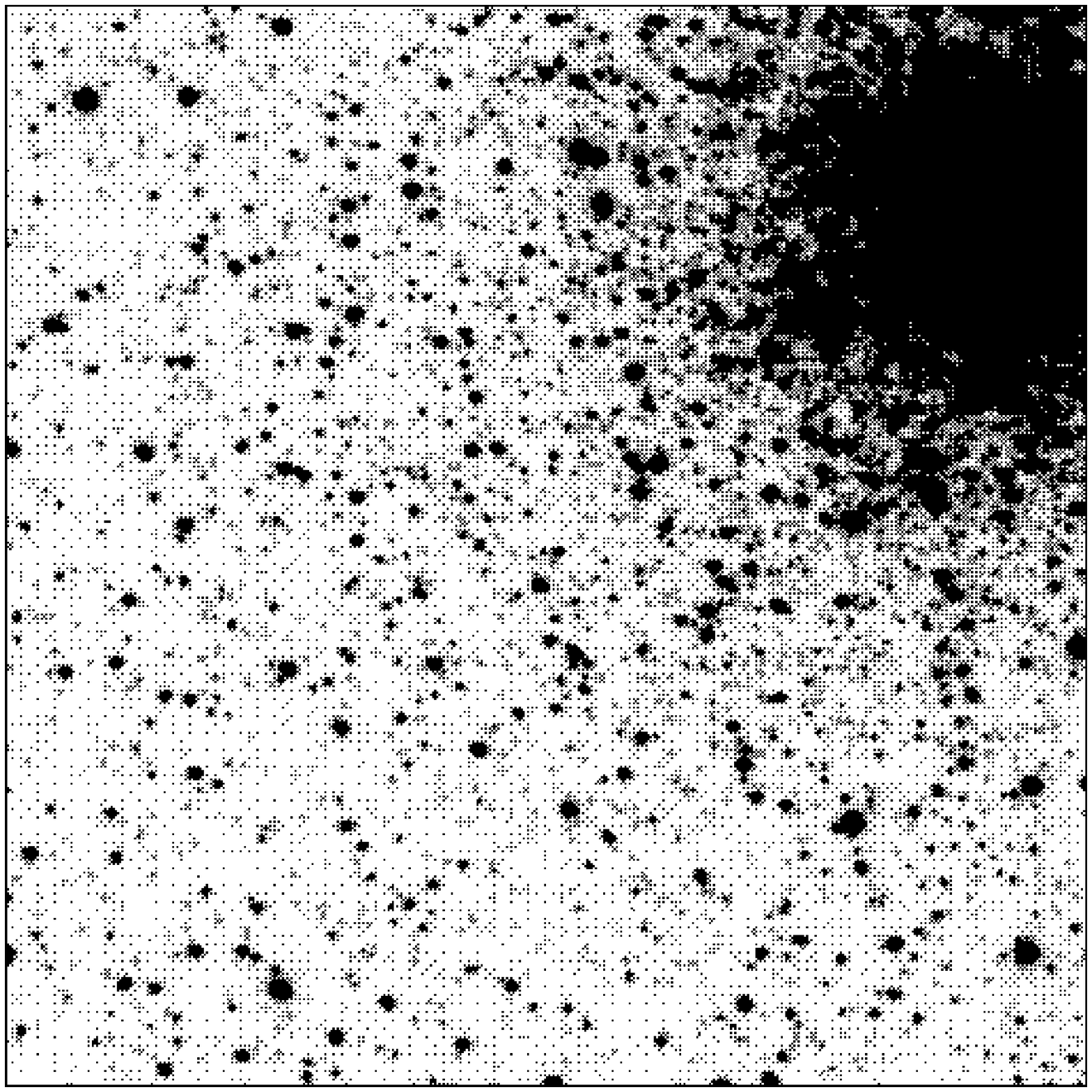,width=4cm}}
\end{tabular}
\end{minipage}
\end{tabular}
\caption[]{CMD and covered field for NGC~6637 (M~69)}
\label{ngc6637}
\end{figure*}

\paragraph {\bf NGC 6637 (M~69).} (Fig.~\ref{ngc6637})

The CMD of NGC 6637 presents the typical distribution in color for the
RGB stars discussed for other metal rich clusters, with the reddest
stars $\sim2.4$~mag redder than the RHB. Previous $B$ and $V$ CCD
photometry is presented by Richtler et al. (\cite{richtler94}), and
the RGB-HB region is also studied by Sarajedini \& Norris
(\cite{sarajedininorris94}) in the same bands.

\begin{figure*}
\begin{tabular}{c@{}c}
\raisebox{-6cm}{
\psfig{figure=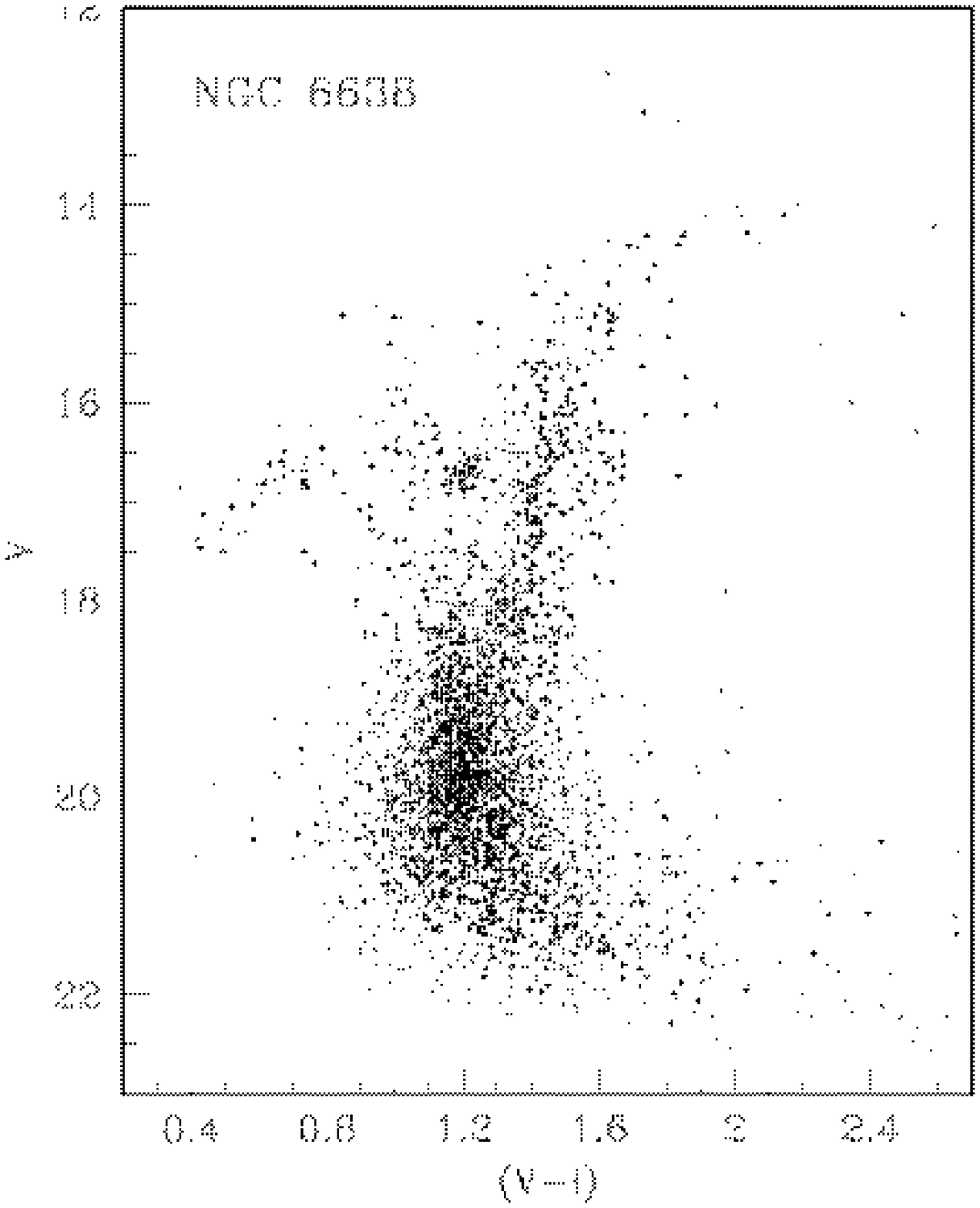,width=8.8cm}
} &
\begin{minipage}[t]{8.8cm}
\begin{tabular}{c@{}c}
\fbox{\psfig{figure=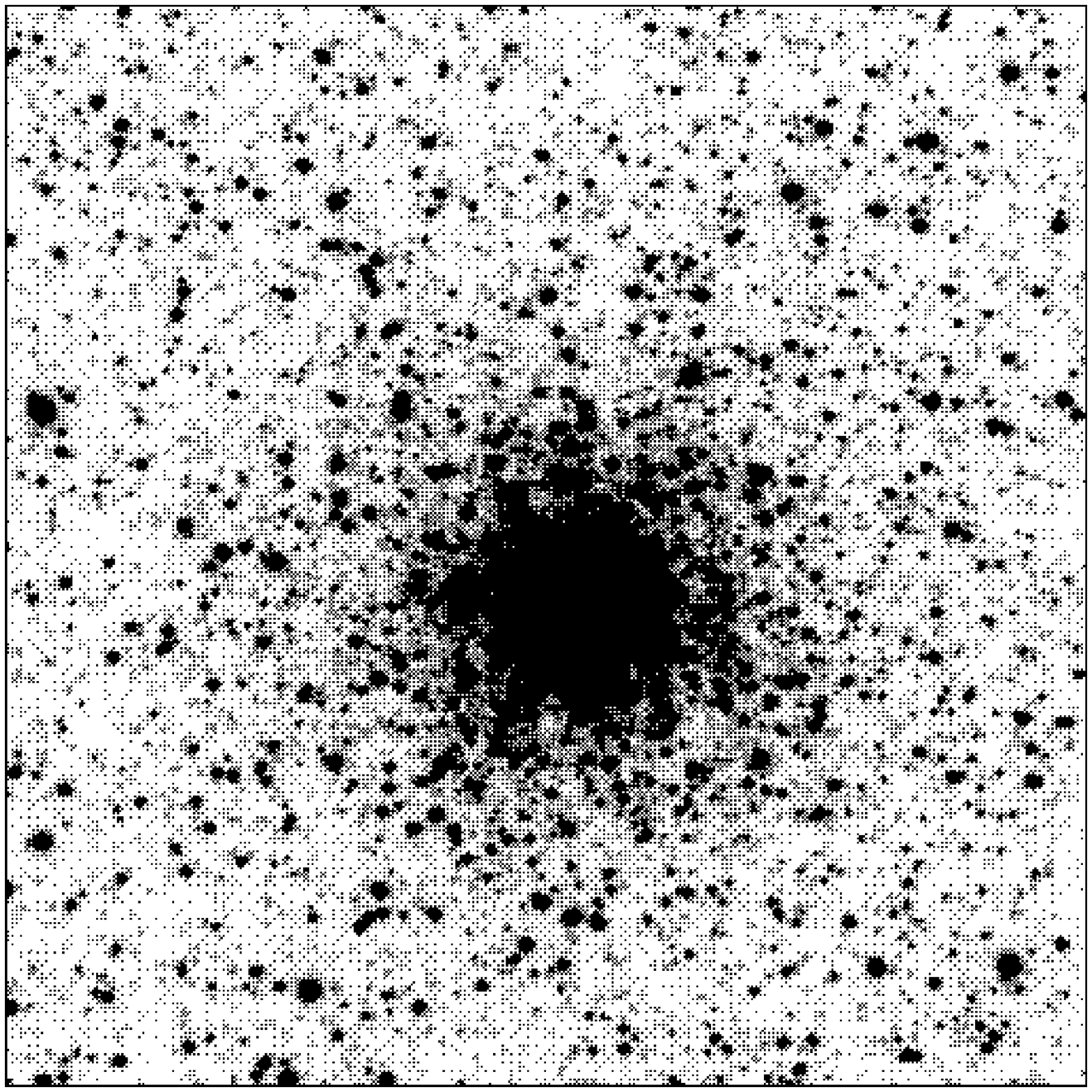,width=4cm}}
\end{tabular}
\end{minipage}
\end{tabular}
\caption[]{CMD and covered field for NGC~6638}
\label{ngc6638}
\end{figure*}

\paragraph {\bf NGC 6638.}  Fig.~\ref{ngc6638}

Affected by high differential reddening, the CMD this cluster is not
very well defined. However, the HB is clearly populated on both sides
of the instability strip, and probably there are many RR-Lyrae. We
have not found any previous CCD photometries of this cluster.

\begin{figure*}
\begin{tabular}{c@{}c}
\raisebox{-6cm}{
\psfig{figure=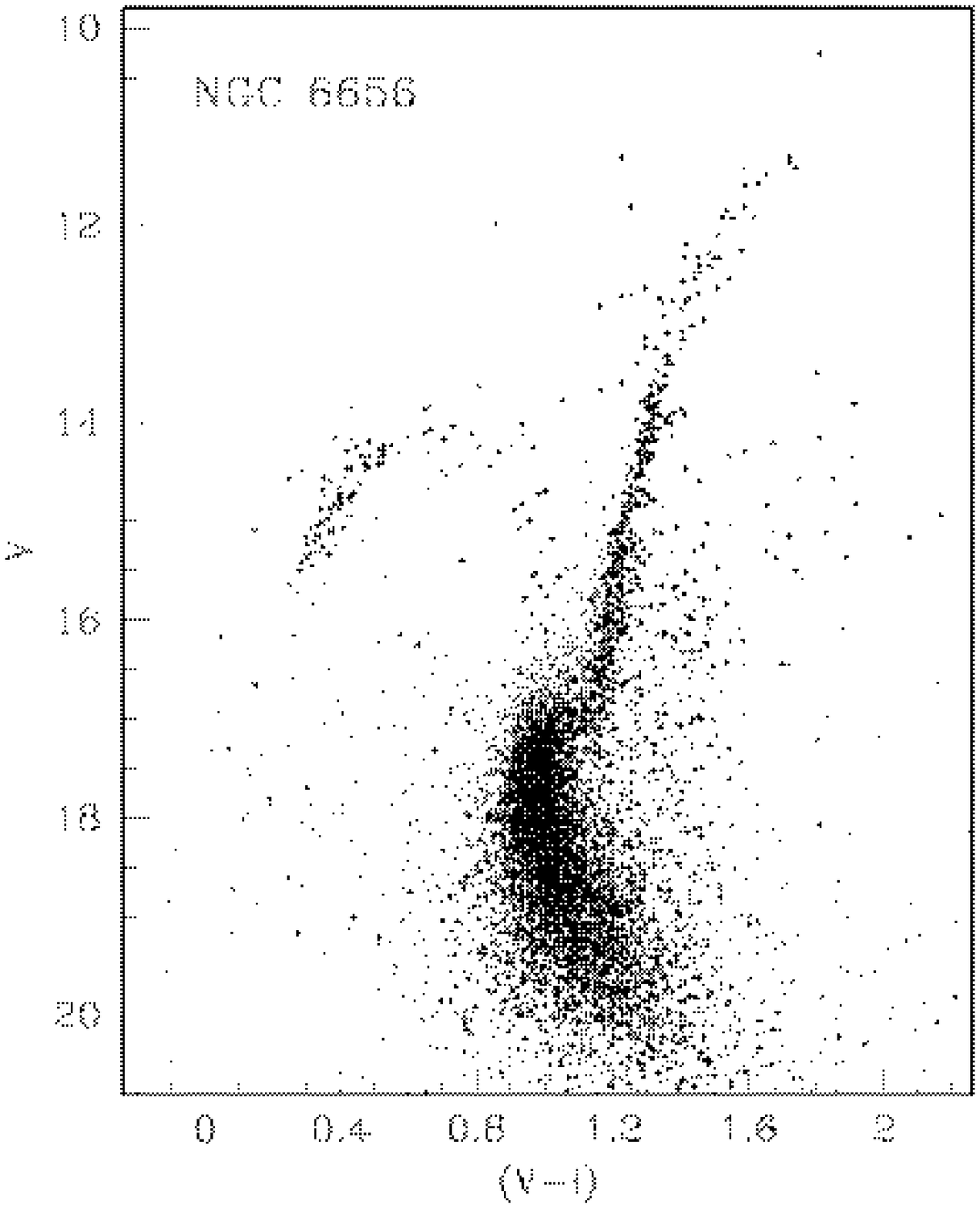,width=8.8cm}
} &
\begin{minipage}[t]{8.8cm}
\begin{tabular}{c@{}c}
\fbox{\psfig{figure=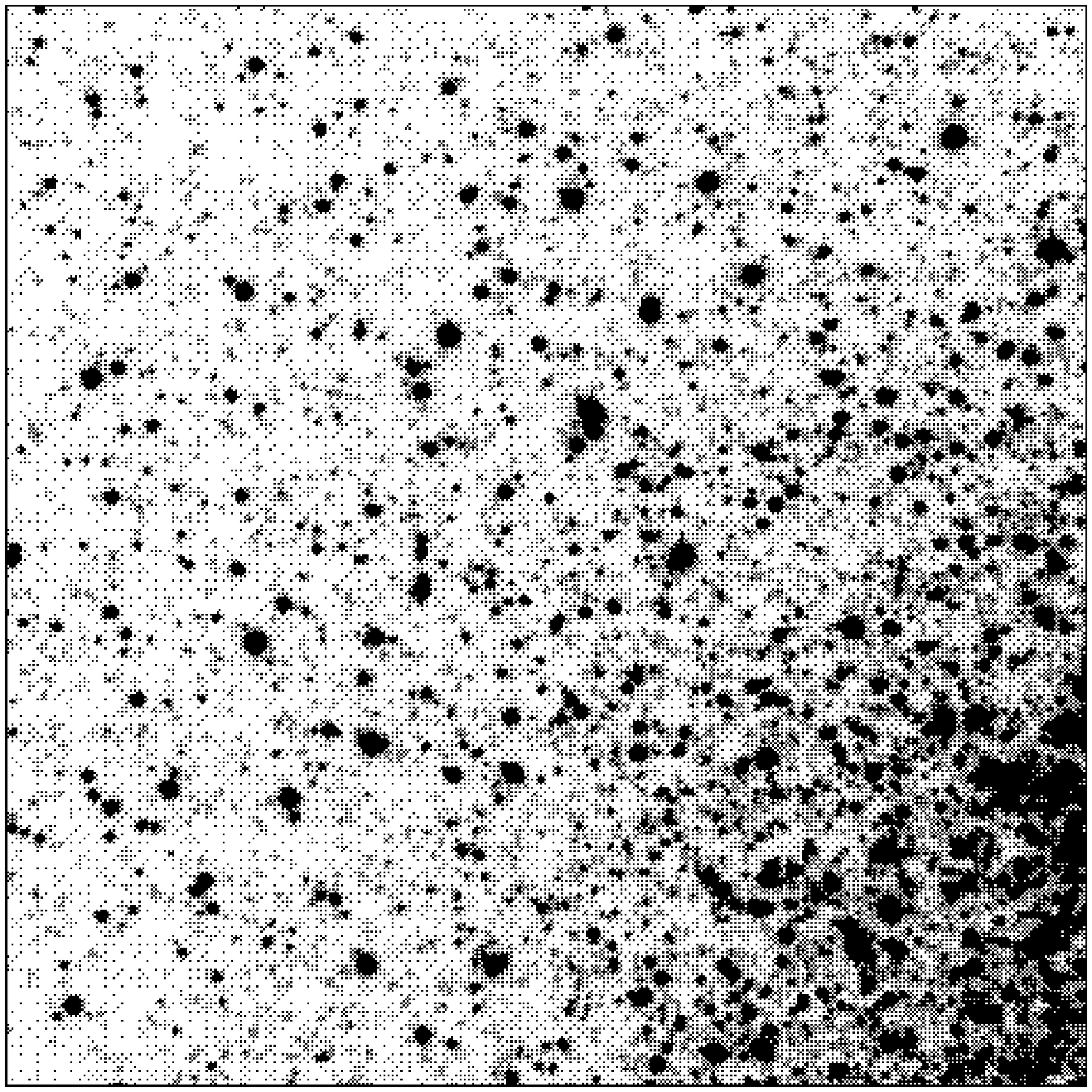,width=4cm}} \\
\fbox{\psfig{figure=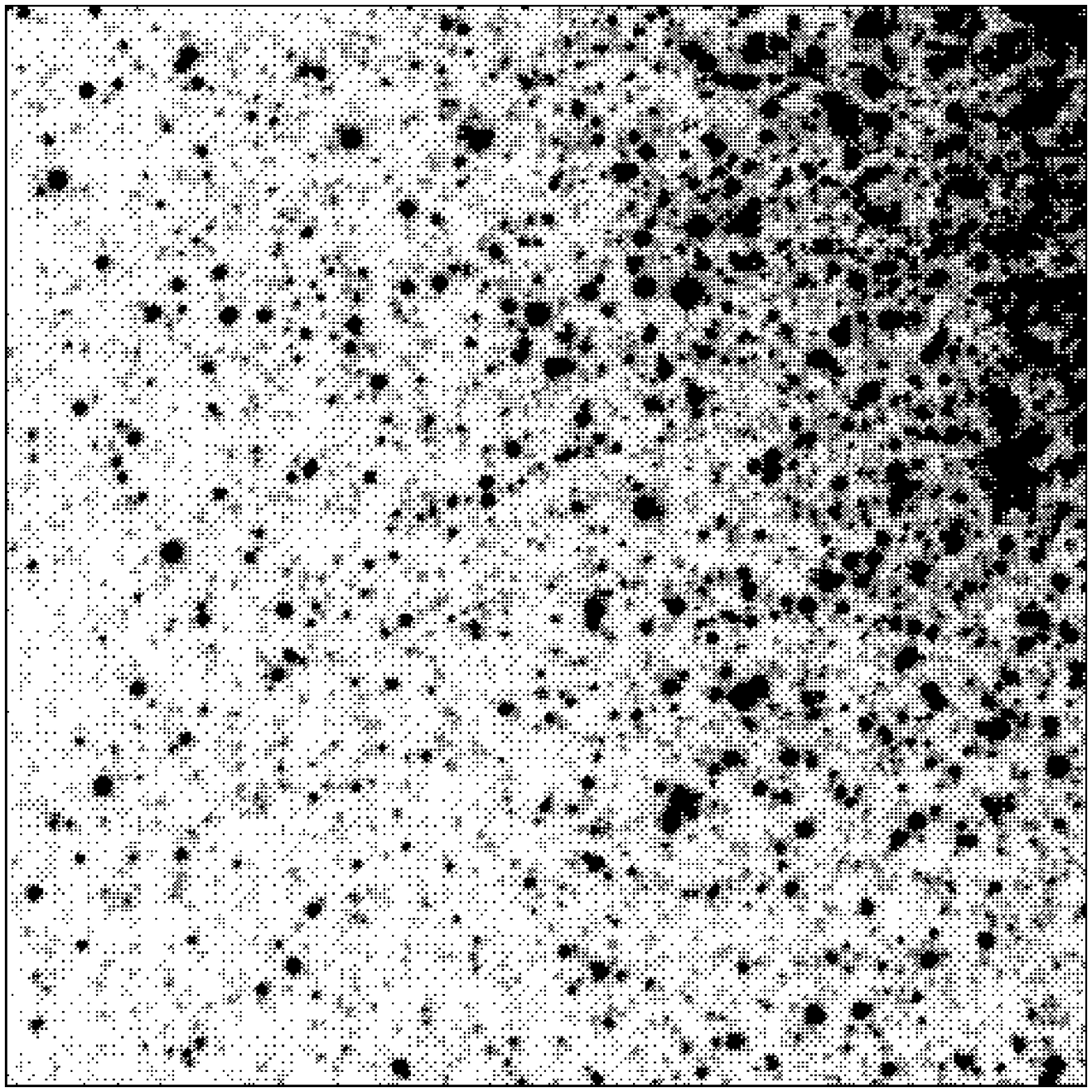,width=4cm}} 
\end{tabular}
\end{minipage}
\end{tabular}
\caption[]{CMD and covered fields for NGC~6656 (M~22)}
\label{ngc6656}
\end{figure*}

\paragraph {\bf NGC 6656 (M~22).} (Fig.~\ref{ngc6656})

A possible internal dispersion in metallicity has been proposed for
M22. It presents an EBHB with some HB stars as faint as the
TO, and several possible RR-Lyrae stars. It is close to the
Galactic center and to the Galactic plane, with a high
reddening.

Piotto \& Zoccali (\cite{piottozoccali99}) published the most recent
study of this cluster. From a combination of HST data and ground based
CCD photometry, they produced a CMD extending from the tip of the RGB
to below $0.2 M_\odot$. Anthony-Twarog et al. (\cite{anthonytwarog95})
present $uvbyCa$ data for over 300 giant and HB stars, while in
Davidge \& Harris (\cite{davidgeharris96}) there is a deep near
infrared study.

\begin{figure*}
\begin{tabular}{c@{}c}
\raisebox{-6cm}{
\psfig{figure=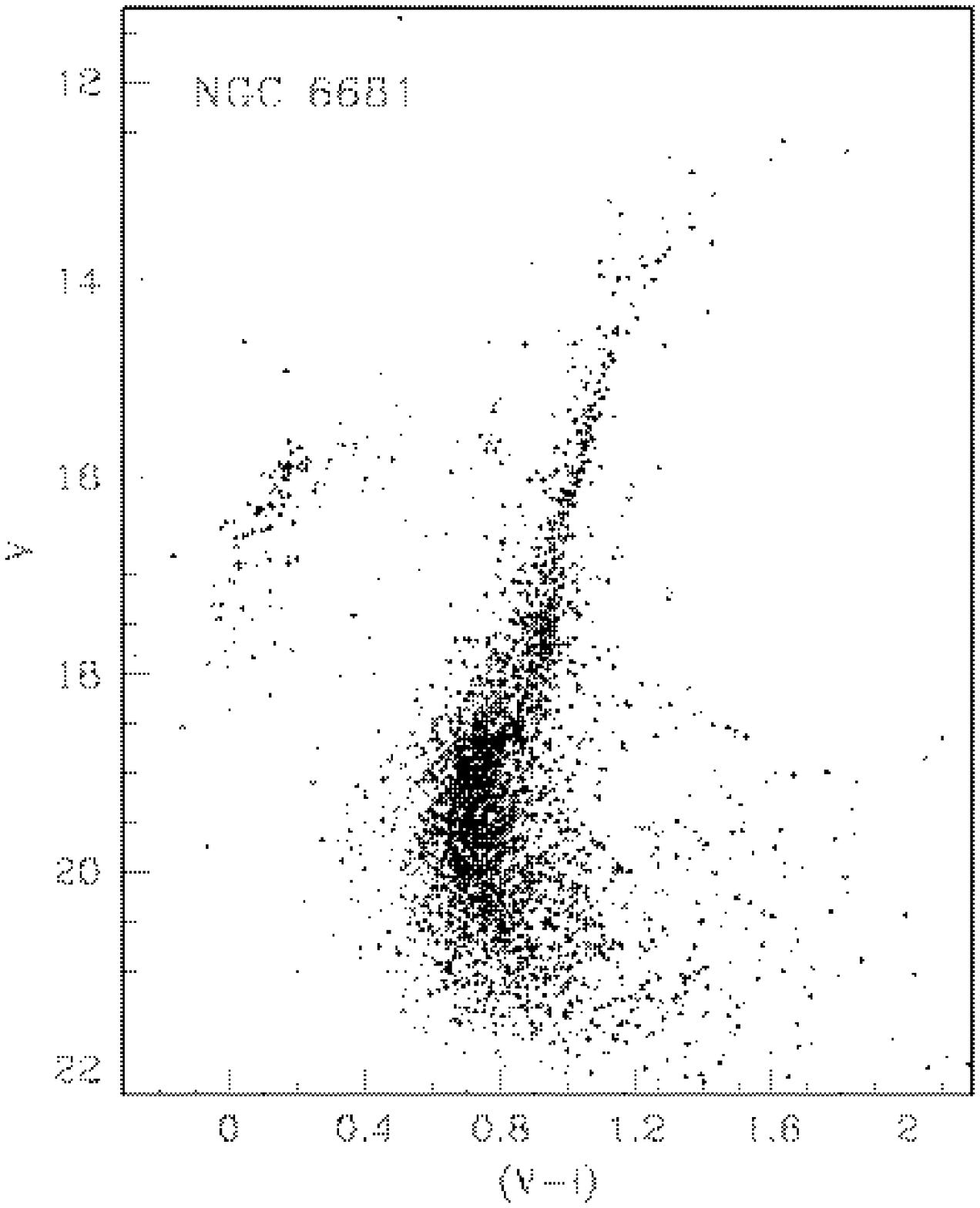,width=8.8cm}
} &
\begin{minipage}[t]{8.8cm}
\begin{tabular}{c@{}c}
\fbox{\psfig{figure=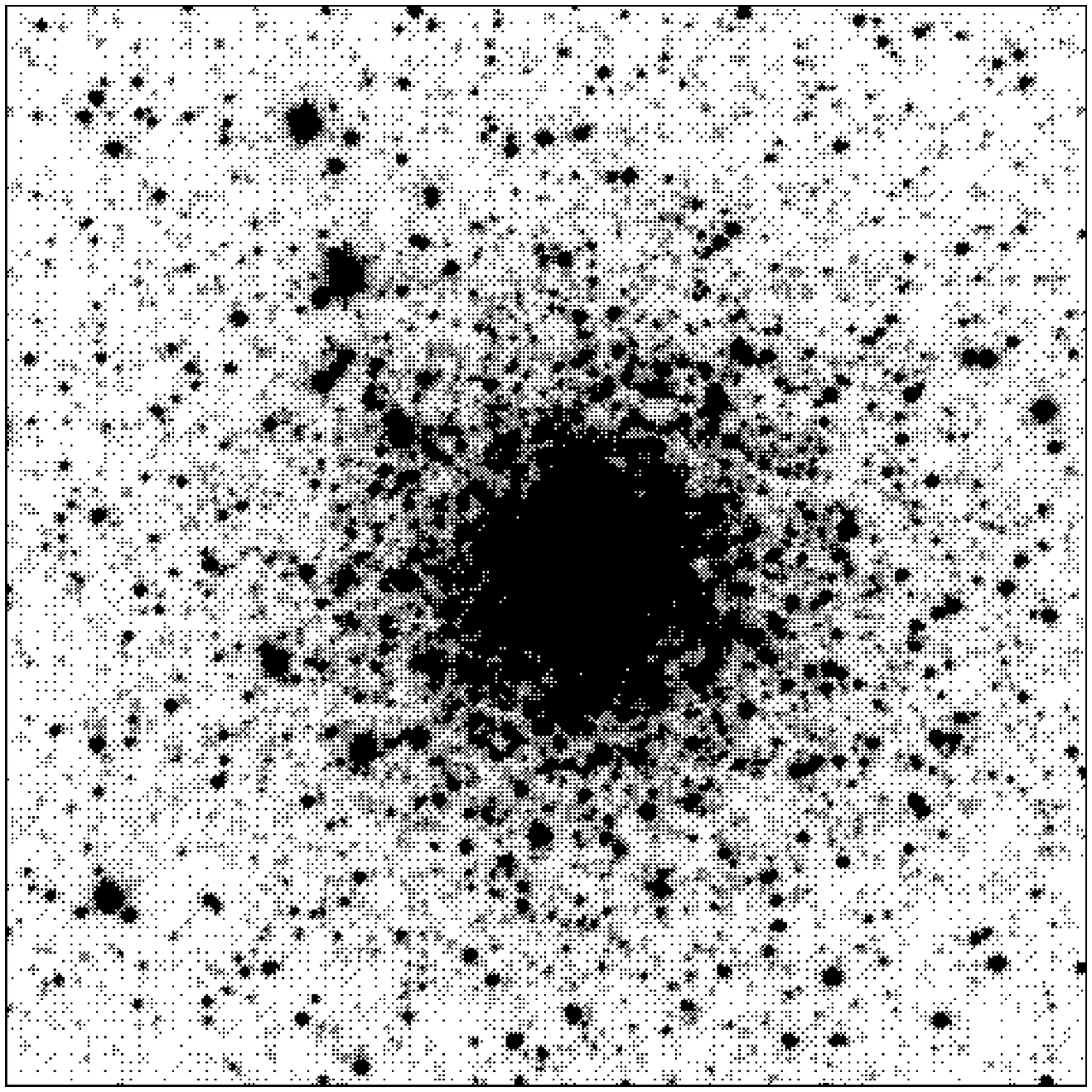,width=4cm}}
\end{tabular}
\end{minipage}
\end{tabular}
\caption[]{CMD and covered field for NGC~6681 (M~70)}
\label{ngc6681}
\end{figure*}

\paragraph {\bf NGC 6681 (M~70).}  (Fig.~\ref{ngc6681})

NGC 6681 has a predominantly blue HB with a few HB stars on the red side
of the instability strip. Brocato et al. (\cite{brocato96}) present
the only other available CCD photometry for this cluster in the $B$
and $V$ bands.

\begin{figure*}
\begin{tabular}{c@{}c}
\raisebox{-6cm}{
\psfig{figure=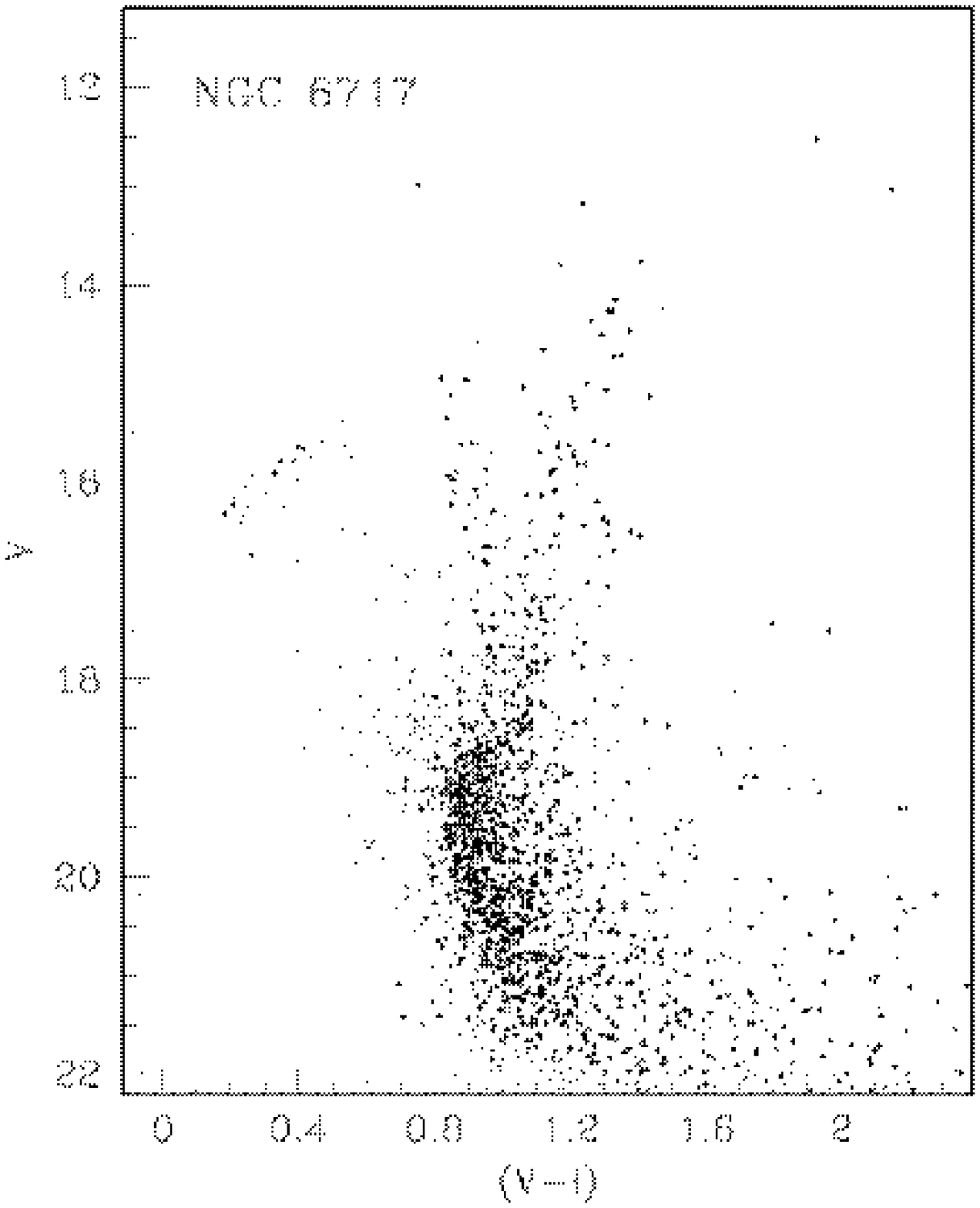,width=8.8cm}
} &
\begin{minipage}[t]{8.8cm}
\begin{tabular}{c@{}c}
\fbox{\psfig{figure=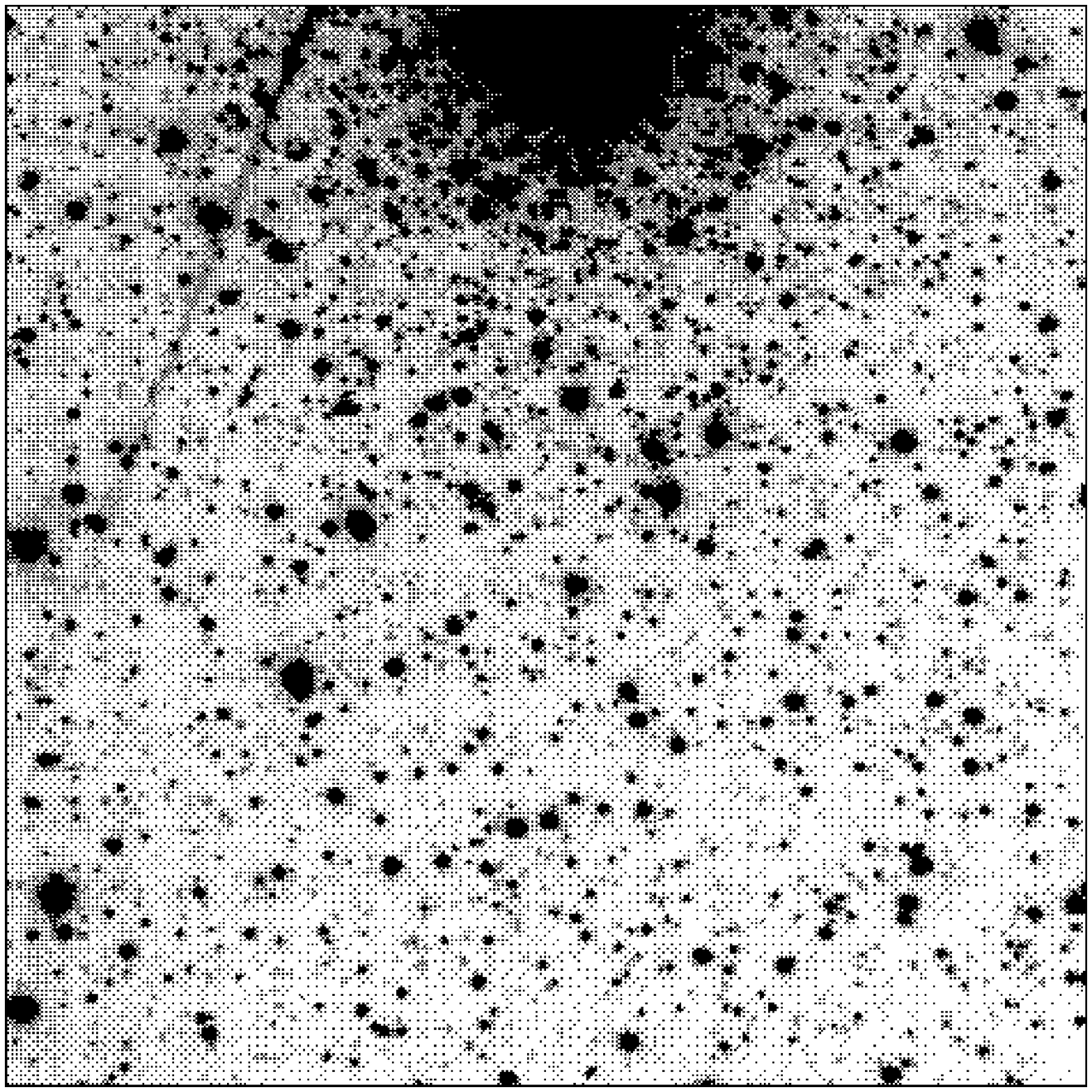,width=4cm}}
\end{tabular}
\end{minipage}
\end{tabular}
\caption[]{CMD and covered field for NGC~6717 (Palomar~9)}
\label{ngc6717}
\end{figure*}

\paragraph {\bf NGC 6717 (Palomar~9).}  (Fig.~\ref{ngc6717})

NGC~6717 is a poorly populated cluster (as most of the
``Palomar-like'' objects), and the CMD is contaminated by bulge
stars. The RGB is difficult to identify, and its HB is blue,
resembling that of NGC 288. Notice that there is a very bright field
star close to the cluster, located at the north side of it. Brocato
et al. (\cite{brocato96}) present the first CCD photometry for this
cluster; their $B$ and $V$ CMD resembles that of
Fig.~\ref{ngc6717}. Recently, Ortolani et al. (\cite{ortolani99})
presented a new CMD, in the same bands, but the CMD branches are more
poorly defined.

\begin{figure*}
\begin{tabular}{c@{}c}
\raisebox{-6cm}{
\psfig{figure=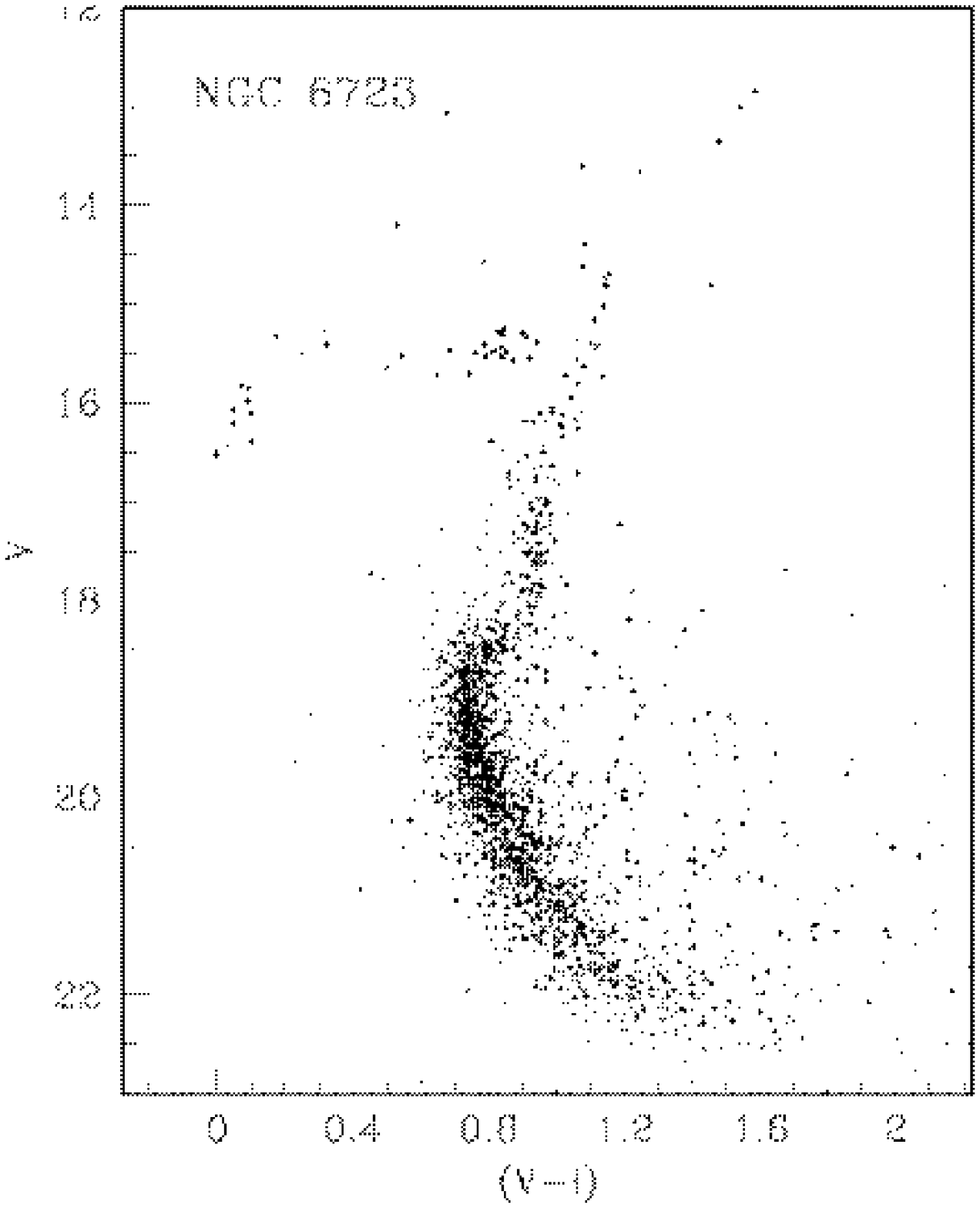,width=8.8cm}
} &
\begin{minipage}[t]{8.8cm}
\begin{tabular}{c@{}c}
\fbox{\psfig{figure=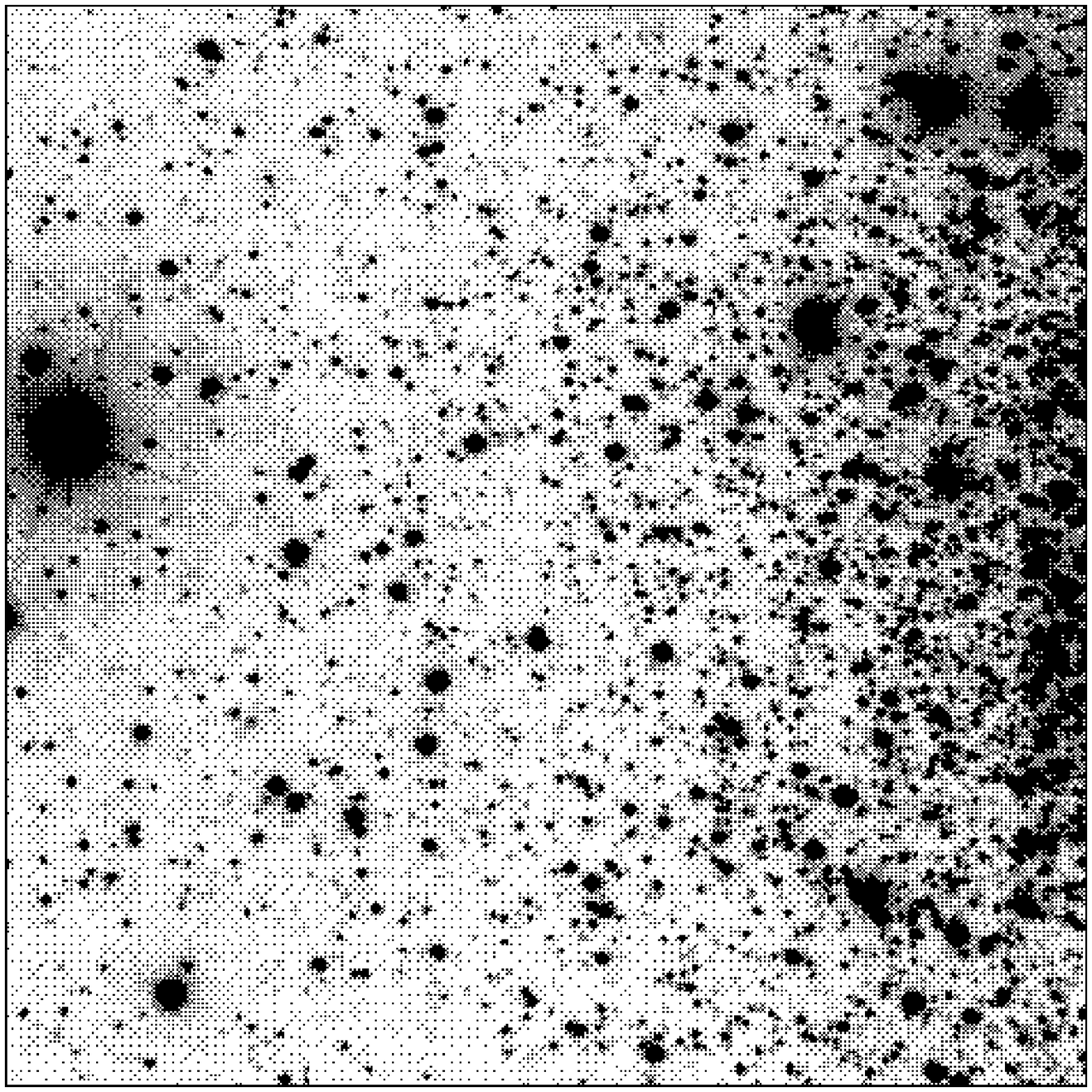,width=4cm}}
\end{tabular}
\end{minipage}
\end{tabular}
\caption[]{CMD and covered field for NGC~6723}
\label{ngc6723}
\end{figure*}

\paragraph {\bf NGC 6723.}  (Fig.~\ref{ngc6723})

NGC 6723 has both a red and blue HB, and the overall morphology is
typical of a cluster of intermediate metallicity. Alcaino et
al. (\cite{alcaino99}) present the most recent CCD study (multicolor
photometry), with a CMD extending down to $V\sim21$. 
Fullton \& Carney (\cite{fulltoncarney96}) have obtained a
deep $B$ and $V$ photometry, extending to $V\sim24$, though the
results of this study have not been completely published, yet.

\begin{figure*}
\begin{tabular}{c@{}c}
\raisebox{-6cm}{
\psfig{figure=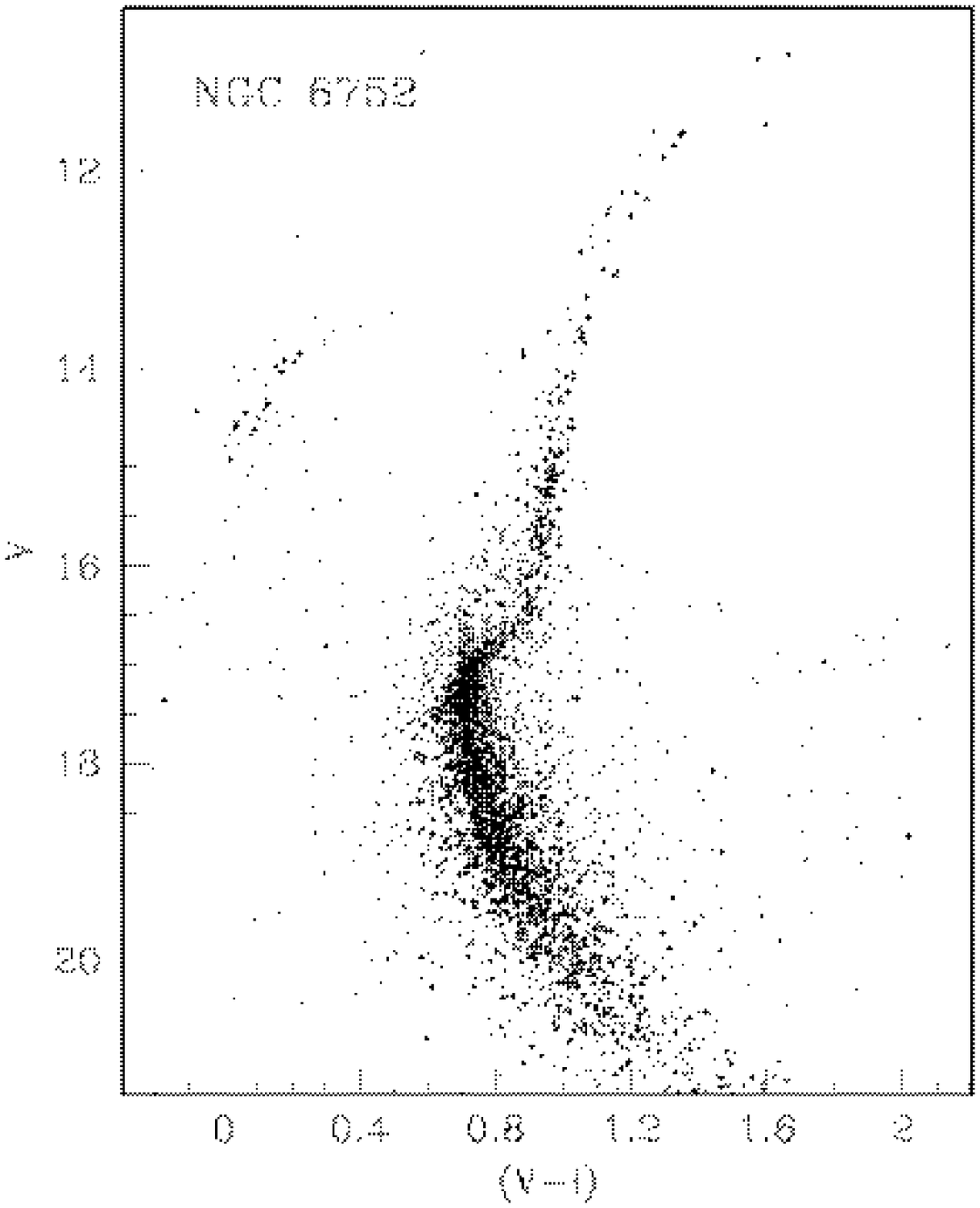,width=8.8cm}
} &
\begin{minipage}[t]{8.8cm}
\begin{tabular}{c@{}c}
\fbox{\psfig{figure=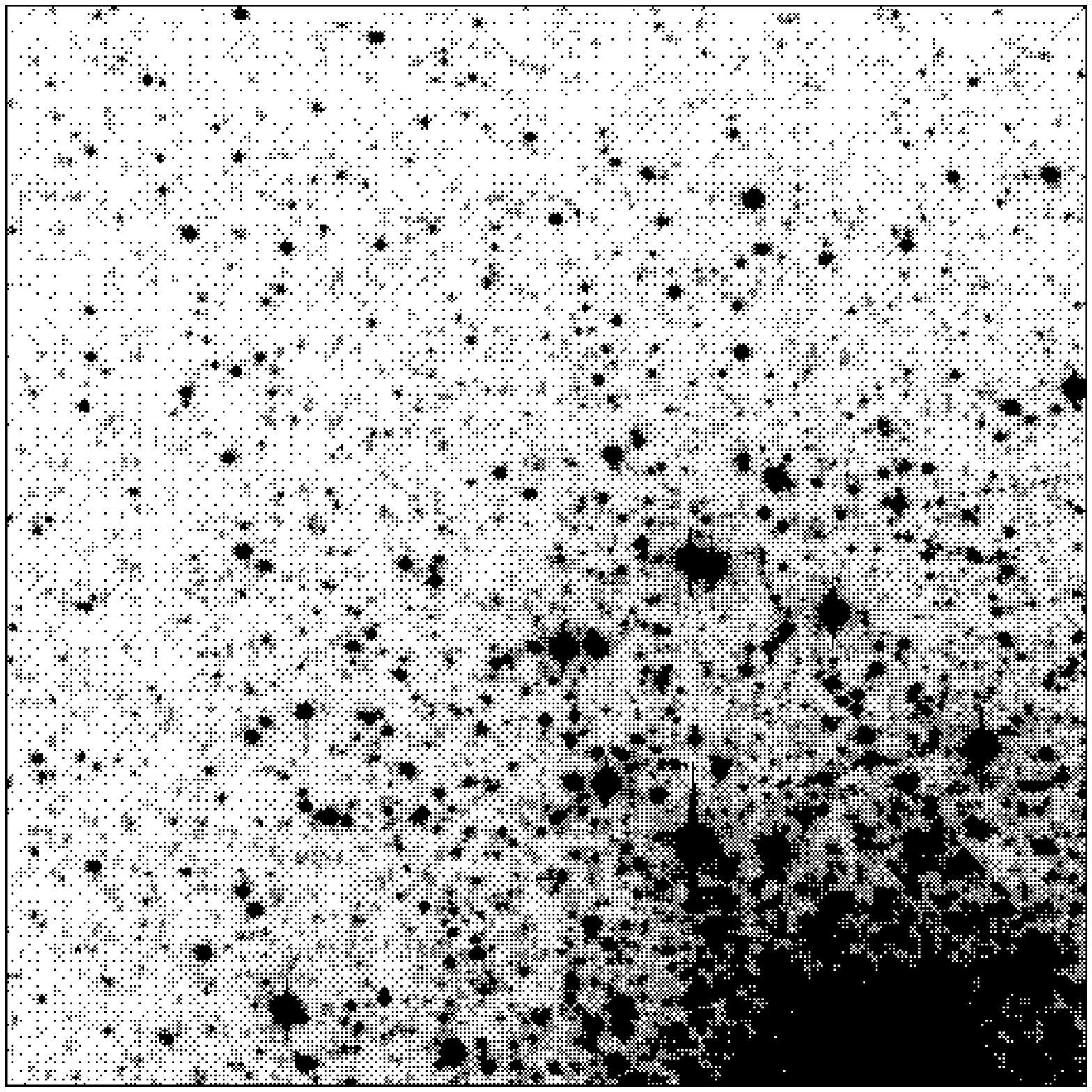,width=4cm}}
\end{tabular}
\end{minipage}
\end{tabular}
\caption[]{CMD and covered field for NGC~6752}
\label{ngc6752}
\end{figure*}

\paragraph {\bf NGC 6752.} (Fig.~\ref{ngc6752})

NGC~6752 has been largely studied in the past. It has a very well
defined EBHB. Penny \& Dickens (\cite{pennydickens86}) presented a
$B$ and $V$ CCD study from a combination of data from two telescopes,
and published a CMD from the RGB tip to $V\sim24$ mag, though with a
small number of measured stars. In the same year, Buonanno et
al. (\cite{buonanno86}) present a CMD in the same bands for stars from
$\sim1$ mag above the TO to $\sim5$ mag below it.  More recently,
Renzini et al. (\cite{renzini96}) and Rubenstein \& Baylin
(\cite{rubensteinbaylin97}) published a CMD from HST data.

\begin{figure*}
\begin{tabular}{c@{}c}
\raisebox{-6cm}{
\psfig{figure=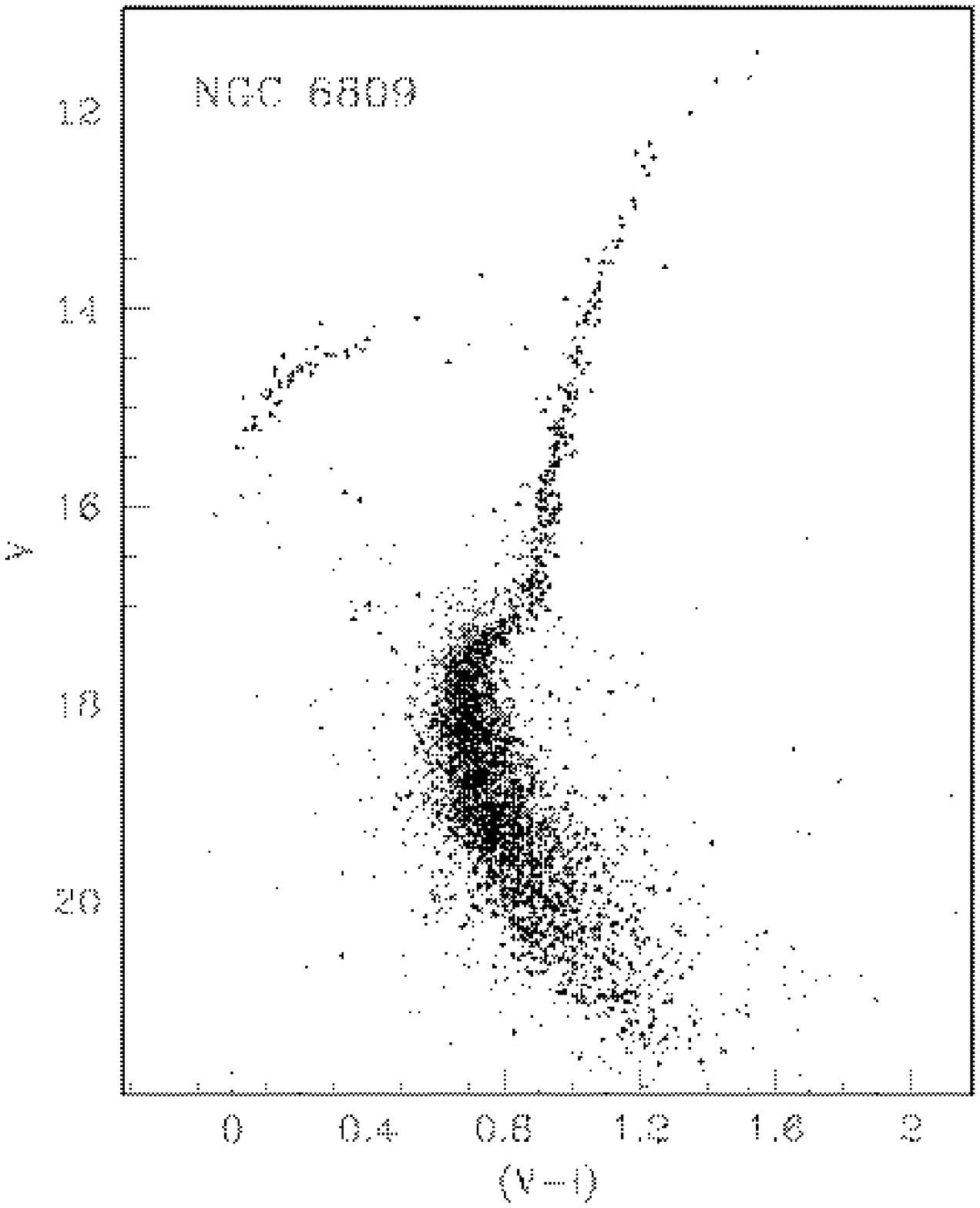,width=8.8cm}
} &
\begin{minipage}[t]{8.8cm}
\begin{tabular}{c@{}c}
\fbox{\psfig{figure=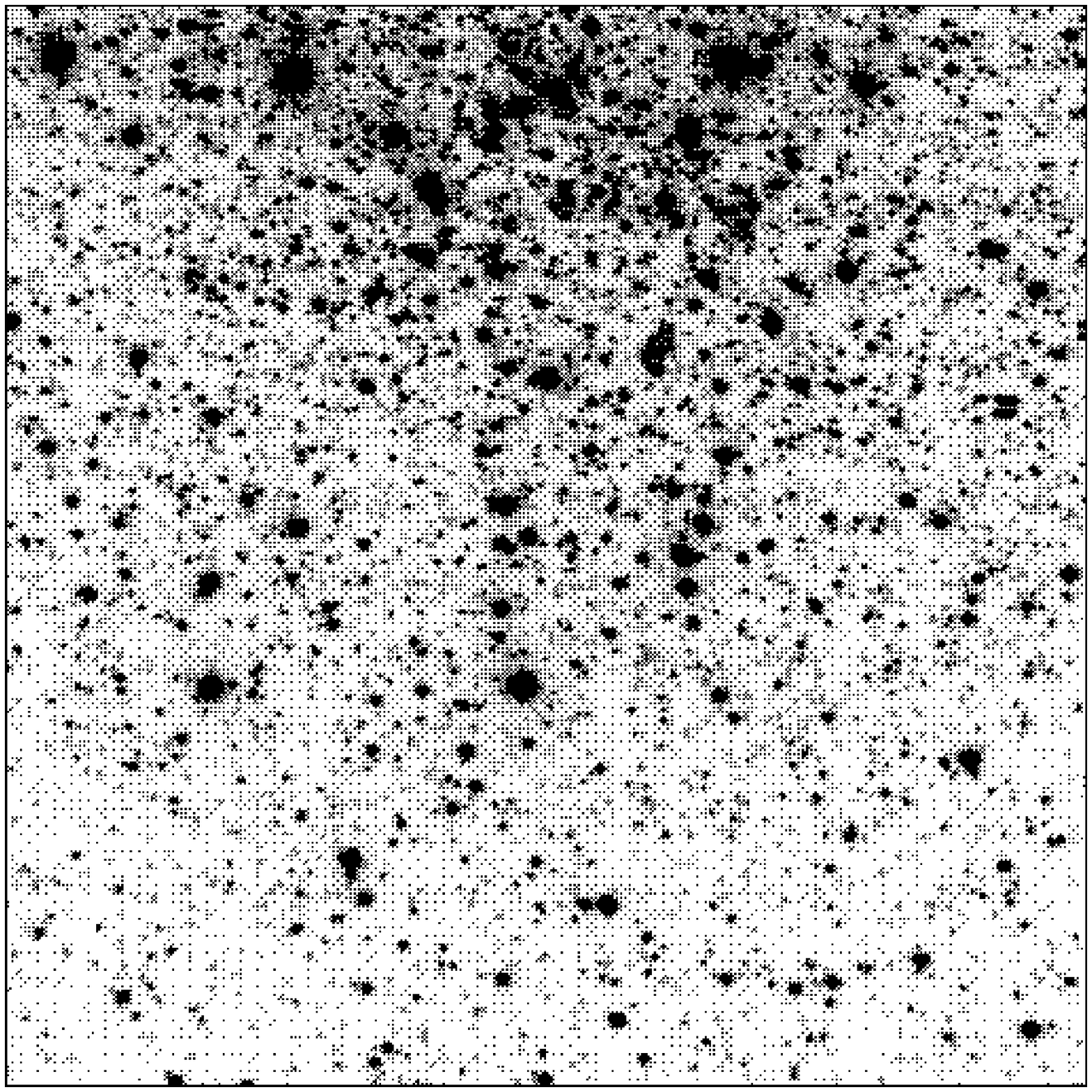,width=4cm}}
\end{tabular}
\end{minipage}
\end{tabular}
\caption[]{CMD and covered field for NGC~6809 (M~55)}
\label{ngc6809}
\end{figure*}

\paragraph {\bf NGC 6809 (M~55).} (Fig.~\ref{ngc6809})

Also the CMD of NGC~6809 is typical for its (low) metallicity. A very
well defined BS sequence is visible in Fig.~\ref{ngc6809}. The
most recent CCD study is in Piotto \& Zoccali
(\cite{piottozoccali99}), who study the cluster luminosity function
based on deep HST data combined with ground-based CCD data for the
evolved part of the CMD. Zaggia et al. (\cite{zaggia97}) present $V$
and $I$ CCD photometry of $\sim34000$ stars covering an entire
quadrant of the cluster (out to $\sim1.5$ times the tidal radius) down
to $V\sim21$. Mateo et al. (\cite{mateo96}) and Fahlman et
al. (\cite{fahlman96}) presented photometric datasets of M~55 that
have been mainly used to study the age and the tidal extension of the
Sagittarius dwarf galaxy. Mandushev et al. (\cite{mandushev96})
published the first deep (down to $V\sim24.5$) photometry of the
cluster.

\clearpage

\clearpage

\end{document}